\documentclass[twoside,numbers,english,a4paper,openright]{ezthesis}
\usepackage{braket}
\usepackage{color}
\usepackage{graphics}
\usepackage{graphicx}
\usepackage{subfigure}
\usepackage{epsfig}
\usepackage{amsmath}
\usepackage{amssymb}
\usepackage{amsthm}
\usepackage{slashed}
\usepackage{float}
\usepackage[mathscr]{eucal}
\usepackage[export]{adjustbox}
\usepackage{emptypage}
\graphicspath{{Figures/}}
\newcommand{\1}{{\text A}}
\newcommand{\2}{{\text B}}

\usepackage[Glenn]{fncychap}
\usepackage[%draft
final,inline,nomargin]{fixme} \fxsetup{theme=color}
\fxsetup{status=draft}
\usepackage{makeidx}
\makeindex

\author{Grecia Guijarro G\'amez}
\degree{DOCTORAL PROGRAM IN COMPUTATIONAL AND APPLIED PHYSICS}
\institution{UNIVERSITAT POLITÈCNICA DE CATALUNYA}
\faculty{DEPARTMENT OF PHYSICS}
\submitdate{Barcelona 2020}

\hyperlinking

\begin{document}
\newgeometry{margin=1in}
\vspace*{-3cm}
\begin{titlepage}
\hspace*{-8cm}
\title{QUANTUM MONTE CARLO STUDY OF FEW- AND MANY-BODY BOSE SYSTEMS
    IN ONE AND TWO DIMENSIONS}
\vspace{2.0cm}
\TitleBlock{\Large \\ \insertinstitution\\ 
    \insertfaculty\\}
\TitleBlock{\Large DOCTORAL THESIS\\}
\vspace{1.0cm}
\rule{13cm}{0.01cm}
\TitleBlock{\bfseries\Large\inserttitle}
\vspace{0.2cm}
\rule{13cm}{0.01cm}
\vspace{1.0cm}
\TitleBlock{\large Author: \bfseries\large\insertauthor}
\vspace{1.0cm}
\TitleBlock{\large Supervisors:\\
    {\bfseries Prof. Dr. Grigory Astrakharchik\\
        Prof. Dr. Jordi Boronat Medico}}
\vspace{1.0cm}
\TitleBlock{\large\insertdegree}
 \vspace*{2cm}
\TitleBlock{\large\insertsubmitdate}
\end{titlepage}

\restoregeometry

\prefacesection{Abstract}
In this Thesis, we report a detailed study of the ground-state properties
of a set of quantum few- and many-body systems by using Quantum Monte Carlo methods.
First, we introduced the Variational
Monte Carlo and Diffusion Monte Carlo methods, these are the
methods used in this Thesis to obtain the properties of the systems. 
The first systems we studied consist of few-body clusters in a
one-dimensional Bose-Bose and Bose-Fermi mixtures.
Each mixture is formed by two different species with attractive
interspecies and repulsive intraspecies contact interactions.
For each mixture, we focused
on the study of the dimer, tetramer, and hexamer clusters. We calculated
their binding energies and unbinding thresholds.
Combining these results with a three-body theory, we
extracted the three-dimer scattering length close to the dimer-dimer zero
crossing. For both mixtures, the three-dimer interaction turns out
to be repulsive. Our results constitute a concrete proposal for obtaining
a one-dimensional gas with a pure three-body repulsion. The next system
analyzed consists of few-body clusters in a two-dimensional
Bose-Bose mixture using two types of interactions. The first case
corresponds to a bilayer of dipoles 
aligned perpendicularly to the planes
and, in the second, we model the
interactions by finite-range Gaussian potentials. 
We find that all the considered clusters are bound states and that their energies
are universal functions of the scattering lengths, for sufficiently large
attraction-to-repulsion ratios. Studying the hexamer energy
close to the corresponding threshold, we discovered an effective three-dimer
repulsion, which can stabilize interesting many-body phases. Once the
existence of the bound states in the dipolar bilayer has been demonstrated,
we investigated whether halos can occur in this system. A halo state is
a quantum bound state whose size is much larger than the range
of the attractive interaction between the atoms that form it, showing universal ratios between
energy and size. 
For clusters composed
from three up to six dipoles, we find two very distinct halo structures. For
large interlayer separation, the halo structure is roughly symmetric. However,
for the deepest bound clusters and as the clusters approach the threshold,
we discover an unusual shape of the halo states, highly anisotropic. Importantly,
our results prove the existence of stable halo states composed of up to six
particles. To the best of our knowledge, this is the first time that halo
states with such a large number of particles have been predicted and observed in 
a numerical simulation. The next system we studied is a two-dimensional
many-body dipolar fluid confined to a bilayer geometry. We calculated the
ground-state phase diagram as a function of the density and the separation
between layers. Our simulations show that the system undergoes a phase
transition from a gas to a stable liquid as the interlayer
distance increases. The liquid phase is stable in a wide range of densities
and interlayer values. 
In the final part of this Thesis, we studied a system of dipolar bosons confined
to a multilayer geometry formed by equally spaced two-dimensional layers. We
calculated the ground-state phase diagram as a function of the density, the
separation between layers, and the number of layers. The key result of our
study in the dipolar multilayer is the existence of three phases: atomic gas,
solid, and gas of chains,
in a wide range of the system parameters. Remarkably, we find that the
density of the solid phase decreases several orders of magnitude as the
number of layers in the system increases. The results reported in this Thesis
show that a dipolar system in a bilayer and multilayer geometries offer 
stable and highly controllable setups for observing interesting phases of
quantum matter, such as halo states, and ultra-dilute liquids and solids. 

\prefacesection{Resumen}
En esta Tesis, presentamos un estudio detallado de las propiedades del estado
fundamental de un conjunto de sistemas cuánticos 
de pocos y muchos cuerpos mediante el uso de
los métodos de Monte Carlo Cuántico.
Primero, introducimos los métodos de Monte Carlo Variacional y Monte Carlo
Difusivo que usamos en esta Tesis para obtener las
propiedades de los sistemas.
Los primeros sistemas que estudiamos son cúmulos de pocos
cuerpos en mezclas unidimensionales de Bose-Bose y Bose-Fermi.
Cada una de las mezclas está formada por dos especies con 
interacciones atractivas para interespecies y repulsivas para
intraespecies.
Para cada una de las mezclas nos enfocamos en el estudio de dímeros, tetrámeros y
hexámeros. Calculamos las energías de ligadura
y los valores umbrales de ruptura 
de los cúmulos.
Combinando estos resultados con una teoría de tres cuerpos,
extraemos la longitud de dispersión de tres dímeros cerca del punto
de ruptura dímero-dímero. Para ambas mezclas
la interacción de tres dímeros resulta
ser repulsiva. 
El siguiente sistema analizado son cúmulos de pocos cuerpos en
una mezcla bidimensional de Bose-Bose con dos tipos de interacciones. 
El primer caso
corresponde a una bicapa de dipolos
con momentos dipolares orientados perpendicularmente a los planos
y, en el segundo, modelamos las
interacciones con potenciales gaussianos de rango finito.
Encontramos que para relaciones de atracción-repulsión suficientemente
grandes todos los cúmulos considerados son estados ligados
y sus energías
son funciones universales de las longitudes de dispersión.
Estudiando la energía del hexámero
cerca del punto umbral correspondiente, descubrimos una repulsión
efectiva de tres dímeros,
que puede estabilizar fases interesantes de muchos cuerpos. Después de demostrar
la existencia de los estados ligados en la bicapa dipolar,
investigamos si pueden ocurrir estados de halo en este sistema. 
Un estado de halo es
un estado ligado cuántico cuyo
tamaño es mucho mayor que el rango
de la interacción atractiva entre los átomos que lo forman.
Para cúmulos compuestos
de tres hasta seis dipolos encontramos dos estructuras de halo muy distintas. 
Para separaciones
grandes entre las capas, la estructura de halo es aproximadamente simétrica.
Sin embargo, para los estados más ligados y a medida que los cúmulos
se acercan al punto umbral,
descubrimos una estructura de halo inusual, altamente anisotrópica. 
Nuestros resultados demuestran la existencia de estados de halo estables
compuestos de hasta seis dipolos. Hasta donde sabemos, esta es la primera vez
que estados de halo con un número tan grande de partículas se predicen y observan en
una simulación numérica. El siguiente sistema estudiado es un fluido
bidimensional dipolar de muchos cuerpos confinado a una geometría de bicapa. 
Calculamos el diagrama de fases del estado fundamental como función de la densidad y
de la separación entre las capas. 
Nuestras simulaciones muestran que en el sistema ocurre una
transición de fase, de un gas a un líquido 
a medida que se incrementa la distancia entre las capas. El líquido es estable en
una región amplia de densidades y de la distancia entre las capas.
En la parte final de esta Tesis se estudia un sistema de bosones dipolares
confinados a una geometría multicapa formada por capas bidimensionales
igualmente espaciadas. Calculamos el diagrama de fases del estado fundamental como
función de la densidad, la separación entre las capas y el número de capas. 
El resultado clave de nuestro estudio sobre la multicapa 
es la existencia de tres fases: gas atómico,
sólido y gas de cadenas, en una región amplia de los parámetros del sistema. 
Encontramos que la densidad del sólido disminuye
varios órdenes de magnitud a medida que el número de capas en el sistema
aumenta. Los resultados reportados en esta Tesis
muestran que un sistema de dipolos confinados a una bicapa o multicapa
ofrecen
configuraciones estables y altamente controlables para observar fases interesantes de
materia cuántica.

\tableofcontents
\section*{}
\thispagestyle{empty}
%\clearpage\thispagestyle{empty}
%empty page

\clearpage
\pagenumbering{arabic}
\chapter{Introduction}
In this Thesis, we report a detailed study of the ground-state properties
of a set of quantum few- and many-body systems in one and two dimensions
with different types of interactions. Nevertheless, the main focus of this
work is the study of the ground-state properties of an ultracold Bose system
with dipole-dipole interaction between the particles.
We consider the cases where the bosons are confined to a bilayer and 
multilayer geometries, that consist of equally spaced two-dimensional layers.
These layers can be experimentally realized by imposing tight confinement in
one direction. We specifically address the study of new quantum phases, their
properties, and transitions between them. One expects these systems to have a
rich collection of few- and many-body phases because the dipole-dipole
interaction is anisotropic and quasi long-range. We will now present a short
historical review of the experiments and theoretical predictions that motivated
the study of ultracold dipolar bosonic gases.

The Bose-Einstein condensation (BEC) is a quantum phenomenon occurring when a
macroscopic number of bosons occupy the zero momentum state. This
happens when the system reaches a temperature below a critical value.
Although BEC was predicted by Albert Einstein in 1924~\cite{Einstein1924}
based on a previous work by Satyendranath Bose, it was not until 1995
that this phenomenon was experimentally observed in rubidium~\cite{Anderson198}
and sodium~\cite{PhysRevLett.75.3969} gases independently. 
The Nobel Prize in Physics 2001 was awarded to Wolfgang Ketterle, Eric A. Cornell,
and Carl E. Wieman for the achievement of BEC in dilute gases of alkali
atoms~\cite{nobelprize}. Typically the BEC state is reached for temperatures
and densities of the order of $T\sim 10^{-7}$K and $n\sim 10^{13}-10^{14}$cm$^{-3}$,
respectively. Since the experimental observation of the BEC, there have been
intense theoretical and experimental efforts to understand ultracold bosonic
and fermionic gases. Interesting quantum phases have been predicted and
experimentally realized in these systems, for example, quantum droplets in a
mixture of Bose-Einstein 
condensates~\cite{PhysRevLett.115.155302,Cabrera2017,Semeghini2018,Ferioli2019}
and in dipolar bosonic gases~\cite{Kadau2016,Schmitt2016,Ferrier2016,Chomaz2016},
quantum droplets in optical lattices~\cite{PhysRevResearch.2.022008,morera2020universal},
impurities of atoms immersed in a gas of
fermions~\cite{kohstall2012metastability,Massignan_2014,RaulBombi2019}
or bosons~\cite{PhysRevX.8.011024,PhysRevLett.120.050405,PhysRevResearch.2.023405},
the so-called polaron problem, among others. At very low densities, some ultracold
gases can be characterized by the s-wave scattering length, which means that they
can be described by an isotropic, short-range, contact interaction model. However,
there are gases with more complex interactions like dipolar interactions.

Recent experiments have enabled the experimental study of ultracold gases with
dipole-dipole interaction (DDI). The DDI has two main properties that greatly
distinguish it from the contact interactions. Firstly, DDI is long-range in
three dimensions, it falls off with a power-law $1/r^3$ dependence, where $r$ is
the distance between particles. Secondly, DDI is anisotropic which means that the
interaction strength and its sign (repulsive or attractive), depends on the angle
between the polarization direction and the relative distance of the particles.
DDI can be found in magnetic atoms, ground-state heteronuclear molecules, and
Rydberg atoms, among others~\cite{Lahaye_2009}. The first Bose-Einstein condensate
of magnetic atoms was realized in a gas of chromium atoms in 
2005~\cite{PhysRevLett.94.160401,PhysRevLett.95.150406}. The most recent experiments
of dipolar gases are done with Dysprosium~\cite{PhysRevLett.107.190401,PhysRevLett.108.215301}
and Erbium~\cite{Chomaz2018}. Many interesting phenomena have been observed and
predicted in dipolar gases, for example, dipolar Bose supersolid 
stripes~\cite{PhysRevLett.119.250402}, dipolar quantum 
mixtures~\cite{PhysRevLett.121.213601,PhysRevA.99.043609}, formation of a crystal
phase~\cite{Buchler2007,Astrakharchik2007}, and a pair
superfluid~\cite{Macia2014,PhysRevA.94.063630,Nespolo_2017}.

Describing a quantum many-body system is a demanding task, as it involves the
interactions of a large number of particles subject to spatial constraints. 
Only for systems with very simple interactions, and under some assumptions, 
can the Schrödinger equation be solved exactly. As we are studying systems with
dipolar interactions, complementary numerical methods become necessary, like
in our case, quantum Monte Carlo methods.

Quantum Monte Carlo (QMC) methods are a set of stochastic techniques that are
used to calculate the ground-state properties of quantum many-body systems at
zero or finite temperature~\cite{hammond1994monte,toulouse2015introduction,Ceperley}. 
One of the most used QMC techniques for its simplicity is the Variational Monte
Carlo (VMC) method. The VMC technique uses the variational principle of quantum
mechanics to provide an upper bound to the ground-state energy of a quantum system.
The accuracy of this method depends entirely on the accuracy of the trial wave
function used to calculate the expectation value of the Hamiltonian. Another QMC
technique is the Diffusion Monte Carlo (DMC) method that solves the many-body
Schrödinger equation in imaginary time. This method consists of evolving in imaginary
time the wave function of a quantum system, and after enough time has passed, it
projects out the ground state. The DMC method allows one to calculate the exact
ground-state energy of the system, as well as other properties, within controllable
statistical errors. Both VMC and DMC methods have been shown to give an accurate
description of correlated quantum systems~\cite{Ceperley}.
Examples include ultradilute bosonic~\cite{PhysRevA.99.023618,LParisi-Thel-2019} and
fermionic mixtures~\cite{PhysRevA.99.043609},
Bose~\cite{PhysRevA.99.063607,PhysRevResearch.2.023405} and 
Fermi~\cite{RaulBombi2019} polarons, dipolar Bose supersolid
stripes~\cite{PhysRevLett.119.250402}, Bose gas subject to a multi-rod
lattice~\cite{ThesisOmar}, and ultracold quantum gases with spin-orbit
interactions~\cite{PhysRevA.98.053632}.

In this Thesis, we have used QMC methods to study the ground-state
properties of a set of quantum few- and many-body systems. A large part
of this Thesis is focused on the study of dipolar Bose systems confined
to a two-dimensional bilayer and multilayer geometries. This Thesis is
organized in the following way: 

In Chapter~\ref{Chapter:Quantum Monte Carlo methods}, we explain the basics
of the Quantum Monte Carlo methods used in this Thesis. First, we present the
Variational Monte Carlo method, which is used to calculate an approximation to
the ground-state energy of a quantum system. Then, we introduce the Metropolis
algorithm, a method used to generate random numbers from an arbitrary
probability distribution function. Afterwards, we discuss the Diffusion Monte
Carlo method, which allows one to calculate the exact ground-state energy of
bosonic systems at zero temperature. Later, we describe a number of trial wave
functions used for QMC calculations. Finally, we show how several ground-state
properties are evaluated in the Monte Carlo algorithm.  

In Chapter~\ref{Chapter:One-dimensional three-boson problem with two- and three-body
    interactions}, 
we use the DMC method to calculate the ground-state properties
of a one-dimensional Bose-Bose and Bose-Fermi mixtures with attractive
interspecies and repulsive intraspecies interactions. We focus on the study
of the tetramer and hexamer clusters. First, we describe the trial wave
functions for the system and the boundary conditions to be satisfied. Then, we
evaluate the tetramer and hexamer ground-state energies for Bose-Bose and
Bose-Fermi mixtures. Afterwards, we determine the threshold for unbinding
for the tetramer and hexamer, where the clusters break into two and three
dimers, respectively. Then, combining these results with a one-dimensional
three-body theory, we extract the three-dimer scattering length close to the
dimer-dimer zero crossing. Finally, we discuss a mixture of ultracold gases  
for obtaining a one-dimensional gas with a pure three-body repulsion.

In Chapter~\ref{Chapter:Few-body bound states of two-dimensional bosons}, we 
study the ground-state properties of few-body bound states in a two-dimensional
mixture of A and B bosons with two types of interactions. The first case corresponds
to a bilayer of dipoles and, in the second, we model the interactions by non-local
(separable) finite-range Gaussian potentials. First, we show the details of the
numerical techniques used to study the two models. In the dipolar case, we use the
diffusion Monte Carlo and in the Gaussian model, we use the stochastic variational
method. Then, using these methods we evaluate the ground-state binding energies 
of the clusters. Also, we numerically determine the threshold for unbinding of the
bound states in the bilayer geometry. Afterwards, studying the hexamer energy near
to the tetramer threshold allows us to characterize an effective three-dimer interaction,
which may have important implications for the many-body problem, particularly for
observing liquid states of dipolar dimers in the bilayer geometry. Finally, we give
some examples of dipolar molecules as promising candidates for observing the predicted
few-body states within a bilayer setup.

In Chapter~\ref{Chapter:Quantum halo states in two-dimensional dipolar clusters}, we
analyze the ground-state properties of loosely bound dipolar states confined to a
two-dimensional bilayer geometry by using the VMC and DMC methods. We study dipolar
dimers, trimers, tetramers, pentamers, and hexamers. First, we evaluate the pair
distribution functions for the dimer, trimer, and tetramer for different values of
the interlayer separation. Then, we calculate the spatial distributions functions
for the trimer and tetramer for two characteristic interlayer distances. Knowledge
of these structural properties permits us to understand how the size and shape of
the clusters change with the interlayer distance. Finally, the calculations of the
binding energies and sizes of the clusters allow us to investigate whether quantum
halos, bound states with a wave function that extends deeply into the classically
forbidden region, can occur in this system.

In Chapter~\ref{Chapter:Two-dimensional dipolar liquid}, we study a many-body system
of dipolar bosons within a bilayer geometry by using exact many-body quantum Monte
Carlo methods. We consider the case in which the dipoles are aligned perpendicularly
to the parallel layers. First, we describe the trial wave functions for the system.
Then, we calculate the equation of state (energy per particle as a function of the
density) for different values of the interlayer distance. Knowledge of the equation
of state permits us to establish the quantum phases present in the bilayer of dipoles.
Afterwards, we obtain the gas-liquid phase diagram of the dipolar fluid as a function
of the density and the separation between layers. Finally, we show numerical results
for the one-body density matrix, condensate fraction and polarization.

In Chapter~\ref{Multilayer System of Dipolar Bosons}, we use the diffusion Monte Carlo
approach to study the ground-state phase diagram of dipolar bosons in a geometry formed
by equally spaced two-dimensional layers. First, we discuss the trial wave functions to
describe the gas, solid, and gas of chains phases. In particular, for the trial function 
of the chains, we have derived the expressions of the drift force and the local energy,
which are necessary to implement the DMC algorithm. Then, we consider the case where
there are four layers and the same number of dipoles in each layer. In this case, we
calculate the pair distribution functions for the different phases present in the system.
Also, we calculate the ground-state phase diagram as a function of the total density and
the interlayer distance. Finally, we consider the case where the dipoles are confined
to three up to ten layers. Here, we calculate the zero-temperature phase diagram.

In Chapter~\ref{Conclusiones}, we present a summary of the principal results obtained in
this Thesis and the main conclusions achieved.

\subsubsection*{Pubications}
The results of this doctoral research were published in:
\begin{itemize}
 \item G. Guijarro, A. Pricoupenko,  G. E. Astrakharchik, J. Boronat, and D. S. Petrov, One-dimensional three-boson problem with two- and three-body
 interactions, Phys. Rev. A {\bf97}, 061605(R) (2018).
 \item G. Guijarro, G. E. Astrakharchik, J. Boronat, B. Bazak, and D. S. Petrov, Few-body bound states of two-dimensional bosons, Phys. Rev. A {\bf101}, 041602(R) (2020).
\end{itemize}
Manuscripts in process
\begin{itemize}
 \item G. Guijarro, G. E. Astrakharchik, and J. Boronat, Quantum halo states in two-dimensional dipolar clusters. Manuscript submitted for publication.
 \item G. Guijarro, G. E. Astrakharchik, and J. Boronat, Quantum liquid of two-dimensional dipolar bosons. Manuscript in preparation.
 \item G. Guijarro, G. E. Astrakharchik, and J. Boronat, Phases of dipolar bosons confined to a multilayer geometry. Manuscript in preparation.
\end{itemize}

\chapter{Quantum Monte Carlo Methods}\label{Chapter:Quantum Monte Carlo methods}
The term Quantum Monte Carlo (QMC) refers to a set of stochastic techniques
whose objective is to solve as exactly as possible quantum many-body problems,
by determining the expectation values of quantum observables~\cite{hammond1994monte}. 
The QMC methods have been demonstrated to give an accurate description of 
correlated quantum systems at zero and low temperature~\cite{Ceperley}.
Examples include ultracold gases with bosonic~\cite{PhysRevA.99.023618,LParisi-Thel-2019}
and fermionic statistics~\cite{RaulBombi2019,PhysRevA.99.043609}, 
quantum solids~\cite{Foulkes,RevModPhys.89.035003},
and Helium~\cite{WDmowski-Obser-2017,PhysRevLett.124.205301}.

To study systems at zero temperature, one can use the \textit{Variational Monte Carlo} 
method (VMC) or the \textit{Diffusion Monte Carlo} method (DMC). The VMC algorithm was introduced by 
McMillan in 1965 to study liquid Helium~\cite{PhysRev.138.A442}. In contrast, the DMC technique was
developed in several works over the years~\cite{RafaelGuardiola}.
The VMC method uses the variational principle of quantum mechanics to
provide an upper bound to the ground-state energy of a quantum system. 
On the other hand, the DMC method allows one to calculate the 
\emph{exact} ground-state energy, of bosonic systems
by solving the many-body Schr\"odinger equation in imaginary time. 

For fermionic systems, the DMC method provides an upper bound to the 
ground-state energy and not the exact one~\cite{Reynolds1982}.  
This is because the wave function of fermions is
antisymmetric under the exchange of two particles. 
Therefore, there are regions where it is positive and other regions where it is negative.
This leads to the so-called \textit{fermion sign problem}.

To study quantum many-body systems with finite, but low temperature
there exists the \textit{Path Integral Monte Carlo} (PIMC) method. 
This method is based on the
thermal density matrix and Feynman's path-integral formulation of
quantum mechanics~\cite{ThesisGuillem,ThesisRaul}.

In this chapter, we introduce the fundamental concepts of the Variational Monte Carlo (VMC)
and the Diffusion Monte Carlo (DMC) methods, which are the Quantum Monte Carlo methods
used in this Thesis.
First, we discuss the theoretical basis of the VMC method and its algorithm.
Second, we present the DMC method and its stochastic realization.
Third, we discuss different types of trial wave functions used in the method.
Finally, we show how several ground-state properties are evaluated
in the Monte Carlo algorithm.  
%%%%%%%%%%%%%%%%%%%%%%%%%%%%%%%%%%%%%
%%%%%%%%%%%%%%%%%%%%%%%%%%%%%%%%%%%%%
\section{The Quantum Many-Body Problem}
We will consider the generic quantum many-body problem involving $N$ interacting particles
of mass $m$. We restrict ourselves to the case of particles in an external potential
$V_{\rm ext}(\mathbf{r}_i)$ and pairwise interactions $V_{\rm int}(\mathbf{r}_i-\mathbf{r}_j)$.
We can write the Hamiltonian of such problems as
\begin{equation}
    \hat{H}=-\frac{\hbar^2}{2m}\sum_{i=1}^N \nabla^2_{\mathbf{r}_i}
    +\sum_{i=1}^N V_{\rm ext}(\mathbf{r}_i)
    +\sum_{i=1}^N\sum_{j=i+1}^N V_{\rm int}(\mathbf{r}_i-\mathbf{r}_j),
    \label{Eq:2.Hamiltonian}
\end{equation}
where $\mathbf{r}_i$ is the position of a single particle. 
It is difficult, if not impossible, to exactly solve the Schrödinger equation for the many-body 
Hamiltonian, which involves obtaining all its eigenstates.
As the complete analytical solution is unavailable, we use numerical methods to calculate
the wave function and the properties of the ground state.
We would like to calculate the ground-state expectation value of an observable $\hat{O}$
\begin{equation}
    \langle\hat{O}\rangle=
    \frac{\langle\Phi_0| \hat{O} |\Phi_0\rangle}{\langle\Phi_0|\Phi_0\rangle},
\end{equation}
$\Phi_0$ being the ground-state wave function.
In particular, we are interested in obtaining the ground-state energy of the system, which
is defined as
\begin{equation}
E_0=\langle\hat{H}\rangle=
\frac{\langle\Phi_0| \hat{H} |\Phi_0\rangle}{\langle\Phi_0|\Phi_0\rangle}.
\end{equation}
Using Monte Carlo methods we can calculate the exact value of the ground-state
energy of a Bose system at zero temperature, within some statistical errors.
%%%%%%%%%%%%%%%%%%%%%%%%%%%%%%%%%%%%%%%
%%%%%%%%%%%%%%%%%%%%%%%%%%%%%%%%%%%%%%%
\section{Variational Monte Carlo Method}
\subsection{Variational Principle}
The Variational Monte Carlo (VMC) method can be used to obtain an approximated value of
the ground-state energy of a quantum system by using the variational principle of
quantum mechanics.
The variational principle states that the expectation value of
a Hamiltonian, $\hat{H}$, obtained with a trial wave function $|\Psi_{\rm T}\rangle$,
provides an upper bound to the ground-state energy $E_0$ of the system:
\begin{equation}
    \frac{\langle\Psi_{\rm T}| \hat{H} |\Psi_{\rm T}\rangle}{\langle\Psi_{\rm T}
        |\Psi_{\rm T}\rangle}\ge E_0,
\label{Eq:2.1}
\end{equation}
if $|\Psi_{\rm T}\rangle$ is not orthogonal to the ground-state wave function. The
equality in Eq.~(\ref{Eq:2.1}) is fulfilled only when the trial function
$|\Psi_{\rm T}\rangle$ is the exact ground-state wave function. The proof of 
Eq.~(\ref{Eq:2.1}) is as follows. If $|\phi_n\rangle$ is an eigenfunction with
eigenvalue $E_n$ of $\hat{H}$, the following properties are fulfilled 
\begin{equation}
\hat{H}|\phi_n\rangle=E_n|\phi_n\rangle,\qquad
\langle\phi_n|\phi_m\rangle=\delta_{n,m},\quad
{\rm and}\quad
\sum_n|\phi_n\rangle\langle\phi_n|=1.
\label{Eq:2.2}
\end{equation}
Using these relations the expectation value of $\hat{H}$ can be written as
\begin{equation}
\begin{aligned}
\frac{\langle\Psi| \hat{H} |\Psi\rangle}{\langle\Psi|\Psi\rangle}&=\frac{\sum_{n,m}\langle\Psi|
\phi_n\rangle \langle \phi_n| \hat{H} |\phi_m\rangle\langle\phi_m|\Psi\rangle}
{\sum_{n}\langle\Psi|\phi_n\rangle \langle \phi_n|\Psi\rangle}\\
&=\frac{\sum_nE_n\langle\phi_n| \Psi\rangle |^2}{\sum_n|\langle\phi_n| \Psi\rangle |^2}.
\end{aligned}
\label{Eq:2.3}
\end{equation}
Since $E_n\ge E_0$, it follows that
\begin{equation}
\begin{aligned}
\frac{\langle\Psi| \hat{H} |\Psi\rangle}{\langle\Psi|\Psi\rangle}
&=\frac{\sum_nE_n|\langle\phi_n| \Psi\rangle |^2}{\sum_n|\langle\phi_n| \Psi\rangle |^2}\\
&\ge\frac{\sum_nE_0|\langle\phi_n| \Psi\rangle |^2}{\sum_n|\langle\phi_n| \Psi\rangle |^2}\\
&=E_0\frac{\sum_n|\langle\phi_n| \Psi\rangle |^2}{\sum_n|\langle\phi_n| \Psi\rangle |^2}=E_0,
\end{aligned}
\label{Eq:2.4}
\end{equation}
and this proves the upper bound reported in Eq.~(\ref{Eq:2.1}).
In general, the trial wave function $|\Psi_{\rm T}\rangle$ depends on a set of parameters that can be
optimized in order to find the lowest possible value of the energy. The trial wave function
with these optimal parameters is an approximation to the ground-state wave function 
of $\hat{H}$ and the lowest energy is an upper bound to the ground-state energy.
\subsection{The Method}
In the Variational Monte Carlo (VMC) method one defines a normalized probability density function 
$\rho(\mathbf{R})$ 
\begin{equation}
    \rho(\mathbf{R})=\frac{|\Psi_{\rm T}(\mathbf{R})|^2}{\int d\mathbf{R}|\Psi_{\rm T}(\mathbf{R})|^2},
    \label{probabilitydistribution}
\end{equation}
and a local energy $E_{\rm L}(\mathbf{R})$ 
\begin{equation}
    E_{\rm L}(\mathbf{R})=\frac{1}{\Psi_{\rm T}(\mathbf{R})}\hat{H}\Psi_{\rm T}(\mathbf{R}),
\label{Eq:2.5}
\end{equation}
here $\mathbf{R}~=~(\vec{\mathbf{r}}_1, \cdots, \vec{\mathbf{r}}_N)$ is a 
3$N$-dimensional vector specifying the positions 
of $N$ particles. The expectation value of $\hat{H}$ can be written in the integral form
\begin{equation}
    E_{\rm var}=\frac{\langle\Psi_{\rm T}|\hat{H}|\Psi_{\rm T}\rangle}{\langle\Psi_{\rm T}|\Psi_{\rm T}\rangle}=
    \frac{\int d\mathbf{R}\Psi_{\rm T}^{*}(\mathbf{R})\hat{H}\Psi_{\rm T}(\mathbf{R})}
    {\int d\mathbf{R}|\Psi_{\rm T}(\mathbf{R})|^2}=
    \int d\mathbf{R}\rho(\mathbf{R})E_{\rm L}(\mathbf{R}). 
\label{Eq:2.6}
\end{equation}
The estimator of the variational energy $E_{\rm var}$ is then calculated as the
mean value of $E_{\rm L}(\mathbf{R})$:
\begin{equation}
    \overline{E_{\rm var}}=\frac{1}{M}\sum_{k=1}^ME_{\rm L}(\mathbf{R}_k), 
\label{Eq:2.7}
\end{equation}
where $M$ is the number of points $\mathbf{R}_k$ 
sampled from the probability density function $\rho(\mathbf{R})$.
As we mentioned before, $\Psi_{\rm T}$ depends on a set of parameters
that are optimized to minimize the energy.
Therefore, we calculate the variational energy Eq.~(\ref{Eq:2.7})
for several values of the parameters and obtain the minimum.

Other observables can also be calculated in the VMC method. The variational
expectation value of an observable $\hat{O}$ is given by
\begin{equation}
    \langle \hat{O} \rangle_{\rm var}=
    \frac{\int {\rm d}\mathbf{R}\Psi_{\rm T}^{*}(\mathbf{R})\hat{O}\Psi_{\rm T}(\mathbf{R})}
    {\int {\rm d} \mathbf{R}|\Psi_{\rm T}(\mathbf{R})|^2},
\end{equation}
which can be written as
\begin{equation}
    \langle \hat{O} \rangle_{\rm var}=
    \int {\rm d}\mathbf{R}\rho(\mathbf{R})O_{\rm L}(\mathbf{R}),
\end{equation}
where $O_{\rm L}(\mathbf{R})$ is the local observable
\begin{equation}
    O_{\rm L}(\mathbf{R})=\frac{1}{\Psi_{\rm T}(\mathbf{R})}
    \hat{O}\Psi_{\rm T}(\mathbf{R}).
\end{equation}
The variational estimator of any local observable can be computed by
averaging the corresponding local value 
\begin{equation}
    \langle \hat{O} \rangle_{\rm var}=
    \frac{1}{M}\sum_{k=1}^MO_{\rm L}(\mathbf{R}_k).
\end{equation}

In general, the probability density
$\rho(\mathbf{R})=|\Psi_{\rm T}(\mathbf{R})|^2/\int d
\mathbf{R}|\Psi_{\rm T}(\mathbf{R})|^2$ Eq.~(\ref{probabilitydistribution})
is complicated and depends on many variables, thus it cannot be sampled by using other 
methods such as the rejection method~\cite{RafaelGuardiola}. 
The solution to this problem is found 
in the Metropolis algorithm which will be discussed below. This method is used to generate random numbers 
from any probability distribution by constructing a Markov process.
Before presenting the Metropolis algorithm, we are going to introduce the
concepts of \textit{stochastic processes} and \textit{Markov processes}.
%%%%%%%%%%%%%%%%%%%%%%%%%%%%%%%%%%%%%%%%%%%%%%%%%%%%%%%%%%
%%%%%%%%%%%%%%%%%%%%%%%%%%%%%%%%%%%%%%%%%%%%%%%%%%%%%%%%%%
\subsection{Stochastic Processes}
A \textit{stochastic process} describes a time-dependent random variable 
$\mathbf{R}(t)$.
For times $t_1,t_2,\ldots,t_n$ there exist a probability distribution
\begin{equation}
P(\mathbf{R}_1,t_1;\mathbf{R}_2,t_2;\ldots;\mathbf{R}_n,t_n)
\end{equation}
where
$\mathbf{R}_1,\ldots,\mathbf{R}_n$ are random variables 
associated to $\mathbf{R}(t)$. Usually the times are ordered, $t_1\leq t_2\leq\ldots\leq t_n$.
We can write the probability distribution in terms of the conditional probabilities as
\begin{equation}
    \begin{aligned}    
    P(\mathbf{R}_n,t_n;\ldots;\mathbf{R}_2,t_2;\mathbf{R}_1,t_1)=&
    P(\mathbf{R}_n,t_n|\mathbf{R}_{n-1},t_{n-1};\ldots;\mathbf{R}_1,t_1)\ldots\\
    &\times P(\mathbf{R}_2,t_2|\mathbf{R}_1,t_1)P(\mathbf{R}_1,t_1).
   \label{Eq:2.9}    
   \end{aligned}
\end{equation}    
It is therefore clear that $\mathbf{R}_j$ is conditioned to
$\mathbf{R}_{j-1},\ldots,\mathbf{R}_1$.
To calculate the probability distribution of a particular realization of 
$\mathbf{R}_1,\ldots,\mathbf{R}_n$ we need to do it in order, this means,
first calculate $P(\mathbf{R}_1,t_1)$ 
then $P(\mathbf{R}_2,t_2|\mathbf{R}_1,t_1)$ and so on. 
%%%%%%%%%%%%%%%%%%%%%%%%%%%%%%%%%%%%%%%%%%%%%%%%%%%%%%%%%%%%%%%%%
%%%%%%%%%%%%%%%%%%%%%%%%%%%%%%%%%%%%%%%%%%%%%%%%%%%%%%%%%%%%%%%%%
\subsection{Markov Processes}
A \textit{Markov process} is a stochastic process for which the conditional probability
for the transition to a new state $\mathbf{R}_j$ depends only on the previous state $\mathbf{R}_{j-1}$
\begin{equation}
    P(\mathbf{R}_j,t_j|\mathbf{R}_{j-1},t_{j-1};\ldots;\mathbf{R}_1,t_1)
    =P(\mathbf{R}_j,t_j|\mathbf{R}_{j-1},t_{j-1}).
\label{Eq:2.10}
\end{equation}   
Therefore for a Markov process we can rewrite Eq.~(\ref{Eq:2.9}) as
\begin{equation}
\begin{aligned}
    P(\mathbf{R}_n,t_n;\ldots;\mathbf{R}_2,t_2;\mathbf{R}_1,t_1)=&
    P(\mathbf{R}_n,t_n|\mathbf{R}_{n-1},t_{n-1})\ldots\\
    &\times P(\mathbf{R}_2,t_2|\mathbf{R}_1,t_1)P(\mathbf{R}_1,t_1).
\label{Eq:2.11}
\end{aligned}
\end{equation}
From here onwards we will consider Markov processes independent of time 
which are known as stationary Markov processes.
The probability $P(\mathbf{R}_f,|\mathbf{R}_i)$ is called the transition probability (or matrix) of going from an
initial state $\mathbf{R}_i$ to a final state $\mathbf{R}_f$.
The transition probability satisfy the following properties
\begin{equation}
    P(\mathbf{R}_f|\mathbf{R}_i)\ge 0,
\label{Eq:2.12}
\end{equation} 
 \begin{equation}
     \int d\mathbf{R}_f P(\mathbf{R}_f|\mathbf{R}_i)=1.
\label{Eq:2.13}
 \end{equation}
The last property simply means that given an initial state $\mathbf{R}_i$, 
a posterior state (the same or different) will be reached with certainty.
Also, there is not a fully absorbing state where the random walk stops.
 
We want to construct a Markov process that converges to the target probability
distribution $\rho(\mathbf{R})$ Eq.~(\ref{probabilitydistribution}) by repeated applications of the transition probability. 
In order for this to happen several conditions must be met.
The first one is that the distribution $\rho(\mathbf{R})$ must be 
an eigenvector of 
$P(\mathbf{R}_f|\mathbf{R}_i)$ with eigenvalue 1~\cite{toulouse2015introduction}
\begin{equation}
     \int d\mathbf{R}_iP(\mathbf{R}_f|\mathbf{R}_i)\rho(\mathbf{R}_i)=
     \rho(\mathbf{R}_f)= \int d\mathbf{R}_i P(\mathbf{R}_i|\mathbf{R}_f)
     \rho(\mathbf{R}_f)\quad \forall \mathbf{R}_f,
 \label{Eq:2.14}
\end{equation}
this condition is known as \textit{stationarity condition},
which means that is we start from the target distribution $\rho(\mathbf{R})$, after repeted
applications of the transition probability, we will continue to sample the target
distribution $\rho(\mathbf{R})$. In general it is required that starting from any initial 
distribution $\rho_{\rm ini}(\mathbf{R})$,
it should converge to the target distribution $\rho(\mathbf{R})$
after applying the transition probability a finite number of times,
\begin{equation}
\begin{aligned}
      \lim_{n\to\infty}&\int d\mathbf{R}_1 d\mathbf{R}_2 \ldots d\mathbf{R}_n P(\mathbf{R}|\mathbf{R}_n) P(\mathbf{R}_n|\mathbf{R}_{n-1}) \ldots
      P(\mathbf{R}_2|\mathbf{R}_1)\rho_{\rm ini}(\mathbf{R}_1)\\
      &=\rho(\mathbf{R}_f).
\label{Eq:2.15}
\end{aligned} 
\end{equation}
To ensure the convergence to a unique stationary distribution $\rho(\mathbf{R})$
the Markov process must be ergodic, which means that it must be possible to move between any 
pair of states $\mathbf{R}_j$ and $\mathbf{R}_l$ in a finite number of steps, then all the states can be visited.
Another condition that the Markov process must fulfill is the
\textit{detailed balanced} contition
\begin{equation}
     P(\mathbf{R}_f|\mathbf{R}_i)\rho(\mathbf{R}_i)=
     P(\mathbf{R}_i|\mathbf{R}_f)\rho(\mathbf{R}_f),
\label{Eq:2.16}
\end{equation}
for any states $\mathbf{R}_i$ and $\mathbf{R}_f$. This condition imposes that the probability flux between
the states $\mathbf{R}_i$ and $\mathbf{R}_f$ to be the same in both directions.
%%%%%%%%%%%%%%%%%%%%%%%%%%%%%%%%%%%%%%%%%%%%%%%%%%%%%%%%%%%%%%%%
%%%%%%%%%%%%%%%%%%%%%%%%%%%%%%%%%%%%%%%%%%%%%%%%%%%%%%%%%%%%%%%%
\subsection{Metropolis Algorithm}
The Metropolis algorithm consists of a Markov process plus a decision criterium
on the random outcomes. We start with an initial state $\mathbf{R}_i$. Then, we
propose a temporary state $\mathbf{R}_f^\prime$ according to a probability distribution 
$P_\mathrm{prop}(\mathbf{R}_f|\mathbf{R}_i)$,
which is known a priori. After that, we test the temporary state. If the temporary state passes
the test then we accept it as the new initial state. If it does not pass the test then
the initial state remains unchanged. The test consists of accepting the move (the temporary state)
with probability
$P_\mathrm{acc}(\mathbf{R}_f|\mathbf{R}_i)$ or rejecting the move with probability 
$1-P_\mathrm{acc}(\mathbf{R}_f|\mathbf{R}_i)$.
Notice that, the transition probability is given by
\begin{equation}
P(\mathbf{R}_f|\mathbf{R}_i)=
    \begin{cases}
        P_\mathrm{acc}(\mathbf{R}_f|\mathbf{R}_i) 
        P_\mathrm{prop}(\mathbf{R}_f|\mathbf{R}_i)
        &\qquad \mathrm{if}\quad \mathbf{R}_f\neq \mathbf{R}_i \\
        1-\int d\mathbf{R}^\prime_f P_\mathrm{acc}
        (\mathbf{R}^\prime_f|\mathbf{R}_i) P_\mathrm{prop}
        (\mathbf{R}^\prime_f|\mathbf{R}_i)
        &\qquad \mathrm{if}\quad \mathbf{R}_f=\mathbf{R}_i
    \end{cases}
\label{Eq:2.17}
\end{equation}   
where $P_\mathrm{acc}(\mathbf{R}_f|\mathbf{R}_i)$ is the probability of accepting the move.
We are free to choose
the criterium for accepting a move, this means we are free to choose
$P_\mathrm{acc}(\mathbf{R}_f|\mathbf{R}_i)$.
However, $P_\mathrm{acc}(\mathbf{R}_f|\mathbf{R}_i)$ has to fulfill
the detailed balanced condition Eq.~(\ref{Eq:2.16})
\begin{equation}
    \frac{P_\mathrm{acc}(\mathbf{R}_f|\mathbf{R}_i)}{P_\mathrm{acc}(\mathbf{R}_i|\mathbf{R}_f)}=
    \frac{P_\mathrm{prop}(\mathbf{R}_i|\mathbf{R}_f)\rho(\mathbf{R}_f)}{P_\mathrm{prop}(\mathbf{R}_f|\mathbf{R}_i)\rho(\mathbf{R}_i)}.
\label{Eq:2.18}
\end{equation}    

The Metropolis algorithm makes a particular choice of $P_\mathrm{acc}(\mathbf{R}_f|\mathbf{R}_i)$
\begin{equation}
    P_\mathrm{acc}(\mathbf{R}_f|\mathbf{R}_i)=\mathrm{min}\left\{1,
        \frac{P_\mathrm{prop}(\mathbf{R}_i|\mathbf{R}_f)\rho(\mathbf{R}_f)}
        {P_\mathrm{prop}(\mathbf{R}_f|\mathbf{R}_i)\rho(\mathbf{R}_i)}\right\}.
\label{Eq:2.19}
\end{equation}    
An advantage of this choice is that we do not need to calculate the normalization factor
for $\rho(\mathbf{R})$, because it will cancel out. 

To implement the Metropolis algorithm we need
to choose a proposal probability $P_\mathrm{prop}(\mathbf{R}_f|\mathbf{R}_i)$.
A simple choice of $P_\mathrm{prop}(\mathbf{R}_f|\mathbf{R}_i)$ is a normal
Gaussian distribution.

The Metropolis algorithm reads as:
\begin{enumerate}
    \item Start from a random state $\mathbf{R}_i$.
    \item Propose a trial state $\mathbf{R}^\prime$ according to
        \begin{equation}  
        \mathbf{R}^\prime=\mathbf{R}_i+\mathbf{\chi},
        \nonumber
    \end{equation}
        where $\mathbf{\chi}$
          is an N-dimensional random vector sampled from a Gaussian distribution.
      \item Calculate the quotient $\rho(\mathbf{R}^\prime)/\rho(\mathbf{R}_i)$.
  \item Generate a random number $\xi$ from the uniform distribution in $[0,1)$.
  \item If $\rho(\mathbf{R}^\prime)/\rho(\mathbf{R}_i)>\xi$ the move is accepted and 
      $\mathbf{R}_{i+1}=\mathbf{R}^\prime$.
      Otherwise stay in the same state $\mathbf{R}_{i+1}=\mathbf{R}_i$.
\end{enumerate}
After applying the Metropolis algorithm a large enough number of times, the Markov process
will sampled the target distribution $\rho(\mathbf{R})$.

Notice that in step 3 only the quotient
$\rho(\mathbf{R}^\prime)/\rho(\mathbf{R}_i)$
defines the acceptance probabilty because
$P_\mathrm{prop}(\mathbf{R}_i|\mathbf{R}^\prime)=
P_\mathrm{prop}(\mathbf{R}^\prime|\mathbf{R}_i)$,
since the Gaussian probability distribution is symmetric.
%%%%%%%%%%%%%%%%%%%%%%%%%%%%%%%%%%%%%%%%%%%%%%%
%%%%%%%%%%%%%%%%%%%%%%%%%%%%%%%%%%%%%%%%%%%%%%%
\subsection{VMC Stochastic Realization}
Here we present the VMC algorithm:
\begin{enumerate}
    \item We start with a random point $\mathbf{R}_1$ that represents the
        initial distribution $\rho_{\rm ini}(\mathbf{R})$ given by
        \begin{equation}
            \rho_{\rm ini}(\mathbf{R})=\delta(\mathbf{R}-\mathbf{R}_1).
            \nonumber
        \end{equation}
    \item Using the Metropolis algorithm we construct the Markov process given by
        $\{\mathbf{R}_{1},\mathbf{R}_2,\ldots,\mathbf{R}_B,\ldots,\mathbf{R}_{B+M}\}$.
    \item We remove the first $B$ elements of the Markov process. The remaining elements
        $\{\mathbf{R}_{1},\mathbf{R}_2,\ldots,\mathbf{R}_{M}\}$ (with the corresponding
        relabeling) are sampled from the target distribution $\rho(\mathbf{R})$.
    \item Now we can calculate the variational estimator of the Hamiltonian 
        \begin{equation}
            \overline{E_{\rm var}}=\frac{1}{M}\sum_{k=1}^ME_{\rm L}(\mathbf{R}_k). 
\label{Eq:EnergyVMC}
\end{equation}
\end{enumerate}        
%%%%%%%%%%%%%%%%%%%%%%%%%%%%%%%%%%%%%%%%%%%%%%%
%%%%%%%%%%%%%%%%%%%%%%%%%%%%%%%%%%%%%%%%%%%%%%%
\section{Diffusion Monte Carlo Method}
In the VMC method, the accuracy of the energy Eq.~(\ref{Eq:EnergyVMC})
depends entirely on the accuracy of the trial wave function.
The larger the overlap between the trial wave function and the ground-state
wave function the better the estimation of the ground-state energy.
To overcome the limitations of the VMC method, we introduce the
\textit{Diffusion Monte Carlo} (DMC) method. 
This method provides a practical way of
evolving in imaginary time the wave function of a quantum
system and obtaining, ultimately, the ground-state 
energy~\cite{doi:10.1119/1.18168}.

The starting point of the DMC method is the time-dependent many-body Schrödinger equation
with an energy shift $E_{\rm T}$, which is equivalent to replacing $\hat{H} \to \hat{H}-E_{\rm T}$
\begin{equation}
\begin{aligned}
    i\hbar\frac{\partial\Psi(\mathbf{R},t)}{\partial t}&=\left[\hat H-E_{\rm T}\right]\Psi(\mathbf{R},t)\\
    &=\left[-\frac{\hbar^2}{2m}\nabla^2_{\mathbf{R}} + V(\mathbf{R}) - E_{\rm T}\right]\Psi(\mathbf{R},t),
\end{aligned}
\label{Eq:2.23}
\end{equation}   
where $\mathbf{R}~=~(\vec{\mathbf{r}}_1, \cdots, \vec{\mathbf{r}}_N)$ is a 
3$N$-dimensional vector specifying the coordinates of all $N$ particles,
$\Psi(\mathbf{R},t)$ is the many-body wave function of the system, which depends on
the particle coordinates and the time,
and $\hat H$ is the many-body Hamiltonian
Eq.~(\ref{Eq:2.Hamiltonian})
\begin{equation}
    \nabla^2_{\mathbf{R}}=\sum_{i=1}^N \nabla^2_{\mathbf{r_i}},\quad 
 V(\mathbf{R})=\sum_{i=1}^NV_{\rm ext}(\mathbf{r_i}) + 
   \sum_{i=1}^N\sum_{j=i+1}^NV_{\rm int}(\mathbf{r_i}-\mathbf{r_j}).
\label{Eq:2.24}
\end{equation}
Let us now perform a transformation from real time to imaginary time
by introducing the new variable $\tau=it/\hbar$. After this, the Schrödinger
equation Eq.~(\ref{Eq:2.23}) becomes
\begin{equation}
\begin{aligned}
    -\frac{\partial\Psi(\mathbf{R},\tau)}{\partial\tau}&=\left[\hat H-E_{\rm T}\right]\Psi(\mathbf{R},\tau)\\
    &=\left[-D\nabla^2_{\mathbf{R}} + V(\mathbf{R}) - E_{\rm T}\right]\Psi(\mathbf{R},\tau),
\end{aligned}
\label{Eq:2.25}
\end{equation}  
where $D=\hbar^2/2m$. Eq.~(\ref{Eq:2.25}) can be identified as a modified
diffusion equation in the 3$N$- dimensional space. If the 
$\left[V(\mathbf{R}) - E_{\rm T}\right]$ term were removed, Eq.~(\ref{Eq:2.25})
becomes the usual diffusion equation with a difussion constant $D$. On the other hand,
if the term with the Laplacian were removed, Eq.~(\ref{Eq:2.25}) would be a
rate equation, describing and exponential growth or decrease of the function
$\Psi(\mathbf{R},\tau)$.

The objective is to solve Eq.~(\ref{Eq:2.25}) to access the 
ground state of the system.
Using the spectral descomposition
\begin{equation}
    e^{-(\hat{H}-E_{\rm T})\tau}=\sum_{i}|\Phi_i\rangle
    e^{-(E_i-E_{\rm T})\tau}\langle\Phi_i|,
\end{equation}
the formal solution of Eq.~({\ref{Eq:2.25}})
\begin{equation}
    |\Psi(\mathbf{R},\tau)\rangle=e^{-(\hat{H}-E_{\rm T})\tau}
    |\Psi(\mathbf{R},0)\rangle,
\end{equation}    
can be expressed as
\begin{equation}
    |\Psi(\mathbf{R},\tau)\rangle=\sum_{i=0}
    e^{-(E_i-E_{\rm T})\tau}
|\Phi_i\rangle
\langle\Phi_i
|\Psi(\mathbf{R},0)\rangle,
\label{Eq:2.250}
\end{equation}   
where $\{\Phi_i\}$ and $\{E_i\}$,
with $\hat{H}|\Phi_i\rangle=E_i|\Phi_i\rangle$,
denote a complete sets of 
eigenfunctions and eigenvalues of $\hat{H}$, respectively.
We consider that the eigenvalues are ordered 
\begin{equation}
    E_0<E_1\leq E_2\leq\ldots
\label{Eq:2.260}
\end{equation}
The amplitudes of each one of the terms in Eq.~(\ref{Eq:2.250}) can increase or decrease
in time depending on the sign of ($E_n-E_{\rm T}$).  
Notice that, for sufficiently long times $\tau\to\infty$ the operator
$e^{-(\hat{H}-E_{\rm T})\tau}$ projects out the lowest eigenstate $|\Phi_0\rangle$
that has non-zero overlap with $|\Psi(\mathbf{R},0)\rangle$
\begin{equation}
    \begin{aligned}
 \lim_{\tau\to\infty}|\Psi(\mathbf{R},\tau)\rangle=&
 \lim_{\tau\to\infty}\sum_{i=0}
 e^{-(E_i-E_{\rm T})\tau}
|\Phi_i\rangle
\langle\Phi_i
|\Psi(\mathbf{R},0)\rangle\\
&=\lim_{\tau\to\infty}
e^{-(E_0-E_{\rm T})\tau}
|\Phi_0\rangle
\langle\Phi_0
|\Psi(\mathbf{R},0)\rangle.
\label{Eq:2.27}
\end{aligned}
\end{equation}    
The higher terms will decay exponentially faster since $E_n>E_0~~\forall n\neq0$.
For $E_{\rm T}=E_0$ the function $|\Psi(\mathbf{R},\tau)\rangle$ converges to the ground-state wave
function $|\Phi_0(\mathbf{R})\rangle$ regardless of the choice of the initial wave function
$|\Psi(\mathbf{R},0)\rangle$ 
\begin{equation}
    \lim_{\tau\to\infty}|\Psi(\mathbf{R},\tau)\rangle
    \propto |\Phi_0(\mathbf{R})\rangle.
\label{Eq:2.28}
\end{equation}
This fundamental property of the projector $e^{-(\hat{H}-E_{\rm T})\tau}$
is the basis of the DMC technique~\cite{Foulkes}.
The DMC method follows the evolution of an initial many-body state
$|\Psi(\mathbf{R},0)\rangle$
in imaginary time, until long enough time passes and only the contribution of the ground state
to the many-body wave function dominates according to Eq.~(\ref{Eq:2.27}).
%%%%%%%%%%%%%%%%%%%%%%%%%%%%%%%%%%%%%%%%%%%
%%%%%%%%%%%%%%%%%%%%%%%%%%%%%%%%%%%%%%%%%%%
\subsection{Green{\rq}s Function}
To follow the evolution of the Schrödinger equation in imaginary time we will use the 
Green{\rq}s function formalism.

The solution of the imaginary-time Schrödinger 
equation Eq.~(\ref{Eq:2.25}) in integral form is given by
\begin{equation}
    \langle\mathbf{ R}|\Psi(\tau)\rangle=
    \int d\mathbf{R^\prime} \langle\mathbf{ R}|e^{-(\hat{H}-E_{\rm T})\tau}|\mathbf{R^\prime}\rangle
    \langle\mathbf{R^\prime}|\Psi(0)\rangle,
\label{Eq:G1}
\end{equation}
and it can be written as
\begin{equation}
    \Psi(\mathbf{R},\tau)=
    \int d\mathbf{R^\prime}G(\mathbf{R}|\mathbf{R^\prime};\tau)
    \Psi(\mathbf{R^\prime},0).
\label{Eq:G2}
\end{equation}
Here, $\Psi(\mathbf{R^\prime},0)$ is the wave function at the initial time $\tau=0$
and we have introduced the \textit{Green{\rq}s function} $G(\mathbf{R}|\mathbf{R^\prime};\tau)$,
also known as the imaginary-time
propagator from $\mathbf{R^\prime}$ to $\mathbf{R}$
\begin{equation}
G(\mathbf{R}|\mathbf{R^\prime};\tau)=
\langle\mathbf{ R}|e^{-(\hat{H}-E_{\rm T})\tau}|\mathbf{R^\prime}\rangle.
\label{Eq:G3}
\end{equation}
The Green{\rq}s function is subject to the boundary condition at the initial time $\tau=0$
\begin{equation}
    G(\mathbf{R}|\mathbf{R^\prime};0)=
    \delta(\mathbf{ R}-\mathbf{R^\prime}).
\label{Eq:G4}
\end{equation}
In general, we do not know
the exact Green{\rq}s function
for all times $\tau$. However, the Green{\rq}s function is known in the
limit of a short propagation time, $G(\mathbf{R}|\mathbf{R^\prime};\Delta\tau)$, where
$\Delta\tau$ is a small imaginary time-step
\begin{equation}
    \Psi(\mathbf{R},\tau+\Delta\tau)=
    \int d\mathbf{R^\prime}G(\mathbf{R}|\mathbf{R^\prime};\Delta\tau)
    \Psi(\mathbf{R^\prime},\tau),
\label{Eq:G5}
\end{equation}
and then Eq.~(\ref{Eq:G2}) can be solved in a step by step process 
\begin{equation}
    \begin{aligned}
    \Psi(\mathbf{R},\tau)=&\lim_{M\to\infty}
    \int d\mathbf{R_1}d\mathbf{R_2}\cdots d\mathbf{R_M}
    G(\mathbf{R}|\mathbf{R_M};\Delta\tau)G(\mathbf{R_M}|\mathbf{R_{M-1}};\Delta\tau)\\
    &\cdots G(\mathbf{R_2}|\mathbf{R_1};\Delta\tau) \Psi(\mathbf{R_1},0).
\label{Eq:G6}
\end{aligned}
\end{equation}
According to Eq.~(\ref{Eq:G6}) an approximation to the final state
$\Psi(\mathbf{R},\tau)$ is obtained by applying $M$ times the short-time
Green{\rq}s function to the intial state $\Psi(\mathbf{R_1},0)$.

Before giving an explicit expression for the short-time Green{\rq}s function 
we are going to introduce 
the \textit{importance sampling} technique. In this technique, we introduce a guiding wave function
that is independent of the imaginary time.
%%%%%%%%%%%%%%%%%%%%%%%%%%%%%%%%%%%%%%%%%%%%%%%%
%%%%%%%%%%%%%%%%%%%%%%%%%%%%%%%%%%%%%%%%%%%%%%%%
\subsection{Importance Sampling}
Solving Eq.~(\ref{Eq:2.23}) is usually inefficient, mainly because of the presence
of the potential $V(\mathbf{R})$, which can diverge when to particles are very close.
This leads to large variance and low convergence when calculating the
expectation values of observables. To overcome these limitations one can use the
importance sampling technique.

In the importance sampling procedure one consider the imaginary-time evolution
of the \textit{mixed distribution} $f(\mathbf{R},\tau)$, which is given by the product,
\begin{equation}
    f(\mathbf{R},\tau)=\Psi_{\rm T}(\mathbf{R})\Psi(\mathbf{R},\tau),
\label{Eq:2.32}
\end{equation}
of the wave function $\Psi(\mathbf{R},\tau)$, which satisfies the Schrödinger equation
Eq.~(\ref{Eq:2.25}), and a trial wave function $\Psi_{\rm T}(\mathbf{R})$, which is imaginary-time
independent. The trial wave function $\Psi_{\rm T}(\mathbf{R})$ is 
designed from the available
knowledge of the exact ground-state wave function. 

The imaginary-time evolution of
$f(\mathbf{R},\tau)$ can be obtained by multiplying Eq.~(\Ref{Eq:2.24})
by $\Psi_{\rm T}(\mathbf{R})$. After rearranging terms, one obtains
\begin{equation}
\begin{aligned}
-\frac{\partial f(\mathbf{R},\tau)}{\partial\tau}=&
-D\nabla^2_{\mathbf{R}} f(\mathbf{R},\tau)+
D\nabla_{\mathbf{R}}\cdot[\mathbf{F}(\mathbf{R})f(\mathbf{R},\tau)]\\
&+\left[E_{\rm L}(\mathbf{R}) - E_{\rm T}\right]f(\mathbf{R},\tau).
\label{Eq:2.33}
\end{aligned}
\end{equation}   
Here, $\mathbf{F}(\mathbf{R})$ denotes the \textit{drift force}, also called the \textit{drift velocity}
\begin{equation}
    \mathbf{F}(\mathbf{R})=2\frac{\nabla_{\mathbf{R}}\Psi_{\rm T}(\mathbf{R})}{\Psi_{\rm T}(\mathbf{R})},
\label{Eq:2.34}
\end{equation}
and $E_{\rm L}(\mathbf{R})$ is the local energy Eq.~(\ref{Eq:2.5})
\begin{equation}
    E_{\rm L}(\mathbf{R})=\frac{\hat{H}\Psi_{\rm T}(\mathbf{R}) }{\Psi_{\rm T}(\mathbf{R})}.
\label{Eq:2.35}
\end{equation}

Eq.~(\ref{Eq:2.33}) describes a modified difussion process for the mixed distribution
$f(\mathbf{R},\tau)$. Notice that, the rate term is now proportional to
$[E_{\rm L}(\mathbf{R})-E_{\rm T}]$, unlike the rate term in Eq.~(\ref{Eq:2.25}) which
depends on the potential $V(\mathbf{R})$. 
With a good choice of $\Psi_{\rm T}(\mathbf{R})$, the local energy $E_{\rm L}(\mathbf{R})$
remain finite even if $V(\mathbf{R})$ diverges~\cite{Reynolds1982}.
Also, notice that there is an 
additional term $\nabla_{\mathbf{R}}\cdot[\mathbf{F}(\mathbf{R})f(\mathbf{R},\tau)]$
in Eq.~({\ref{Eq:2.33}}). This new term imposes a drift on the difussion
process guided by $\Psi_{\rm T}(\mathbf{R})$.

The mixed distribution $f(\mathbf{R},\tau)$ becomes proportional to the 
ground-state wave function in the limit of large enough time
\begin{equation}
    f(\mathbf{R},\tau)\propto{\lim_{\tau\to\infty}}
    \Psi_{\rm T}(\mathbf{R})\Phi_0(\mathbf{R}).
\label{Eq:2.40}
\end{equation}
%%%%%%%%%%%%%%%%%%%%%%%%%%%%%%%%%%
%%%%%%%%%%%%%%%%%%%%%%%%%%%%%%%%%%
\subsection{Importance-Sampling Green{\rq}s Function and 
    Short-Time Approximation}
The evolution described by Eq.~(\Ref{Eq:2.33}) can be written as the sum of
three different operators acting on $f(\mathbf{R},\tau)$
\begin{equation}
-\frac{\partial f(\mathbf{R},\tau)}{\partial\tau}=
(\hat{O}_K + \hat{O}_D+ \hat{O}_B )f(\mathbf{R},\tau)
\equiv \hat{O}f(\mathbf{R},\tau),
\label{Eq:2.36}
\end{equation} 
where
\begin{equation}
    \begin{aligned}
        \hat{O}_K&=-D\nabla^2_{\mathbf{R}},\\
        \hat{O}_D&=D[\nabla_{\mathbf{R}}\cdot \mathbf{F}(\mathbf{R})
        +\mathbf{F}(\mathbf{R})\cdot\nabla_{\mathbf{R}}],\\
        \hat{O}_B&=E_{\rm L}(\mathbf{R}) - E_{\rm T}.
\label{Eq:2.37}
\end{aligned}
\end{equation} 
Here, $\hat{O}_K$, $\hat{O}_D$ and $\hat{O}_B$ are the kinetic, drift and branching operators, respectively.
This division will make easier to solve the Schrödinger equation for $f(\mathbf{R},\tau)$ Eq.~(\ref{Eq:2.33}).

Analogously to Eq.~(\ref{Eq:G5}),
the formal solution of the evolution equation for the mixed distribution
$f(\mathbf{R},\tau)$ 
\begin{equation}
    f(\mathbf{R},\tau+\Delta\tau)=
    \int d\mathbf{R^\prime}\tilde G(\mathbf{R}|\mathbf{R^\prime};\Delta\tau)
    f(\mathbf{R^\prime},\tau),
\label{Eq:2.38}
\end{equation}
where $\tilde G(\mathbf{R}|\mathbf{R^\prime};\tau)$ is the \textit{importance sampling
    Green{\rq}s function}. $\tilde G(\mathbf{R}|\mathbf{R^\prime};\tau)$ satisfies the boundary
condition
\begin{equation}
    \tilde G(\mathbf{R}|\mathbf{R^\prime};0)=
    \delta(\mathbf{R}-\mathbf{R^\prime}).
\label{Eq:2.39}
\end{equation}
The importance sampling Green{\rq}s function is given in terms of the
operator $\hat{O}$, 
\begin{equation}
    \tilde G(\mathbf{R}|\mathbf{R^\prime};\Delta\tau)=
    \langle\mathbf{ R}|e^{-\hat{O}\Delta\tau}|\mathbf{R^\prime}\rangle.
    \label{Eq:2.390}
\end{equation}

Now, we focus on giving an explicit expression for the short-time Green{\rq}s
function.
A short-time approximation of the Green{\rq}s function 
to first order in $\Delta\tau$ is given by
\begin{equation}
    e^{-\hat{O}\Delta\tau}=e^{-(\hat{O}_K+\hat{O}_D+\hat{O}_B)\Delta\tau}
    =e^{-\hat{O}_K\Delta\tau}e^{-\hat{O}_D\Delta\tau}
    e^{-\hat{O}_B\Delta\tau} + \mathcal{O}((\Delta\tau)^2).
\label{Eq:2.391}
\end{equation}
A second order descomposition is given by
\begin{equation}
e^{-\hat{O}\Delta\tau}=
e^{-\hat{O}_B \frac{\Delta\tau}{2}}
e^{-\hat{O}_D \frac{\Delta\tau}{2}}
e^{-\hat{O}_K\Delta\tau}
e^{-\hat{O}_D \frac{\Delta\tau}{2}}
e^{-\hat{O}_B \frac{\Delta\tau}{2}}
     + \mathcal{O}((\Delta\tau)^3).
\label{Eq:2.392}
\end{equation}
Observe that, as $\Delta\tau\rightarrow0$ this will be a valid approximation.
Introducing Eq.~(\ref{Eq:2.392}) into Eq.~(\ref{Eq:2.38}) we obtain
an integral equation of the mixed distribution $f(\mathbf{R},\tau)$ in terms of
the individual Green{\rq}s functions $\tilde G_i$, each one asociated to a single 
operator $\hat{O}_i$
\begin{equation}
\begin{aligned}
f(\mathbf{R},\tau+\Delta\tau)=&
\int d\mathbf{R_1}d\mathbf{R_2}d\mathbf{R_3}d\mathbf{R_4}d\mathbf{R^{\prime}}    
[
\tilde G_B(\mathbf{R}|\mathbf{R_1};\frac{\Delta\tau}{2})\\
&\times\tilde G_D(\mathbf{R_1}|\mathbf{R_2};\frac{\Delta\tau}{2})
\tilde G_K(\mathbf{R_2}|\mathbf{R_3};\Delta\tau)\\
&\times\tilde G_D(\mathbf{R_3}|\mathbf{R_4};\frac{\Delta\tau}{2})
\tilde G_B(\mathbf{R_4}|\mathbf{R^{\prime}};\frac{\Delta\tau}{2})
]
f(\mathbf{R^{\prime}},\tau).
\end{aligned}
\end{equation}
The next step is to solve 
three diferential equations, each corresponding to a Green{\rq}s function
$\tilde G_i$.
The first diferential equation is associated with 
the kinetic operator $\tilde{G}_K$
\begin{equation}
    -\frac{\partial \tilde G_K(\mathbf{R}|\mathbf{R^{\prime}};\tau)}
    {\partial \tau}=-D\nabla^2_{\mathbf{R}}
    \tilde G_K(\mathbf{R}|\mathbf{R^{\prime}};\tau).
    \label{GreenFunctionK}
\end{equation}
This is a diffusion equation which diffusion constant $D$.
The evolution given by $\tilde{G}_K$ corresponds to an 
isotropic Gaussian movement
\begin{equation}
    \tilde G_K(\mathbf{R}|\mathbf{R^{\prime}};\tau)=
    \left(4\pi D\tau \right)^{-3N/2}
    \mathrm{exp}\left[ -\frac{(\mathbf{R}-\mathbf{R^{\prime}})^2}
        {4D\tau}\right].
\end{equation}
The second diferential equation corresponds to the drift operator
$\tilde{G}_D$
\begin{equation}
    -\frac{\partial \tilde G_D(\mathbf{R}|\mathbf{R^{\prime}};\tau)}
    {\partial \tau}=D\nabla_{\mathbf{R}}\cdot
    \left[\mathbf{F}(\mathbf{R})\tilde G_D(\mathbf{R}|\mathbf{R^{\prime}};\tau)\right].
    \label{GreenFunctionD}
\end{equation}
The Green{\rq}s function $\tilde{G}_D$ describes the movement due to the drift force and its solution is
\begin{equation}
   \tilde G_D(\mathbf{R}|\mathbf{R^{\prime}};\tau)=
   \delta(\mathbf{R}-\mathcal{R}^\prime(\tau)),
\end{equation}
where $\mathcal{R}(\tau)$ is defined by the following equations 
\begin{equation}
\begin{aligned}
\frac{d\mathcal{R}(\tau)}{d\tau}&=D\mathbf{F}(\mathcal{R}(\tau)),\\
\mathcal{R}(0)&=\mathbf{R}.
\end{aligned}
\label{Eq:DifferentialEquation}
\end{equation}
The last diferential equation is associated with the branching
operator $\tilde{G}_B$
\begin{equation}
    -\frac{\partial \tilde G_B(\mathbf{R}|\mathbf{R^{\prime}};\tau)}
    {\partial\tau}=(E_{\rm L}(\mathbf{R})-E_{\rm T}) 
\tilde G_B(\mathbf{R}|\mathbf{R^{\prime}};\tau),
\label{GreenFunctionB}
\end{equation}
and its solution is given by
\begin{equation}
    \tilde G_B(\mathbf{R}|\mathbf{R^{\prime}};\tau)=
    \mathrm{exp}\left[-(E_{\rm L}(\mathbf{R})-E_{\rm T})\tau \right]   
    \delta\left(\mathbf{R}-\mathbf{R}^{\prime}  \right).
\end{equation}
The Green{\rq}s function $\tilde{G}_B$ assigns a weight to $\mathbf{R}$
depending on its local energy.

Now that we have found the solutions
to the equations of the Green{\rq}s functions we can 
describe completely the stochastic realization of the DMC algorithm.

In the stochastic realization of the DMC algorithm, the mixed distribution and its
imaginary-time evolution are represented by a set of random walkers.  
Walkers evolve through repeated applications of the propagators $G_i$,
until one obtains convergence to the ground state in the limit $\tau\to\infty$.
%%%%%%%%%%%%%%%%%%%%%%%%%%%%%%%%%%%%%%%
%%%%%%%%%%%%%%%%%%%%%%%%%%%%%%%%%%%%%%%
\subsection{DMC Stochastic Realization}
In this section, we use the concepts exposed previously to
give a basic version of the DMC algorithm with importance sampling. 

In the DMC method, the probability distribution at the initial time $f(\mathbf{R},0)$ 
and its evolution in imaginary-time $f(\mathbf{R},\tau)$ is represented by a set
of \textit{random walkers}. 
A \textit{walker} is defined by the positions of all the particles in
the system in the configuration space of 3$N$ dimensions 
$\mathbf{R}=\{\vec{\mathbf{r}}_1,\vec{\mathbf{r}}_2,\ldots,\vec{\mathbf{r}}_N\}$.
The set of random walkers can be written as~\cite{ThesisOmar}
\begin{equation}
    \mathcal{R}_k=\{\mathbf{R}_{k,\alpha}|\alpha=1,2,\ldots,N_{w,k}\}.
\end{equation}
Here, $k$ is the time step index, $\tau_k=k\Delta\tau$ is the current time,
and $N_{w,k}$ is the number of walkers which may change between steps.

The initial configuration for the DMC algorithm is drawn from some arbitrary probability distribution.
In most cases the initial configuration
will be the output from the VMC algorithm.

At the time-step $k=0$, we start with an initial configuration $N_{w,0}$
of random walkers $\mathcal{R}_0$ drawn from
\begin{equation}
    f_{\rm ini}(\mathbf{R})=f(\mathbf{R},0)=|\Psi_{\rm T}(\mathbf{R})|^2,
\end{equation}
after passing a large enough Metropolis steps. An initial estimate of $E_{\rm T}$
is obtained from the mean of the local energies of the walkers
\begin{equation}
    E_{T,0}=\frac{1}{N_{w,0}}\sum_{\alpha=1}^{N_{w,0}}
    E_{\rm L}(\mathbf{R}_{0,\alpha}).
\end{equation}
The following algorithm is iterated $M$ times.
\begin{enumerate}
    \item Starting from the random walker $\mathbf{R}_{k-1,\alpha}$ we obtain a temporary configuration
        by appliyng the Green{\rq}s function $\hat{G}_K$ Eq.~(\ref{GreenFunctionK}). This is done for all random walkers in $\mathcal{R}_{k-1}$.
        This means that if we start with the configuration $\mathbf{R}_{k-1,\alpha}$ we obtain a
        temporary configuration
        $\mathbf{R}^\prime_{k-1,\alpha}$ as
        \begin{equation}
            \mathbf{R}^\prime_{k-1,\alpha} = \mathbf{R}_{k-1,\alpha} + \mathbf{\chi}.
        \end{equation}
        Here, $\mathbf{\chi}$ is an N-dimensional random vector sampled from a multivariate
        Gaussian distribution with zero mean and variance $\sigma^2=2D\Delta\tau$. 
    \item Now we will apply a second Green{\rq}s function $\hat{G}_D$ Eq.~(\ref{GreenFunctionD}),
        which corresponds to the action of the drift force.
        From the temporary configuration $\mathbf{R^\prime}_{k-1,\alpha}$ we obtain a new
        configuration $\mathbf{R}_{k,\alpha}$ by doing the following steps 
        \begin{itemize}
         \item $\mathbf{R}^{(1)}_{k-1,\alpha}=\mathbf{R}^\prime_{k-1,\alpha}+\mathbf{F}(\mathbf{R}^\prime_{k-1,\alpha})\frac{\Delta\tau}{2}$
         \item $\mathbf{R}^{(2)}_{k-1,\alpha}=\mathbf{R}^\prime_{k-1,\alpha}+
             \left[\mathbf{F}(\mathbf{R}^\prime_{k-1,\alpha})+\mathbf{F}(\mathbf{R}^{(1)}_{k-1,\alpha})\right]\frac{\Delta\tau}{4}$
         \item $\mathbf{R}^{(3)}_{k-1,\alpha}=\mathbf{R}^\prime_{k-1,\alpha}+\mathbf{F}(\mathbf{R}^{(2)}_{k-1,\alpha})\Delta\tau$
         \item $\mathbf{R}_{k,\alpha}=\mathbf{R}^{(3)}_{k-1,\alpha}$
        \end{itemize}
        The new configuration form a new set $\mathcal{R}_k$.

        We used a second order integration method called Runge-Kutta~\cite{PhysRevA.42.6991} to integrated
        the differential equation Eq.~(\ref{Eq:DifferentialEquation}) in order to do the 
        displacement from $\mathbf{R^\prime}_{k-1,\alpha}$ to $\mathbf{R}_{k,\alpha}$.
    \item Calculate the \textit{branching probability} for each walker in $\mathcal{R}_k$:
        \begin{equation}
            w_{\alpha}=e^{-\left(
                    \frac{E_{\rm L}(\mathbf{R}_{k,\alpha})-E_{\rm L}(\mathbf{R}_{k-1,\alpha})}{2}
                    -E_{\rm T}\right)\Delta\tau}.
        \end{equation}
    \item Calculate the \textit{branching factor} for each walker in $\mathcal{R}_k$:
        \begin{equation}
            n_{\alpha}=int(w_{\alpha}+\eta).
        \end{equation}
        Here, $int$ denotes the integer part of a real number 
        and $\eta$ is a random number drawn from the uniform
        distribution on the interval $[0,1)$. If $n_\alpha=0$,
        removed $\mathbf{R}_{k,\alpha}$ from $\mathcal{R}_k$.
        If $n_\alpha\ge1$, replace $\mathbf{R}_{k,\alpha}$ with $n_\alpha$
        copies of iftsel in $\mathcal{R}_k$.
    \item Update the estimators of energy and other observables of interest.
    \item Repeat steps 1 to 5 until $M$ time steps are reached. 
\end{enumerate}
For sufficiently long times the ground-state energy is given by
\begin{equation}
    E_0=\lim_{\tau\to\infty}
    \frac{\int {d}\mathbf{R}f(\mathbf{R},\tau)E_{\rm L}(\mathbf{R})}
    {\int d \mathbf{R}f(\mathbf{R},\tau)}.
\end{equation}
The result of the stochastic process describe above is a set of walkers
representing the distribution $f(\mathbf{R},\tau)$. Therefore,
the estimator for $\overline{E_0}$ after $M$ times steps is~\cite{toulouse2015introduction}
\begin{equation}
    \overline{E_0}=\frac{\sum_{k=1}^M\sum_{\alpha=1}^{N_{w,k}}E_{\rm L}(\mathbf{R}_{k,\alpha})}
{\sum_{k=1}^M\sum_{\alpha=1}^{N_{w,k}} w_{k,\alpha}}.
\end{equation}

The value of $E_{\rm T}$ is adjusted during the iterations to keep the size of the walker population
within a desired value. A simple formula for adjusting $E_{\rm T}$ for the iteration $k + 1$ 
is~\cite{toulouse2015introduction}
\begin{equation}
    E_{T,k+1}=\overline{E_{0,k}}-C\mathrm{ln}\left( \frac{N_{w,k}}{N_{w,ave}} \right),
\end{equation}
where $C$ is a constant, and $N_{k,ave}$ is a desired average number of walkers.
%%%%%%%%%%%%%%%%%%%%%%%%%%%%%%%%%%%%%%%
%%%%%%%%%%%%%%%%%%%%%%%%%%%%%%%%%%%%%%%
\subsection{Convergence Analysis}
The DMC algorithm gives exact results for the ground-state energy when
simultaneously the time step $\Delta\tau\to 0$ and the number of walkers
$N_w\to\infty$.
The use of a finite time step $\Delta\tau$
to approximate the Green{\rq}s function introduces a systematic error bias in the calculation.
To overcome this problem one can consider a short-time Green{\rq}s function
accurate to order $(\Delta\tau)^2$ according to Eq.~(\ref{Eq:2.391})
or a more precise algorithm accurate to order $(\Delta\tau)^3$ as stated by Eq.~(\ref{Eq:2.392}).
In the first case, the energy has a linear dependence 
when the time step is sufficiently small.
Then, one can use several values of the time step to extrapolate the value 
of the energy to the $\Delta\tau\to 0$ limit. In the second case, the energy
depends quadratically on the time step. Here, the extrapolation procedure is
not completely necessary because for small $\Delta\tau$ the energy converges
fast to the exact value and the time step can be chosen such that the systematic
error is smaller than the statistical error.

\begin{figure}[b!]
	\centering
    \includegraphics[width=0.7\textwidth]{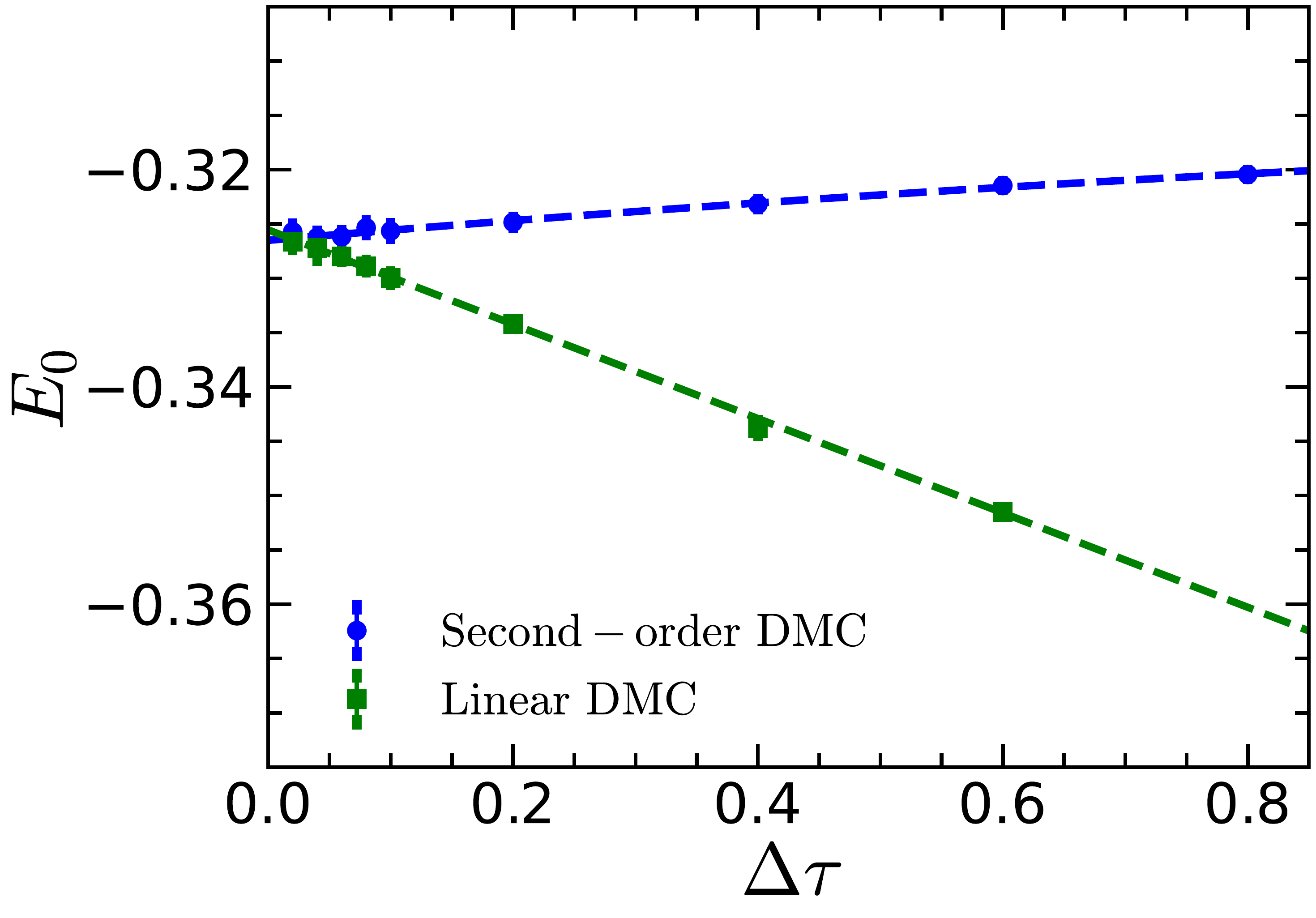}
    \caption{Ground-state energy $E_0$ of a dipolar gas vs the time step $\Delta\tau$
        computed by using the linear and second-order DMC algorithms. 
        The dashed lines correspond to polynomial fits: 
        $E_0(\Delta\tau)=a_0+a_1\Delta\tau$ for linear DMC meyhod and
        $E_0(\Delta\tau)=b_0+b_1\Delta\tau+b_2(\Delta\tau)^2$ for second-order DMC method.
        The unit of energy used is $\hbar^2/(mr_0^2)$, with $r_0=md^2/\hbar^2$ the dipolar length and $d$
        is the dipole moment of an atom of mass $m$.
        The units of $\Delta\tau$ are inverse of energy.}
    \label{Fig:TimeStep}
\end{figure}

In Fig~\ref{Fig:TimeStep} we show an example of the ground-state energy $E_0$ 
dependence on the time step $\Delta\tau$ for a dipolar gas. We compare a linear DMC method with a 
second-order DMC method. For the linear DMC algorithm, in order to obtain the exact energy, the extrapolation
to zero time step is required. In Fig~\ref{Fig:TimeStep} we observe
that the slope of the line that joins the green dots 
is pronounced with respect to the scale that we are using.
In contrast, for the second-order DMC technique, we notice that
the changes in energy are statistically indistinguishable in the range 
$\Delta\tau=0.01-0.1$,  
the slope of the line that joins the blue dots is less pronounced.
This is a very useful feature of the second-order DMC method which
allows one to obtain exact values of the energy with less computational effort.
Besides the time step bias, the DMC algorithm presents
a dependence on the number of walkers $N_w$, which requires
additional analysis.

\begin{figure}[H]
	\centering
    \includegraphics[width=0.7\textwidth]{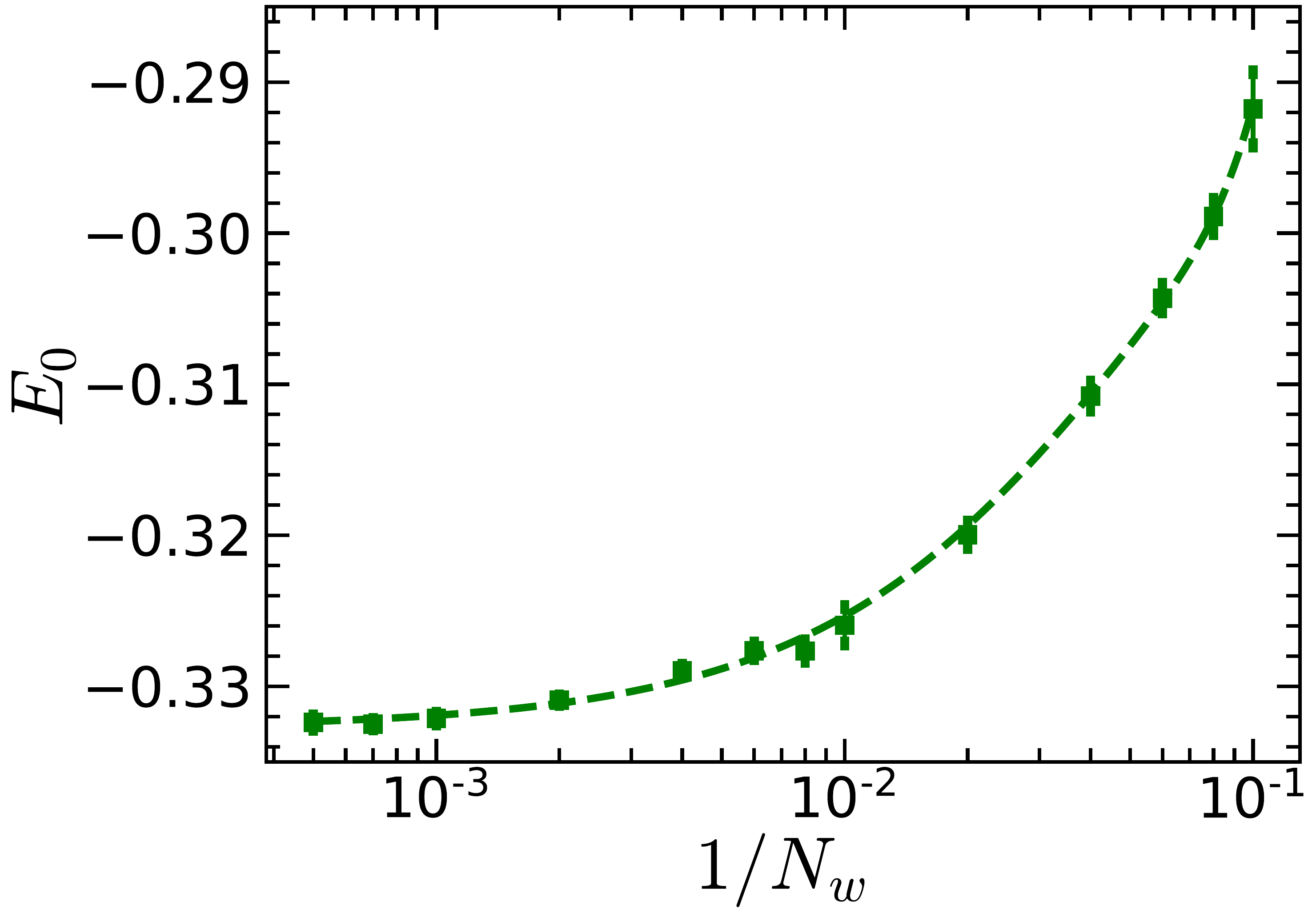}
    \caption{Ground-state energy $E_0$ of a dipolar gas vs the the inverse of the number
        of walkers $1/N_w$
        computed by using a second-order DMC algorithm with time step
        $\Delta\tau=0.01$. The dashed line correspond to the polynomial fit: 
        $E_0(N_w)=a_0+a_1/N_w+a_2/(N_w)^2+a_3/(N_w)^3.$
        The unit of energy used is $\hbar^2/(mr_0^2)$.
        The units of $\Delta\tau$ are inverse of energy.}
    \label{Fig:Nw}
\end{figure}

Fig~\ref{Fig:Nw} shows an example of the ground-state energy $E_0$ dependence on the
number of walkers $N_w$ for a dipolar gas. We notice that the energy changes very little for
$N_w\gtrsim 1000$. Therefore, using $N_w\approx1000$ to estimate the exact ground-state energy
is typically sufficient. This depends on the interaction potential and mainly
on the quality of the trial wave function. The improvement of $\Psi_{\rm T}$ makes
$N_w$ decrease.
%%%%%%%%%%%%%%%%%%%%%%%%%%%%%%%%%%%%%%%
%%%%%%%%%%%%%%%%%%%%%%%%%%%%%%%%%%%%%%%
\section{Trial Wave Functions}
An important part of the VMC and DMC methods is the choice of the
trial wave function. In the VMC technique, the expectation values of all observables
are evaluated with the trial wave function, therefore it determines completely
the accuracy of the results. In the DMC algorithm, the trial function affects the
efficiency of the estimations by increasing or decreasing the
variance.
The DMC technique is based on energy projection, therefore, generally, the energy converges
faster as compared to other quantities. However, the non-diagonal properties can be
more sensitive to the quality of the trial wave function.

The trial wave function $\Psi_{\rm T}(\mathbf{R})$ should be a good approximation
of the ground state of the system. Also, for better computational efficiency
$\Psi_{\rm T}(\mathbf{R})$ and its gradient
and Laplacian should have simple expressions since they are repeatedly evaluated in
the calculation.

The trial wave functions usually used in Quantum Monte Carlo methods are of the
form
\begin{equation}
    \Psi_{\rm T}(\mathbf{R})=\mathcal{F}_1(\mathbf{R})\mathcal{F}_2(\mathbf{R})
   \mathcal{S}(\mathbf{R}).
    \label{Eq:TrialWaveFunction}
\end{equation}
The factor $\mathcal{F}_1(\mathbf{R})$ is constructed as a product of
one-body terms
\begin{equation}
    \mathcal{F}_1(\mathbf{R})=
    \prod_{i=1}^{N}f_1(\mathbf{r_i}).
\end{equation}
The one-body term $f_1(\mathbf{r_i})$ depends only on the position of a single particle
$\mathbf{r_i}$. The choice of
$f_1(\mathbf{r_i})$
is based on the characteristics
of the system under study. In general, the one body functions are taken as
the solution of the one-body problem with an external potential $V_{ext}(\mathbf{r_i})$.

The interparticle correlations are commonly described by a pair-product form
$\mathcal{F}_2(\mathbf{R})$
known as the Bijl-Jastrow term
and it is constructed as a product of two-body terms
\begin{equation}
    \mathcal{F}_2(\mathbf{R})=
    \prod_{j<k}^{N}f_2(|\mathbf{r_j}-\mathbf{r_k}|).
    \label{BijlJastrow}
\end{equation}
The two-body term depends on the distance between a pair of particles
$f_2(|\mathbf{r_j}-\mathbf{r_k}|)$. Typically, for short
distances the two-body function is constructed
as the solution of the two-body problem with
an interaction potential $V_{int}(|\mathbf{r_j}-\mathbf{r_k}|)$.

The factor $\mathcal{S}(\mathbf{R})$ defines the symmetry or antisymmetry of
$\Psi_{\rm T}(\mathbf{R})$ under the exchange of two particles. 

In Chapters ~\ref{Chapter:Few-body bound states of two-dimensional bosons},
~\ref{Chapter:Quantum halo states in two-dimensional dipolar clusters},
and~\ref{Chapter:Two-dimensional dipolar liquid}
we study a mixture
of bosons of types A and B with attractive interspecies AB interactions and 
equally repulsive intraspecies AA and BB interactions. In this case,
we use $\Psi_{\rm S}(\mathbf{R})$ as a trial wave function
\begin{equation}               
\begin{aligned}
\Psi_{\rm S}(\mathbf{R})=&
\prod_{i<j}^{N_\1}f_{\1\1}(|\mathbf{r}_i-\mathbf{r}_j|)
\prod_{\alpha<\beta}^{N_{\2}}f_{\2\2}(|\mathbf{r}_{\alpha}-\mathbf{r}_{\beta}|)\\
&\times\left[\prod_{i=1}^{N_{\1}}\sum_{\alpha=1}^{N_{\2}}f_{\1\2}(|\mathbf{r}_i-\mathbf{r}_{\alpha}|)+
\prod_{\alpha=1}^{N_{\2}}\sum_{i=1}^{N_{\1}}f_{\1\2}(|\mathbf{r}_i-\mathbf{r}_{\alpha}|)\right],               
\label{Eq:SymTrialWaveFunction}
\end{aligned}
\end{equation}      
where $N_\1$ and $N_\2$ are the number of bosons of the species A and B, respectively.
We denote with Latin letters the bosons of the species A and with Greek letters
the bosons of the species B. In Eq.~(\ref{Eq:SymTrialWaveFunction}) 
we have removed the one-body terms since there is no a external potential.
The terms in the first row of Eq.~(\ref{Eq:SymTrialWaveFunction}) are of the
Bijl-Jastrow form and the term in the second row corresponds
to the factor $\mathcal{S}(\mathbf{R})$, in this case it is symmetric because our
system consists of bosons. The advantage of using $\Psi_{\rm S}(\mathbf{R})$
is that it is suitable for describing systems with pairing. In particular,
$\Psi_{\rm S}(\mathbf{R})$ describes well the dimer-dimer problem.

Once the trial wave function has been chosen it is necessary to calculate its gradient
and Laplacian in order to implement the QMC method. The expressions for
the gradient and the Laplacian of $\Psi_{\rm T}(\mathbf{R})$ and $\Psi_{\rm S}(\mathbf{R})$
can be found in the 
Appendices~\ref{Appendix:Jastrow Trial wave function} and~\ref{Appendix:Sym Trial wave function},
respectively.
%%%%%%%%%%%%%%%%%%%%%%%%%%%%%%%%%%%%%%%%%%%%%%%%%%%%%%%%%%%%%%%%%%%%%%%%%%%%%%%%%%%%%%%
%%%%%%%%%%%%%%%%%%%%%%%%%%%%%%%%%%%%%%%%%%%%%%%%%%%%%%%%%%%%%%%%%%%%%%%%%%%%%%%%%%%%%%%
\section{Quantum Monte Carlo Estimators}
The aim of this section is to show how the ground-state properties are computed in the
Monte Carlo algorithm.
\subsection{Pair Distribution Function}
\label{Sec:PairDistribution}
The pair distribution function $g(\mathbf{r}_1,\mathbf{r}_2)$ is proportional
to the probability of finding two particles at the positions $\mathbf{r}_1$
and $\mathbf{r}_2$, simultaneously. In coordinate representation, for a homogeneous system
$g(\mathbf{r}_1,\mathbf{r}_2)$ is given by
\begin{equation}
    g(\mathbf{r}_1,\mathbf{r}_2)=\frac{N(N-1)}{n^2}
    \frac{\int|\Psi(\mathbf{R})|^2d\mathbf{r}_3\cdots d\mathbf{r}_N}
    {\int|\Psi(\mathbf{R})|^2d\mathbf{R}},
\end{equation}
where $n$ is the density of the system. In a homogeneous system the pair distribution
function $g(\mathbf{r}_1,\mathbf{r}_2)$ depends only on the relative distance
$\mathbf{r}=\mathbf{r}_1-\mathbf{r}_2$, with this assumption $g(\mathbf{r}_1,\mathbf{r}_2)$
becomes
\begin{equation}
    g(\mathbf{r})=\frac{N(N-1)}{n^2L^d}
    \frac{\int|\Psi(\mathbf{R})|^2 \delta(\mathbf{r}_{12}-\mathbf{r})  d\mathbf{R}}
    {\int|\Psi(\mathbf{R})|^2d\mathbf{R}},
\end{equation}
where $L$ is the size of the simulation box and $d$ is the dimensionality of the system. To
improve the efficiency of the calculation we sum over all pair of particles
\begin{equation}
    g(\mathbf{r})=\frac{2}{nN}
    \frac{\int|\Psi(\mathbf{R})|^2\sum_{i<j}
        \delta(\mathbf{r}_{ij}-\mathbf{r})  d\mathbf{R}}
    {\int|\Psi(\mathbf{R})|^2d\mathbf{R}},
\end{equation}
where $\mathbf{r}_{ij}=\mathbf{r}_i-\mathbf{r}_j$. In Monte Carlo,
the pair distribution function is determined by making a histogram
of the distance between all pair of particles in the system.
\subsection{One-Body Density Matrix}
For a homogeneous system described by the many-body wave function
$\Psi(\mathbf{r}_1,\cdots,\mathbf{r}_N)$ the one body
density matrix $\rho(\mathbf{r}_1,\mathbf{r}_1^\prime)$ is defined as
\begin{equation}
\rho(\mathbf{r}_1,\mathbf{r}_1^\prime)=N
\frac{\int d\mathbf{r}_2\cdots d\mathbf{r}_N
\Psi^*(\mathbf{r}_1,\mathbf{r}_2,\cdots,\mathbf{r}_N) 
\Psi(\mathbf{r}_1^\prime,\mathbf{r}_2,\cdots,\mathbf{r}_N)}
{\int d\mathbf{r}_1\cdots d\mathbf{r}_N
|\Psi(\mathbf{r}_1,\mathbf{r}_2,\cdots,\mathbf{r}_N) |^2}.
\end{equation}

In the VMC calculations, we sample the trial wave function $\Psi_{\rm T}(\mathbf{R})$. 
Thus, a variational estimation of the one body density matrix is given by
\begin{equation}
\rho(\mathbf{r}_1,\mathbf{r}_1^\prime)=N
\frac{\int d\mathbf{r}_2\cdots d\mathbf{r}_N
    \frac{\Psi_{\rm T}^*(\mathbf{R})}{\Psi_{\rm T}^*(\mathbf{R}^\prime)}
    |\Psi_{\rm T}(\mathbf{R}^\prime) |^2}
{\int d\mathbf{R}|\Psi_{\rm T}(\mathbf{R}) |^2},
\end{equation}
 where $\mathbf{R}=\{\mathbf{r}_1,\mathbf{r}_2,\cdots,\mathbf{r}_N\}$ and
$\mathbf{R}^\prime=\{\mathbf{r}^\prime_1,\mathbf{r}_2,\cdots,\mathbf{r}_N\}$.

Instead, in the DMC method we sample the mixed distribution
$f(\mathbf{R})=\Psi_{\rm T}(\mathbf{R})\Psi(\mathbf{R})$. Thus, a mixed
estimation of the one body density matrix is given by
\begin{equation}
\rho(\mathbf{r}_1,\mathbf{r}_1^\prime)=N
\frac{\int d\mathbf{r}_2\cdots d\mathbf{r}_N
    \frac{\Psi_{\rm T}^*(\mathbf{R})}{\Psi_{\rm T}^*(\mathbf{R}^\prime)}
    f(\mathbf{R}^\prime)}
{\int d\mathbf{R}f(\mathbf{R})}.
\end{equation}

For a homogeneous Bose system, the condensate fraction $N_0/N$ is
obtained from the asymptotic behavior of the one body density matrix 
\begin{equation}
\lim_{|\mathbf{r}-\mathbf{r'}|\to\infty}
\frac{\rho(\mathbf{r}_1,\mathbf{r}_1^\prime)}{n}
=\frac{N_0}{N},
\end{equation}
where $N_0$ is the number of particles in the condensate.
\subsection{Mixed Estimators and Extrapolation Technique}
\label{Extrapolation Technique}
The expectation value of a given observable $\hat{O}$ is obtained
from
\begin{equation}
    \langle\hat{O}\rangle=\frac{\langle\Psi|\hat{O}|\Psi\rangle}
    {\langle\Psi|\Psi\rangle},
    \label{Eq:2.61}
\end{equation}
where $\Psi$ is the wave function of the system.
In a VMC calculation, the expectation values are evaluated with 
the trial wave function $\Psi_{\rm T}$, therefore we obtain a
variational estimator
\begin{equation}
    \langle\hat{O}\rangle_{\rm var}=\frac{\langle\Psi_{\rm T}|\hat{O}|\Psi_{\rm T}\rangle}
    {\langle\Psi_{\rm T}|\Psi_{\rm T}\rangle}.
    \label{Eq:2.62}
\end{equation}
A more precise estimator is obtained from a DMC calculation, where
the expectation values are sample for the
mixed distribution $f=\Psi_{\rm T}\Psi$. After long enough imaginary time
propagation we have $f\approx\Psi_{\rm T}\Phi_0$, and the expectation value is
obtained from 
\begin{equation}
    \langle\hat{O}\rangle_{\rm mix}=\frac{\langle\Psi_{\rm T}|\hat{O}|\Phi_0\rangle}
    {\langle\Psi_{\rm T}|\Phi_0\rangle},
\label{Eq:2.63}
\end{equation}
where $\Phi_0$ is the ground-state wave function. The last equation is known
as the \textit{mixed estimator} since it is calculated over two different
states. The Eq.~(\ref{Eq:2.63}) gives the exact expectation value for the
Hamiltonian (i.e. the calculation of the ground-state energy is exact)
and for observables that commute with it. In the case of
operators that do not commute with $\hat{H}$, the result obtained
from Eq.~(\ref{Eq:2.63}) will be biased by $\Psi_{\rm T}$.
In this case, it is possible to improve the description by employing a 
first order correction in $\Psi_{\rm T}$
using the extrapolation method. In this method, one assumes that the
difference between the trial wave function $\Psi_{\rm T}$ and the ground-state 
wave function $\Phi_0$ is small: $\delta \Psi=\Phi_0-\Psi_{\rm T}$.
The approximated value of the exact estimator with a second-order error
in $\delta\Psi$ can be written in two forms
\begin{equation}
\begin{aligned}
    \langle\hat{O}\rangle_{\rm ext_1}=&2\langle\hat{O}\rangle_{\rm mix}
    -\langle\hat{O}\rangle_{\rm var} + \mathcal{O}(\delta\Psi^2),\\
    \langle\hat{O}\rangle_{\rm ext_2}=&\frac{\langle\hat{O}\rangle_{\rm mix}^2}
    {\langle\hat{O}\rangle_{\rm var}} + \mathcal{O}(\delta\Psi^2).
    \label{Eq:2.64}
\end{aligned}
\end{equation}
The main limitation of using the \textit{extrapolated estimators} $\langle\hat{O}\rangle_{\rm ext_1}$ 
and $\langle\hat{O}\rangle_{\rm ext_2}$ is that they depend on the quality of trial wave function
$\Psi_{\rm T}$ used for importance sampling. 
However, it is usefull to have two different estimators, as if they differ among themselves,
the difference will show the typical difference with the exact result.
%%%%%%%%%%%%%%%%%%%%%%%%%%%%%%%%%%%%%%%%%%%%%%
%%%%%%%%%%%%%%%%%%%%%%%%%%%%%%%%%%%%%%%%%%%%%%
\subsection{Pure Estimators}\label{Pure Estimators}
To overcome the limitations of the extrapolation method, one can use
\textit{forward walking} techniques or similar methods to calculate pure estimators for
local observables that do not commute with the Hamiltonian.
The pure estimator of a local observable $\hat{O}$ is given by
\begin{equation}
    \langle\hat{O}\rangle_{\rm pure}=
    \frac{\langle\Phi_0|\hat{O}|\Phi_0\rangle}
    {\langle\Phi_0|\Phi_0\rangle}.
    \label{Eq:PE}
\end{equation}
The natural outcome of the DMC method is instead a mixed estimator Eq.~(\ref{Eq:2.63}),
which differs from Eq.~(\ref{Eq:PE}) by the presence of the trial wave function
$\Psi_{\rm T}$ on one of the sides. Nevertheless, the pure
estimator can be related to the mixed one by reweighting the observable
with the quotient $\Phi_0/\Psi_{\rm T}$ calculated for the same coordinates
as the local observable
\begin{equation}
    \langle\hat{O}\rangle_{\rm pure}=
    \frac{\langle\Psi_{\rm T}| \frac{\Phi_0}{\Psi_{\rm T}} \hat{O}|\Phi_0\rangle}
    {\langle\Psi_{\rm T}| \frac{\Phi_0}{\Psi_{\rm T}} |\Phi_0\rangle}
    =\left\langle\frac{\Phi_0}{\Psi_{\rm T}}\hat{O}\right \rangle_{\rm mix}.
\label{Eq:PE.1}
\end{equation}
According to Lie \textit{et. al.}~\cite{PhysRevA.10.303} the quotient 
$\Phi_0/\Psi_{\rm T}$ can be computed from the asymptotic number of descendants
of each of the walkers
\begin{equation}
    W(\mathbf{R})=\lim_{\tau\to\infty}n\left(\mathbf{R}(\tau)\right).
    \label{Eq:PE.2}
\end{equation}

Usually, in the Monte Carlo algorithm the local observables are
calculated by taking block averages. Each block consists of $M$ time
steps or iterations. Inside one of these blocks, after one iteration,
when a walker is replicated, we replicate its coordinates and its weight
Eq.~(\ref{Eq:PE.2}), and computed the observables associated with 
it~\cite{CasullerasBoronat1995,doi:10.1063/1.1446847}:
\begin{equation}
    \begin{aligned}   
        O_{k,\alpha}&=\langle\hat{O}
        \left(\mathbf{R}_{k,\alpha}\right)\rangle_{\rm mix},\\
    W_{k,\alpha}&=n\left(\mathbf{R}_{k,\alpha}\right),
\end{aligned}
\end{equation}
where $k$ is the time step index and $\alpha$ is an index
over the number of walkers $N_{w,k}$
After a block is completed, the estimator of the observable
is calculated as
\begin{equation}
    \langle\hat{O}\rangle^{\rm block}=
    \frac{\sum_{k=0}^M\sum_{\alpha=1}^{N_{w,k}} 
        W_{k,\alpha}O_{k,\alpha}}
    {\sum_{k=0}^M\sum_{\alpha=1}^{N_{w,k}} W_{k,\alpha}}.
\end{equation}
After $N_{block}$ blocks, the pure estimator is given by
\begin{equation}
    \langle\hat{O}\rangle_{\rm pure}=
    \frac{1}{N_{\rm block}}\sum_{j=1}^{N_{\rm block}}
    \langle\hat{O}\rangle^j.
\end{equation}
The pure estimator depends on the size $M$ of a block. $M$
has to be large enough to reach the asymptotic regime given
by Eq.~(\ref{Eq:PE.2}).

Although the calculation of the pure estimators for the potential
energy, density profile, pair distribution function, static structure
factor, and other correlation functions are routinely done to our
best knowledge, the calculation of pure coordinates never has been done.
We found it convenient to store the coordinates of the walker and
replicate them during the branching process. After long enough
propagation time, the pure coordinates are stored and at the end of
the simulation, we have a large number of pure coordinates. At this
point an average of a local observable over them automatically becomes
pure. In particular, we find this trick to be very flexible and
especially useful for the pure estimation of all sorts of complicated
correlation functions common for few-body analysis and often involving
calculations of hyperradius. For example, Fig.~\ref{Fig:SpatialDistributions},
Fig.~\ref{Fig:ABHaloStates} and Fig.~\ref{Fig:HaloStates},
were obtained by this method.

\chapter{One-Dimensional Three-Boson Problem with Two- and Three-Body Interactions}
\label{Chapter:One-dimensional three-boson problem with two- and three-body interactions}
In this chapter, we study the three-boson problem with contact two- and three-body
interactions in one dimension.
By using the diffusion Monte
Carlo technique we calculate the binding energy of two and three dimers
formed in a
Bose-Bose or Fermi-Bose mixture with attractive
interspecies and repulsive intraspecies interactions. Combining these
results with a three-body theory~\cite{Guijarro2018},
we extract the three-dimer
scattering length close to the dimer-dimer zero crossing. In both
considered cases the three-dimer interaction turns out to be
repulsive. Our results constitute a concrete proposal for obtaining
a one-dimensional gas with a pure three-body repulsion.
%%%%%%%%%%%%%%%%%%%%%%%%%%%%%%%%%%%%%%%%%%%%%%%%%%%%%%%%%%%%%%%%
\section{Introduction}
The one-dimensional $N$-boson problem with the two-body contact
interaction $g_2\delta(x)$ is exactly solvable. Lieb and 
Liniger~\cite{LiebLiniger} have shown that for $g_2>0$ the system is in the
gas phase with positive compressibility. McGuire~\cite{McGuireBosons}
has demonstrated that for $g_2<0$ the ground state is a soliton with
the chemical potential diverging with $N$. In the case $N=\infty$ the
limits $g_2\rightarrow +0$ and $g_2\rightarrow -0$ are manifestly
different: The former corresponds to an ideal gas whereas the latter
corresponds to collapse. Accordingly, the behavior of a realistic one-
or quasi-one-dimensional system close to the two-body zero crossing
strongly depends on higher-order terms not included in
the Lieb-Liniger or McGuire zero-range models. Sekino and 
Nishida~\cite{Nishida} have considered one-dimensional bosons with a pure
zero-range three-body attraction and found that the ground state of
the system is a droplet with the binding energy exponentially
increasing with $N$, which also means collapse in the thermodynamic limit. 
In Ref.~\cite{PricoupenkoPetrov}, the authors have argued that in a
sufficiently dilute regime the three-body interaction is effectively
repulsive, providing a mechanical stabilization against collapse for
$g_2<0$. The competition between the two-body attraction and
three-body repulsion leads to a dilute liquid state similar to the
one discussed by Bulgac~\cite{Bulgac} in three dimensions.

The three-body scattering in one dimension is kinematically equivalent
to a two-dimensional two-body scattering~\cite{Petrov3body,Nishida}.
Therefore, the corresponding interaction shift depends logarithmically
on the product of the scattering momentum and three-body scattering
length $a_3$. An important consequence of this fact is that, in
contrast to higher dimensions, the one-dimensional three-body
interaction can become noticeable even if $a_3$ is exponentially small
compared to the mean interparticle distance. Therefore, three-body
effects can be studied in the universal dilute regime essentially in
any one-dimensional system that preserves a finite residual three-body
interaction close to a two-body zero crossing. Universality means that
the effective-range effects are exponentially small and the relevant
interaction parameters are the two- and three-body scattering lengths
$a_2$ and $a_3$, respectively.

In this chapter, we consider a
two-component Bose-Bose mixture with attractive interspecies and
repulsive intraspecies interactions. In this system, the interspecies
attraction binds atoms into dimers while the dimer-dimer interaction
is tunable by changing the intraspecies 
repulsion~\cite{PricoupenkoPetrov}. 
Analytical predictions~\cite{Guijarro2018} are complemented
by diffusion Monte Carlo calculations of the hexamer energy, permitting
to determine the three-dimer scattering length close to the
dimer-dimer zero crossing. We perform this procedure for equal
intraspecies coupling constants and in the case where their ratio is
infinite. In the latter limit, one of the components is in the
Tonks-Girardeau regime and the system is equivalent to a Fermi-Bose
mixture. We find that the three-dimer interaction is repulsive in both
cases.
%%%%%%%%%%%%%%%%%%%%%%%%%%%%%%%%%%%%%%%%%%%%%%%%%%%%%%%%%%%%%%%%
\section{The System}
In Ref.~\cite{Guijarro2018}, the authors considered a one-dimensional system of
three bosons of mass $m$ interacting via contact two- and
three-body forces characterized by the scattering lengths $a_2$ and
$a_3$, respectively.
They obtained the following analytical expression for the ground and
excited trimer energies
\begin{equation}\label{Relation}
\ln\frac{a_3\kappa
e^\gamma}{a_2}=\frac{2}{\kappa^2-1}\left[\frac{\pi}{3\sqrt{3}}+\frac{3\kappa^2-1}{\sqrt{4\kappa^2-1}}\arctan\sqrt{\frac{2\kappa+1}{2\kappa-1}}\right],
\end{equation}
where $\kappa=\sqrt{-mE}a_2/(2\hbar)$ and $\gamma=0.577$ is Euler's constant.
The Eq.~(\ref{Relation}) relates the trimer binding energy $E=E_3<0$
with $a_2$ and $a_3$. Considering the dimer binding energy as $|E_2|=\hbar^2/ma_2^2$
we obtain the following relation $E_3/E_2=4\kappa^2$.
In the following we are going to use Eq.~(\ref{Relation}) to extract the three-dimer
scattering length close to the dimer-dimer zero crossing. 

Systems where two- and three-body effective interactions can be
controlled independently are difficult to produce or engineer
(see~\cite{Petrov3body} and references therein). We now discuss a model
tunable to the regime of pure three-body repulsion. Namely, we consider
a mixture of one-dimensional pointlike bosons $\1$ and $\2$ of unit
mass characterized by the coupling constants
\begin{equation}
g_{\1\2}=-\frac{2\hbar^2}{ma_{\1\2}}<0,
\end{equation}
for the interspecies attraction and 
\begin{equation}
g_{\sigma\sigma}=-\frac{2\hbar^2}{ma_{\sigma\sigma}}>0,
\end{equation}
for the intraspecies repulsions. 
The interspecies attraction leads to the formation
of $\1\2$ dimers of size $a_{\1\2}$ and energy 
\begin{equation}
E_{\1\2}=-\frac{\hbar^2}{ma_{\1\2}^2}.
\end{equation}

One can show~\cite{PricoupenkoPetrov} that the two-dimer interaction changes
from attractive to repulsive with increasing $g_{\sigma\sigma}$. In
particular, the two-dimer zero crossing is predicted to take place
for the Bose-Bose (BB) case at
\begin{equation}
g_{\1\1}=g_{\2\2}=2.2|g_{\1\2}|,
\label{Eq:3.5}
\end{equation}
and fo the Fermi-Bose (FB) case at
\begin{equation}
g_{\1\1}=0.575|g_{\1\2}|,
\label{Eq:3.6}
\end{equation}
if $g_{\2\2}=\infty$.

Here, we consider three such dimers and characterize their three-dimer interaction
by calculating the hexamer energy $E_{\1\1\1\2\2\2}$ and by comparing it with
the tetramer energy $E_{\1\1\2\2}$ on the attractive side of the two-dimer
zero crossing where the tetramer exists. The idea is that sufficiently close
to this crossing the dimers behave as pointlike particles weakly bound to
each other. One can then extract the three-dimer scattering length $a_3$
from the zero-range three-boson formalism [Eq.~(\ref{Relation})], 
with $m\to2m$
\begin{equation}
\begin{aligned}
E_2&=E_{\1\1\2\2}-2E_{\1\2},\\
E_3&=E_{\1\1\1\2\2\2}-3E_{\1\2},
\end{aligned}
\end{equation}
and using the asymptotic expression for the dimer-dimer scattering length
\begin{equation}
    E_2=-\frac{\hbar^2}{2ma_2^2},
\end{equation}
we obtain 
\begin{equation}
    a_2=\frac{\hbar}{\sqrt{2m|E_2|}}.
\end{equation}
%%%%%%%%%%%%%%%%%%%%%%%%%%%%%%%%%%%%%%%%%%%%%%%%%%%%%%%%%%%%%%%%
\section{Details of the Methods}
In order to calculate $E_2$ and $E_3$, we resort to the diffusion Monte
Carlo (DMC) technique,, which was explained in
Chapter~\ref{Chapter:Quantum Monte Carlo methods}.
The importance sampling
is used to reduce the statistical noise and also to impose the Bethe-Peierls
boundary conditions stemming from the $\delta$-function interactions. We
construct the guiding wave function $\psi_{\rm T}$ in the pair-product form
\begin{equation}\label{trial wf}
    \psi_{\rm T}
=
\prod\limits_{i<j} f^{\1\1}(x_{ij}^{\1\1})
\prod\limits_{i<j} f^{\2\2}(x_{ij}^{\2\2})
\prod\limits_{i,j}f^{\1\2}(x_{ij}^{\1\2})\;,
\end{equation}
where $x_{ij}^{\sigma\sigma'} = x_i^\sigma-x_j^{\sigma'}$ is the distance
between particles $i$ and $j$ of components $\sigma$ and $\sigma'$,
respectively. The intercomponent correlations are governed by the dimer
wave function 
\begin{equation}
    f^{\1\2}(x) = \exp  \left(-\frac{|x|}{a_{\1\2}}\right) ,
\end{equation}
and the intracomponent terms are 
\begin{equation}
f^{\sigma\sigma}(x) = 
\sinh\left(\frac{|x|}{a_{\1\2}}-\frac{|x|}{2a_{\rm dd}}\right)
-\left(\frac{a_{\sigma\sigma}}{a_{\1\2}}-\frac{a_{\sigma\sigma}}
    {2a_{\rm dd}}\right).
\end{equation}
These functions satisfy the Bethe-Peierls boundary
conditions, 
\begin{equation}
    \frac{\partial f^{\sigma\sigma'}(x)}{\partial x}
    |_{x=+0}=-\frac{f^{\sigma\sigma'}(0)}{a_{\sigma\sigma'}},
\end{equation}
which, because of the
product form, also ensures the correct behavior of the total guiding function
$\psi_{\rm T}$ at any two-body coincidence.
At the same time, the long-distance behavior of $f^{\sigma\sigma}(x)$
is chosen such that $\psi_{\rm T}$ allows dimers to be at distances larger
than their size. When the distance $x$ between pairs $\{x_1^\1,
x_1^\2\}$ and $\{x_2^\1, x_2^\2\}$ is much larger than the dimer
size $a_{\1\2}$, Eq.~(\ref{trial wf}) reduces to 
\begin{equation}
    \psi_{\rm T}\propto
f^{\1\2}(x_{11}^{\1\2})f^{\1\2}(x_{22}^{\1\2})
\exp\left(-\frac{|x|}{a_{\rm dd}}\right).
\label{Eq:longerange}
\end{equation}
For $a_{\rm dd}\gg a_{\1\2}$, this wave function describes two dimers
weakly-bound to each other.
It can be noted that the choice of the same spin Jastrow terms might seem
rather unusual as $f^{\sigma\sigma}$ become exponentially large at large
distances due to divergence of $\sinh(x)$ function. 
This divergence is cured by multiplication of exponentially small
opposite-spin Jatsrow terms $f^{\1\2}(x)$. 
Thus, the use of the $\sinh(x)$ function allows both to impose the
physically correct long-range properties, Eq.~(\ref{Eq:longerange})
and the correct Bethe-Peierls boundary condition at short distances
where the expansion $\sinh(x)\to x$ results in 
$f^{\sigma\sigma} \propto |x| - a^{\sigma\sigma}$.

While $a_{\sigma\sigma'}$ are fixed by the Hamiltonian, we
treat $a_{\rm dd}$ as a free parameter in Eq.~(\ref{trial wf}). Close to
the dimer-dimer zero crossing $a_{\rm dd}\approx a_2$ and this parameter is
related self-consistently to the tetramer energy while far from the crossing
its value is optimized according to the variational principle. It is useful
to mention that in case FB, where $a_{\2\2}=0$, the $\2$ component is in the
Tonks-Girardeau limit and can be mapped to ideal fermions by Girardeau's
mapping~\cite{Girardeau60}. Replacing $|x|$ by $x$ in the definition of
$f^{\2\2}(x)$ makes $\psi_{\rm T}$ antisymmetric with respect to permutations of
$\2$ coordinates.
%%%%%%%%%%%%%%%%%%%%%%%%%%%%%%%%%%%%%%%%%%%%%%%%%%%%%%%%%%%%%%%%
\section{Results}
\subsection{Binding Energies}
\begin{figure}[ht]
  \centering
  \subfigure{\includegraphics[width=0.484\textwidth]{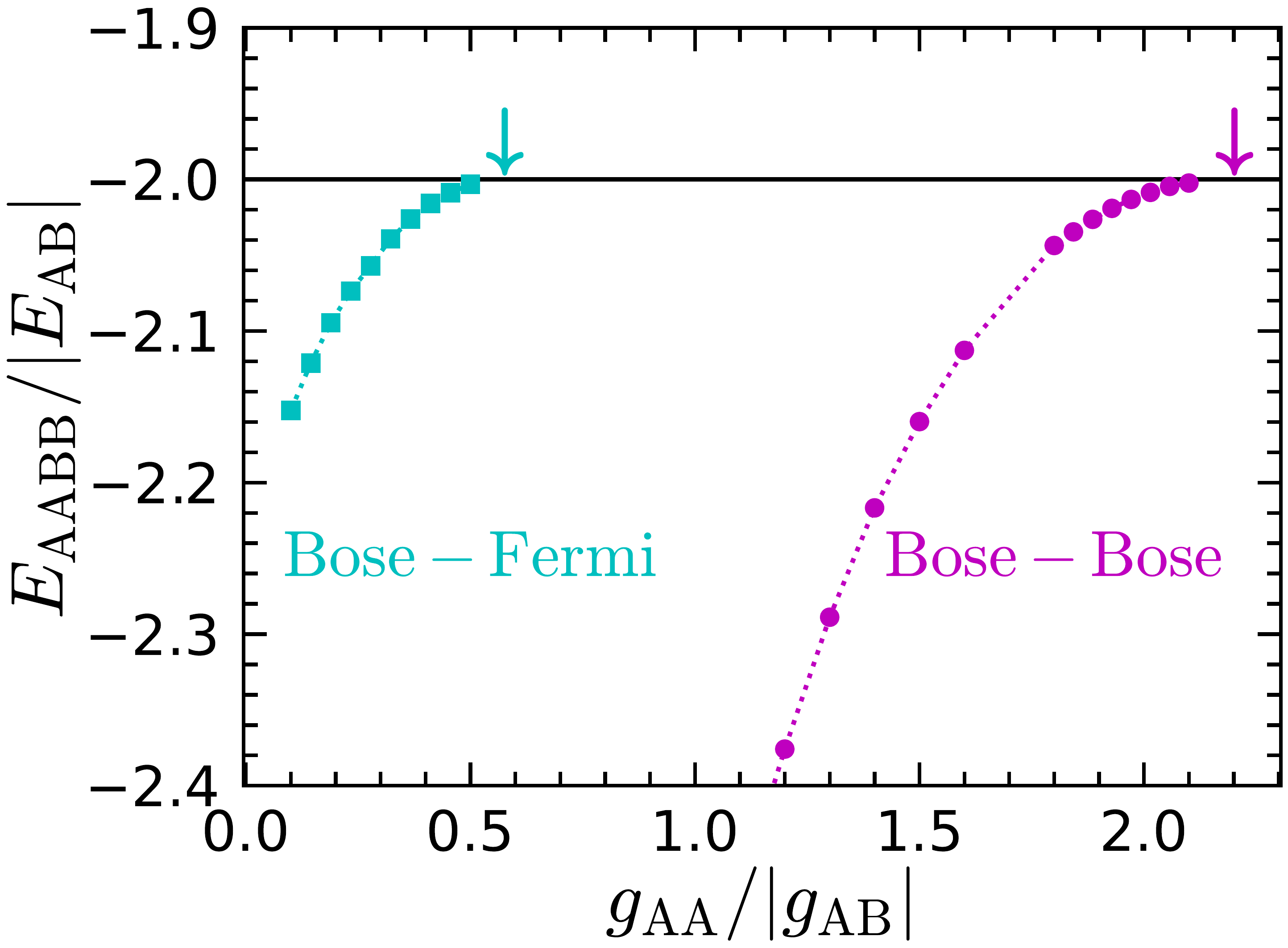}}\quad
  \subfigure{\includegraphics[width=0.484\textwidth]{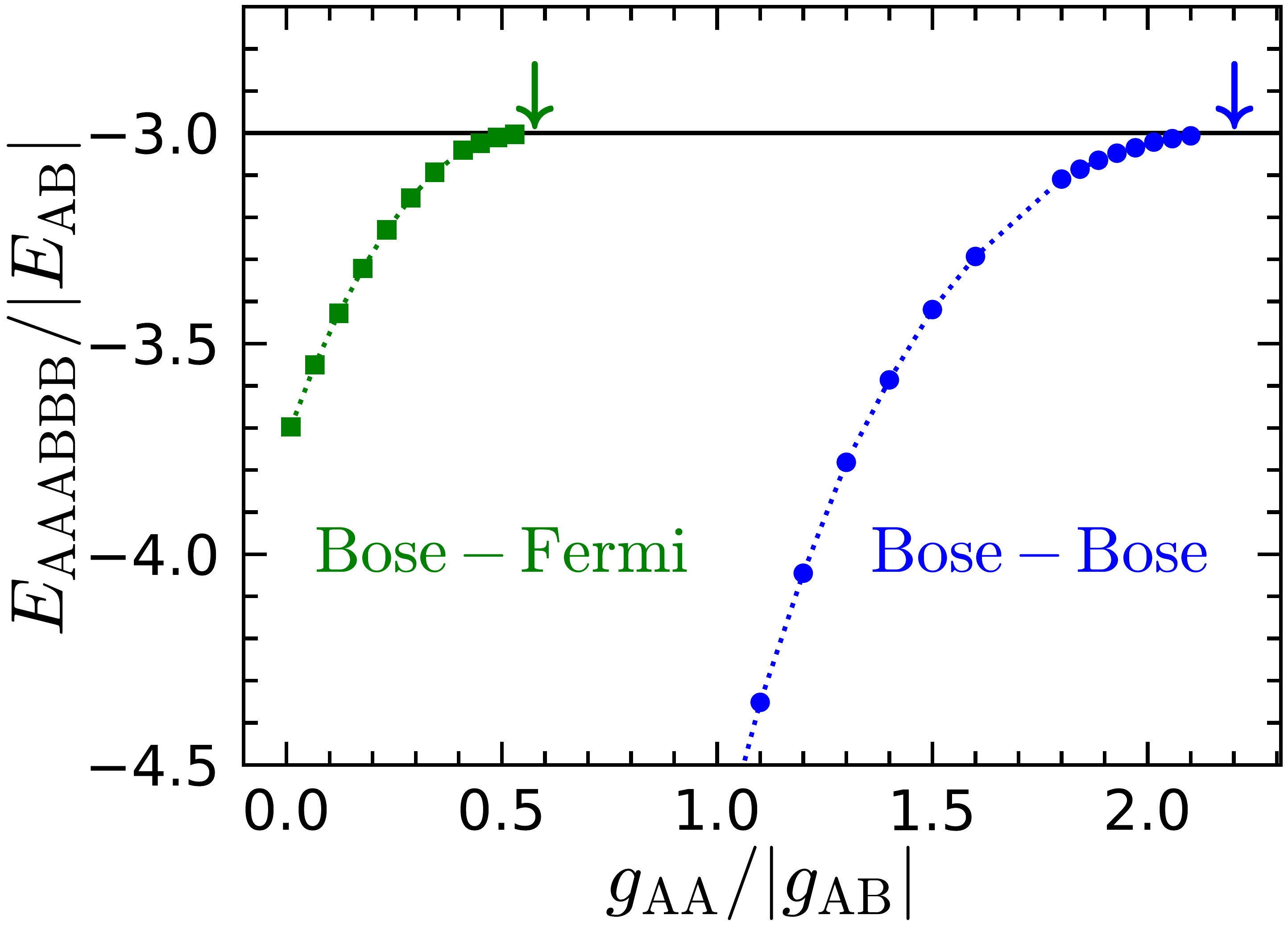}}
  \caption{Tetramer $E_{\1\1\2\2}$ (left) and
      hexamer $E_{\1\1\1\2\2\2}$ (right) energies in units of the dimer
        energy $|E_{\1\2}|$ for Bose-Fermi and Bose-Bose mixtures
        as a function of the ratio $g_{\1\1}/|g_{\1\2}|$.
    The arrows show the positions of the thresholds for binding.}
\label{Fig:Ch3_Energies}
\end{figure}

In Fig.~\ref{Fig:Ch3_Energies} (left) we show the
tetramer energy $E_{\1\1\2\2}$ as a function of the ratio
$g_{\1\1}/|g_{\1\2}|$ for the Bose-Fermi and Bose-Bose mixtures.
The thresholds for binding are shown by arrows. 

In Fig.~\ref{Fig:Ch3_Energies} (right) we show the
hexamer energy $E_{\1\1\1\2\2\2}$ as a function of the ratio
$g_{\1\1}/|g_{\1\2}|$ for the Bose-Fermi and Bose-Bose mixtures.
We find that for sufficiently strong intercomponent
repulsion (larger $g_{\1\1}/|g_{\1\2}|$) the hexamer gets unbound,
first for the Bose-Fermi case and then for the Bose-Bose mixture.
%%%%%%%%%%%%%%%%%%%%%%%%%%%%%%%%%%%%%%%%%%%%
\subsection{Threshold Determination}
\begin{figure}
  \centering
  \subfigure{\includegraphics[width=0.484\textwidth]{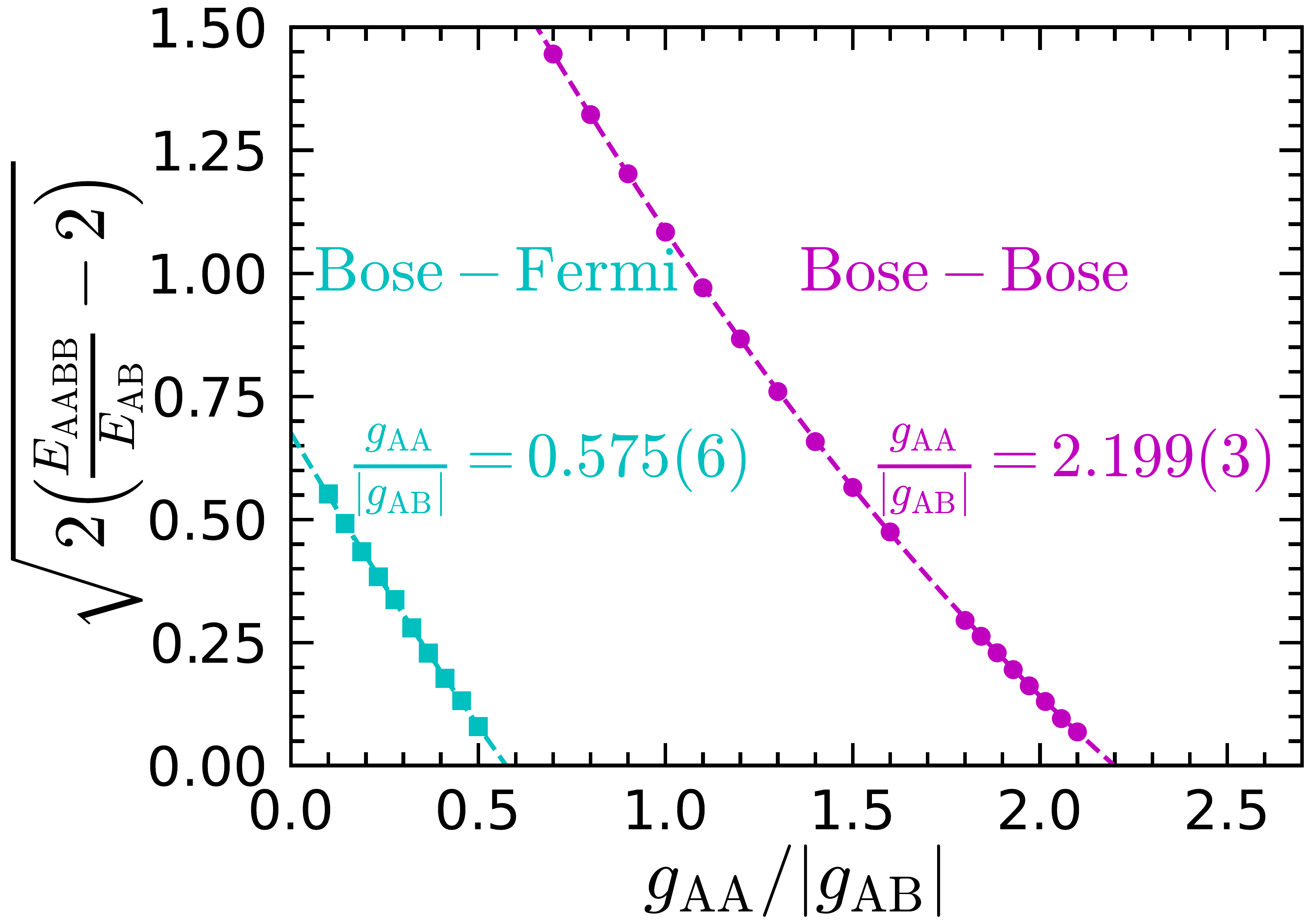}}\quad
  \subfigure{\includegraphics[width=0.484\textwidth]{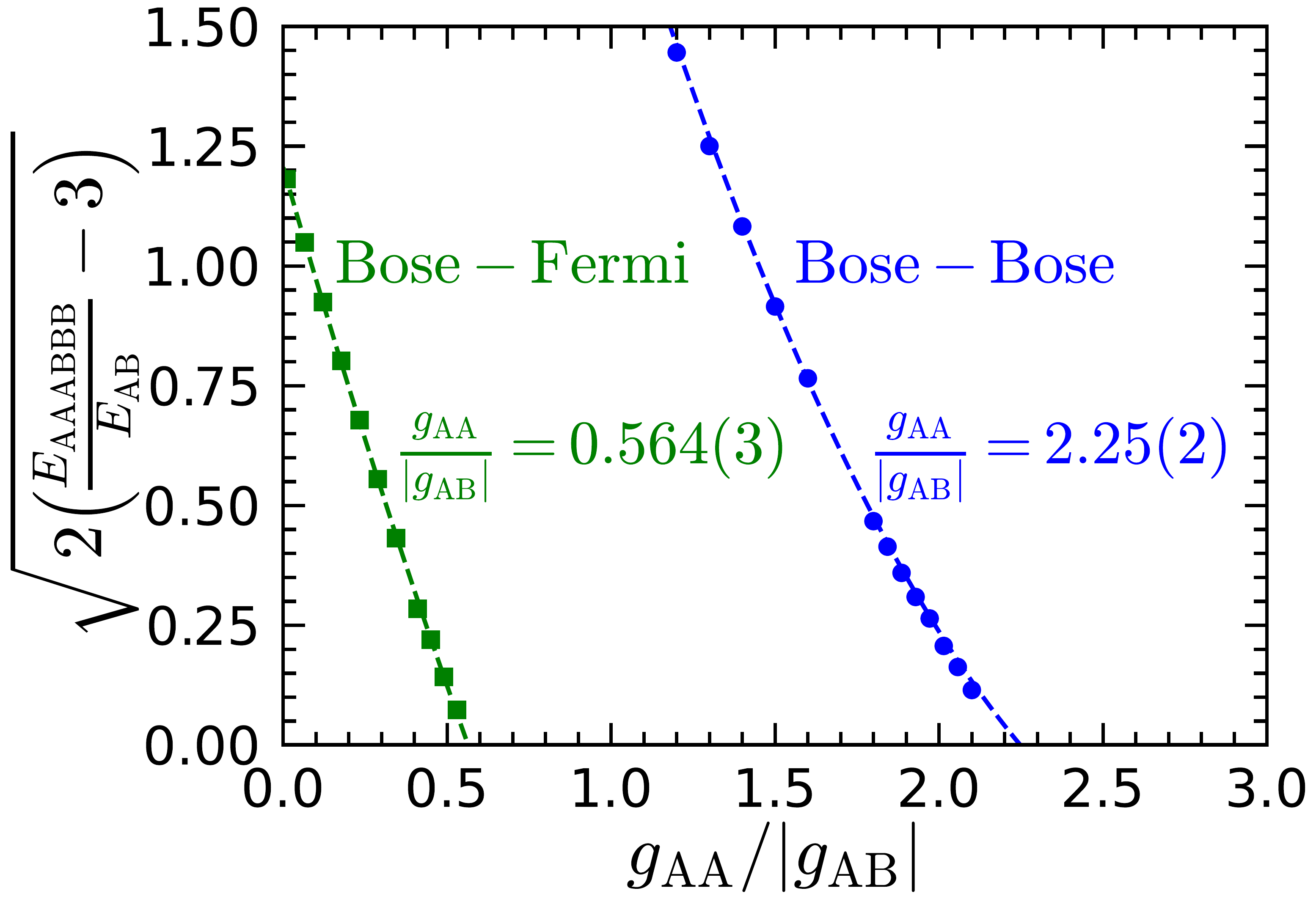}}
  \caption{Fitting procedure for the determination of the threshold
      values of the AABB tetramer and AABBB hexamer for Bose-Fermi and Bose-Bose mixtures.
 The dashed curves
 correspond to the fit $c_1\left(\frac{g_\mathrm{AA}}
     {|g_\mathrm{AB}|}-\frac{g_\mathrm{AA}^c}{|g_\mathrm{AB}^c|}\right)+
 c_2\left(\frac{g_\mathrm{AA}}{|g_\mathrm{AB}|}
     -\frac{g_\mathrm{AA}^c}{|g_\mathrm{AB}^c|}\right)^2$.}
    \label{Fig:Ch3_Threshold}
\end{figure}

In Fig.~\ref{Fig:Ch3_Threshold} we show the numerical
threshold determination for the tetramer and hexamer for the Bose-Fermi and
Bose-Bose mixtures.
Our numerical results for the tetramer threshold values
are consistent with the predictions of Ref.~\cite{PricoupenkoPetrov}
(Eq.~\ref{Eq:3.5} and Eq.~\ref{Eq:3.6}).
We find that the hexamer threshold for the Bose-Fermi mixture
is located at $g_{\1\1}/|g_{\1\2}|\approx 0.575$.
In the case of the Bose-Bose mixture, the hexamer threshold
occurs at $g_{\1\1}/|g_{\1\2}|\approx 2.2$. In both cases, the hexamer
thresholds coincide with the tetramer thresholds.
%%%%%%%%%%%%%%%%%%%%%%%%%%%%%%%%%%%%%%%%%%%%%%%%%%%%%%%%%%%%%%%%
\subsection{Three-Dimer Repulsion}
\begin{figure}[ht]
    \centering
    \vskip 0 pt \includegraphics[clip,width=0.7\columnwidth]{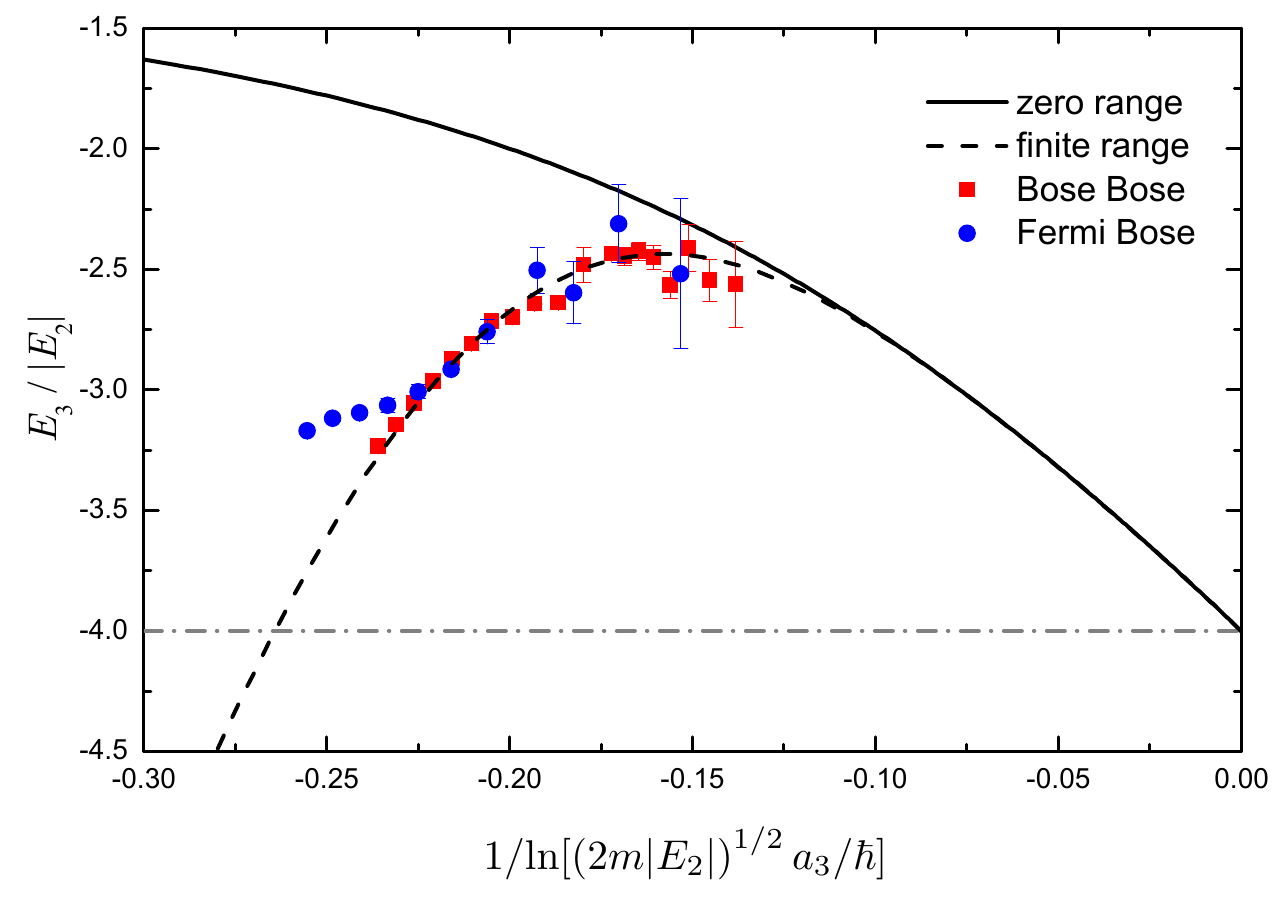}
\caption{
$E_3/|E_2|$ vs $1 / \ln(\sqrt{2m|E_2|}a_3/\hbar)$
for one-dimensional dimers. Here
$E_2$ and $E_3$ are the tetramer and hexamer energies measured relative
to the two- and three-dimer thresholds, respectively. The solid curve is
the prediction of Eq.~(\ref{Relation}) and the dashed curve is a fit, which
includes finite-dimer-size effects into account. The dash-dotted
line is the McGuire result $E_3=4E_2$ for three pointlike bosons with no
three-body interaction. The red squares are the DMC data for case BB plotted
using $a_3 = 0.01 a_{\1\2}$ and the blue circles stand for case FB with $a_3
= 0.03 a_{\1\2}$. The error bars are larger in the latter case because of
the larger statistical noise induced by the nodal surface imposed by the
Fermi statistics.}
\label{Fig:DMC}
\end{figure}

In Fig.~\ref{Fig:DMC}, we show $E_3/|E_2|$ for cases BB (red squares) FB
(blue circles) as a function of 
$\delta=1/\ln(\sqrt{2m|E_2|}a_3/\hbar)$ 
along with
the prediction of Eq.~(\ref{Relation}) (solid black). The quantity $a_3$
is a fitting parameter to the DMC results; changing it essentially shifts
the data horizontally. We clearly see that in both cases the three-dimer
interaction is repulsive since $E_3/|E_2|$ is above the McGuire trimer 
limit~\cite{McGuireBosons} (dash-dotted line). For rightmost data points the hexamer
is about ten times larger than the dimer and the data align with the universal
zero-range analytics. For the other points we observe significant effective
range effects related to the finite size of the dimer. In the universal
limit $a_{\1\2}\ll a_2$, the leading effective-range correction to the
ratio $E_3/|E_2|$ is expected to be proportional to $a_{\1\2}/a_2\propto
e^{1/\delta}$~\cite{PricoupenkoPetrov}. Indeed, adding the term $C
e^{1/\delta}$ to the zero-range prediction well explains deviations of our
results from the universal curve and we have checked that other exponents
do not work that well. We thus treat $a_3$ and $C$ as fitting parameters; in
case BB we obtain $a_3=0.01a_{\1\2}$ and in case FB $a_3=0.03 a_{\1\2}$. Both
cases are fit with $C=-100$ (dashed curve in Fig.~\ref{Fig:DMC}). We emphasize
that we are dealing with the true ground state of three dimers. The lower
``attractive'' state formally existing for these values of $a_2$ and
$a_3$ in the zero-range model is an artifact since it does not satisfy
the zero-range applicability condition. The three-dimer interaction is an
effective finite-range repulsion which supports no bound states.
%%%%%%%%%%%%%%%%%%%%%%%%%%%%%%%%%%%%%%%%%%%%%%%%%%%%%%%%%%%%%%%%
\section{Summary}
In conclusion, 
we argue that since in one dimension the three-body
energy correction scales logarithmically with the three-body scattering length
$a_3$, three-body effects are observable even for exponentially small $a_3$,
which significantly simplifies the task of engineering three-body-interacting
systems in one dimension. We demonstrate that Bose-Bose or Fermi-Bose dimers,
previously shown to be tunable to the dimer-dimer zero crossing, exhibit
a noticeable three-dimer repulsion. We can now be certain that the ground
state of many such dimers slightly below the dimer-dimer zero crossing is
a liquid in which the two-body attraction is compensated by the three-body
repulsion~\cite{Bulgac,PricoupenkoPetrov}.

Our results have implications for quasi-one-dimensional mixtures. We
mention particularly the $^{40}$K-$^{41}$K Fermi-Bose mixture which
emerges as a suitable candidate for exploring the liquid state of fermionic
dimers. Here the intraspecies $^{41}$K-$^{41}$K background interaction is
weakly repulsive (the triplet $^{41}$K-$^{41}$K scattering length equals
3.2nm~\cite{Falke}) and the interspecies one features a wide Feshbach
resonance at 540G~\cite{Zwierlein}. Let us identify $\1$ with $^{40}$K, $\2$
with $^{41}$K, and assume the radial oscillator length $l_0=56$nm, which
corresponds to the confinement frequency $2\pi\times 80$kHz. Under these
conditions the effective coupling constants equal $g_{\sigma\sigma'}\approx
2a_{\sigma\sigma'}^{\rm (3D)}/l_0^2$~\cite{Olshanii} and the dimer-dimer zero
crossing at $g_{\1\1}=0.575|g_{\1\2}|$ is realized for the three-dimensional
scattering lengths $a_{\2\2}^{\rm (3D)}\approx 3.2$nm and $a_{\1\2}^{\rm
(3D)}\approx-5.6$nm. The dimer size is then $\approx 560$nm and dimer binding
energy corresponds to $\approx 2\pi \times 800$Hz placing the system in the
one-dimensional regime. For the rightmost (next to rightmost) blue circle
in Fig.~\ref{Fig:DMC}, the tetramer is approximately 20 (10) times larger
than the dimer and 800 (200) times less bound. Moving left in this figure is
realized by increasing $|a_{\1\2}^{\rm 3D}|$ and thus getting deeper in the
region $g_{\1\1}<0.575|g_{\1\2}|$. Note, however, that this also pushes the
system out of the one-dimensional regime and effects of transversal 
modes~\cite{Gora,Kovrizhin,Mazets} become important.

\chapter{Few-Body Bound States of Two-Dimensional Bosons}
\label{Chapter:Few-body bound states of two-dimensional bosons}
In this chapter, we study clusters of the type A$_N$B$_M$ with $N\leq M\leq 3$
in a two-dimensional mixture of A and B bosons, with attractive AB and equally
repulsive AA and BB interactions. In order to check universal aspects of the
problem, we choose
two very different models: dipolar bosons in a bilayer
geometry (this work) and particles interacting via separable Gaussian
potentials (reported in Ref.~\cite{PhysRevA.101.041602}). 
We find that all the considered clusters are bound and that their energies are
universal functions of the scattering lengths 
$a_{\1\2}$ and $a_{\1\1}=a_{\2\2}$, for sufficiently large
attraction-to-repulsion ratios $a_{\1\2}/a_{\2\2}$. When $a_{\1\2}/a_{\2\2}$
decreases below $\approx 10$, the dimer-dimer interaction changes from
attractive to repulsive and the population-balanced AABB and AAABBB clusters
break into AB dimers. Calculating the AAABBB hexamer energy just below this
threshold, we find an effective three-dimer repulsion which may have important
implications for the many-body problem, particularly for observing liquid and
supersolid states of dipolar dimers in the bilayer geometry. The
population-imbalanced ABB trimer, ABBB tetramer, and AABBB pentamer remain
bound beyond the dimer-dimer threshold. In the dipolar model, they break up at
$a_{\1\2}\approx 2 a_{\2\2}$ where the atom-dimer interaction switches to
repulsion.
The work presented in this chapter was a 
collaboration~\cite{PhysRevA.101.041602}.
I did the calculations for the dipolar clusters.
%%%%%%%%%%%%%%%%%%%%%%%%%%%%%%%%%%%%%%%%%%%%%%%%%%%%%%%%%%%%%%%%
\section{Introduction}
Recent experiments on dilute quantum droplets in dipolar
bosonic gases~\cite{Kadau2016,Schmitt2016,Ferrier2016,Chomaz2016} and in
Bose-Bose mixtures~\cite{Cabrera2017,Semeghini2018,Ferioli2019} with
competing interactions have exposed the important role of
beyond-mean-field effects in weakly-interacting systems. A natural
strategy to boost these effects and enhance exotic behaviors is to
make the interactions stronger while keeping the attraction-repulsion
balance for mechanical stability. The most straightforward way of
getting into this regime is to increase the gas parameter $na_s^3$.
However, this leads to enhanced three-body losses which results in
very short lifetimes (as it has been observed in
experiments~\cite{Kadau2016,Schmitt2016,Ferrier2016,Chomaz2016,
Cabrera2017,Semeghini2018,Ferioli2019}).
Nevertheless, this regime is achievable in reduced geometries. It has
been shown that a one-dimensional Bose-Bose mixture with
strongly-attractive interspecies interaction becomes dimerized and,
by increasing the intraspecies repulsion, the dimer-dimer interaction
can be tuned from attractive to repulsive~\cite{Pricoupenko2018}.
Then, an effective three-dimer repulsion has been found in this
system and predicted to stabilize a liquid phase of attractive
dimers~\cite{Guijarro2018}. 

In two dimensions, a particularly interesting realization of such a
strongly-interacting, tunable, and long-lived Bose-Bose mixture is a
system of dipolar bosons confined to a bilayer 
geometry~\cite{WangLukinDemler2006,Wang2007,Trefzger_2011}.
When the dipoles are oriented perpendicularly to the plane, there is
a competing effect between repulsive intralayer and partially
attractive interlayer interactions, interesting from the viewpoint of
liquid formation. 
In addition, the quasi-long range character of the dipolar interaction
can produce the rotonization of its spectrum and a supersolid 
behavior~\cite{ODell2003,Santos2003,Lu2015,Tanzi2019,Chomaz2019,
Bottcher2019,Tanzi2019Modes,Guo2019,Ferlaino2019},
formation of a crystal phase~\cite{Buchler2007,Astrakharchik2007},
and a pair superfluid~\cite{Macia2014,PhysRevA.94.063630,Nespolo_2017}
(see also lattice calculations of Ref.~\cite{Safavi2017}). A peculiar
feature of bilayer model is the vanishing Born integral for the
interlayer interaction~\cite{bitube},
\begin{equation}
\int V_{\1\2}(\rho)d^2\rho = 0,
\label{Eq:4.1}
\end{equation}
which has led to controversial claims about the existence of a
two-body bound state~\cite{Yudson1997} till it has finally been
established that this bound state always exists, although its energy
can be exponentially small~\cite{Shih2009,Armstrong2010,Klawunn2010,
Baranov2011,Volosniev2011},
consistently with Ref.~\cite{Simon1976}. Interestingly, a similar
controversy seems to continue at the few-body level; it has been
claimed~\cite{Volosniev2012} that the repulsive dipolar tails will
never allow for three- or four-body bound states in this geometry.

In this chapter, we investigate few-body bound states in a 
two-dimensional mass-balanced mixture of A and B bosons with two types of
interactions characterized by the two-dimensional scattering lengths
$a_{\rm AB}$ and $a_{\rm AA}=a_{\rm BB}$. The first case corresponds to the
bilayer of dipoles discussed above and, in the second, the interactions
were modeled by non-local (separable) finite-range Gaussian 
potentials~\cite{PhysRevA.101.041602}. By
using the diffusion Monte Carlo (DMC) technique in the first case, and the
Stochastic Variational Method (SVM) in the second, we find that for
sufficiently weak BB repulsion compared to the AB attraction,
$a_{\rm AB}\gg a_{\rm BB}$, all clusters of the type A$_N$B$_M$ with
$1 \leq N \leq M \leq 3$ are bound. We then locate thresholds for their
unbinding with decreasing $a_{\1\2}/a_{\2\2}$. By looking at the AAABBB hexamer
energy close to the corresponding threshold, we discover an effective
three-dimer repulsion, which can stabilize interesting many-body phases.
%%%%%%%%%%%%%%%%%%%%%%%%%%%%%%%%%%%%%%%%%%%%%%%%%%%%%%%%%%%%%%%%%
\section{The Hamiltonian}
\begin{figure}[b!]
	\centering
    \includegraphics[width=0.7\textwidth]{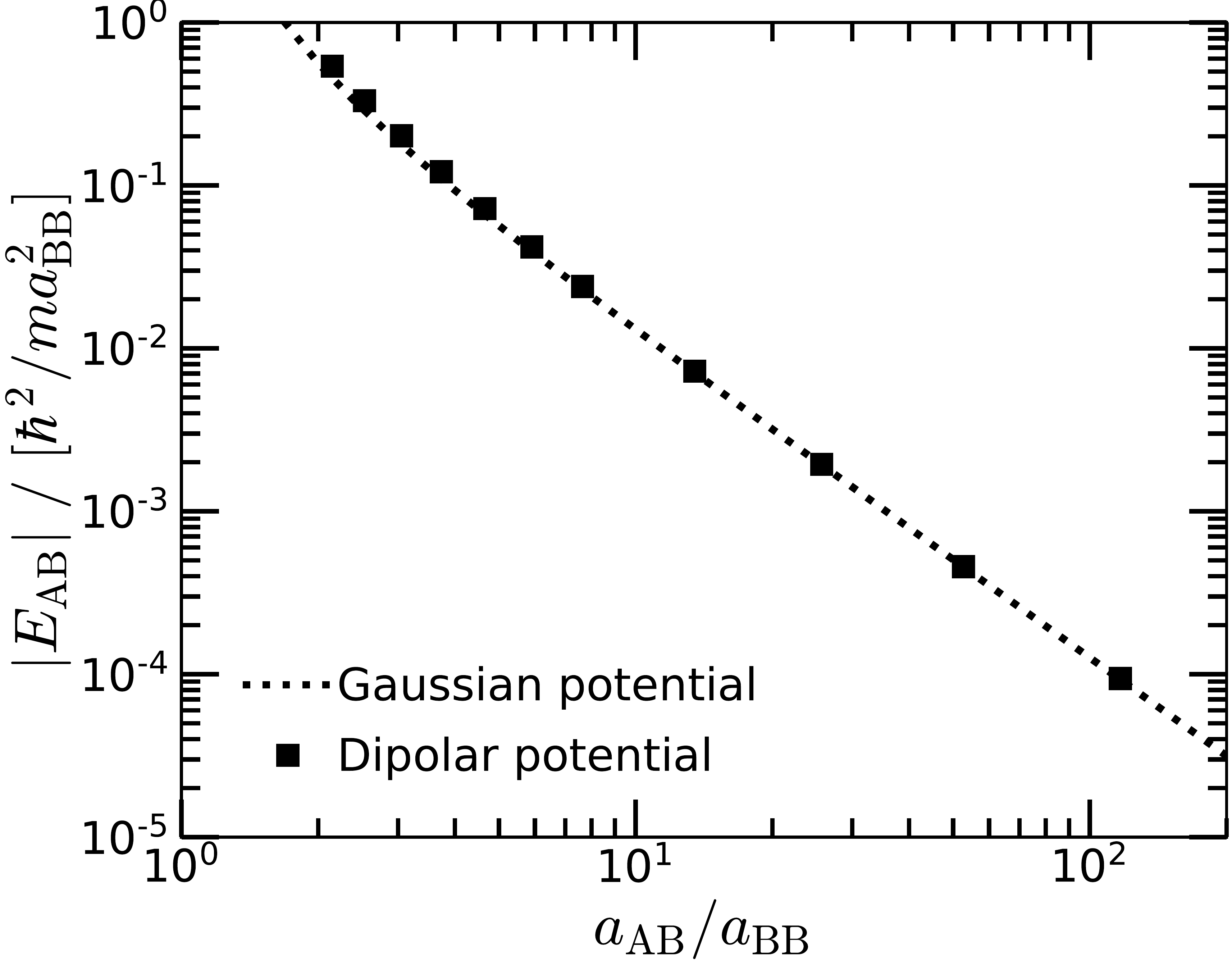}
    \caption{Dimer energy $E_{\1\2}$ in units of
        $\hbar^2/ma_{\2\2}^2$ for Gaussian (curve) and dipolar
        (symbols) potentials as a function of the scattering length ratio
        $a_{\1\2}/a_{\2\2}$.}
    \label{Fig:DimerEnergy}
\end{figure}

The Hamiltonian of the system is
\begin{equation}
\begin{aligned}
\label{Eq:4.2}
\hat H=&-\frac{\hbar^2}{2m}\sum_{i=1}^{N}\nabla^2_i-\frac{\hbar^2}{2m}
\sum_{\alpha=1}^{M}\nabla_\alpha^2
+\sum_{i<j}\hat V_{\1\1}(r_{ij})\\
&+\sum_{\alpha<\beta} \hat V_{\2\2}(r_{\alpha\beta})
+\sum_{i,\alpha} \hat V_{\1\2}(r_{i\alpha}) \ ,
\end{aligned}
\end{equation}
where the two-dimensional vectors ${\bf r}_i$ and ${\bf r}_\alpha$
denote particle positions of species A and B containing, respectively,
$N$ and $M$ atoms, $\hat V_{\1\2}$ and $\hat V_{\1\1}=\hat V_{\2\2}$
are the interspecies and intraspecies interaction potentials, and $m$
is the mass of each particle. For the bilayer setup, we have
\begin{equation}
    V_{\1\1}(r) = V_{\2\2}(r) = \frac{d^2}{r^3}, 
\label{Eq:4.3}
\end{equation}
and
\begin{equation}
   V_{\1\2}(r) = \frac{d^2(r^2-2h^2)}{(r^2+h^2)^{5/2}},
\end{equation}
where $d$ is the dipole moment and $h$ is the distance between the
layers. Dipoles are aligned perpendicularly to the layers and there
is no interlayer tunneling. The potential $V_{\2\2}(r)$ is purely
repulsive and is characterized by the $h$-independent scattering
length $a_{\2\2}=e^{2\gamma}r_0$~\cite{Ticknor2009}, where $\gamma\approx 0.577$
is the Euler constant and $r_0=md^2/\hbar^2$ is the dipolar length. 
The interlayer potential $V_{\1\2}(r)$ always supports at least one dimer state. 
Its energy reported in the Fig.~\ref{Fig:DimerEnergy}
diverges for $h\to 0$ and exponentially vanishes in the opposite
limit~\cite{Shih2009,Armstrong2010,Klawunn2010,Baranov2011}. 
The scattering length $a_{\1\2}$, which is a function of $r_0$ and $h$,
is $\sim a_{\2\2}\sim r_0$ for $h\sim r_0$, and exponentially large for
$h \gg r_0$. In the following, we parametrize the system by specifying
$a_{\2\2}$ and $a_{\1\2}$ rather than $h$ and $r_0$.

In the more academic case of Gaussian interactions,
the following potential was used~\cite{PhysRevA.101.041602}
\begin{equation}
\hat{V}_{\1\2}(r_{i\alpha})\psi(r_{i\alpha}) =
\int V_{\1\2}(r_{i\alpha},r'_{i\alpha})\psi(r'_{i\alpha})d^2 r'_{i\alpha},
\label{Eq:4.4}
\end{equation}
and similarly for $V_{\1\1}$ and $V_{\2\2}$, where 
\begin{equation}
\begin{aligned}
V_{\sigma\sigma'}(r,r') &= C_{\sigma\sigma'}G_\xi(r)G_\xi(r'),\\
G_\xi(r) &= (2\pi \xi^2)^{-1}\exp(-r^2/2\xi^2),
\label{Eq:4.5}
\end{aligned}
\end{equation}
and $\xi$ is the characteristic range of the potential. An advantage of
this non-local potential is that the two-body problem can be solved
analytically, giving 
\begin{equation}
C_{\sigma\sigma'}^{-1}=\frac{m}{4\pi\hbar^2}
\left[2\ln \frac {2\xi}{a_{\sigma\sigma'}}-\gamma\right].
\label{Eq:4.6}
\end{equation}
In the following, the ratio is varied $a_{\1\2}/a_{\2\2}$, with $a_{\2\2}=1.4\xi$
fixed. Note that the available ratio is limited to $a_{\1\2}/a_{\2\2}>1.1$.
%%%%%%%%%%%%%%%%%%%%%%%%%%%%%%%%%%%%%%%%%%%%%%%%%%%%%%%%%%%%%%%%
\section{Details of the Methods}\label{Section:Details of the methods}
In order to calculate the energies of the different few-body clusters
with dipolar interactions we use the diffusion Monte Carlo (DMC) method 
(see Chapter~\ref{Chapter:Quantum Monte Carlo methods}),
which leads to the exact ground-state energy of the system, within a
statistical error. This stochastic technique solves the Schr\"odinger
equation in imaginary time using a trial wave function for importance
sampling. We choose it to be 
\begin{equation}
\begin{aligned}
    \Psi_{\rm S}(\mathbf{r}_1,\dots,\mathbf{r}_{N+M})&=\prod_{i<j}^{N}f_{\1\1}(r_{ij})
\prod_{\alpha<\beta}^{M}f_{\2\2}(r_{\alpha\beta})\\
&\times\left[\prod_{i=1}^{N}\sum_{\alpha=1}^{M}f_{\1\2}(r_{i\alpha}) +
\prod_{\alpha=1}^{M}\sum_{i=1}^{N}f_{\1\2}(r_{i\alpha})\right],
\end{aligned}
\end{equation}
which takes into account a possible formation of AB dimers.
\begin{figure}[b!]
	\centering
    \includegraphics[width=0.7\textwidth]{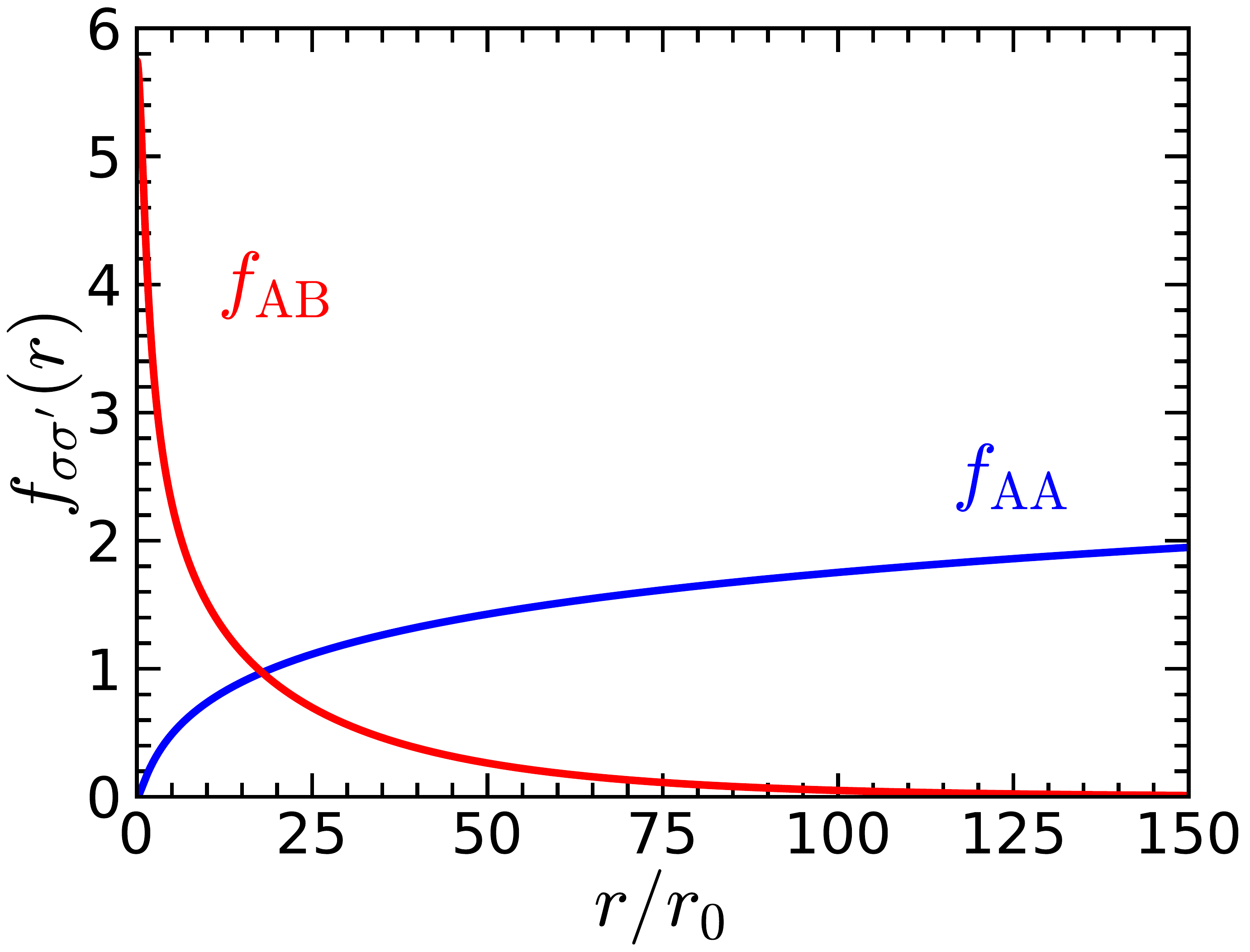}
    \caption{Intraspecies $f_{\1\1}(r)$ and interspecies $f_{\1\2}(r)$ wave
        functions.}
    \label{Fig:WaveFunction}
\end{figure}

The intraspecies Jastrow factors are chosen as the zero-energy
two-body scattering solution,
\begin{equation}
f_{\1\1}(r)=f_{\2\2}(r)=K_0(2\sqrt{r_0/r}),
\label{Eq:4.7}
\end{equation}
with $K_0$ the modified Bessel function.
The interspecies interactions are described by the dimer wave function
$f_{\1\2}(r)$ up to $R_0$, calculated numerically.
The variational parameter $R_0$ is chosen to be large enough that for
distances larger than $R_0$ we neglected the dipolar potential and took
the free scattering solution $f_{\1\2}(r)=CK_0(\sqrt{-mE_{\1\2}}r/\hbar)$.
We impose continuity of the logarithmic derivative at the matching point
$R_0$, this condition yields to the following equality
\begin{equation}
    \frac{ f^{'}_{\1\2}(R_0)}{f_{\1\2}(R_0)}=-\frac{\sqrt{-mE_{\1\2}}}{\hbar} 
    \frac{K_1(\sqrt{-mE_{\1\2}}R_0/\hbar)}{K_0(\sqrt{-mE_{\1\2}}R_0/\hbar)}.  
\label{Eq:4.8}
\end{equation} 
In Fig.~\ref{Fig:WaveFunction} we show the intraspecies $f_{\1\1}(r)$ and
interspecies $f_{\1\2}(r)$ wave functions.

In the Gaussian model, the stochastic variational method (SVM) was used.
Details of the SVM method can be found in 
Refs.~\cite{SuzVar98,BazEliKol16}. 
%%%%%%%%%%%%%%%%%%%%%%%%%%%%%%%%%%%%%%%%%%%%%%%%%%%%%%%%%%%%%%%%
\section{Results}
%%%%%%%%%%%%%%%%%%%%%%%%%%%%%%%%%%%%%%%%%%%%%%%%%%%%%%%%%%%%%%%%
\subsection{Binding Energies}
\label{Ch4:Binding Energies}
\begin{figure}[b!]
	\centering
    \includegraphics[width=0.7\textwidth]{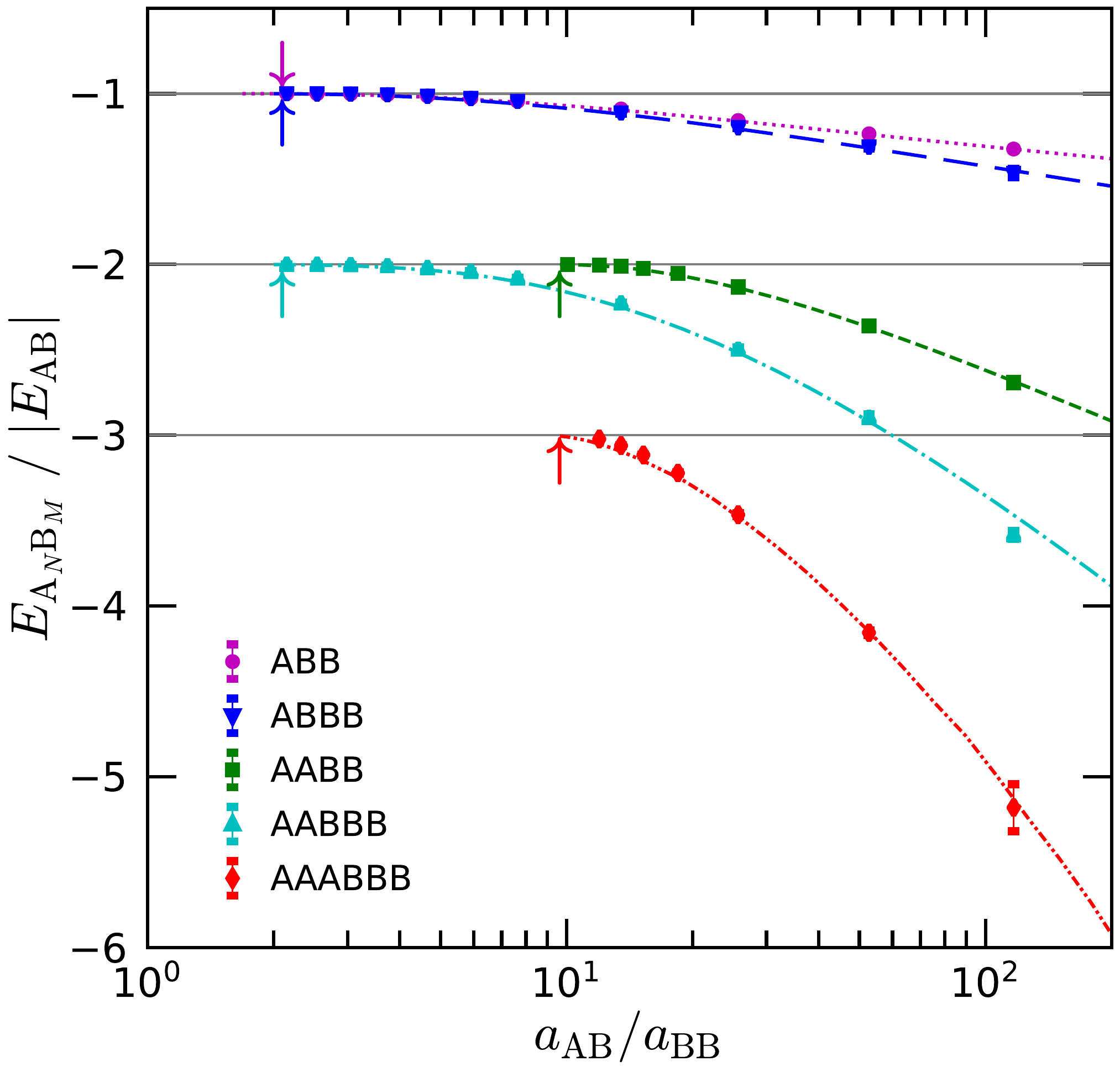}
	\caption{Energies of ${\rm A}_N{\rm B}_M$ clusters in units of the dimer
        energy $E_{\1\2}$ for Gaussian (curves) and dipolar
        (symbols) potentials as a function of the scattering length ratio
        $a_{\1\2}/a_{\2\2}$. The arrows show the positions of the thresholds
        for binding in the dipolar case.}
  	\label{Fig:RatioEnergies}
\end{figure}

We first discuss the limit of very large $a_{\1\2}$ (large dimer size)
when the interaction range and the intraspecies interactions can be
neglected. In this case, the problem can be treated in the zero-range
approximation giving for the ABB trimer
$E_{\rm ABB}^{a_{\2\2}=0}=2.39 E_{\rm AB}$ 
~\cite{Brodsky2006,PricoupenkoPedri2010,Bellotti2011} and for the
tetramers $E_{\rm ABBB}^{a_{\2\2}=0}=4.1 E_{\rm AB}$ and
$E_{\rm AABB}^{a_{\2\2}=0}=10.6 E_{\rm AB}$~\cite{Brodsky2006}.
The other A$_N$B$_M$ clusters
(with $1\leq N\leq M\leq 3$) are also bound in absence of the
intraspecies repulsion. 
In Ref.~\cite{BazakPetrov2018}, the authors calculated
their binding energies (and they also updated the energies of
smaller clusters), which are reported in Table~\ref{Table:ClustersEnergies}. 
\begin{table}[b!]
\centering
\begin{tabular}{||c c||} 
    \hline
 A$_N$B$_M$ &  $E^{a_{\2\2}=0}_{\1_N\2_M}/E_{\1\2}$\\ [0.5ex] 
 \hline\hline
 ABB & 2.3896(1) \\[0.5ex]
 ABBB & 4.1364(2) \\[0.5ex]
 AABB & 10.690(2) \\[0.5ex]
 AABBB & 28.282(5) \\ [0.5ex]
 AAABBB & 104.01(5) \\ [0.5ex]
 \hline
\end{tabular}
\caption{Energies of ${\rm A}_N{\rm B}_M$ clusters in units of the dimer
    energy $E_{\1\2}$ in absence of the
    intraspecies repulsion $a_{\2\2}=0$~\cite{BazakPetrov2018}.
The number between parenthesis indicates the error in the
last digits of the corresponding value.}
\label{Table:ClustersEnergies}
\end{table}
\clearpage
The intraspecies repulsion shifts the cluster energies upwards as has
been seen for the ABB trimer~\cite{PRB95_045401,BazakPetrovPRL2018}
and for the ABBB tetramer~\cite{BazakPetrovPRL2018}. In
Fig.~\ref{Fig:RatioEnergies}, we report the energies of these and
bigger clusters for both the dipolar and Gaussian interactions. Note
that, even for the weakest BB repulsion shown in this figure
($a_{\1\2}/a_{\2\2} = 200$), the clusters are significantly less
bound compared to the case of no BB repulsion. This happens since the
small parameter that controls the weakness of the intraspecies
interaction relative to the interspecies one is actually
$\lambda=1/\ln(a_{\1\2}/a_{\2\2}) \ll 1$. By contrast, effective-range
corrections contain powers of 
$r_0\sqrt{mE}/\hbar$ or $\xi \sqrt{mE}/\hbar$ for dipolar or Gaussian
interactions, respectively, which are exponentially small in terms of
$\lambda$. This explains why the two interaction models lead to almost
indistinguishable results for large $a_{\1\2}/a_{\2\2}$. 
\begin{figure}[b!]
	\centering
	\includegraphics[width=0.64\textwidth]{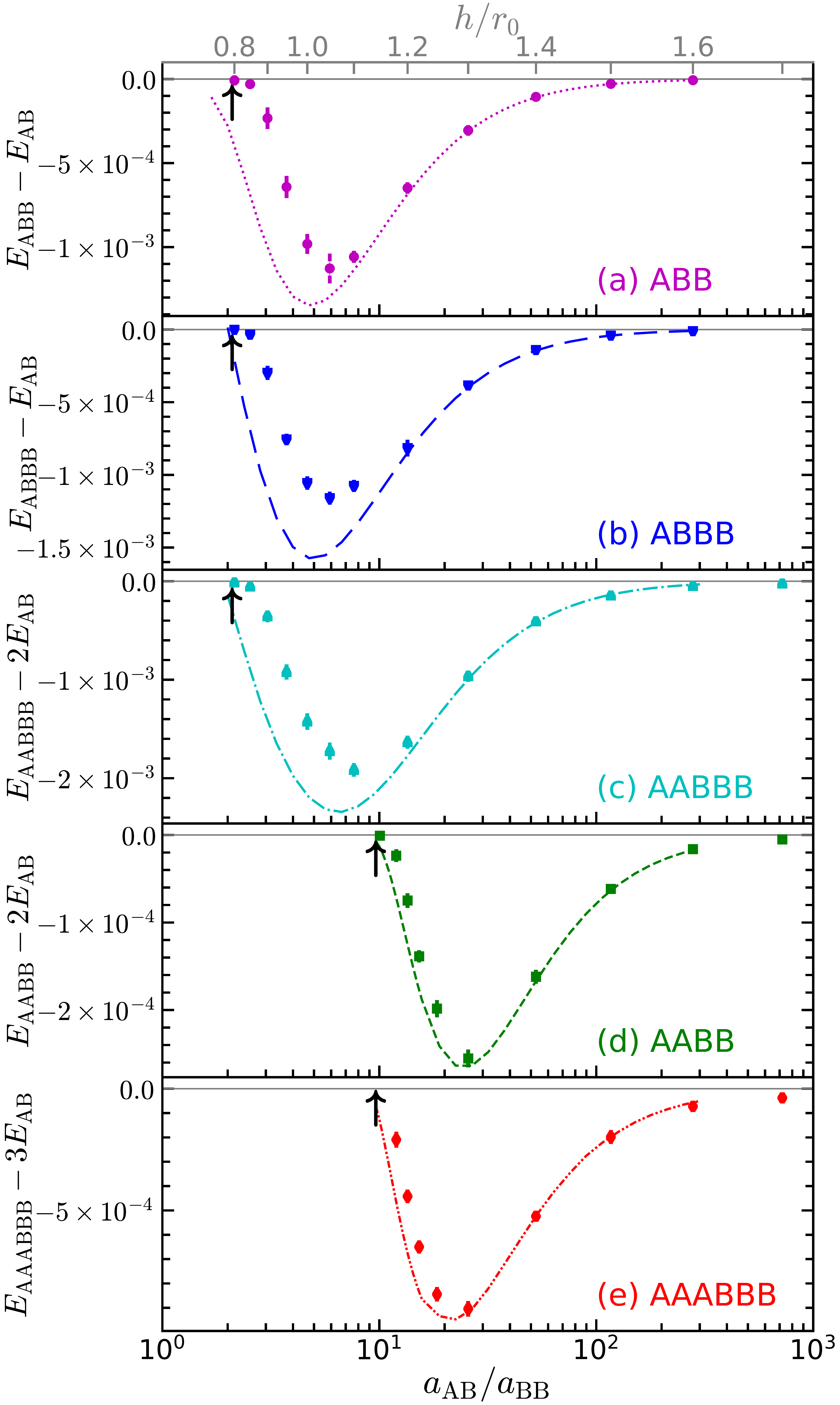}
	\caption{Binding energies of the few-body clusters 
        $E_{{\rm A}_N{\rm B}_M}-NE_{\rm AB}$, in units of
        $\hbar^2/ma_{\2\2}^2$, versus $a_{\1\2}/a_{\2\2}$,
        for Gaussian (curves) and dipolar (symbols) potentials.
        The arrows show the positions of the thresholds
        for binding in the dipolar case.
        The uppper axis is labeled in dipolar units.}
	\label{BindingEnergies}
\end{figure}

We find that for sufficiently strong intraspecies repulsion (smaller
$a_{\1\2}/a_{\2\2}$) the trimer and all higher clusters get unbound.
In Fig.~\ref{Fig:RatioEnergies}, the thresholds for binding in the
dipolar model are shown by arrows. We find that the tetramer threshold
is located at $a_{\1\2}/a_{\2\2} \approx 10$ ($h/r_0\approx 1.1$) and
the trimer threshold, corresponding to the atom-dimer zero crossing,
occurs in the regime where all relevant length scales (scattering
lengths, dimer sizes, interaction ranges) are comparable to one
another; $a_{\1\2}/a_{\2\2} \approx 2$ ($h/r_0\approx 0.8$) for the
dipolar model. The positions of the threshold and differences between
the results of the two models are better visible in 
Fig.~\ref{BindingEnergies} where we plot the cluster energies in
units of $\hbar^2/ma_{\2\2}^2$. 

Our numerical calculations for larger clusters indicate that,
depending on whether they are balanced ($M=N$) or not, their
unbinding thresholds coincide, respectively, with the tetramer or
with the trimer ones. To understand these results note that close to
these thresholds the clusters are much larger than the dimer. Treating
there the latter as an elementary boson D, the AABBB pentamer and the
ABBB tetramer can be thought of as weakly bound DDB or DBB ``trimers''
characterized by a large $a_{{\rm D}{\rm B}}$ value and repulsive DD
and BB interactions (the DD interaction is repulsive since we are
above the tetramer AABB threshold). In the limit
$a_{{\rm D}{\rm B}}\rightarrow \infty$ the DD and BB interactions can
be neglected and the binding energies of the DDB and DBB composite
trimers are asymptotically fractions of 
$E_{\1\2\2}-E_{\1\2}$~\cite{PricoupenkoPedri2010}. The ABB trimer,
ABBB tetramer, and AABBB pentamer thresholds are therefore the same
[see Fig.~\ref{BindingEnergies}~(a,b,c)]. In the same reasoning, close
to the AABB tetramer crossing, the hexamer AAABBB is a weakly-bound
DDD state which splits into three dimers when the dimer-dimer
attraction changes to repulsion resulting in the same threshold value.
%%%%%%%%%%%%%%%%%%%%%%%%%%%%%%%%%%%%%%%%%%%%%%%%%%%%%%%%%%%%%%%%
\subsection{Threshold Determination}
In this section, we numerically determine the threshold
values of the few-body clusters in the bilayer setup. To do
this we need to know how the energy depends on the
interaction potential close to the threshold for unbinding.
To find out this energy dependency, let us review the
principal properties of the two-body bound state in 
one 1D, two 2D, and three dimensions 3D.

\begin{figure}[t!]
  \centering
  \subfigure{\includegraphics[width=0.484\textwidth]{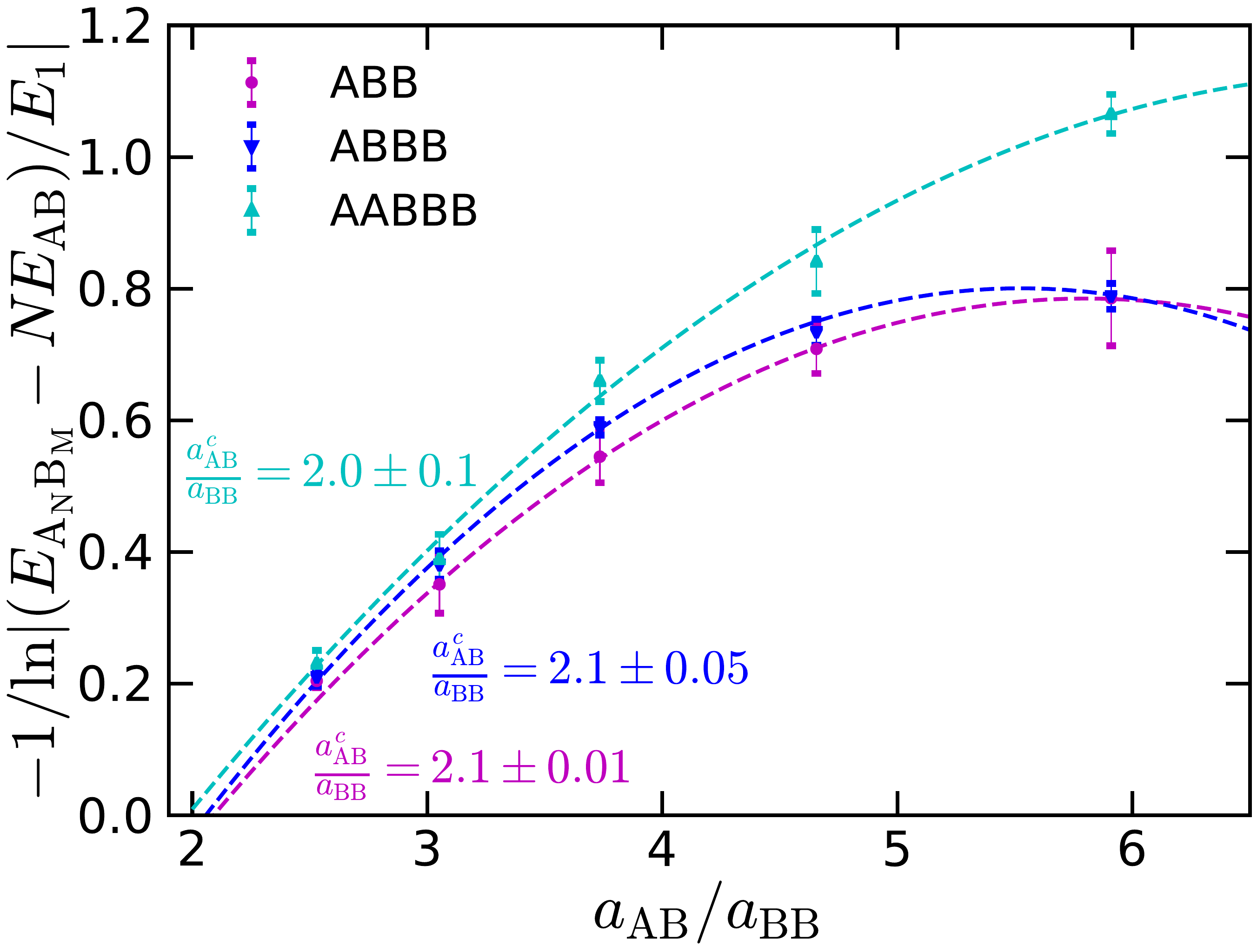}}\quad
  \subfigure{\includegraphics[width=0.484\textwidth]{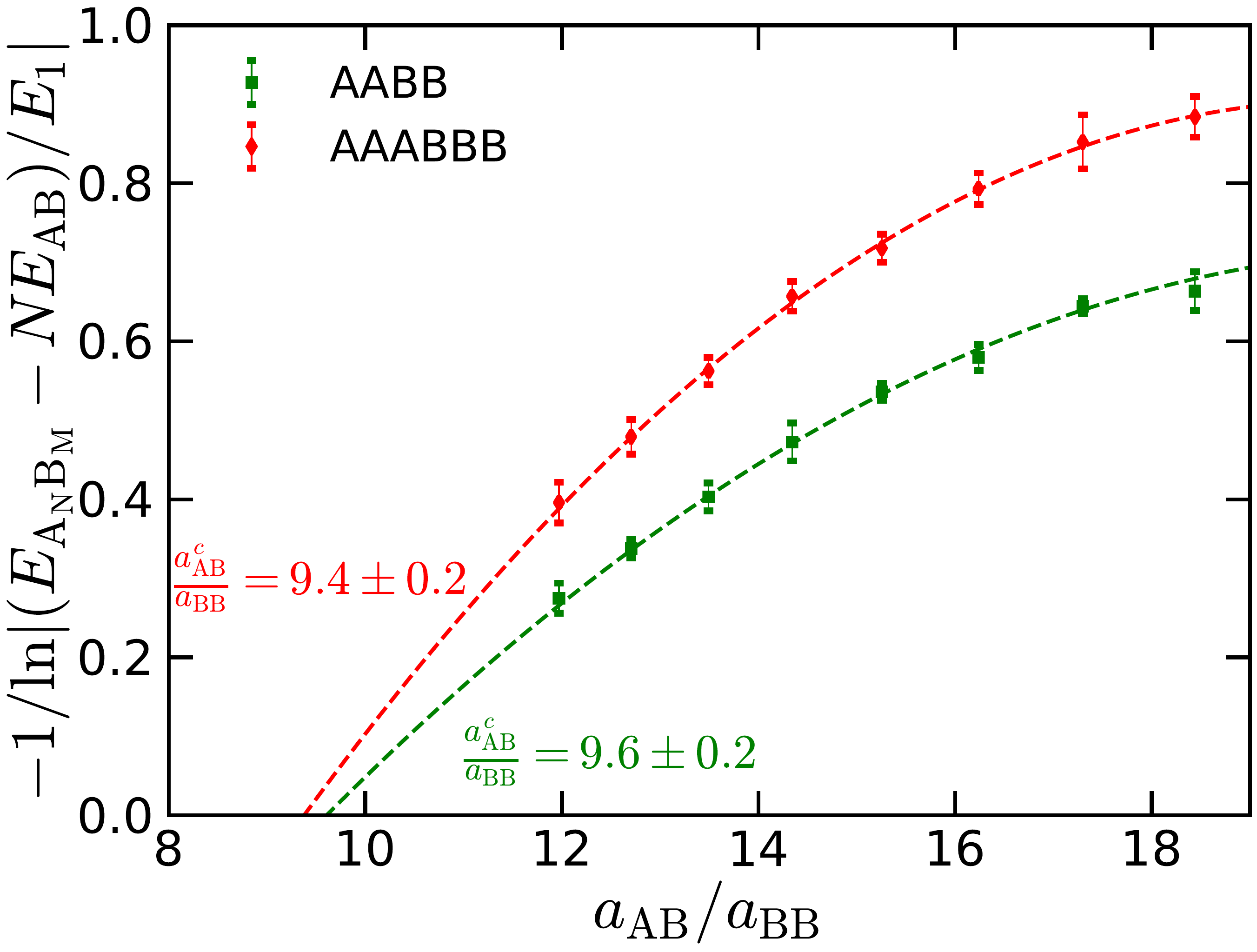}}
  \caption{Fitting procedure for the determination of the threshold values
      of the A$_N$B$_M$ clusters with dipolar interactions. The dashed curves
      correspond to the fit $c_1(\frac{a_\mathrm{AB}}
      {a_\mathrm{BB}}-\frac{a_\mathrm{AB}^c}{a_\mathrm{BB}})+
      c_2(\frac{a_\mathrm{AB}}{a_\mathrm{BB}}-\frac{a_\mathrm{AB}^c}{a_\mathrm{BB}})^2$.
      The fitting parameters are reported in Table~\ref{Table:ThresholdsFittingParameters}.}
    \label{Fig:Thresholds}
\end{figure}

According to Quantum mechanics, a symmetric attractive well in 3D 
supports a bound state of two particles only if the potential well
depth $V$ is larger than a \textit{critical depth} $V_c$~\cite{kagan2013modern}.
Thus there is
a threshold for a two-body bound state in 3D. This is in contrast with
the 1D and 2D cases where the dimer state is formed even for infinitely
small attraction between the two particles. Therefore, in 1D and 2D the
threshold for the formation of the two-body bound state is 
absent~\cite{kagan2013modern}.
In Table~\ref{Table:BoundStates} we present a summary of the principal
properties of the dimer state in 1D, 2D, and 3D~\cite{inguscio2008ultra}.
We notice that for a symmetric attractive well in 2D the
dimer state is weakly bound, with its energy depending exponentially
on the shallow potential $-V$, according to
\begin{equation}
    E_B\approx E_Re^{-2cE_R/V}, \quad \mathrm{with} 
    \quad E_R=\frac{\hbar^2}{mR^2},
\label{Eq:4.9}
\end{equation}
with c on the order of 1. This is the energy dependency we were looking for. 
\begin{table}
\centering
\begin{tabular}{||c c c c||} 
 \hline
  & & &\\
 & 1D & 2D & 3D \\ [2.0ex] 
 \hline\hline
  & & &\\
 $V$ & $\ll E_R$ & $\ll E_R$ & $>V_c\approx E_R$\\ [2ex]
 \hline
 & & &\\
 $\psi(r>R)$ & $e^{-r/r_B}$ &
 $K_0(\frac{r}{r_B})
\begin{cases}
    -\mathrm{log}(\frac{r}{r_B}),&\!R\ll r\ll r_B\\
    e^{-r/r_B},&\!r\gg r_B
\end{cases}$
& $\frac{e^{-r/r_B}}{r}$\\ [3ex]
\hline
 & & &\\
$r_B$ & $R\frac{E_R}{V}$ & $Re^{cE_R/V}$ & $R\frac{E_R}{V-V_c}$ \\ [2ex]
\hline
 & & &\\
$E_B=-\frac{\hbar^2}{mr_B^2}$ & $-\frac{V^2}{E_R}$ & $-E_Re^{-2cE_R/V}$ & $-\frac{(V-V_c)^2}{E_R}$\\  [2ex]
 \hline
\end{tabular}
\caption{Bound-states in 1D, 2D and 3D for a potential well of size $R$
    and depth $V$. $\psi(r>R)$ is the wave function outside the well,
    $r_B$ is the size of the bound state, and $E_B$ its energy
    ($E_R=\hbar^2/(mR^2$))~\cite{inguscio2008ultra}.}
\label{Table:BoundStates}
\end{table}
Although Eq.~(\ref{Eq:4.9}) is for two-body bound states we are going to
use it for larger clusters and let see if it works.
Using the above result we propose to fit the
DMC binding energies with the function
\begin{equation}
E_{{\rm A}_N{\rm B}_M}-NE_{\rm AB}=E_1\exp 
\left[ \frac{-1}{c_1(a_{\1\2}-a_{\1\2}^c)
        + c_2(a_{\1\2}-a_{\1\2}^c)^2} \right],
\label{Eq:4.10}
\end{equation}
for $a_{\1\2}>a_{\1\2}^c$, where $a_{\1\2}^c,E_1,c_1,c_2$ are free parameters.
The Eq.~(\ref{Eq:4.10}) can be rewritten as 
\begin{equation}
    -\frac{1}{\mathrm{ln}|(E_{{\rm A}_N{\rm B}_M}-NE_{\rm AB})/E_1|}= 
    c_1(a_{\1\2}-a_{\1\2}^c)
        + c_2(a_{\1\2}-a_{\1\2}^c)^2 ,
\label{Eq:4.11}
\end{equation}
which is more convenient to fit the energies.
In Fig.~\ref{Fig:Thresholds} we show the numerical threshold determination for the
dipolar clusters. On the left panel of Fig.~\ref{Fig:Thresholds} we show
the threshold fitting for the trimer ABB, tetramer ABBB, and pentamer AABBB.
On the right panel of Fig.~\ref{Fig:Thresholds} we show the
tetramer AABB and hexamer AAABBB thresholds.
The threshold values and the fitting parameters are reported in
Table~\ref{Table:ThresholdsFittingParameters}.
Our numerical results are consistent with our conclusions of the previous
secction, the bilayer setup have two thresholds, one for population-imbalanced
clusters at $a^c_{\1\2}/a_{\2\2}\approx 2$ and the second one for 
population-balanced cluster at $a^c_{\1\2}/a_{\2\2}\approx 10$.  
\begin{table}[t!]
\centering
\begin{tabular}{||c c c c c||} 
 \hline
 A$_N$B$_M$ & $E_1$ & $c_1$ & $c_2$ & $a_{\1\2}^c/a_{\2\2}$\\ [0.5ex] 
 \hline\hline
 ABB    & 0.0004    & 0.42307  & -0.0569969  & 2.1 $\pm$ 0.01\\[0.5ex] 
 ABBB   & 0.00041   & 0.461707 & -0.0665619  & 2.1 $\pm$ 0.05\\[0.5ex]
 AABBB  & 0.00046   & 0.437039 & -0.042321   & 2.0 $\pm$ 0.1\\[0.5ex]
 AABB   &  0.000089 & 0.125298 & -0.00547874 & 9.4 $\pm$ 0.2\\[0.5ex]
 AAABBB & 0.00026   & 0.170496 & -0.00803067 & 9.6 $\pm$ 0.2\\ [0.5ex] 
 \hline
\end{tabular}
\caption{Fitting parameters for the threshold determination.}
\label{Table:ThresholdsFittingParameters}
\end{table}
%%%%%%%%%%%%%%%%%%%%%%%%%%%%%%%%%%%%%%%%%%%%%%%%%%%%%%%%%%%%%%%%
\subsection{Three-Dimer Repulsion}
In the last part of Subsection~\ref{Ch4:Binding Energies},
we have integrated out the internal degrees
of freedom of the dimers, replacing them by elementary point-like
bosons. In fact, the DD zero crossing that we observe for
$a_{\1\2}\approx 10 a_{\2\2}$ is a nonperturbative phenomenon
resulting from a competition between strong repulsive and
attractive interatomic forces among four individual atoms. These
interactions are strong since the corresponding scattering lengths
are comparable to the typical atomic de Broglie wave lengths
$\sim 1/a_{\rm AB}$. We emphasize that this cancellation is achieved
only for two dimers. For three dimers it is incomplete and there is a
residual effective three-dimer force of range $\sim a_{\rm AB}$ 
(distance, where the dimers start touching one another). In the
many-body problem, this higher-order force may compete with the
dimer-dimer interaction (if it is not completely zero) or even become
dominant. In principle, one can also discuss higher-order effects of
this type at the DB zero crossing in a DB mixture, but they are
expected to be subleading since the DD and BB interactions remain
finite. In the remainder of this section we thus
concentrate on the population-balanced case.

\begin{figure}[t!]
	\centering
    \includegraphics[width=0.7\textwidth]{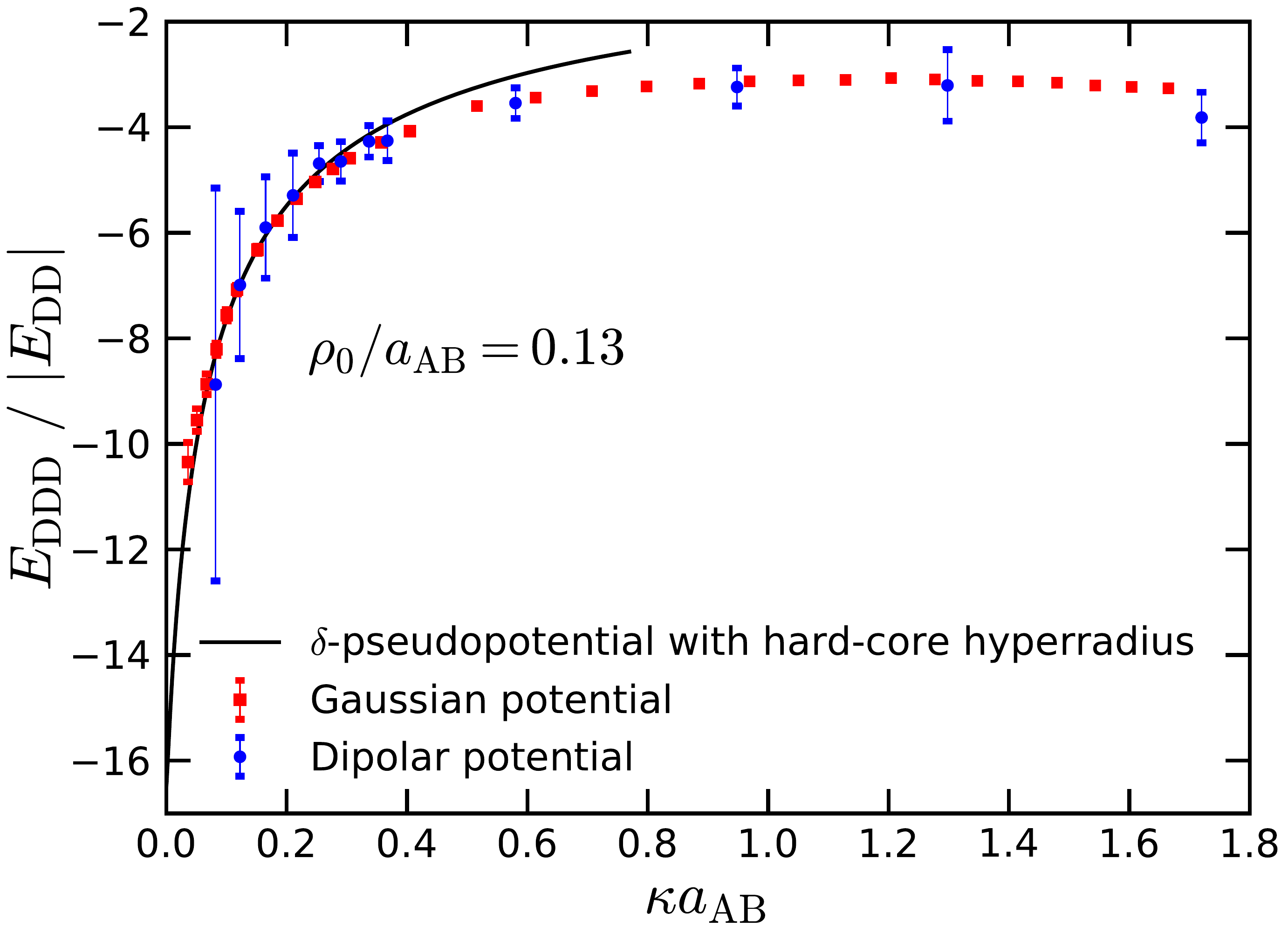}
	\caption{The hexamer-to-tetramer binding energy ratio 
    $E_{\rm DDD}/|E_{\rm DD}|$ as a function of $\kappa a_{\rm AB}$.
    The solid line is the result of the zero-range model with the
    hard-core hyperradius constraint at 
    $\rho_0=0.13a_{\rm AB}$~\cite{PhysRevA.101.041602}.}
	\label{Fig:ThreeBodyForce}
\end{figure}

In order to characterize the effective three-dimer interaction, we
follow the method developed previously in one
dimension (Chapter~\ref{Chapter:One-dimensional three-boson problem with two- and three-body
interactions}). Namely, we analyze the behavior of the
hexamer energy just below the tetramer threshold. If the tetramer
binding energy 
\begin{equation}
E_{\rm DD}=E_{\rm AABB}-2E_{\rm AB},
\end{equation}
is much smaller
than $E_{\1\2}$, the dimer-dimer interaction can be considered
point-like and the relative DD wave function can be approximated by
\begin{equation}
\phi(r)\propto K_0(\kappa r),
\end{equation}
where
$\kappa=\sqrt{-2mE_{\rm DD}/\hbar^2}$ is the inverse size of the
tetramer. Similarly, the AAABBB hexamer under these conditions reduces
to the well-studied problem of three point-like
bosons~\cite{BruchTjon1979,Adhikari1988,Nielsen1997,Nielsen1999,
HammerSon2004,KartavtsevMalykh2006,Brodsky2006}, according to which
the ground-state hexamer binding energy
\begin{equation}
E_{\rm DDD}=E_{\rm AAABBB}-3E_{\rm AB},
\end{equation}
should satisfy~\cite{HammerSon2004,KartavtsevMalykh2006}
\begin{equation}\label{E3zr}
E_{\rm DDD}/|E_{\rm DD}|=-16.5226874.
\end{equation}
We expect the ratio $E_{\rm DDD}/|E_{\rm DD}|$ to reach the zero-range
limit~(\ref{E3zr}) as we approach the dimer-dimer zero crossing, i.e.,
as $\kappa a_{\rm AB}\rightarrow 0$. In Fig.~\ref{Fig:ThreeBodyForce},
we plot $E_{\rm DDD}/|E_{\rm DD}|$ versus $\kappa a_{\rm AB}$ and
indeed see a tendency towards the value~(\ref{E3zr}) although the
effects of the finite size of the dimers and their internal degrees
of freedom, that we have neglected in the zero-range model, are
obviously important. The fact that the hexamer energy lies above the
limit~(\ref{E3zr}) points to an effective three-dimer repulsive force.
We note again that the values of the ratio $E_{\rm DDD}/|E_{\rm DD}|$
obtained for Gaussian and dipolar potentials are quite close to each
other for all values of $a_{\1\2}$ suggesting a certain universality
of this problem and a relative unimportance of the long-range
interaction tails.

In order to quantify the three-dimer interaction observed in 
Fig.~\ref{Fig:ThreeBodyForce}, we compared our results with
a zero-range model with hard-core hyperradius constraint $\rho_0$,
developed in Ref.~\cite{PhysRevA.101.041602}.

The authors in Ref.~\cite{PhysRevA.101.041602} extended the model of three point-like
dimers by requiring that the three-dimer wave function vanishes at a
hyperradius $\rho_0$. For three dimers, with coordinates 
${\bf r}_1$, ${\bf r}_2$, and ${\bf r}_3$, the hyperradius is defined
as
\begin{equation}
\rho=\sqrt{x^2+y^2},
\end{equation}
where 
\begin{equation}
    {\bf x}=\frac{2{\bf r}_3-{\bf r}_1-{\bf r}_2}{\sqrt{3}} \quad \mathrm{and}
    \quad{\bf y}={\bf r}_1-{\bf r}_2,
\end{equation}
are the Jacobi coordinates. For
this minimalistic model $E_{\rm DDD}/|E_{\rm DD}|$ is a function of
the ratio $\kappa \rho_0$, relating the three- and two-dimer
interaction strengths. Kartavtsev and Malykh~\cite{KartavtsevMalykh2006}
discussed the adiabatic hyperspherical method in detail and
applied it to the $\rho_0=0$ limit, i.e., the case of
no three-body interaction. The authors in Ref.~\cite{PhysRevA.101.041602}
accounted for finite $\rho_0$, by set the hyperradial channel
functions to zero at $\rho=\rho_0$. In this way, they obtained the ratio
$E_{\rm DDD}/|E_{\rm DD}|$ as a function of $\kappa \rho_0$ (for more
details see Ref.~\cite{PhysRevA.101.041602}). We then
treat $\rho_0$ as a constant (independent of $E_{\rm DD}$) determined
by fitting the DMC and SVM data in $\kappa a_{\rm AB} < 0.4$ range. By
minimizing $\chi^2$ we obtain $\rho_0=0.13 a_{\rm AB}$.

The inclusion of the three-body hard-core constraint, even
corresponding to numerically very small $\kappa\rho_0$, leads to a
spectacular deviation from Eq.~(\ref{E3zr}). This interesting effect
is due to an enhancement of the three-dimer interaction by strong
two-dimer correlations. 

Promising candidates for observing the predicted cluster states are
bosonic dipolar molecules characterized by large and tunable dipolar
lengths which, at large electric fields, tend to
$r_0=5\times 10^{-6}$m for 
$^{87}$Rb$^{133}$Cs~\cite{PhysRevLett.113.205301,PhysRevLett.113.255301},
$r_0 = 2\times10^{-5}$m for
$^{23}$Na$^{87}$Rb~\cite{PhysRevLett.116.205303,PhysRevA.97.020501}
and $r_0=6\times10^{-5}$m for $^7$Li$^{133}$Cs~\cite{Deiglmayr2010}.
Fermionic $^{87}$Rb$^{40}$K~\cite{Moses2015,DeMarco2019} and
$^{23}$Na$^{40}$K~\cite{Park2015,Park2017,Yang2019} molecules
($r_0=7\times 10^{-7}$m and $r_0=7\times 10^{-6}$, respectively)
could be turned into bosons by choosing another isotope of K. The
interlayer distance, fixed by the laser wavelength, has typical values
of $h\approx (2-5)\times 10^{-7}$m, which is thus sufficient for
observing the few-body bound states that we predict for ratios
$h/r_0 > 0.8$. 

The next step in this work is to generalize these findings to the
many-body problem when a new scale (density $n$) comes into play.
It is important to understand how the two- and three-body effects
correlate with each other as one passes through the dimer-dimer
zero crossing. Although we find no qualitative difference between the
dipolar and Gaussian models in our few-body results, the long-range
tails will be important when the quantity $nr_0$ becomes comparable
to the inverse healing length (which is where the dipolar condensate
becomes rotonized). For bilayer dipoles the relevant region of
parameters is close to the dimer-dimer zero crossing, which we predict
to be at $h/r_0\approx 1.1$.
In Chapter~\ref{Chapter:Two-dimensional dipolar liquid}, we studied
the many-body problem of dipolar bosons in a bilayer geometry.
%%%%%%%%%%%%%%%%%%%%%%%%%%%%%%%%%%%%%%%%%%%%%%%%%%%%%%%%%%%%%%%%
\section{Summary}
To summarize, we have studied few-body clusters A$_N$B$_M$ with
$N\leq M\leq 3$ in a two-dimensional Bose-Bose mixture using different
(long-range dipolar and short-range Gaussian) intraspecies repulsion
and interspecies attraction models. In both cases, the intraspecies
scattering length $a_{\1\1}=a_{\2\2}$ is of the order of the potential
ranges, whereas we tune $a_{\1\2}$ by adjusting the AB attractive
potential (or the interlayer distance in the bilayer setup). We find
that for $a_{\1\2}\gg a_{\2\2}$ all considered clusters are (weakly)
bound and their energies are independent of the interaction model. As
the ratio $a_{\1\2}/a_{\2\2}$ decreases, the increasing intraspecies
repulsion pushes the clusters upwards in energy and eventually breaks
them up into $N$ dimers and $M-N$ free B atoms. In the population
balanced case ($N=M$) this happens at $a_{\1\2}/a_{\2\2}\approx 10$
where the dimer-dimer attraction changes to repulsion. By studying
the AAABBB hexamer near the dimer-dimer zero crossing we find that
it very much behaves like a system of three point-like particles
(dimers) characterized by an effective three-dimer repulsion. A
dipolar system in a bilayer geometry can thus exhibit the tunability
and mechanical stability necessary for observing dilute liquids and
supersolid phases.

\chapter{Quantum Halo States in Two-Dimensional Dipolar Clusters}
\label{Chapter:Quantum halo states in two-dimensional dipolar clusters}
The purpose of the present chapter is to
study the ground-state properties of loosely bound dipolar
clusters composed of two to six particles
in a two-dimensional bilayer geometry.
We investigate whether halos, bound states with a wave function that
extends deeply into the classically forbidden region~\cite{Jensen2004,Riisager_2013},
can occur in this system. The dipoles are confined
to two layers, A and B, with dipolar moments aligned perpendicularly to the
planes. The binding energies, pair correlation functions, spatial distributions,
and sizes are calculated at different values of the interlayer distance
by using the diffusion Monte Carlo method. We find that for large interlayer
separations the AB dimers are halo states and following a universal scaling
law relating the energy and size of the bound state. For ABB trimers and
AABB tetramers, we find two very distinct halo structures. For large values of
the interlayer separation, such halo states are weakly bound and the typical
distances between BB and AB dipoles are similar. However, for the
deepest bound ABB and AABB clusters, and as the clusters approach the unbinding
energy threshold, we find a highly anisotropic structure, in which the AB
pair is on average several times closer than the BB pair.
Similar symmetric and asymmetric structures are observed in pentamers and
hexamers, both being halo states, thus providing halos with the largest
number of particles ever observed or predicted before.  
%%%%%%%%%%%%%%%%%%%%%%%%%%%%%%%%%%%%%%%%%%%%%%%%%%%%%%%%%%%%%%%%
\section{Introduction}
One of the most remarkable aspects of ultracold quantum
gases is their versatility,
which permits to bring ideas from other areas of physics and
implement them in a clean and highly controllable manner. Some of
the examples of fruitful interdisciplinary borrowings include Efimov,
states originally introduced in nuclear physics and observed in alkali
atoms~\cite{EFIMOV1970563,kraemer2006evidence,Naidon_2017}, lattices
created with counter-propagating laser beams~\cite{sachdev_2009,Bloch2005} as models
for crystals in condensed matter physics,
Bardeen-Cooper Schrieffer (BCS) pairing theory first introduced to
explain superconductivity and later used to describe two-component Fermi
gases~\cite{ReviewFermiGases,Zwerger2012book}. In the present chapter, 
we exploit the tunability of ultracold gases to create halo states with a
number of atoms never achieved before. Originating
in nuclear physics~\cite{RevModPhys.66.1105,TANIHATA1985,PhysRevLett.55.2676},
halo dimer states have been studied and experimentally observed in ultracold
gases~\cite{RevModPhys.82.1225}.

A halo is an intrinsically quantum object and it is defined as a bound state
with a wave function that extends deeply into the classically forbidden
region~\cite{Jensen2004,Riisager_2013}. These states are characterized by
two simultaneous features: a large spatial size, due to that extension of the bound
state, and a binding energy which is much smaller than the typical energy of
the interaction. One of the most
dramatic examples of a  halo system, experimentally known, is the Helium dimer
($^4\mathrm{He}_2$), which is about ten times more extended than the size
of a typical diatomic molecule~\cite{RevModPhys.82.1225}.

While most of the theoretical and
experimental studies of halos have been carried out in three
dimensions~\cite{RIISAGER1992393,PhysRevC.49.201,PhysRevLett.113.253401,Stipanovi2017QuantumHS,Stipanovic2019},
there is an increasing interest in halos in two dimensions 
(2D)~\cite{Nielsen1997,Nielsen1999,Jensen2004}. In fact, two dimensions
are especially interesting as halos in 2D have different
properties~\cite{Jensen2004} of the 3D ones.
A crucial difference between 3D and 2D geometries is that lower dimensionality
dramatically enhances the possibility of forming bound states.
If the integral of the interaction potential $V(r)$ over all the space is finite
and negative, 
\begin{equation}
V_{k=0} = \int V({\bf r})d{\bf r}<0,
\end{equation}
this is always sufficient
to create a two-body bound state in 2D but not necessarily in 3D, where the
potential depth should be larger than a critical value. Furthermore,
the energy of the bound-state is exponentially small in 2D and it can
be expressed as~\cite{LandauLifshitz_iii}
\begin{equation}
    E = - \frac{\hbar^2}{(2ma^2)} \exp\left[-\frac{\hbar^2|V_{k=0}|}{2\pi m}\right],
\end{equation}
where $a$ is the typical size of the bound state.
An intriguing possibility arises when such integral is exactly equal
to zero,
\begin{equation}
V_{k=0} = 0,
\end{equation}
since a priori it is not clear if a bound state exists.
This situation exactly happens in a dipolar bilayer in which atoms or
molecules with a dipolar moment $d$ are confined to two layers
separated by a distance $h$ and the interaction between particles
of different layer is given by
\begin{equation}
    V(r)=\frac{d^2(r^2-2h^2)}{(r^2+h^2)^{5/2}}.
\end{equation}
The vanishing Born integral has first lead to conclusions
that the two-body bound state disappears when the distance
between the layers is large~\cite{Yudson1997} although later
it was concluded that the bound state exists for any 
separation~\cite{Shih2009,Armstrong2010,Klawunn2010,Baranov2011,Volosniev2011},
consistently with Ref.~\cite{Simon1976}.
A peculiarity of this system is that the bound state is extremely weakly
bound in the limit $h\to\infty$.
That is, a potential with depth $V(r=0) = -d^2/h^3$ and width
$h$ would be expected to have binding energy equal to 
\begin{equation}
E =-\hbar^2/(2ma^2)\exp(-\text{const}\cdot r_0/h),
\end{equation}
where $r_0 = md^2/\hbar^2$ is
the characteristic distance associated with the dipolar interaction and $m$
is the particle mass.
Instead, the correct binding energy~\cite{Klawunn2010,Baranov2011}  
\begin{equation}
E = - 4\hbar^2/mh^2\exp(-8r_0^2/h^2 + O(r_0/h)),
\end{equation}
is much smaller as it has $h^{-2}$
in the exponent and not the usual $h^{-1}$.
This suggests that the bilayer configuration is very promising for the formation
of two-body halo states.
Moreover, the peculiarity of the bilayer problem has resulted in the
controversial claim that the three- and four-body~\cite{Volosniev2012}
bound states never exist in this system, and only very recently it has been shown
that
actually they are formed~\cite{PhysRevA.101.041602}.

In this chapter, we analyze the ground-state properties of dipolar few-body bound states
within a two-dimensional bilayer setup, as candidates for halo states. In
particular, we study the ground state of 
up to six particles occupying the A and B layers, with A and B denoting
particles in different planes. 
To find the exact system properties we rely on the diffusion
Monte Carlo (DMC) method 
(see Chapter~\ref{Chapter:Quantum Monte Carlo methods}) with
pure estimators ( see Subsection~\ref{Pure Estimators}), which has been
used previously to get an accurate description of quantum halo
states in Helium dimers~\cite{PhysRevLett.113.253401}, trimers and
tetramers~\cite{Stipanovi2017QuantumHS,Stipanovic2019}. Also, we report
relevant structure properties of the clusters, such as the spatial density distributions
and the pair distribution functions for characteristic interlayer separations.
%%%%%%%%%%%%%%%%%%%%%%%%%%%%%%%%%%%%%%%%%%%%%%%%%%%%%%%%%%%%%%%%
\section{The Hamiltonian}
We consider two-dimensional systems consisting from two to six
dipolar bosons of mass $m$ and dipole moment $d$ confined to a bilayer
setup. All the dipole moments are oriented perpendicularly to the layers
making the system always stable.
The Hamiltonian of this system is
\begin{equation}
H=-\frac{\hbar^2}{2m}\sum_{i=1}^{N_\1}\nabla^2_i-\frac{\hbar^2}{2m}
\sum_{\alpha=1}^{N_\2}\nabla_\alpha^2
+\sum_{i<j}\frac{d^2}{r^3_{ij}}+\sum_{\alpha<\beta}\frac{d^2}
{r^3_{\alpha\beta}}+\sum_{i\alpha}\frac{d^2(r_{i\alpha}^2-2h^2)}
{(r_{i\alpha}^2+h^2)^{5/2}}\;,
\label{Hamiltonian}
\end{equation}
where $h$ is the distance between the layers.
The terms of the Hamiltonian~(\ref{Hamiltonian}) are
the kinetic energy of $N_\1$ dipoles in the bottom layer and $N_\2$ dipoles
in the top layer; the other two terms correspond
to the intralayer dipolar interactions of $N_A$ and $N_B$ bosons; and
the last accounts for the interlayer interactions. The in-plane
distance between pairs of bosons in the bottom (top) layer is denoted
by $r_{ij(\alpha\beta)}=|{\bf{r}}_{i(\alpha)}-{\bf{r}}_{j(\beta)}|$,
and $r_{i \alpha}=|{\bf{r}}_{i}-{\bf{r}}_{\alpha}|$ stands for
the distance between the projections onto any of the layers of
the positions of the $\alpha$-th and $i$-th particles.
We use the characteristic dipolar length $r_0=md^2/\hbar^2$ and energy
$E_0=\hbar^2/(mr_0^2)$ as units of length and energy, respectively.

Dipoles in the same layer are repulsive, with an interaction decaying 
as $1/r^3$. However, for dipoles in different layers the interaction is 
attractive for small in-plane distance $r$ and repulsive for larger $r$.
In other words, a dipole in the bottom layer induces 
attractive and  repulsive zones for a dipole in the top layer. Importantly, the 
area of the attractive cone increases with the distance between layers $h$, 
making the formation of few-body bound states more efficient.
%%%%%%%%%%%%%%%%%%%%%%%%%%%%%%%%%%%%%%%%%%%%%%%%%%%%%%%%%%%%%%%%
\section{Details of the Methods}
To investigate the structural properties of the dipolar clusters
we use a second-order DMC method (see Chapter~\ref{Chapter:Quantum Monte Carlo methods})
with pure estimators (see Subsection~\ref{Pure Estimators}).
We use the same trial wave function as in Section~\ref{Section:Details of the methods}.
%%%%%%%%%%%%%%%%%%%%%%%%%%%%%%%%%%%%%%%%%%%%%%%%%%%%%%%%%%%%%%%%
\section{Results}
\subsection{Structure of the Bound States}
We first analyze the structure of few-body clusters, composed by up four 
particles, as a function of the interlayer separation $h$. To this end, we 
calculate the pair distribution function $g_{\sigma\sigma^{'}}(r)$, which 
is proportional to the probability of finding two particles at a relative 
distance $r$ (see Subsection~\ref{Sec:PairDistribution}). In the case of the ABB trimers  
and AABB tetramers, we also determine the ground-state density 
distributions for different values of the interlayer separation.
\begin{figure}[t!]
\centering
\includegraphics[width=0.7\textwidth]{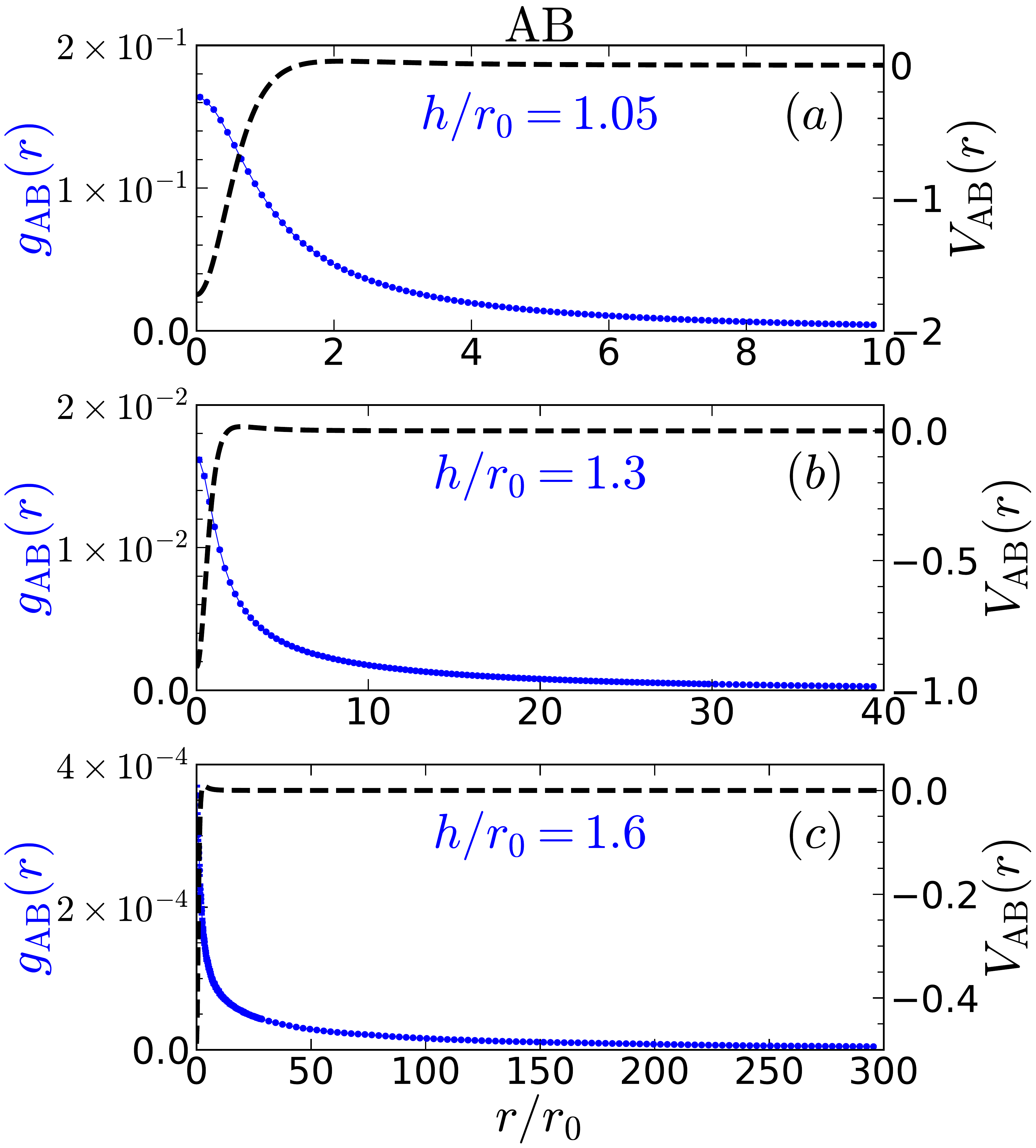}
\caption{
Interlayer pair distributions $g_{\1\2}(r)$ (left) and dipolar potentials
$V_{\1\2}(r)$ (right) for AB and for three values of the interlayer distance
$h/r_0=1.05$, $1.3$ and $1.6$. Notice the different scales in the $x$ axis.}
\label{Fig:grAB}
\end{figure}
\subsubsection{AB Dimer}
The AB dimer is strongly bound for $h\lesssim r_0$ and its energy decays 
exponentially in the opposite limit~\cite{Baranov2011} of large interlayer separation.
In order to understand 
how the cluster size changes with $h/r_0$, we show in 
Fig.~\ref{Fig:grAB} the interlayer pair 
distributions $g_{\1\2}(r)$ (left) and the dipolar potential $V_{\1\2}(r)$ (right),
for three values of $h/r_0$. The strong-correlation 
peak of $g_{\1\2}$ at $r/r_0=0$ is due to the interlayer attraction
$V_{\1\2}(r)$ at short distances. For the 
cases shown in Fig.~\ref{Fig:grAB} we notice that $g_{\1\2}$ are very wide in 
comparison to the dipolar potential and interlayer distance $h/r_0$
reflecting the exponential decay of the bound state. The tail at large distances
becomes longer as the interlayer distance increases.
%%%%%%%%%%%%%%%%%%%%%%%%%%%%%%%%%%%%%%%%%%%%%%%%%%%%%%%%%%%%%%%%
\subsubsection{ABB Trimer}
\begin{figure}[t!]
  \centering
  \subfigure{\includegraphics[width=0.48\textwidth]{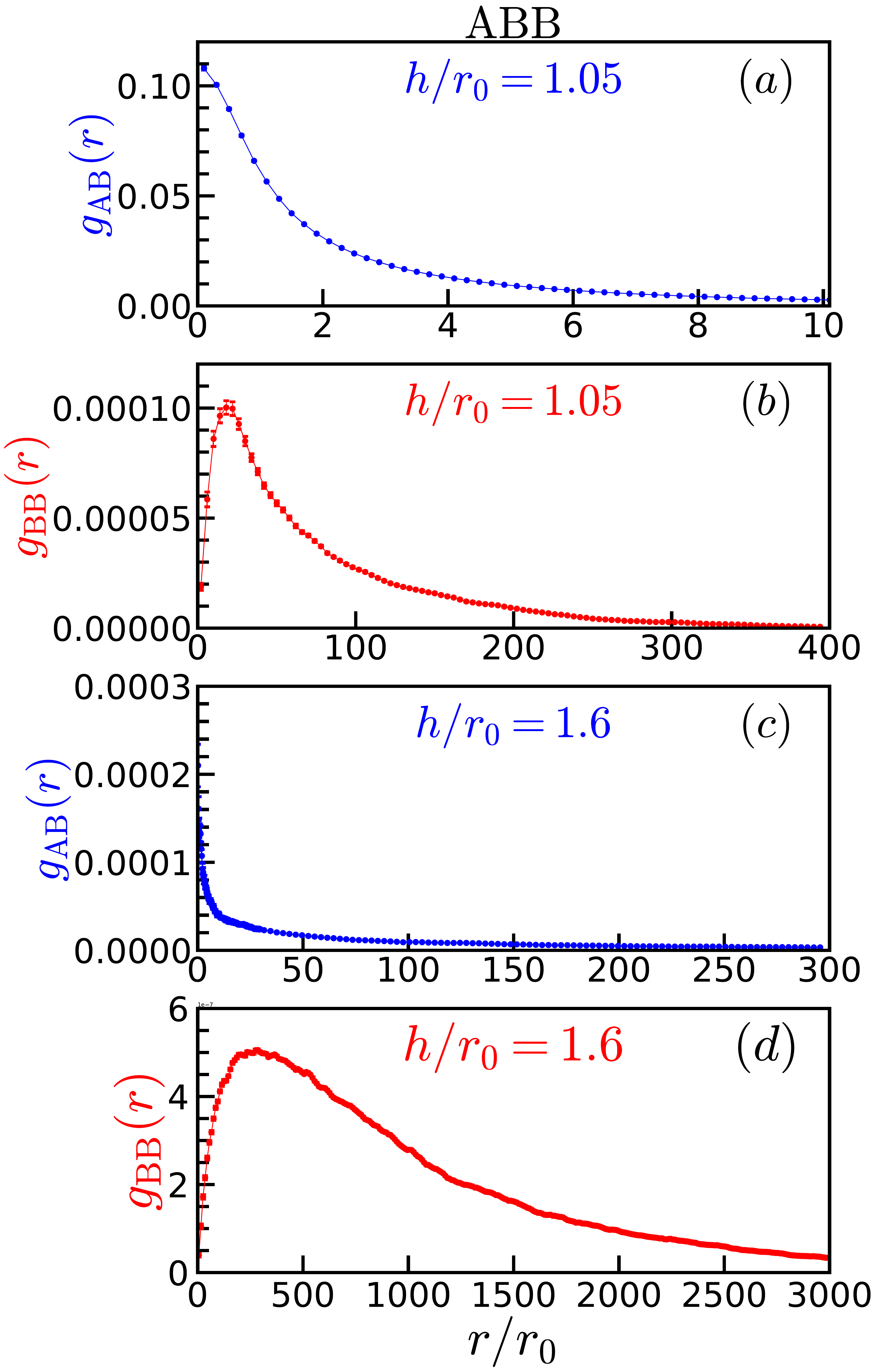}}\quad
  \subfigure{\includegraphics[width=0.48\textwidth]{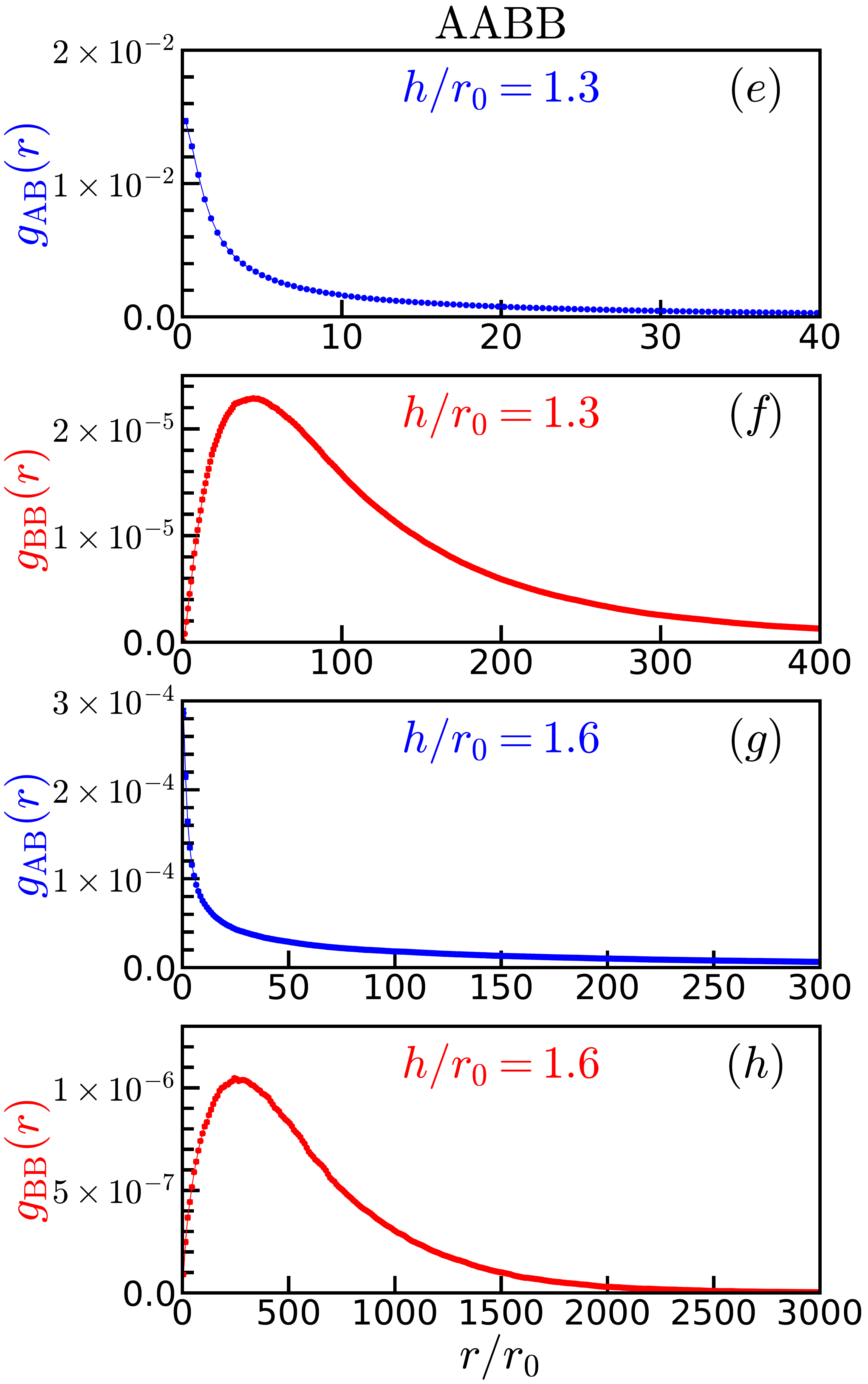}}
\caption{Interlayer and intralayer pair distributions, $g_{\1\2}(r)$
    and $g_{\2\2}(r)$, for ABB (a, b, c, d) and AABB (e, f, g, h) clusters, and for
    different values of
         the interlayer distance $h/r_0$.}  
	\label{Fig:PairDistributions}
\end{figure}

The ABB trimer is bound for large enough separation 
between the layers $h/r_0>0.8$ while, for smaller separations, it breaks 
into a dimer and an isolated atom ~\cite{PhysRevA.101.041602}.  The trimer 
binding  energy is vanishingly small for $h\approx h_c$, with $h_c \simeq 0.8 
r_0$, and it  becomes larger as $h$ is increased, reaching its maximum 
absolute value at $h/r_0\approx  1.05$. Then, it vanishes again in the 
limit of $h\to\infty$ ~\cite{PhysRevA.101.041602}.
We report the intralayer and
interlayer pair distributions, $g_{\2\2}(r)$ and $g_{\1\2}(r)$, respectively, in
Fig.~\ref{Fig:PairDistributions} for strongly- (a, b) and weakly-bound (c, d)
trimers. We observe that the $g_{\1\2}$ distributions are very wide in
comparison to $h$, similarly to what has been observed in
Fig.~\ref{Fig:grAB} for dimers. The same-layer distribution $g_{\2\2}$
vanishes when $r/r_0\to0$ as a consequence of the strongly repulsive dipolar
intralayer potential at short distances. As $r$ increases, $g_{\2\2}$ exhibits
a maximum,
next it monotonically decreases with $r/r_0$. For a weakly-bound trimer
($h/r_0=1.6$), both $g_{\1\2}$ and $g_{\2\2}$ produce long tails at large distances.

\begin{figure}[t!]
  \centering
  \subfigure{\includegraphics[width=0.48\textwidth]{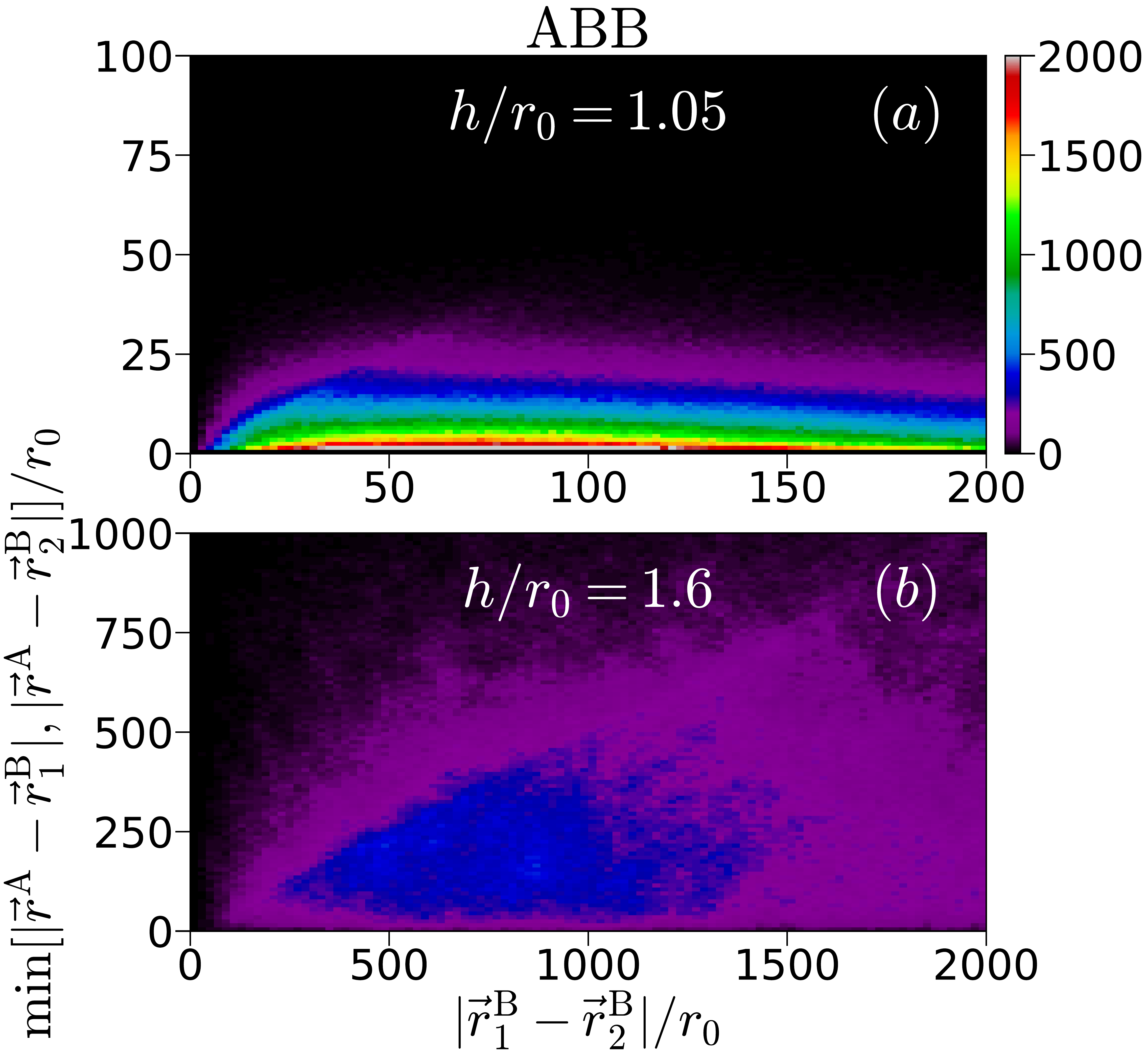}}\quad
  \subfigure{\includegraphics[width=0.48\textwidth]{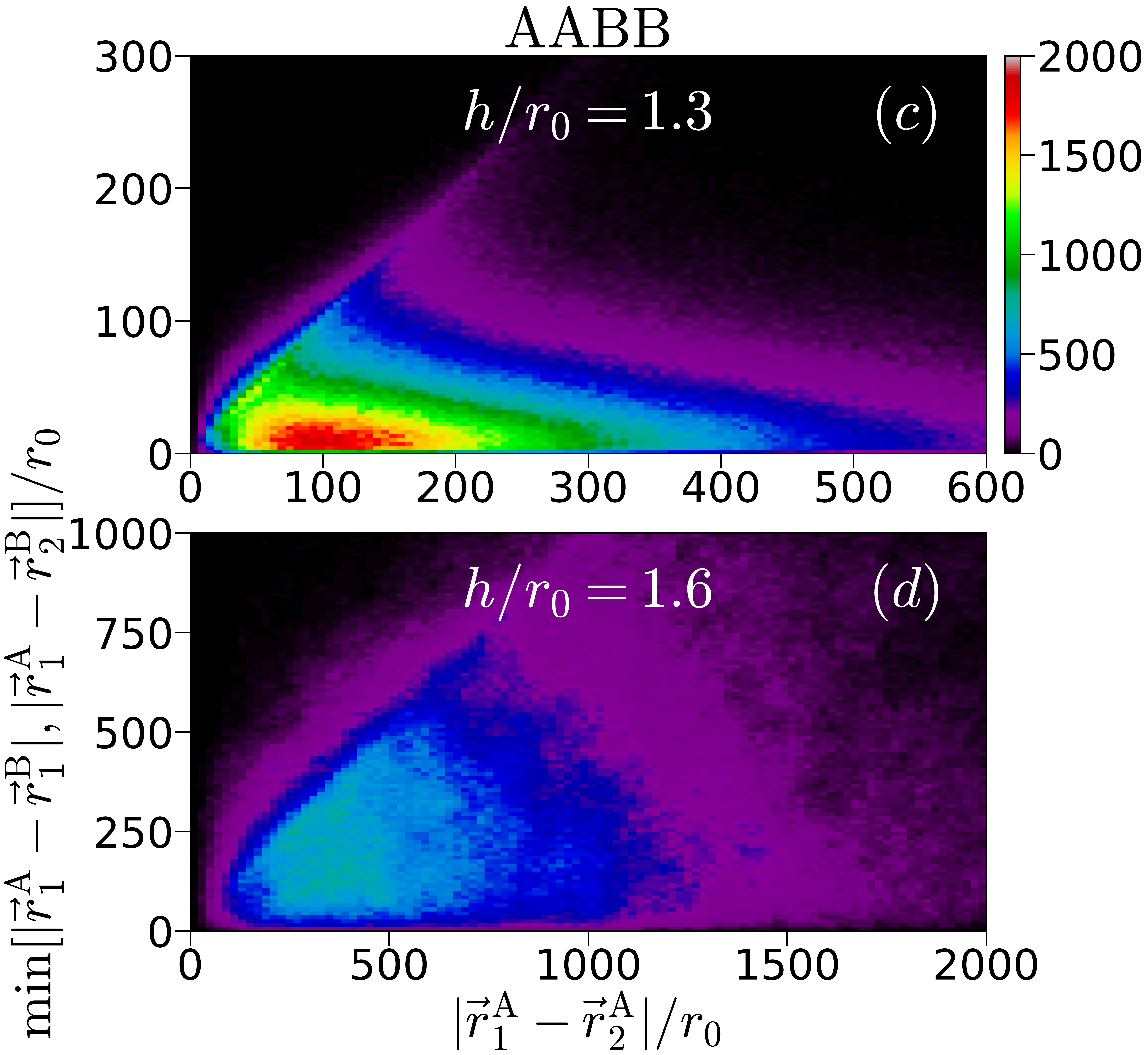}}
  \caption{Heatmap plot representing the spatial structure of the ground state for ABB
      trimer (a, b) and AABB tetramer (c ,d) for different values of the interlayer
      distance. The distance between two dipoles in the same layer is plotted
      in the horizontal axis and the minimum distance between dipoles in different
      layers is shown in the vertical axis. 
}  
\label{Fig:SpatialDistributions}
\end{figure}

The trimer is weakly bound close to the threshold, $h\to h_c$,
and for large interlayer separation, $h\to\infty$, but its
internal structure in those two limits is significantly different.
This can be seen in Fig.~\ref{Fig:SpatialDistributions} (a , b), where we
plot the trimer ground-state spatial distribution for $h/r_0=1.05$ and 1.6.
The heat-map plot is shown as a function of the distance between two dipoles
in the same layer $|\vec{r}_1^{\2}-\vec{r}_2^{\2}|$ (horizontal axis)
and the minimal 3D-distance between dipoles in different layers
$\mathrm{min}\{|\vec{r}_1^{\1}-\vec{r}_1^{\2}|,
|\vec{r}_1^{\1}-\vec{r}_2^{\2}|\}$ (vertical axis).
For large separation between layers, shown in Fig.~\ref{Fig:SpatialDistributions} 
(b) for $h/r_0=1.6$, the distances between AB and BB atoms are all of the same order,
revealing an approximately symmetric structure. However, by decreasing
the distance between layers the particle distribution becomes significantly asymmetric.
For $h/r_0=1.05$ (Fig.~\ref{Fig:SpatialDistributions} (a)), we observe that the trimer 
spatial distribution is elongated:
two dipoles in different layers are close to each other while the third one is far away.
Regardless of the interlayer separation, the pair $\1\2$ is, on average, closer than
the $\2\2$ pair. As the system approaches the threshold value $(h/r_0=0.8)$, the
trimer becomes more extended and eventually breaks into a dimer and an single atom.
%%%%%%%%%%%%%%%%%%%%%%%%%%%%%%%%%%%%%%%%%%%%%%%%%%%%%%%%%%%%%%%%
\subsubsection{AABB Tetramer}
As we have shown in the previous section, an
ABB trimer breaks into a dimer and an atom
when $h\approx h_c$. Here, we address the structure properties of the balanced
case for a tetramer, in which the number of A and B atoms is the same. 
The AABB tetramer is
weakly bound for large values of $h/r_0$. When the distance between layers
decreases, the tetramer becomes unbound at $h/r_0\approx1.1$ and splits into two
AB dimers~\cite{PhysRevA.101.041602}.

The pair distributions $g_{\1\2}$ and $g_{\2\2}$ for AABB are shown in
Fig.~\ref{Fig:PairDistributions} (e, f, g, h) for two characteristic values 
of the interlayer distance
$h/r_0$. We observe a behavior that is similar to that previously reported
for ABB. That is, both $g_{\1\2}$ and $g_{\2\2}$ are compact for the deepest bound
state ($h/r_0=1.3$ for AABB), and become diffuse, showing long tails at large
distances when it turns to a weakly-bound state ($h/r_0=1.6$).   

The ground-state spatial distributions for the symmetric tetramer are shown in
Fig.~\ref{Fig:SpatialDistributions} (c, d). We observe that for large
separation $h$,
i.e., when the tetramer is weakly bound, it
has large spatial extension
and the distances between AA and AB pairs are of the same order. As the
interlayer separation is progressively decreased, the tetramer size decreases and its
structure becomes anisotropic. In this case, the distance between dipoles
in the same layer is several times larger than the distance between dipoles
in different layers. When the tetramer approaches the threshold for
unbinding ($h/r_0\approx1.1$) the cluster becomes even more elongated
and eventually, it breaks into two AB dimers.
%%%%%%%%%%%%%%%%%%%%%%%%%%%%%%%%%%%%%%%%%%%%%%%%%%%%%%%%%%%%%%%%
\subsection{Quantum Halo Characteristics}
A halo is a quantum bound state in which particles have a high probability
to be found in the classically forbidden region, outside the range of the
interaction potential. The key characteristic of a halo is its extended
size and tiny binding energies. To classify a system as a halo state,
one typically introduces two scaling parameters with which the size and
the energy are compared. The first parameter is the scaling length $R$.
For two-body systems, one commonly chooses $R$ as the outer classical
turning point. The second parameter is the scaling energy
$\mu BR^2/\hbar^2$, where $\mu$ is the reduced mass and $B$ is the
absolute value of the ground-state energy of the cluster. The size of
a cluster is usually quantified through its mean-square radius
$\langle r^2 \rangle$, where $r$ is the interparticle distance.
A two-body quantum halo is then defined by the condition
\begin{equation}
    \frac{\langle r^2 \rangle}{R^2}>2,
\end{equation}
which means that the system has a probability to be in the classically
forbidden region larger than 50$\%$.
\begin{figure}[t!]
  \centering
  \includegraphics[width=0.3\textwidth]{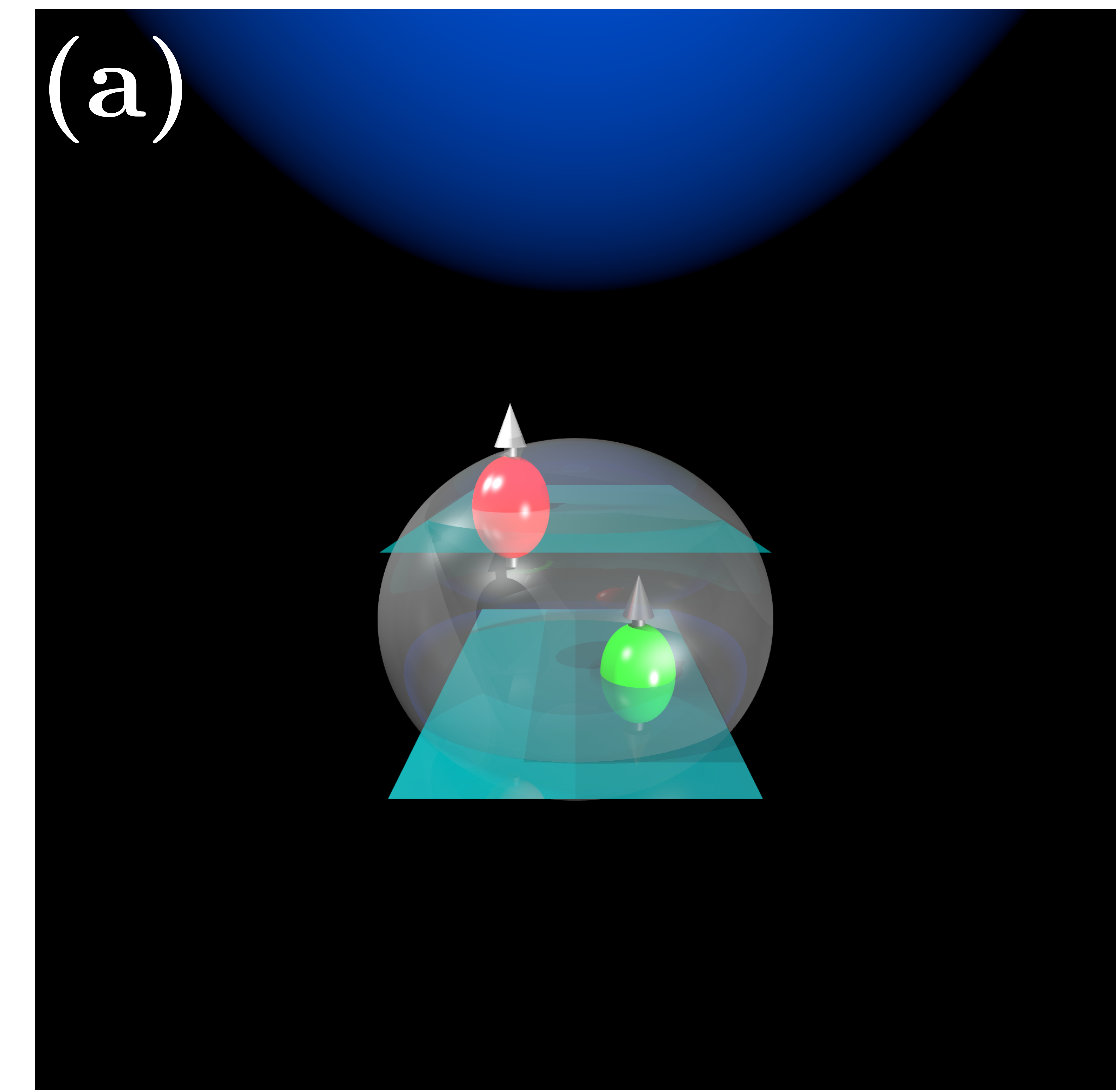}
  \includegraphics[width=0.3\textwidth]{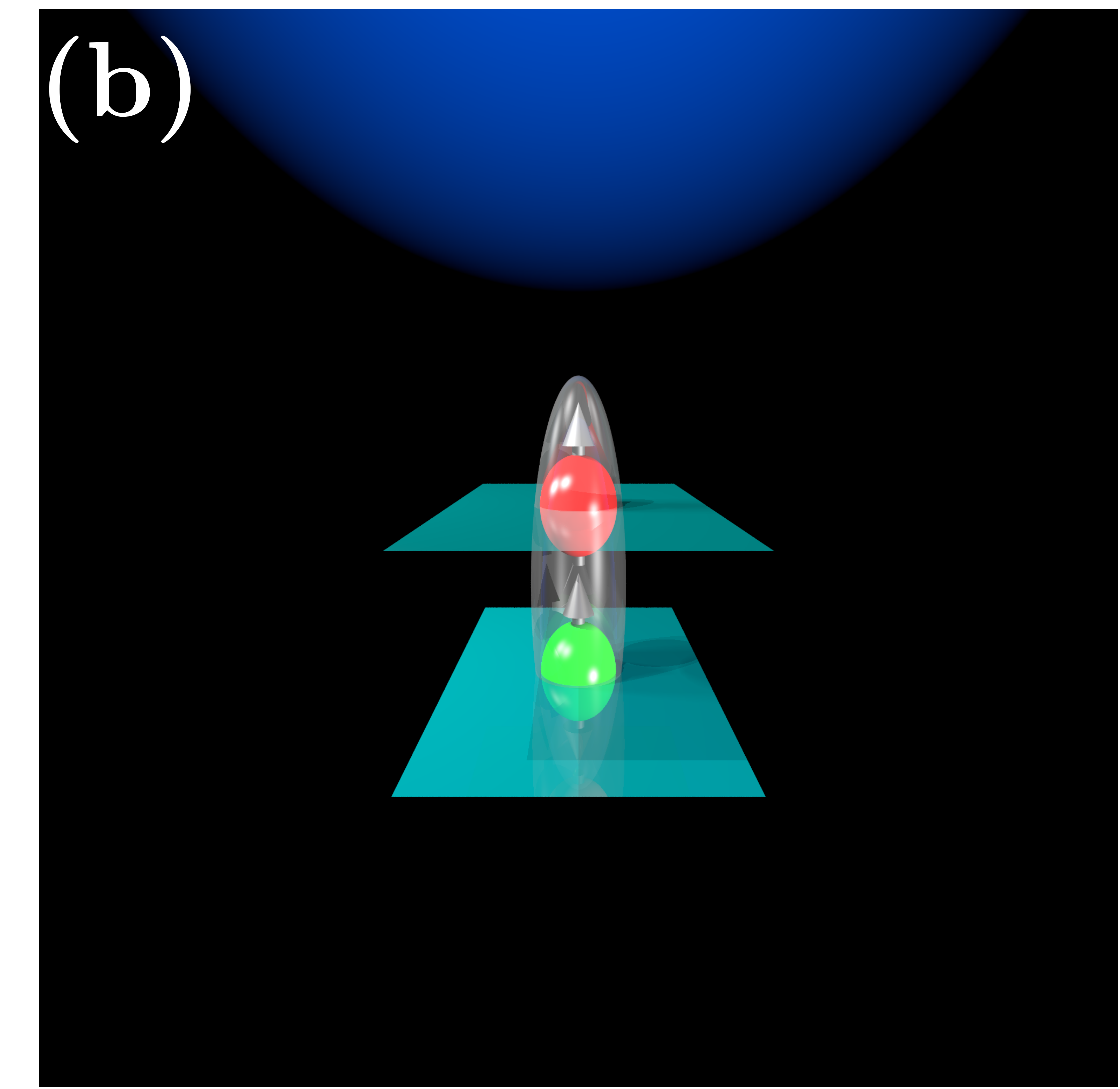}
  \includegraphics[width=0.7\textwidth]{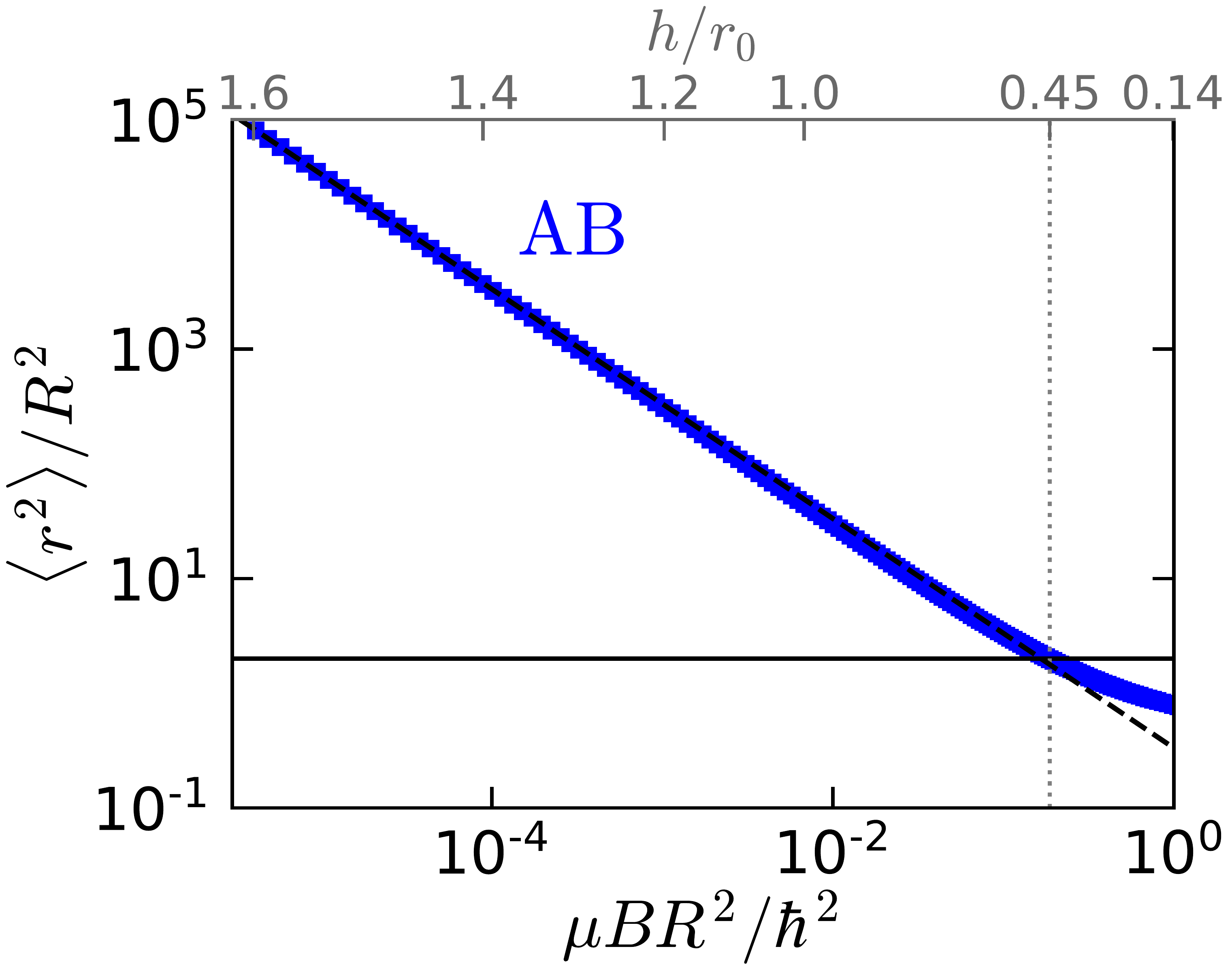}
 \caption{Bottom panel: Size vs ground-state energy scaling plot for two-body halos.
      The horizontal line is the quantum
      halo limit and the dashed one is 
      $\langle r^2\rangle/R^2=\hbar^2/(3\mu BR^2)$, which is a
      zero-range approximation for two-body halos in
      two-dimensions~\cite{Jensen2004}. Top panel: Schematic representation
      of the AB dimer state in two limits: (a) AB is a halo state;
      (b) AB is not a halo state.}  
  \label{Fig:ABHaloStates}
\end{figure}

\begin{figure}[t!]
    \centering
  \includegraphics[width=0.3\textwidth]{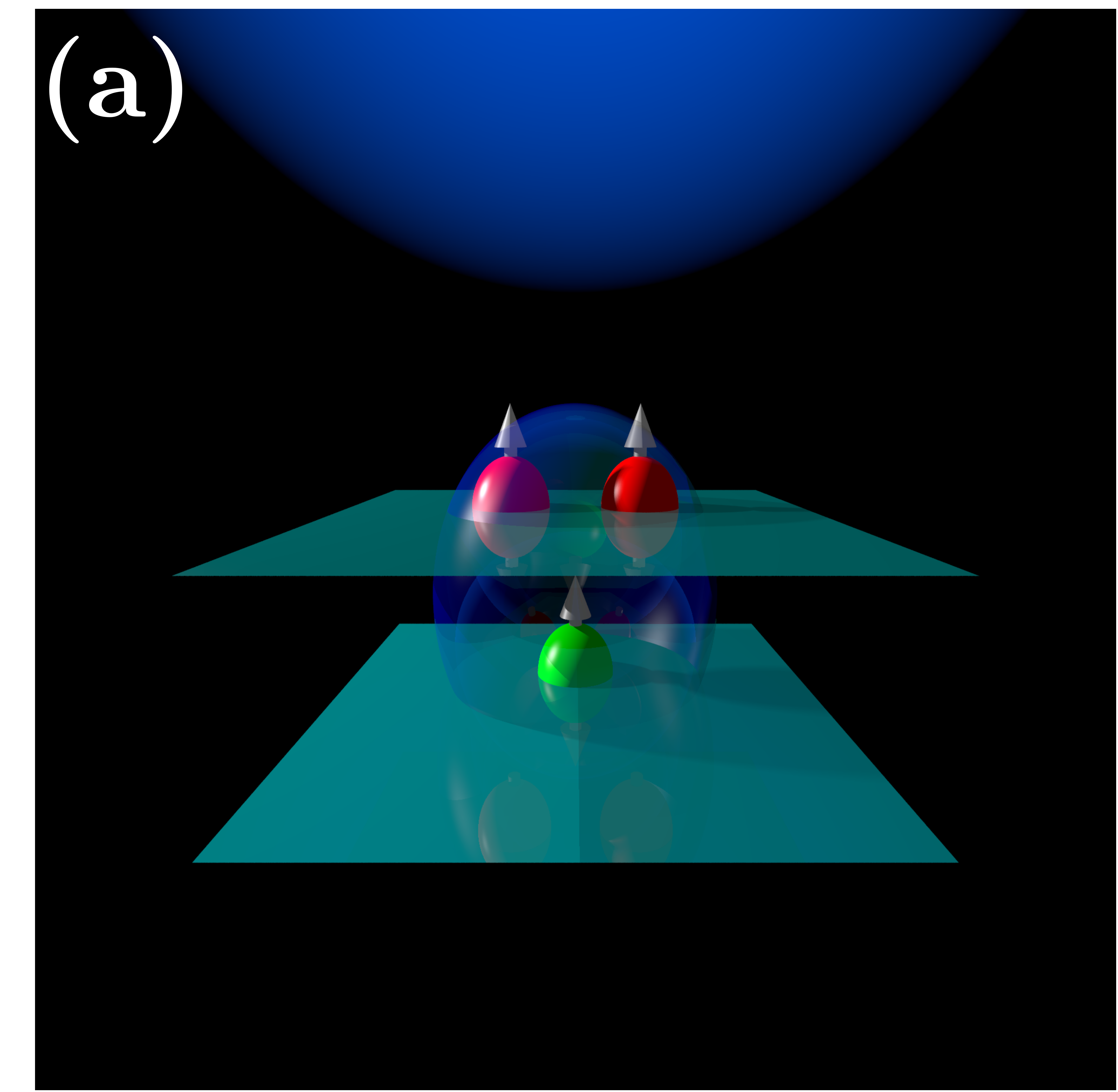}
  \includegraphics[width=0.3\textwidth]{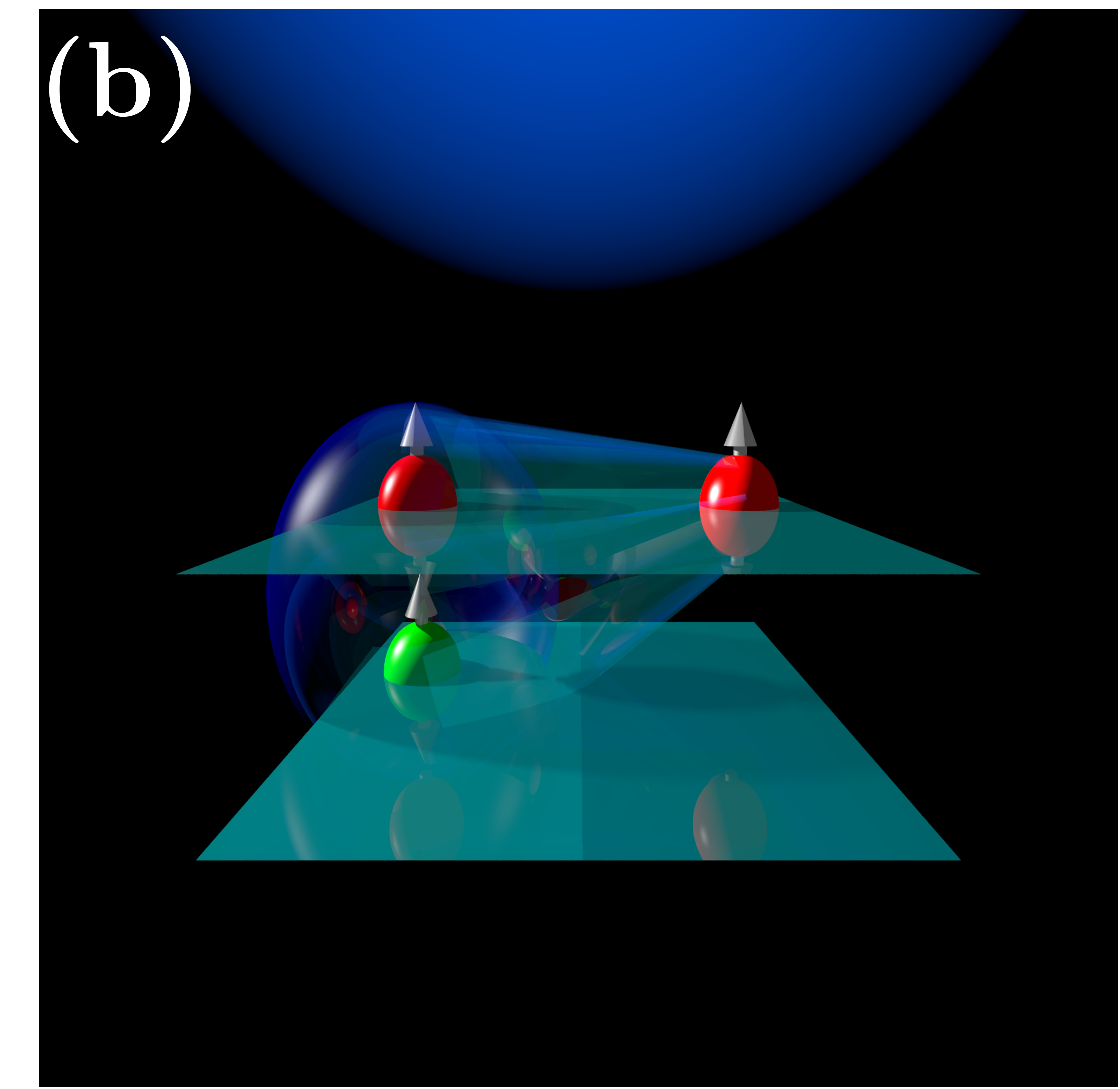}
  \includegraphics[width=0.7\textwidth]{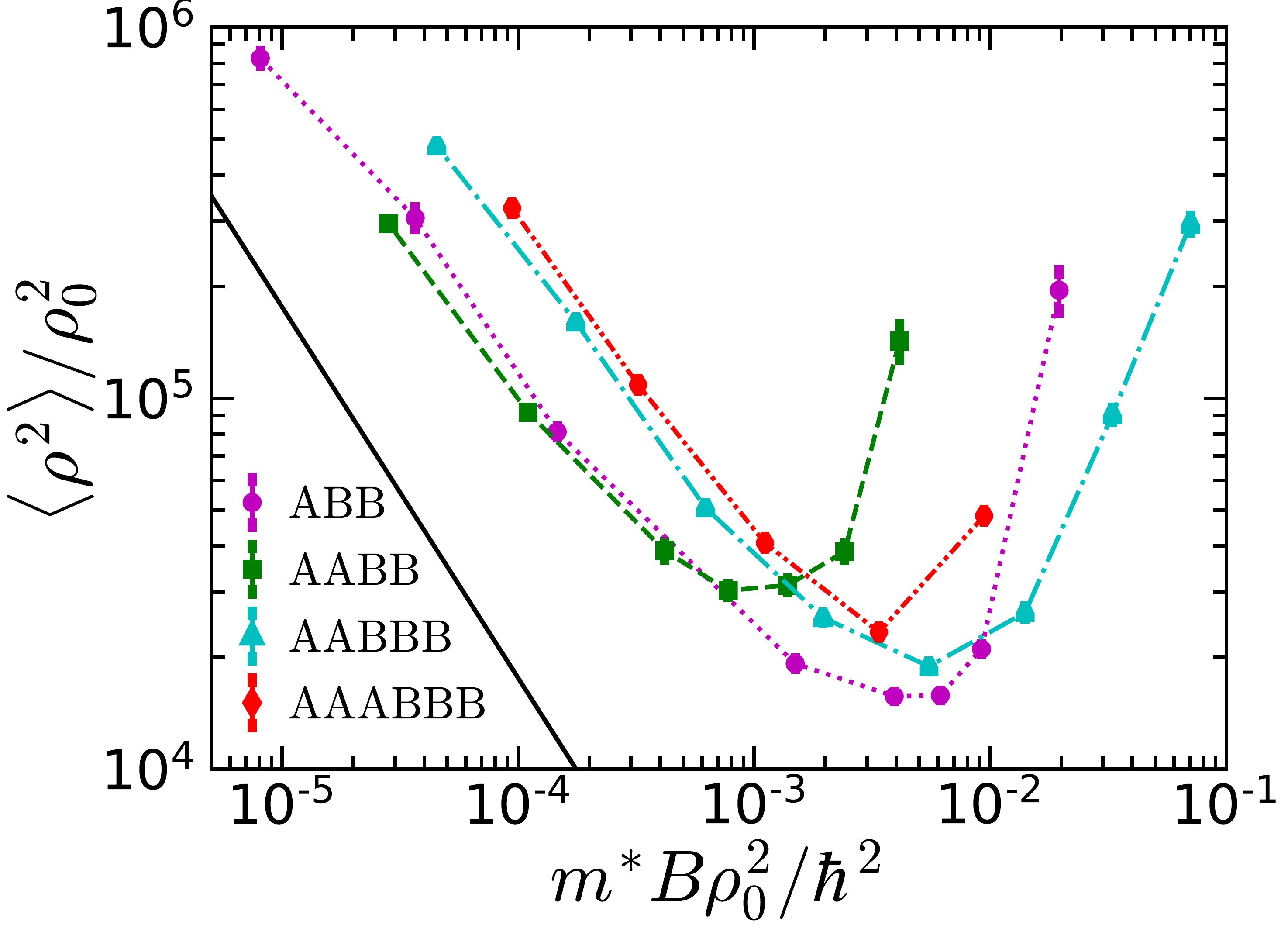}
  \caption{Bottom panel: Size vs ground-state energy scaling plot for
      three- up to six-body halos. The solid line corresponds to
      the case of a ABB trimer with delta interactions and without 
      intraspecies repulsion.
      Top panel: Schematic
      representacion of the ABB trimer state in two limits:
      (a) $h\to \infty$; (b) $h\to h_c$. }  
\label{Fig:HaloStates}
\end{figure}

The dipolar interaction in the bilayer geometry has vanishing Born integral
and thus, the AB dimer can show enhanced halo properties.
In Fig.~\ref{Fig:ABHaloStates}, we show the
scaling plot for the dipolar dimers, corresponding to interlayer
distance from $h/r_0=0.14$ to 1.6, as indicated on the upper axis. 
All dimers which lie above the halo limit $\langle r^2\rangle/R^2=2$
(horizontal line in Fig.~\ref{Fig:HaloStates}) are halo states and
follow a universal scaling law
$\langle r^2\rangle/R^2=\hbar^2/(3\mu BR^2)$,
shown with a dashed line in the figure~\cite{Jensen2004}.
This is exactly the case for all dimers with interlayer separations
$h/r_0 > 0.45$. This threshold value is close to the characteristic value,
$h/r_0=0.5$, for which the dimer binding energy is approximately equal to
the typical energy of the dipolar interaction $E_{\1\2} \approx \hbar^2/(mr_0^2)$.
\begin{figure}[h]
  \centering
  \includegraphics[width=0.7\textwidth]{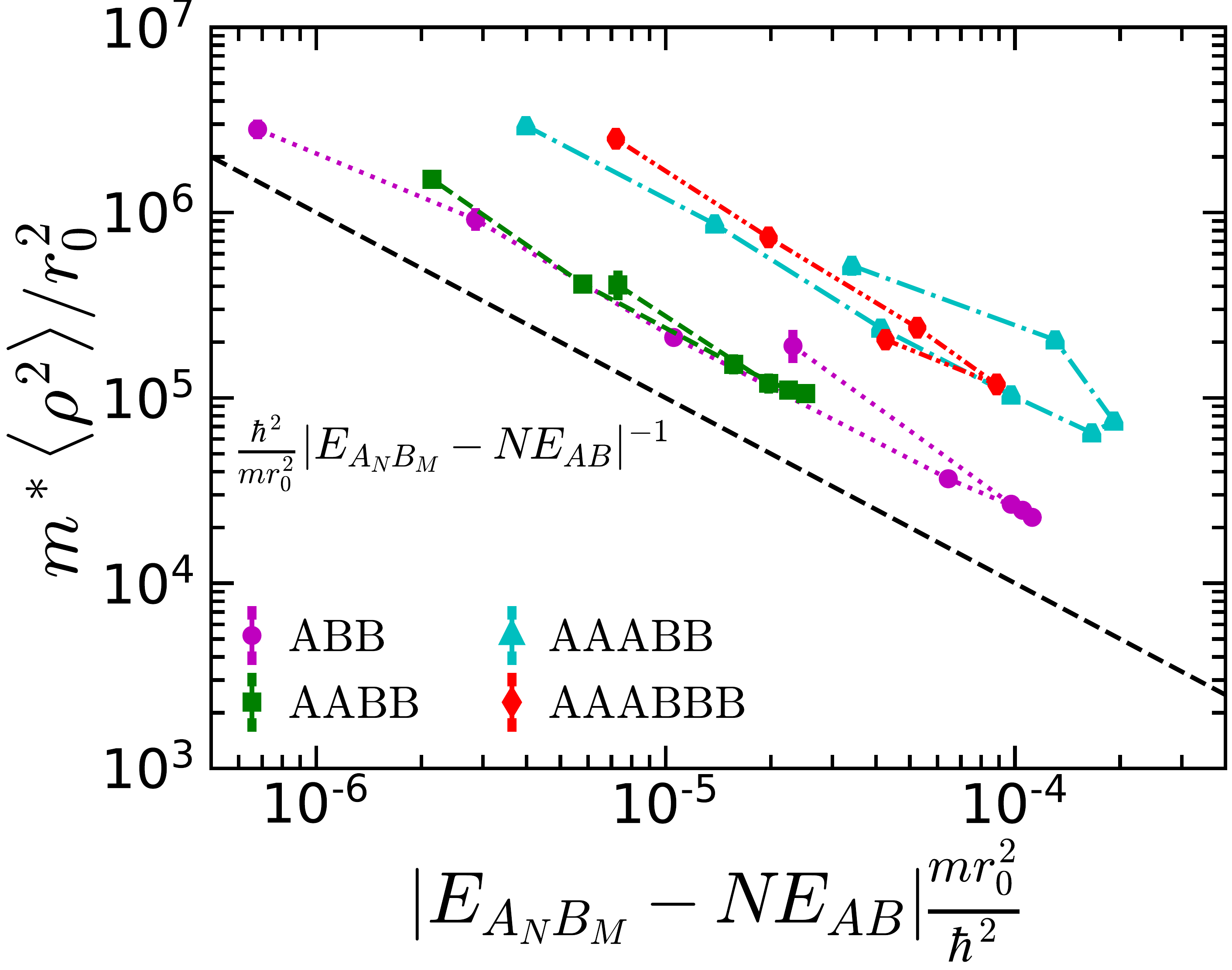}
  \caption{Cluster size vs ground-state binding energy (in dipolar units) for
      three- up to six-body. The dashed line corresponds to 
      $\frac{\hbar^2}{mr_0^2}|E_{\1_N\2_M}-NE_{\1\2}|^{-1}$.}  
  \label{Fig:HaloStatesBindingEnergies}
\end{figure}   

While AB dimers exist for any interlayer separation, ABB trimers and
AABB tetramers
are self-bound  for large $h$ values, where AB dimers are in fact 
halo states. Thus, it can be anticipated that these
few-body bound states are also halos. The sizes of 
three- and four-body systems are
measured in terms of the mean-square hyperradius~\cite{Jensen2004},
\begin{equation}
m^*\rho^2=\frac{1}{M}\sum_{i<k}m_im_k({\bf r_i}-{\bf r_k})^2,
\end{equation}
where $m^*$ is an arbitrary mass unit, $M$ is the total mass of the system,
and $m_i$ is the mass of particle $i$. The scaling size parameter $\rho_0$
is given by
\begin{equation}
m^*\rho^2_0=\frac{1}{M}\sum_{i<k}m_im_kR_{ik}^2,
\end{equation}
with $R_{ik}$ the two-body scaling length of the $i-k$ system, which
is calculated as the outer classical turning point for the $i-k$ potential.
We choose $R_{ik}$ equal
to zero for repulsive potentials. The condition for three- and four-body
quantum halos is now 
\begin{equation}
    \frac{\langle \rho^2 \rangle} { \rho_0^2}>2.
\end{equation}

The dependence of the scaled size on the scaled energy 
for ABB and AABB
are shown in
Fig.~\ref{Fig:HaloStates}.
We find a non-monotonic behavior, in clear contrast with the
dependence observed in the dimer case (see Fig.~\ref{Fig:ABHaloStates}).
That is, the cluster size decreases with increasing energy and reaches a
minimum and then it starts to grow again. The minima correspond to the deepest
bound states~\cite{PhysRevA.101.041602}. This 
resurgence appears
as the clusters approach to the thresholds, where trimers eventually break into
a dimer and an atom, and tetramers into two dimers.  
We want to emphasize that all the trimers and tetramers analized in
Fig.~\ref{Fig:HaloStates} are halo states, although they are organized in significantly 
different spatial structures. On the left side of the minima, the clusters are almost
radially symmetric and all the interparticle distances are of the same order. 
However, at the minima and on the right
side of the minima, the cluster structures are elongated and highly asymmetric.

The solid line in Fig.~\ref{Fig:HaloStates} corresponds to the case of a
trimer with contact interactions and without intraspecies repulsion.
We observe that the size of the trimer with contact interactions is much smaller than
the dipolar trimer, 
which indicates that repulsion has an important role in the size and energy of the clusters.
In the limit of exponentially small energy, large interlayer distance, the dipolar trimer
should follow the solid line.

The AABBB pentamer and AAABBB hexamer are self-bound and are manifestly halo states.
Their mean square size has a similar behavior to
the one observed before for the trimer and tetramer, that is a minimum 
corresponding to the larger binding energy which separates a regime of nearly 
symmetric particle distribution form another one, more elongated, and thus 
asymmetric. 
\begin{figure}[h]
  \centering
  \includegraphics[width=0.7\textwidth]{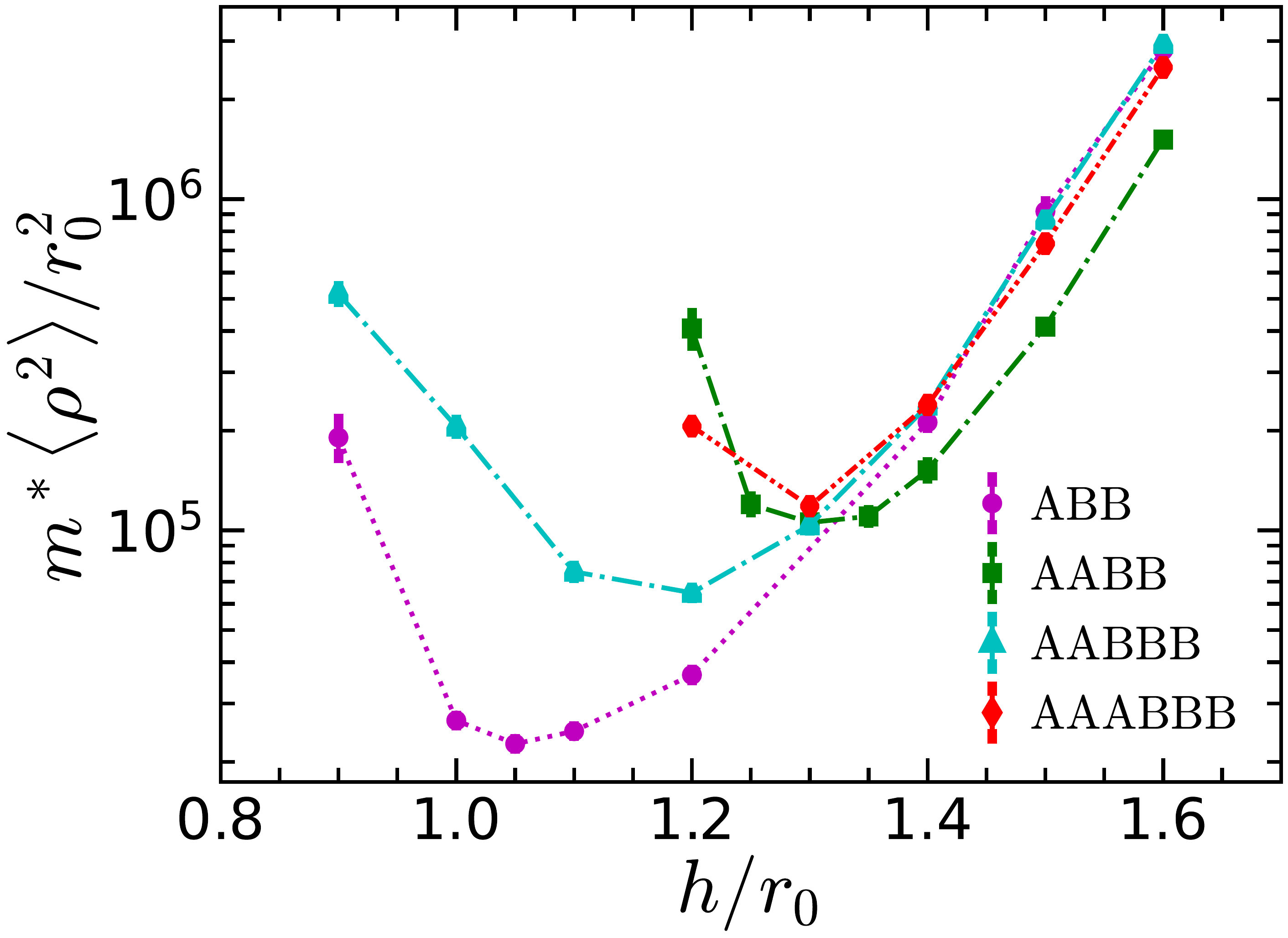}
  \caption{Cluster size (in dipolar units) vs $h/r_0$ interlayer distance for
      three- up to six-body halos.}  
  \label{Fig:HaloStatesvsh}
\end{figure}   

In Fig.~\ref{Fig:HaloStatesBindingEnergies} we plot the cluster size
$m^*\langle \rho^2 \rangle/r_0^2$ as a function of the binding energy
$|E_{A_NB_M}-NE_{AB}|\frac{mr_0^2}{\hbar^2}$ in dipolar units.
We notice that the cluster size is a function of the binding energy
$m^*\langle \rho^2 \rangle/r_0^2 \sim|E_{A_NB_M}-NE_{AB}|^{-1}$,
as indicated by the black dashed line,
and not of the total energy.

Until now, we have shown the size of the clusters as a function of the
total and binding energies. Another possibility is to plot the
size of the clusters as a function of the interlayer separation 
$h/r_0$, as we can see in Fig.~\ref{Fig:HaloStatesvsh}.
Here, for small values of $h/r_0$ we see a clear separation between
population-imbalanced ($M\neq N$) and population-balanced clusters ($M=N$).
This is a direct consequence of what we found in the previous chapter.
That is, in Chapter~\ref{Chapter:Few-body bound states of two-dimensional bosons}
we find that the few-body bound states in the bilayer geometry have two
unbinding thresholds, depending on whether they are balanced or not.
The first one is at $h/r_0\approx 0.8$ for population-imbalanced ABB trimer and
AABBB pentamer. The second one is at $h/r_0\approx 1.1$ for population-balanced
AABB tetramer and AAABBB hexamer.
%%%%%%%%%%%%%%%%%%%%%%%%%%%%%%%%%%%%%%%%%%%%%%%%%%%%%%%%%%%%%%%%
\section{Summary}
We used the diffusion Monte Carlo method to study the ground-state properties of
few-body dipolar bound states in a two-dimensional bilayer setup. 
We have studied 
clusters composed by up to six particles,
for different values of the interlayer distance, as candidates for quantum halo states. 
In the case of dimers, we find that for values of the interlayer separation
larger than $h/r_0 = 0.45$ the clusters are halo states and they
follow a universal scaling law.
In the cases of trimers up to hexamers, we find two very
different halo structures. 
For large values of the interlayer separation, the halo structures are almost radially
symmetric and the distances between dipoles are all of the same scales. In contrast,
in the vicinity of the threshold for unbinding, the clusters are elongated and highly anisotropic.
Importantly, our results prove the existence of stable halo states composed 
of up to six particles. 
To the best of our knowledge this is the first time that
halo states with such a large number of particles are predicted and observed in 
a numerical experiment. Indeed, the addition of particles to a two or 
three body halo states typically makes them shrink towards a more compact liquid 
structure.
This particular bilayer geometry is the reason of our distinct results.
We hope that these results will stimulate experimental activity in this setup,
composed by atoms with dominant dipolar interaction, to bring evidence of these
remarkably quantum halo states.

\chapter{Quantum Liquid of Two-Dimensional Dipolar Bosons}\label{Chapter:Two-dimensional dipolar liquid}
In this chapter, we investigate the ground-state phase diagram of two-dimensional dipoles
confined to a bilayer geometry by using many-body quantum Monte Carlo methods. 
The dipoles are considered to be aligned perpendicularly to the parallel layers.
We find a rich phase diagram that contains quantum phase transitions between liquid, solid,
atomic gas, and molecular gas phases. We predict the formation of a novel liquid phase in which
the bosons interact via purely dipolar potential and no contact potential is required to
stabilize the system. The liquid phase, which is formed due to the balance between an
effective dimer-dimer attraction and an effective three-dimer repulsion, is manifested by
the appearance of a minimum in the equation of state. The equilibrium density is given by
the position of the minimum of the energy and it can be controlled in a wide range by the interlayer
distance. From the equation of state, we extract the spinodal density, below which the
homogeneous system experiences a negative pressure and breaks into droplets. Our results
offer a new example of a two-dimensional interacting dipolar liquid in a clean and highly
controllable setup.
%%%%%%%%%%%%%%%%%%%%%%%%%%%%%%%%%%%%%%%%%%%%%%%%%%%%%%%%%%%%%%%%
\section{Introduction}
Quantum liquids are self-bound systems, in which competing repulsive and attractive
interparticle interactions mechanically balance the system. The effects of quantum
mechanics and quantum statistics, such as the indistinguishability of elementary
particles, play an important role~\cite{leggett_2015} in the description of these systems. 
One of the most known quantum liquids is superfluid Helium, which played a revolutionary
role in low-temperature quantum physics. The interaction between Helium atoms is
characterized by a finite-range potential that has two main features. At large distances,
the potential has an attractive long-range van der Waals tail that tends to hold the atoms together.
Instead, at short distances, the potential has a strongly repulsive core that prevents the
liquid from collapse.

Recent experiments, motivated by a previuos theoretical proposal~\cite{PhysRevLett.115.155302},
have enabled the experimental observation of a qualitatively different type of liquid, 
quantum droplets in a mixture of Bose-Einstein condensates~\cite{Cabrera2017,Semeghini2018,Ferioli2019}
and in dipolar bosonic gases~\cite{Kadau2016,Schmitt2016,Ferrier2016,Chomaz2016}. These quantum
droplets are self-bound clusters of atoms possessing a density that is several orders of magnitude
more dilute than liquid Helium. Both in two-component and dipolar droplets, the system collapses
according to the mean-field theory, thus the stability of the liquid state is a genuinely quantum
many-body effect.

Dipolar liquids were first experimentally observed. However, the precise description
of the system is complicated because their stability is an
interplay between dipolar attraction and short-range repulsion. 
Therefore, there is a strong dependence on the short-range details of the
interaction potential. 
Instead, dipoles in a bilayer geometry may serve as a simpler and
cleaner system in which no short-range repulsion needs to be used. If the dipolar moments of
the bosons are oriented perpendicularly to the parallel layers, there is a competing effect
between repulsive intralayer and partially attractive interlayer interactions, which can produce
interesting few- and many-body states, in particular liquids. For example, a solid and a pair
superfluid phases were characterized in Refs.~\cite{Macia2014,PhysRevA.94.063630} using exact
Monte Carlo simulations.

Our results in Chapter~\ref{Chapter:Few-body bound states of two-dimensional bosons}
for few-body bound states of dipolar bosons confined to a bilayer geometry
predicts that a dimer-dimer attraction plus an effective three-dimer repulsion can stabilize a
many-body liquid state. It is therefore an open challenge to determine
the existence, formation mechanism, and properties of the self-bound many-body system
in this geometry.

In this chapter, we study a two-dimensional system of dipolar bosons confined to the bilayer geometry. 
We calculate the ground-state phase diagram as a function of the density and the separation
between layers by using exact quantum Monte Carlo methods. The key result of our work is the
prediction of a homogeneous liquid in this system. The liquid is stable
in a wide range of densities and interlayer values. We find that the critical interlayer separation
at which the liquid to gas transition happens is the same as the threshold value at which the
effective interaction between dimers changes from attractive to repulsive and a tetramer is formed
in a four-body problem. We characterize the liquid by calculating its equation of state, the
condensate fraction, and the equilibrium and spinodal densities.
%%%%%%%%%%%%%%%%%%%%%%%%%%%%%%%%%%%%%%%%%%%%%%%%%%%%%%%%%%%%%%%%
\section{The Hamiltonian}
We consider $N$ bosons of mass $m$ and dipole moment $d$ confined to two parallel layers separated
by a distance $h$. It is assumed that the dipolar moment of each boson is aligned perpendicularly
to the planes by the external field. Also, we suppose that the confinement to each plane is so
tight that there is no interlayer tunneling and no excitations of the higher levels of the tight
confinement are possible. The Hamiltonian of this system is given by
\begin{equation}
H=-\frac{\hbar^2}{2m}\sum_{i=1}^{N_\1}\nabla^2_i-
\frac{\hbar^2}{2m}\sum_{\alpha=1}^{N_\2}\nabla_\alpha^2
+\sum_{i<j}\frac{d^2}{r^3_{ij}}+\sum_{\alpha<\beta}
\frac{d^2}{r^3_{\alpha\beta}}+\sum_{i\alpha}
\frac{d^2\left(r_{i\alpha}^2-2h^2\right)}{\left(r_{i\alpha}^2+h^2\right)^{5/2}},
\label{Eq:6.1}
\end{equation} 
where Latin (Greek) indices run over each of $N_\1$ ($N_\2$) dipoles in the bottom (top) layer. 
The first two terms in the Hamiltonian~(\ref{Eq:6.1}) correspond to the boson kinetic energy,
the next two terms are the intralayer dipolar interaction, which is always repulsive and falls off
with a power-law $1/r^3$. The last term is the interlayer potential which is attractive for small
values of $r$ but repulsive for large values of $r$, where $r$ is the in-plane distance between
dipoles. The interlayer potential always supports at least one dimer state. The binding energy
diverges for $h\to 0$ and exponentially vanishes in the opposite 
limit~\cite{Shih2009,Armstrong2010,Klawunn2010,Baranov2011}. The dipolar length
$r_0=md^2/\hbar^2$ is used as a unit of length.
%%%%%%%%%%%%%%%%%%%%%%%%%%%%%%%%%%%%%%%%%%%%%%%%%%%%%%%%%%%%%%%%
\section{Details of the Methods}
To investigate the ground-state properties of 
Hamiltonian~(\ref{Eq:6.1}) we employ the diffusion Monte Carlo 
(DMC) method, which was explained in
Chapter~\ref{Chapter:Quantum Monte Carlo methods}. 
In this work, we use two guiding trial wave functions, the first one 
describes two-body correlations between individual dipoles
\begin{equation}
    \Psi_{\rm J}(\mathbf{r}_1,\dots,\mathbf{r}_N)=\prod_{i<j}f_{\1\1}(r_{ij})
\prod_{\alpha<\beta}f_{\2\2}(r_{\alpha\beta})
\prod_{i,\alpha}f_{\1\2}(r_{i\alpha}),
\label{Eq:6.2}
\end{equation}
and the second one takes into account a possible formation of AB dimers,
\begin{equation}
\begin{aligned}
    \Psi_{\rm S}(\mathbf{r}_1,\dots,\mathbf{r}_N)&=\prod_{i<j}^{N_\1}f_{\1\1}(r_{ij})
\prod_{\alpha<\beta}^{N_\2}f_{\2\2}(r_{\alpha\beta})\\
&\times\Bigg[\prod_{i=1}^{N_\1}\sum_{\alpha=1}^{N_\2}f_{\1\2}(r_{i\alpha})+
\prod_{\alpha=1}^{N_\2}\sum_{i=1}^{N_\1}f_{\1\2}(r_{i\alpha})\Bigg].
\label{Eq:6.3}
\end{aligned}
\end{equation}
Both choices result in a comparable DMC energy, while the convergence is different.

\begin{figure}[t!]
	\centering
    \includegraphics[width=0.7\textwidth]{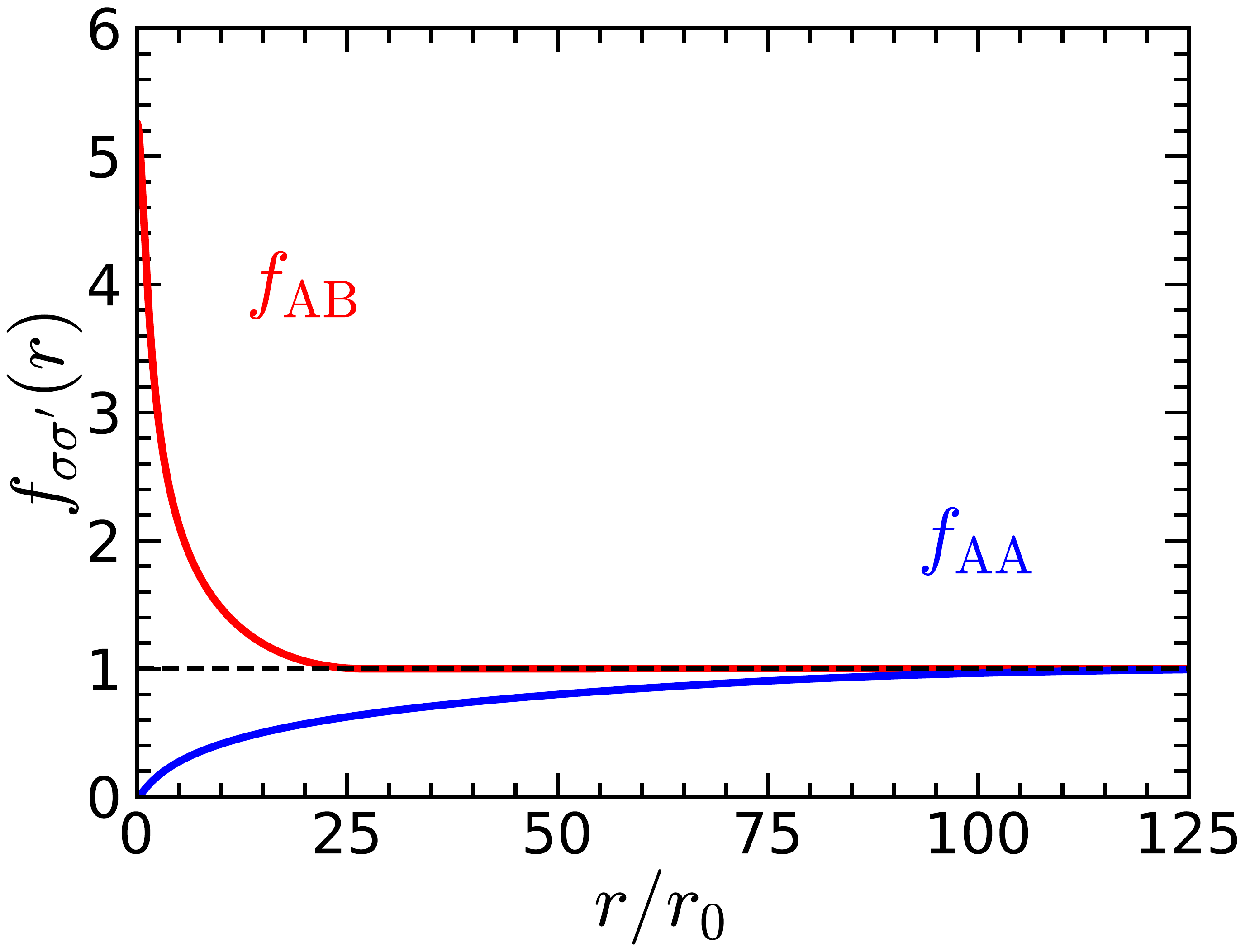}
    \caption{Intraspecies $f_{\1\1}(r)$ and interspecies $f_{\1\2}(r)$ wave
        functions.}
    \label{Fig:Ch6_WaveFunction}
\end{figure}

The intraspecies correlations at short distances, $r<R_{\rm match}$, are modeled by
the zero-energy two-body scattering solution
\begin{equation}
    f_{\1\1}(r)=f_{\2\2}(r)=C_0 K_0\left(2\sqrt{r_0/r}\right),
\end{equation}
with $K_0(r)$ the modified Bessel function and $R_{\rm match}$ a 
variational parameter~\cite{PhysRevLett.98.060405}. For distances
larger than $R_{\rm match}$ we choose 
\begin{equation}
    f_{\1\1}(r)=f_{\2\2}(r)=C_1 \text{exp}\left[ -\frac{C_2}{r} - \frac{C_2}{L-r}\right],
\end{equation}
which describes phonons~\cite{PhysRev.155.88}. The constants $C_0$, $C_1$ and $C_2$ are fixed
by imposing continuity of the function and its first derivative at the
matching distance $R_{\rm match}$, and also that $f_{\1\1}(L/2)=1$,
with $L$ the length of the squared simulation box.
The interspecies correlations $f_{\1\2}(r)$ are taken as 
the solution of the two-body problem
up to $R_1$ and imposing the boundary condition
$f^{'}_{\1\2}(R_1)=0$. For distances larger than the variational
parameter $0<R_1<L/2$ we set $f_{\1\2}(r)=1$.
In Fig.~\ref{Fig:Ch6_WaveFunction} we show the intraspecies $f_{\1\1}(r)$ and
interspecies $f_{\1\2}(r)$ wave functions.

For simplicity, we assume a population-balanced system $N_\1=N_\2=N/2$,
where $N$ is the total number of dipoles. In order to approximate the properties
of extended systems, we perform the simulations in a square box with side length
$L$ and impose periodic boundary conditions. The total density of the system
is defined as $n=N/L^2$. 
%%%%%%%%%%%%%%%%%%%%%%%%%%%%%%%%%%%%%%%%%%%%%%%%%%%%%%%%%%%%%%%%
\section{Results}
%%%%%%%%%%%%%%%%%%%%%%%%%%%%%%%%%%%%%%%%%%%%%%%%%%%%%%%%%%%%%%%%
\subsection{Finite Size Effects}
The dipolar potential is quasi-long ranged in two dimensions,
therefore its truncation at $L/2$ produces large finite-size corrections.
It is possible to diminish them by adding the contribution of the tail
energy to the output of the DMC energy.

For a bilayer system of dipoles the potential energy tail is given by the expression
\begin{equation}
\begin{aligned}
    E_{\rm tail}=&\frac{1}{2} L^2\int_{L/2}^{\infty}V_{\1\1}(r)g_{\1\1} (r)2\pi rdr 
    +\frac{1}{2} L^2\int_{L/2}^{\infty}V_{\2\2}(r)g_{\2\2} (r)2\pi rdr \\
    &+ L^2\int_{L/2}^{\infty}V_{\1\2}(r)g_{\1\2} (r)2\pi rdr, 
\label{Eq:6.4}
\end{aligned}
\end{equation}
we denote by $g_{\1\1}=g_{\2\2}$ and
$g_{\1\2}$ the intraspecies and interspecies pair distribution functions,
respectively. Substituting the interaction potential expressions 
\begin{equation}
    V_{\1\1}(r)=V_{\2\2}(r)=\frac{d^2}{r^3} \quad \mbox{and}
    \quad V_{\1\2}(r)=\frac{d^2\left(r^2-2h^2\right)}{\left(r^2+h^2\right)^{5/2}},
\label{Eq:6.5}
\end{equation}
in Eq.~(\ref{Eq:6.4}) we obtain
\begin{equation}
    \frac{E_{\rm tail}(n,L)}{L^2}=\int_{L/2}^{\infty} \bigg[ 
\frac{d^2}{r^3}g_{\1\1}(r)+\frac{d^2}{r^3}g_{\2\2}(r)+ 
2\frac{d^2\left(r^2-2h^2\right)}{\left(r^2+h^2\right)^{5/2}}g_{\1\2}(r) \bigg]\pi rdr.
\label{Eq:6.6}
\end{equation}
An approximate value of Eq.~(\ref{Eq:6.6}) is obtained by substituting
$g_{\1\1}(r)\to n_{\1}^2$, $g_{\2\2}(r)\to n_{\2}^2$ and
$g_{\1\2}(r)\to n_{\1}n_{\2}$, which leads to
\begin{equation}
    E_{\rm tail}=2\pi d^2L\left[ n_\1^2 + n_\2^2 + 
    \frac{2n_\1n_\2L^3}{\left(4h^2+L^2\right)^{3/2}}\right].
\label{Eq:6.7}
\end{equation}
The total density of the system is $n=N/L^2$, with $N=N_\1 +N_\2$
the total number of particles. The density of each component is one
half of the total density, $n_\1=n_\2=n/2$. We now substitute these
relations into Eq.~(\ref{Eq:6.7})  
\begin{equation}
    \frac{E_{\rm tail}}{N}=\frac{\pi d^2 n^{3/2}}{\sqrt{N}}+\frac{\pi d^2N}
{\left(4h^2+\frac{N}{n}\right)^{3/2}}.
\label{Eq:6.8}
\end{equation}
In units of $\hbar^2/mr_0^2$ we get
\begin{equation}
    \frac{E_{\rm tail}}{N}\frac{mr_0^2}{\hbar^2}= 
\frac{\pi \left(nr_0^2\right)^{3/2}}{2\sqrt{N}} + \frac{\pi N}{\left[4\left(\frac{h}{r_0}\right)^2+\frac{N}{nr_0^2}\right]^{3/2}}.
\label{Eq:6.9}
\end{equation}
This significantly reduces the finite size dependence on the simulation box length
in the energy, while the residual dependence is eliminated by a fitting procedure. 

After adding the tail energy $E_{\rm tail}$ to the DMC energy $E_{\rm DMC}$,
we extrapolate the energy $E(N)=E_{\rm DMC}+E_{\rm tail}$ to the thermodynamic limit
value $E_{\rm th}$ using the fitting formula
\begin{equation}
    E(N)=E_{\rm th}+\frac{C}{N^{1/2}}, 
\label{Eq:6.10}
\end{equation}
where C is a fitting parameter.

\begin{figure}[t!]
	\centering
    \includegraphics[width=0.7\textwidth]{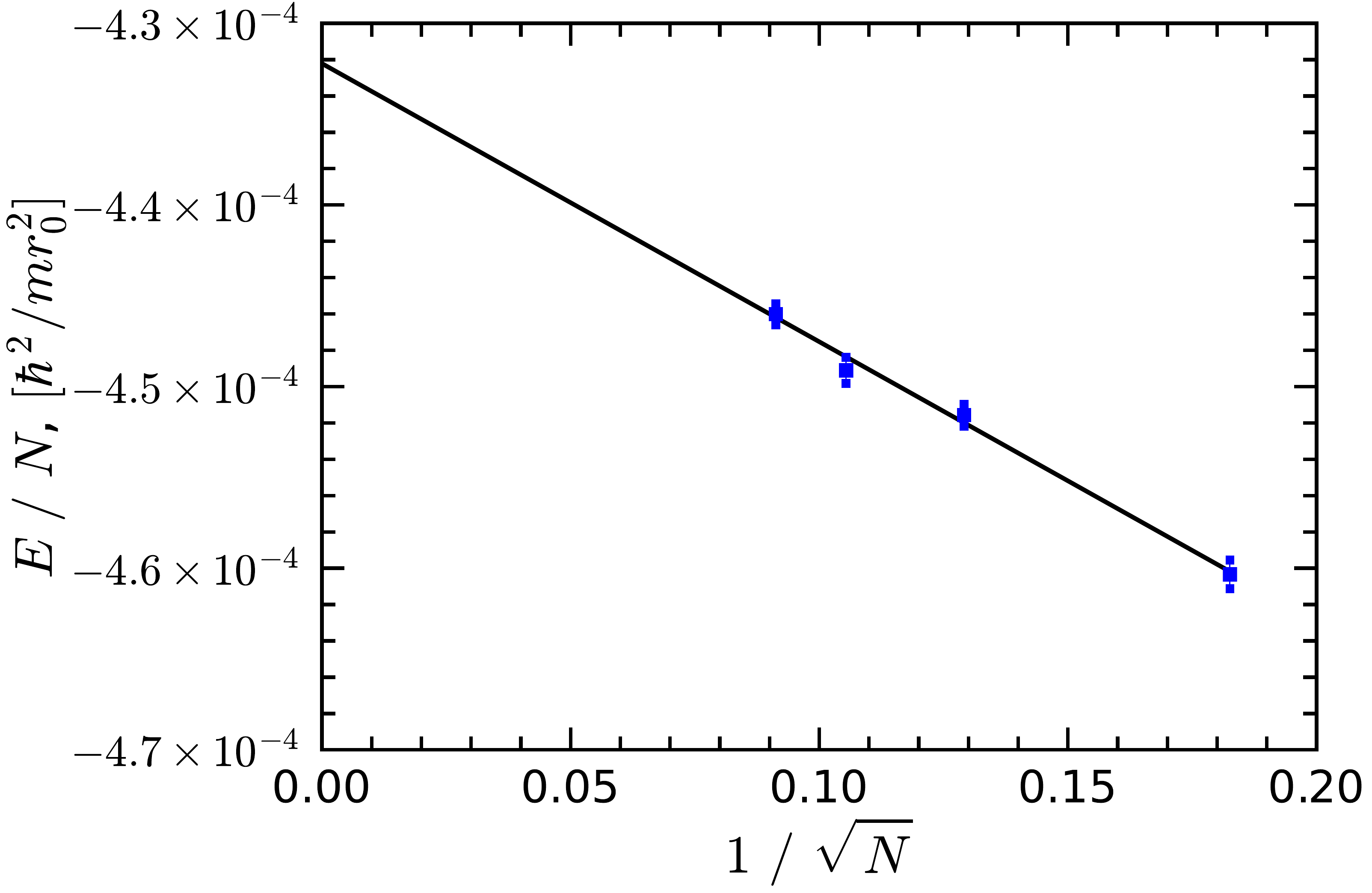}
    \caption{An example of the finite-size dependence for the energy
        in the liquid phase at the dimensionless density $nr^2_0=0.001033$
        and $h/r_0=1.2$.
        Symbols, DMC energy (with added the tail energy, Eq.~(\ref{Eq:6.9}));
        solid line, fit $E_{\rm th}+C/\sqrt{N}$.}
\label{Fig:FiniteSize}
\end{figure}

In Fig.~\ref{Fig:FiniteSize}, we show an example of the finite-size study for the energy.
In it, we consider a liquid phase with density $nr_0^2=0.001033$ and $h/r_0=1.2$.
We observe that the energy dependence on the number of particles scales as
$1/\sqrt{N}$, contrary to the law $1/N$, usual for fast decaying potentials. 
We find that our fitting function describes
well the finite-size dependence. The same procedure is repeated for all the densities
for the gas and liquid phases.
%%%%%%%%%%%%%%%%%%%%%%%%%%%%%%%%%%%%%%%%%%%%%%%%%%%%%%%%%%%%%%%%
\subsection{Equation of State}
\begin{figure}[t!]
  \centering
  \subfigure{\includegraphics[width=0.7\textwidth]{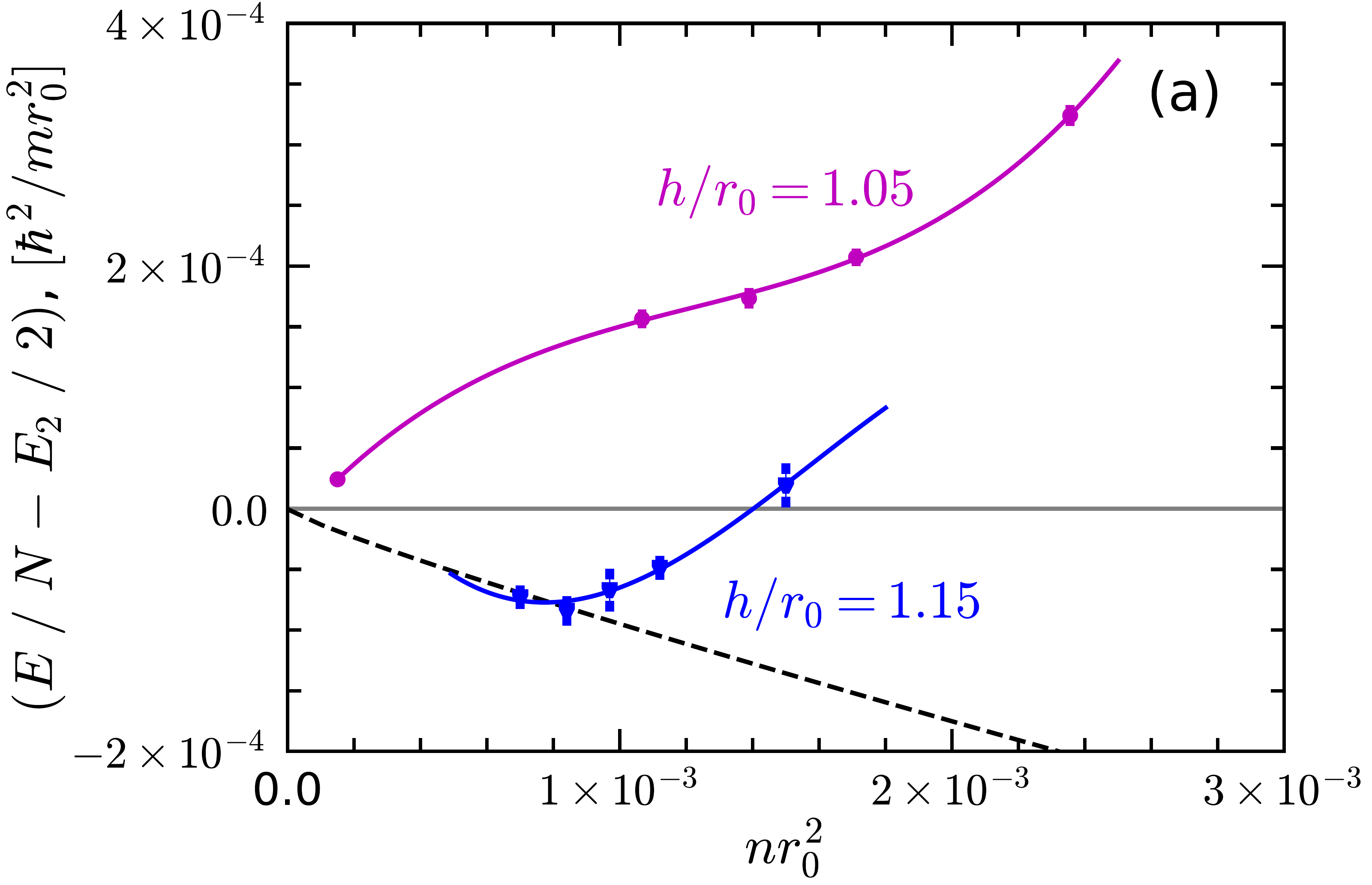}}\quad
  \subfigure{\includegraphics[width=0.7\textwidth]{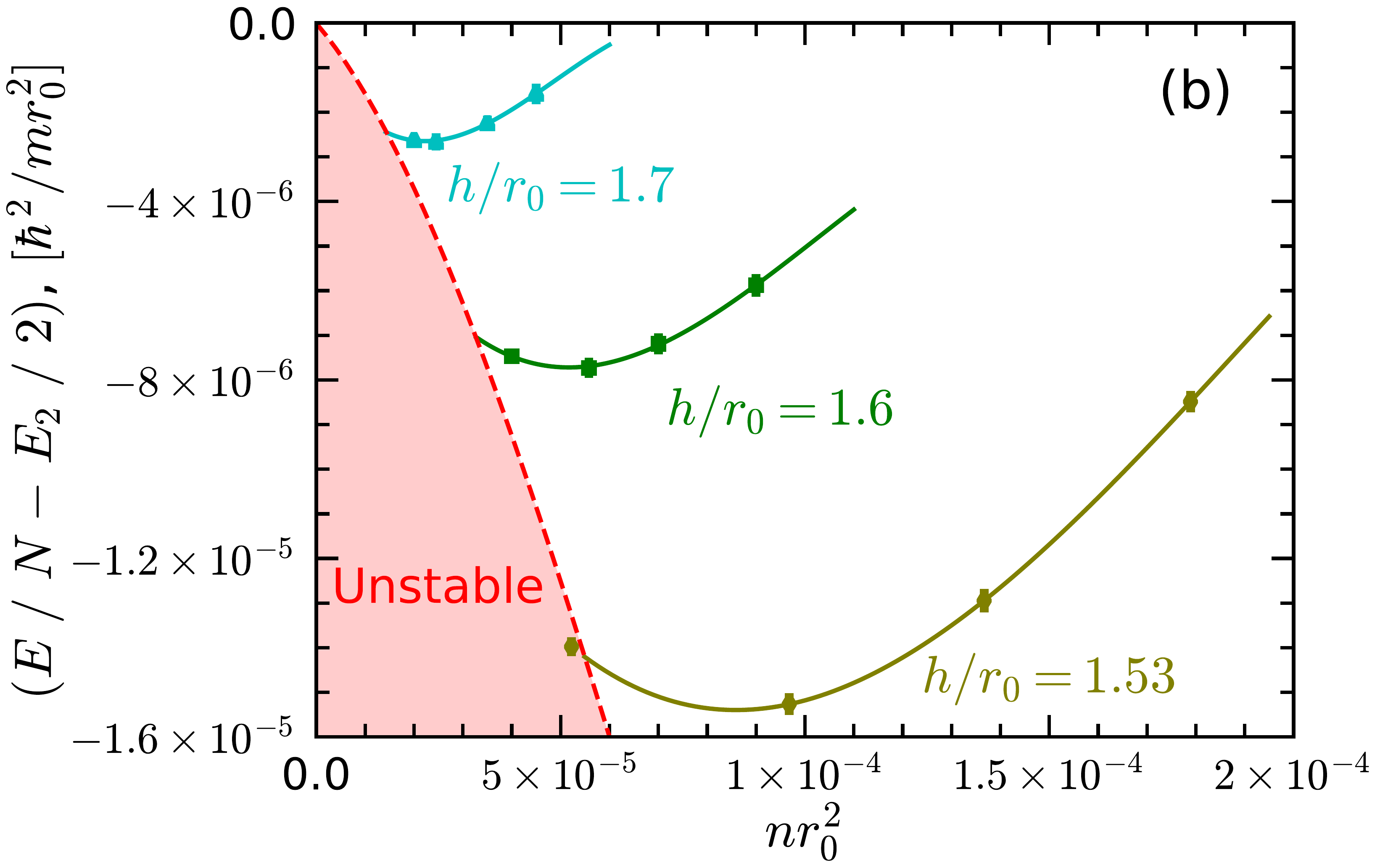}}
  \caption{Energy per particle $E/N$ with half of the dimer binding energy 
$E_2/2$ subtracted as a function of the dimensionless density
$nr_0^2$ for different values of the interlayer 
distance $h/r_0$. The dashed line corresponds to the mean field approximation 
for an attractive molecular gas $E/N=-\pi\hbar^2n/4m\text{ln}[na^2_{\rm dd}]$, 
where $a_{\rm dd}$ is the dimer-dimer scattering length. The solid lines are 
polynomial fits to the energies.}
\label{Fig:EquationOfState}
\end{figure}

A possible existence of a gas or a liquid phase can be determined from
the equation of state. The dependence of the
energy on density is reported in Fig.~\ref{Fig:EquationOfState}
for different values of the interlayer distance. 
We have added the contribution of the tail energy Eq.~(\ref{Eq:6.9})
and performed the extrapolation to the thermodynamic limit. 
We observed that for $h/r_0=1.05$, the energy per particle
monotonically increases as the density is increased. 
Thus, the smallest energy is obtained at vanishing density and this
behavior corresponds to a gas phase. However, a drastically different
behavior is observed as the interlayer separation is increased. That is,
the energy per particle becomes negative and develops a minimum at a
finite density for $h/r_0\ge1.15$. The position of the
minimum corresponds to the equilibrium density.
This behavior
demonstrates the presence of a homogeneous liquid phase.
The balance of forces necessary to stabilize the
liquid comes from a dimer-dimer attraction and an effective
three-dimer repulsion, as proposed in
Chapter~\ref{Chapter:Few-body bound states of two-dimensional bosons}.
Without the repulsive three-dimer force the system would behave as an
attractive gas of dimers as shown by the dashed line in
Fig.~\ref{Fig:EquationOfState}. We found that the interlayer critical
value for the liquid to gas transition ($h/r_0\approx 1.1$) is the
same as the threshold value for the four-body bound state of dipolar
bosons, when the tetramer breaks into two
dimers (Chapter~\ref{Chapter:Few-body bound states of two-dimensional bosons}).
%%%%%%%%%%%%%%%%%%%%%%%%%%%%%%%%%%%%%%%%%%%%%%%%%%%%%%%%%%%%%%%%
\subsection{Phase Diagram}
\begin{figure}[t!]
  \centering
  \subfigure{\includegraphics[width=0.68\textwidth]{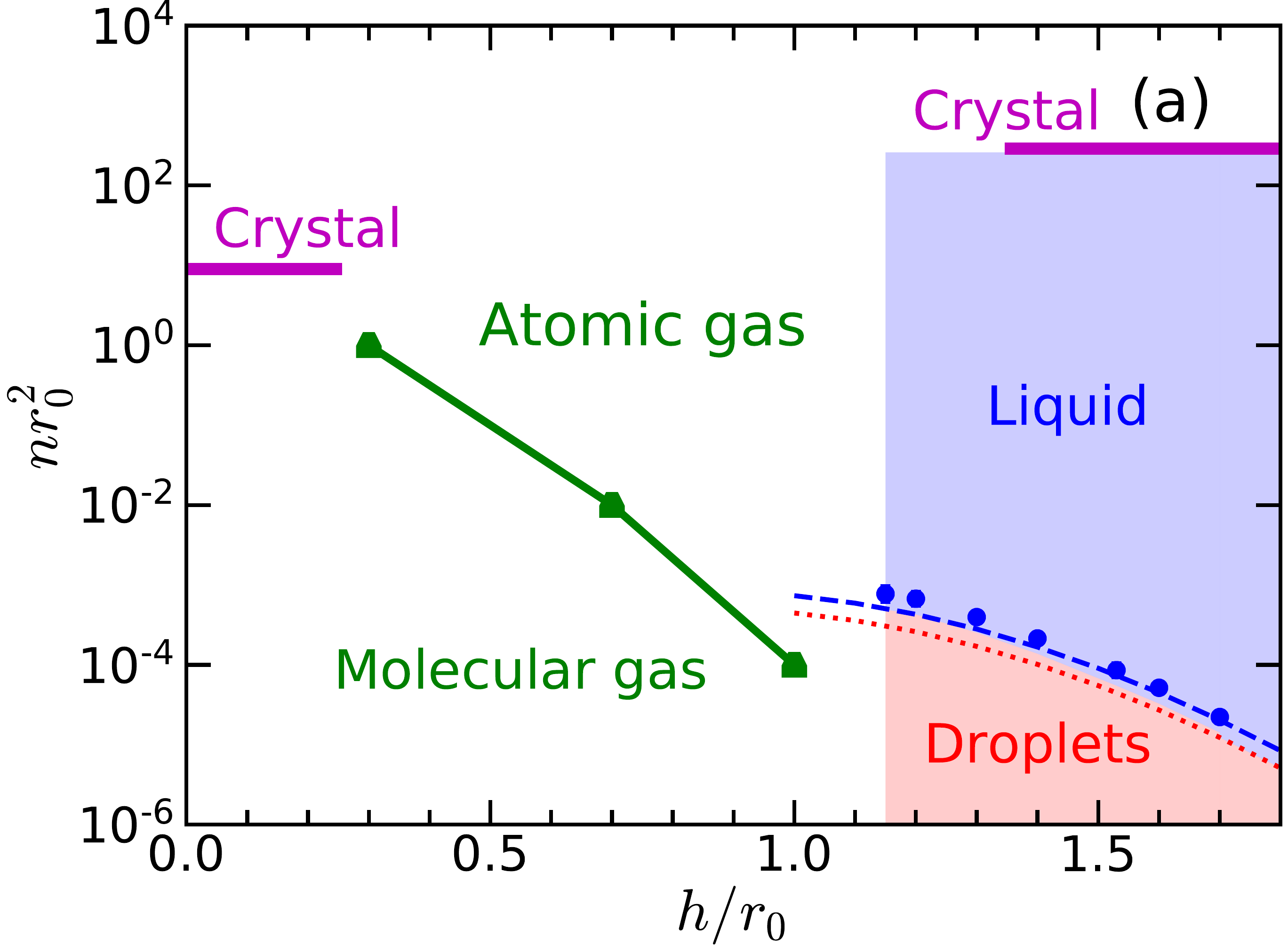}}\quad
  \subfigure{\includegraphics[width=0.73\textwidth]{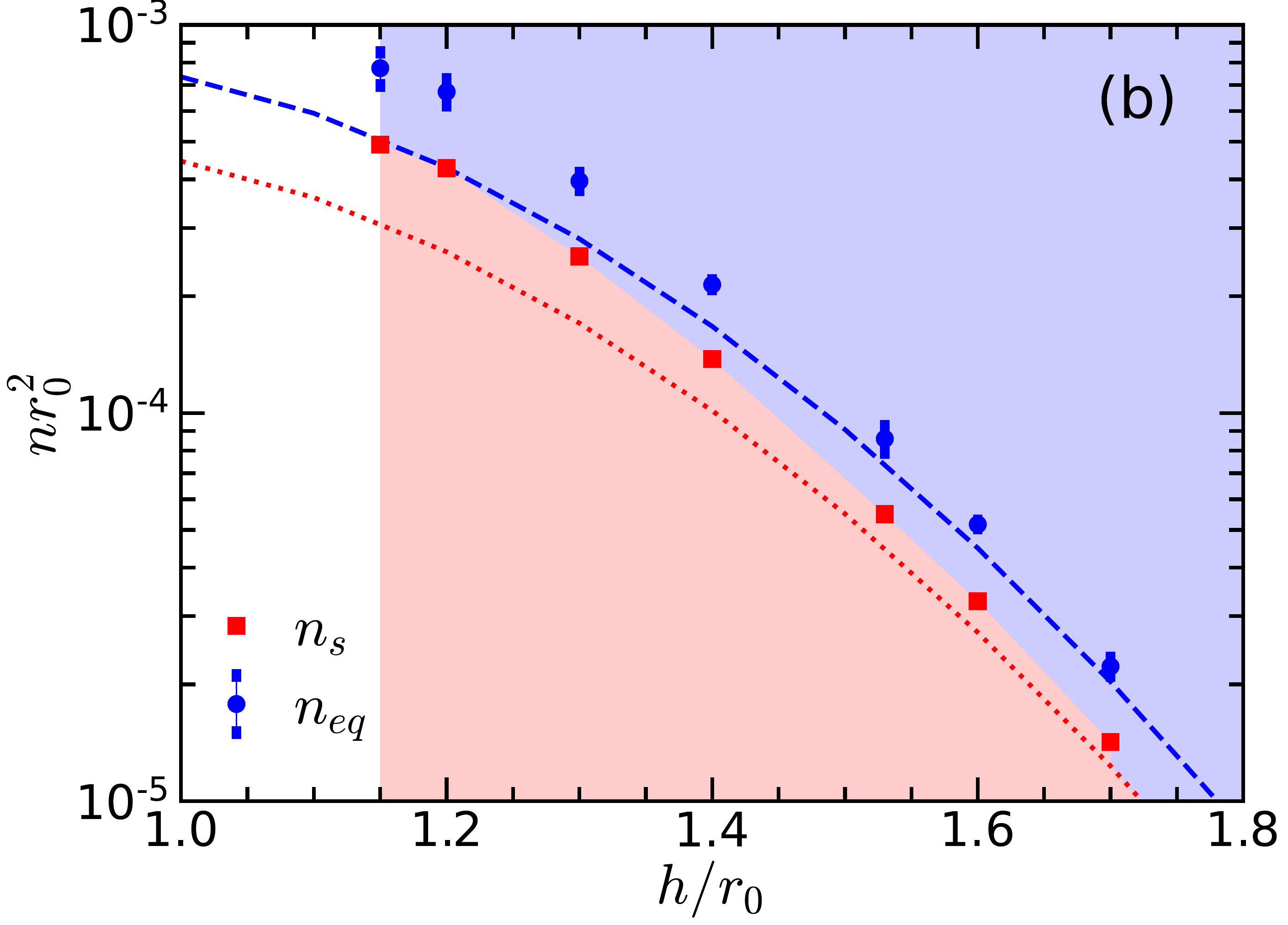}}
\caption{Ground-state phase diagram. The dashed and dotted curves
    are the Bogoliubov approximations for a 2D Bose-Bose mixture with 
		attractive interspecies and repulsive intraspecies short-range 
        interactions~\cite{Petrov2016}.}
    \label{Fig:PhaseDiagram}
\end{figure}
The equations of state are used to extract the equilibrium $n_{eq}$ and
spinodal $n_s$ densities, which are defined by the conditions
\begin{equation}
    \frac{\partial E/N}{\partial n} =0\qquad \mbox{and} \qquad
\frac{\partial P}{\partial n}=0,
    %\frac{\partial^2 E/N}{\partial n^2} =0,
\end{equation}
respectively ($P=n^2 \frac{\partial E/N}{\partial n}$ being the pressure).
The resulting ground-state phase diagram is reported in
Fig.~\ref{Fig:PhaseDiagram} as a function of the total density
$nr_0^2$ and of the interlayer distance $h/r_0$. The self-bound many-body
phases are formed for large interlayer separation, $h/r_0>1.1$. Below the
spinodal curve (dotted line) the homogeneous liquid is unable to sustain
an increasing negative pressure and becomes unstable with respect to droplet formation. 
The stable liquid phase appears above the spinodal curve. Remarkable, this
phase exists in a wide range of densities and interlayer values, which are
experimentally accessible. The equilibrium density (dashed line) can be
adjusted by changing the separation between the layers: $n_{eq}$ decreases
as $h/r_0$ increases. For large separations $h$, the liquid energy is
greatly decreased and in this weakly-interacting regime it is possible to
make a comparison with the predictions of Bogoliubov theory~\cite{Petrov2016}
for zero-range potential (see Fig.~\ref{Fig:PhaseDiagram}b). The best agreement
is found for the smallest equilibrium and spinodal densities
(i.e., for the largest $h$) for which the dipolar potential is better approximated
by the $s$-wave scattering length.
The gaseous and self-bound phases are separated by the threshold $h/r_0\approx1.1$ at
which the effective dimer-dimer interaction switches sign~\cite{PhysRevA.101.041602} 
(Chapter~\ref{Chapter:Few-body bound states of two-dimensional bosons})
from repulsion (gas) to attraction (liquid and droplets). 
The gaseous regime features
a second-order phase transition between atomic and molecular gas phases which on a
qualitative level occurs when the molecule binding energy is similar to the chemical potential. 
In the molecular gas phase, the atomic condensate is absent while the molecular
one is finite~\cite{Macia2014}. In the atomic gas, the atomic condensate is
present and the system features a strong Andreev-Bahskin drag between
superfluids in different layers~\cite{Nespolo_2017}.
The gas phase shows a quantum phase transition from an atomic
to a molecular superfluid as the interlayer distance decreases~\cite{Macia2014}.

As the density is increased, the potential energy starts to dominate and
a triangular crystal is formed. 
For large separation between layers, two
independent atomic crystals are formed and the phase transition occurs when the 
density per layer reaches the same critical value as in a single-layer
geometry, $n_{\1}r_0^2=n_{\2}r_0^2\approx 290$~\cite{PhysRevLett.98.060405,Cinti2017}.
In the limit of small interlayer separations, a single molecular crystal
is formed
at the density $nr_0^2\approx 9$.
%%%%%%%%%%%%%%%%%%%%%%%%%%%%%%%%%%%%%%%%%%%%%%%%%%%%%%%%%%%%%%%%
\subsection{Depletion of the Condensate}
\begin{figure}[b!]
    \centering
    \subfigure{\includegraphics[width=0.7\textwidth]{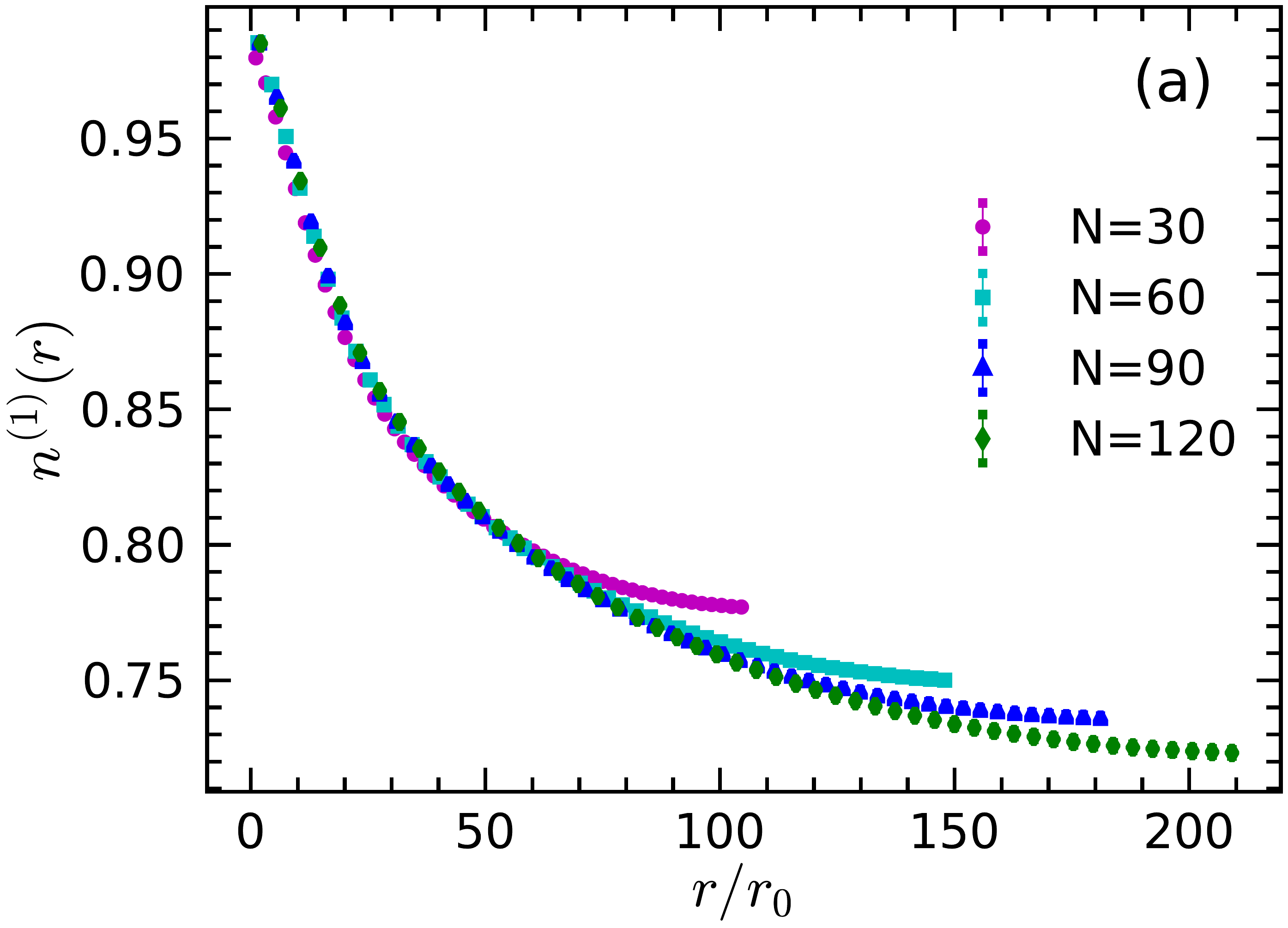}}\quad
    \subfigure{\includegraphics[width=0.7\textwidth]{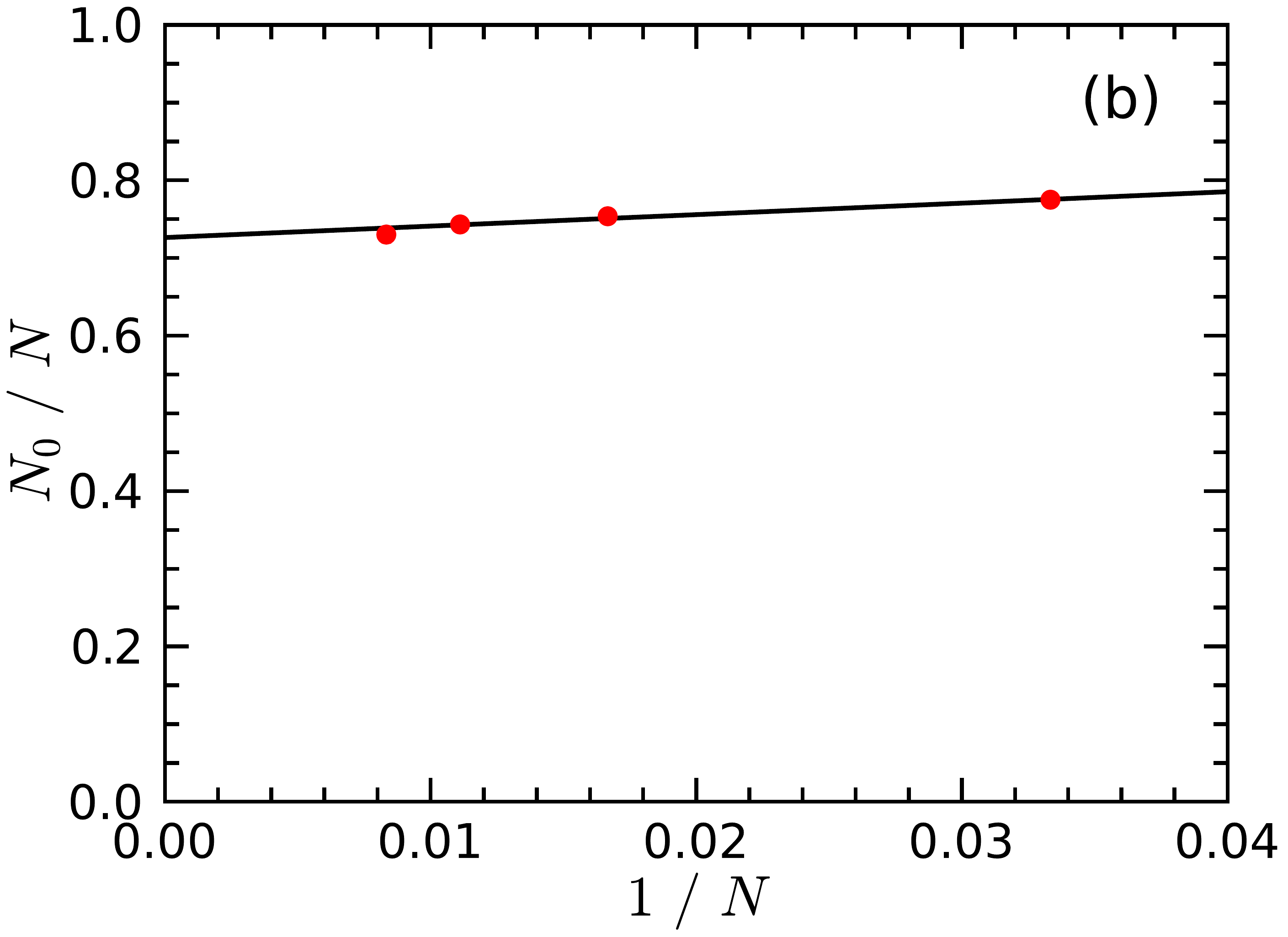}}
    \caption{(a) Example of the one-body density matrix for $h/r_0=1.2$
        at the equilibrium density and for different number of particles.
        (b) An example of the finite-size dependence of $N_0/N$.}
\label{Fig:OBDM}
\end{figure}

In order to have a more complete description of the liquid and gas 
phases we have calculated the condensate fraction using the 
one-body density matrix (OBDM), which is defined as~\cite{Landau:1958:SP}		
\begin{equation}
\label{eq7}
n^{(1)}(\mathbf{r},\mathbf{r'})=\langle \hat{\Psi}^{\dagger}(\mathbf{r})
\hat{\Psi}(\mathbf{r'}) \rangle,
\end{equation}
where $\hat{\Psi}^{\dagger}(\mathbf{r}) (\hat{\Psi}(\mathbf{r}))$ is the 
field operator that creates (annihilates) a particle at the point
$\mathbf{r}$. For a homogeneous system, the condensate fraction is
obtained from the asymptotic behavior of the OBDM
\begin{equation}
n^{(1)}\left(|\mathbf{r}-\mathbf{r'}|\right)_{|\mathbf{r}-\mathbf{r'}|
\to \infty}\to\frac{N_0}{N}.
\label{Eq:CondensateFraction}
\end{equation}
This behaviour is often referred to as off-diagonal long-range order,
since it involves the nondiagonal components
$(\mathbf{r}\neq\mathbf{r'})$ of the OBDM.
In Fig~\ref{Fig:OBDM} (a) we show an example of the OBDM
for different numbers of particles.
We notice an important dependence on the total number of particles,
that is, the $N_0/N$ value decreases as the total number of particles
increases.
From the VMC and DMC results of the OBDM, we obtained the extrapolated
values of $N_0/N$, 
using the extrapolation
technique (Section~\ref{Extrapolation Technique}).
In Fig~\ref{Fig:OBDM} (b) we plot the extrapolated values
of $N_0/N$ as a function of $1/N$.
In order to remove the finite-size effects,
we extrapolate the condensate fraction value $N_0/N$ to the thermodynamic
limit using the fitting formula $b_0+b_1/N$,
where $b_0$ and $b_1$ are fitting parameters.
\begin{figure}[b!]
    \centering
    \subfigure{\includegraphics[width=0.7\textwidth]{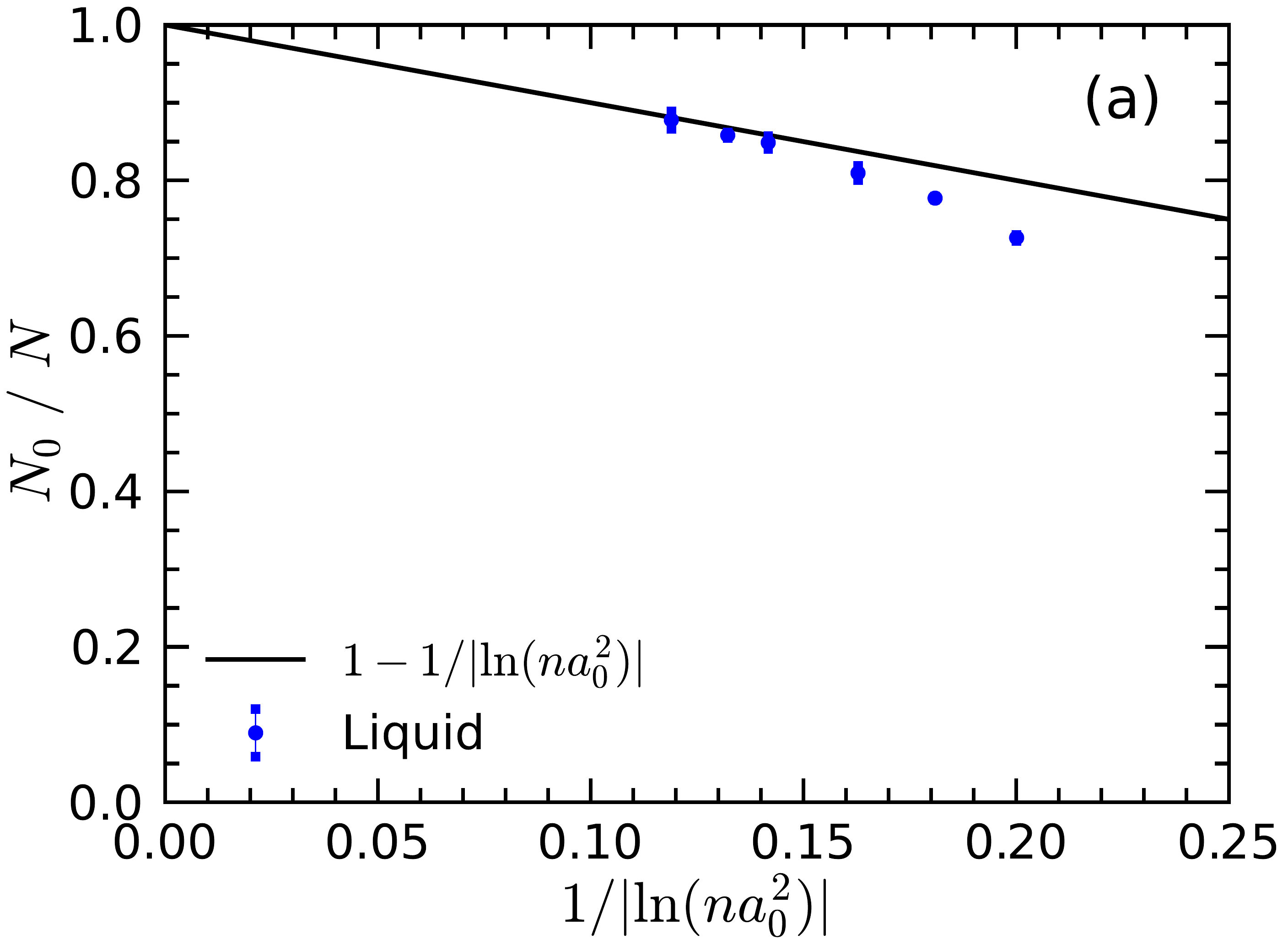}}\quad
    \subfigure{\includegraphics[width=0.7\textwidth]{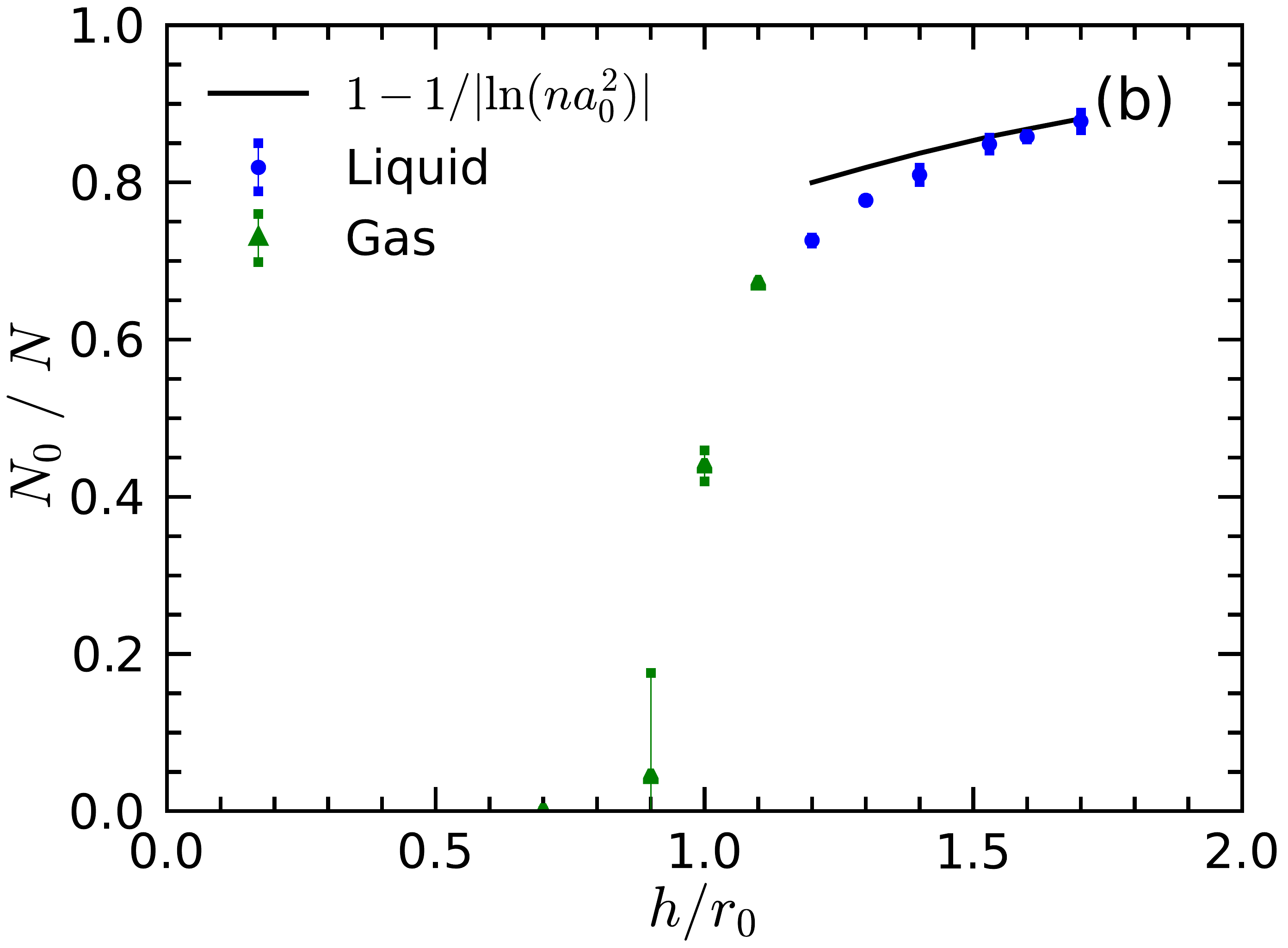}}
    \caption{Quantum depletion of the condensate.
The blue circles correspond to the liquid phase at the equilibrium density. 
The solid line corresponds to the quantum depletion of
short-range potentials having $s$-wave scattering length $a_0$
in two dimensions $1-1/|$ln$(na_0^2)|$. 
(a) Condensate fraction $N_0/N$ as a function of $1/|$ln$(na_0^2)|$. 
(b) Condensate fraction $N_0/N$ as a function of the interlayer distance $h/r_0$, 
the green triangles correspond to the gas phase at the dimensionless 
density $nr_0^2=0.001$.}
\label{Fig:CondensateFraction}
\end{figure}

In Fig.~\ref{Fig:CondensateFraction} (a) we show the condensate fraction
$N_0/N$ as a function of $1/|$ln$(na_0^2)|$, for the dipolar liquid
at the equilibrium density. 
In the dilute limit, we find a good agreement with the quantum
depletion $1/|\ln(na_0^2)|$ calculated in a Bogoliubov theory
for short-range potentials.
The depletion of the condensate is present in interacting
bosonic systems in which, due to the interactions, a portion of bosons
are in a non-zero momentum state even at zero temperature.
The equilibrium density has a strong dependence on the
interlayer separation $h$ (see Fig.~\ref{Fig:PhaseDiagram}).
For liquids formed at separations $h\gtrsim 1.6$ the
perturbative result is expected to hold.
In the Fig.~\ref{Fig:CondensateFraction} (b) we report the
condensate fraction
as a function of the interlayer separation $h$. 
The liquid exists for large separation between the layers $h$.
As $h$ is decreased the equilibrium density grows up until it reaches $nr_0^2\approx 10^{-3}$ at $h/r_0\approx 1.1$ where a phase transition to a gas happens. 
For smaller separations, the liquid does not exist and we show the condensate fraction in the gas with the density fixed to $nr_0^2=10^{-3}$. 
The condensate fraction rapidly drops to zero signaling a phase transition from atomic to molecular gas.
%%%%%%%%%%%%%%%%%%%%%%%%%%%%%%%%%%%%%%%%%%%%%%%%%%%%%%%%%%%%%%%%
\subsection{Polarization}
The different phases present in the system can be further 
characterized by calculated 
the ground-state energy dependence on small polarization.
This dependence can be linear or quadratic depending on the
molecular or atomic nature of the system, respectively.
This can be obtained by slightly unbalanced the number
of particles in the bottom $N_\1$ and top $N_\2$ layers, while keeping fixed
the total number of particles $N_\1+N_\2$.
The polarization is defined as
\begin{equation}
    P=\frac{N_\1-N_\2}{N_\1+N_\2},\qquad\mathrm{with} \qquad |P|\ll 1.
\end{equation}

For an atomic condensate the ground-state energy dependence
on small polarization is quadratic
\begin{equation}
    E(P)=E(0)+N(n/2\chi_s)P^2,
\end{equation}
where $E(0)$ is the ground-state energy of the balanced system and
$\chi_s$ is the spin susceptibility associated with the dispersion of
spin waves of the magnetization density $n_t-n_d$ with speed of
sound $c_s=\sqrt{n/m\chi_s}$. In this case the low-lying excitations
are coupled phonon modes of the two layers~\cite{Macia2014}.
\begin{figure}[t!]
	\centering
    \includegraphics[width=0.7\textwidth]{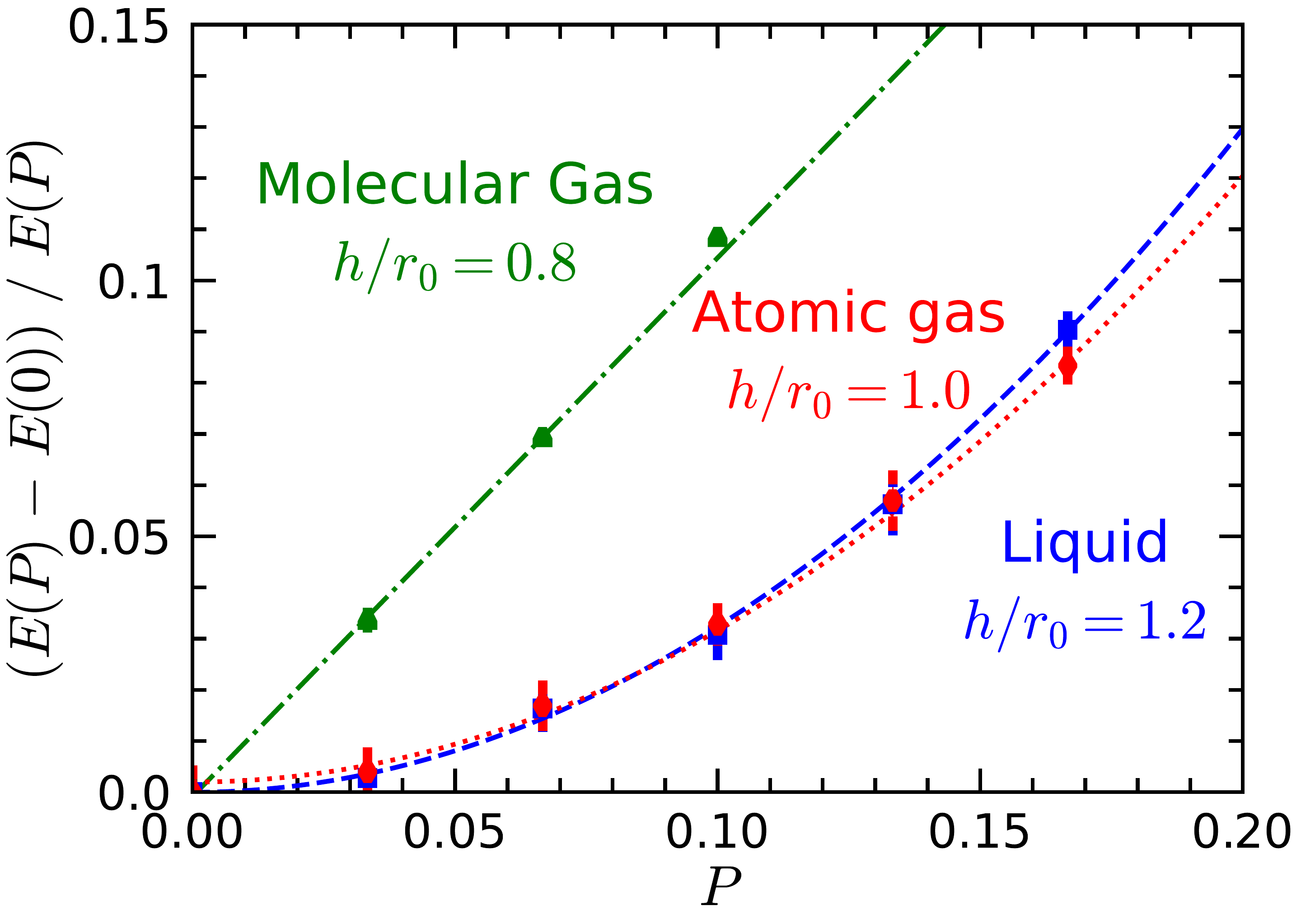}
    \caption{Ground-state energy $E(P)$ as a function of the polarization
        $P$ for three values of $h/r_0$ in the molecular gas ($nr_0^2=0.001$),
        atomic gas ($nr_0^2=0.001$),
        and liquid phase (equilibrium density).}
    \label{Fig:Polarization}
\end{figure}

For a molecular superfluid phase the ground-state energy is a linear
function of the polarization 
\begin{equation}
    E(P)=E(0)+N\Delta_{\rm gap}P,
\end{equation}
in this case an energy $\Delta_{\rm gap}$ is needed to break a pair and
spin excitations are gapped~\cite{Macia2014}.

Examples of the different behaviors of $E(P)$ are reported in
Fig.~\ref{Fig:Polarization} for three values of $h/r_0$ corresponding
to the molecular gas, atomic gas, and liquid phases. We notice that
$E(P)$ is a quadratic function of $P$ for the liquid state, therefore
the liquid is a liquid of atoms and not a liquid of molecules or dimers. 
%%%%%%%%%%%%%%%%%%%%%%%%%%%%%%%%%%%%%%%%%%%%%%%%%%%%%%%%%%%%%%%%
\section{Summary}
We have shown that a dipolar bilayer possesses a rich phase
diagram with quantum phase transitions between gas and solid phases
(known before), and a liquid phase (newly predicted). 
Remarkably, the liquid state, which results from the balance of a
dimer-dimer attraction and an effective three-dimer repulsion, exists
in a wide range of densities and interlayer separations which are
experimentally accessible. From the equations of state, we extracted
the spinodal and equilibrium densities, which are controllable through
the interlayer distance. The equilibrium density decreases as the
interlayer distances increases, allowing access to ultra-dilute liquids
in a stable setup.

\chapter{Phases of Dipolar Bosons Confined to a Multilayer Geometry}
\label{Multilayer System of Dipolar Bosons}
\section{Introduction}
In a classical crystal at low temperature all the atoms are
strongly localized around their equilibrium lattice positions.
In contrast, in a quantum crystal, the atoms move around the
equilibrium lattice positions and exchanges between few particles
occur with frequency~\cite{RevModPhys.89.035003}.
A quantum crystal is then defined as a crystal in which the zero
point motion of an atom
about its equilibrium position is a large fraction of the near
neighbor distance~\cite{GUYER1970413}. This large displacement is
a consequence of the light-weight particles and the weakness of the
long-range forces between the atoms~\cite{RevModPhys.89.035003}.

The most known quantum solids are Hydrogen and Helium.
The solidification of $^4$He in 1926 initiated the experimental
study of quantum solids. $^4$He solidifies at the temperature limit
$T\to0$K under a pressure of $P\simeq25$ bar~\cite{GUYER1970413}.

An interesting property that quantum solids can exhibit is
supersolidity, a state of matter predicted in 1969 for Andreev
and Lifshitz in which crystalline order and Bose-Einstein condensation
coexists~\cite{RevModPhys.89.035003}. A supersolid flows without
friction but its particles form a crystalline lattice.

Ultracold dipolar gases provide a powerful platform to study
highly non-trivial quantum phenomena, 
in particular to study solid and supersolid states of matter.
Recent experimental and theoretical studies have shown that dipolar condensates
within a pancake-~\cite{Santos2003,ronen2007radial,bohn2009does,parker2009structure,
blakie2012roton,jona2013roton,wilson2010critical,natu2014dynamics}
and cigar-like~\cite{Chomaz2018observation} geometries
can undergo a phase transition from a Bose-Einstein condensate (BEC)
to a state that has supersolid properties.

An interesting confined system in two dimensions is a layer of
dipolar bosons. If the dipolar moments of the bosons are oriented
perpendicularly to the layer, the dipole-dipole interaction is
repulsive. When this happens, the system undergoes a quantum phase
transition from a gas to a solid phase as the density
increases~\cite{Buchler2007,Astrakharchik2007}.

Adding a second parallel layer makes the system richer.  
In a bilayer setup, there is a competing effect between repulsive intralayer 
and partially attractive interlayer interactions, which can produce
exotic few- and many-body states. For example, 
a gas and a pair superfluid phases  
were characterized in Refs.~\cite{Macia2014,PhysRevA.94.063630}
using exact quantum Monte Carlo simulations.
Furthermore, when the interlayer distance approaches to zero the system
forms a molecular crystal and for large values of the
interlayer distance an independent atomic crystal is formed in each
layer~\cite{Cinti2017}. 

In this chapter, we study a system of dipolar bosons confined to a
multilayer geometry by using exact many-body quantum Monte Carlo
methods. The multilayer geometry consists of equally spaced
two-dimensional layers. We consider the case in which the dipoles
are aligned perpendicularly to the parallel layers. This system is
predicted to have a rich collection of many-body phases due to the
anisotropic and quasi long-range dipole-dipole interaction between
the bosons. We calculated the
ground-state phase diagram as a function of the density, the
separation between layers, and the number of layers. The key result
of our work is the existence of three phases: atomic gas, solid,
and gas of chains, in a wide range of the system parameters. We find
that the density of the solid phase decreases several orders of
magnitude as the number of layers in the system increases.
Furthermore, we calculated the pair distribution functions for the
three phases.
%%%%%%%%%%%%%%%%%%%%%%%%%%%%%%%%%%%%%%%%%%%%%%%%%%%%%%%%%%%%%%%
\begin{figure}[!b]
	\centering
    \includegraphics[width=0.7\textwidth,trim = {0 5.0cm 0 0.5cm}, clip]{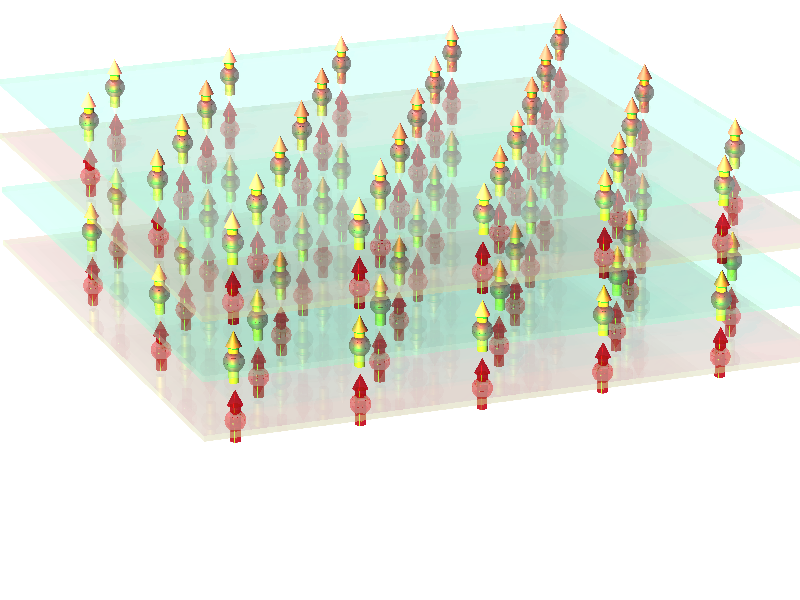} 
\caption{Schematic representation of dipoles confined to a multilayer geometry.}
    \label{Fig:Ch7_Multilayer}
\end{figure}
%%%%%%%%%%%%%%%%%%%%%%%%%%%%%%%%%%%%%%%%%%%%%%%%%%%%%%%%%%%%%%%
\section{The Hamiltonian}
\begin{figure}[!t]
	\centering
    \includegraphics[width=0.7\textwidth]{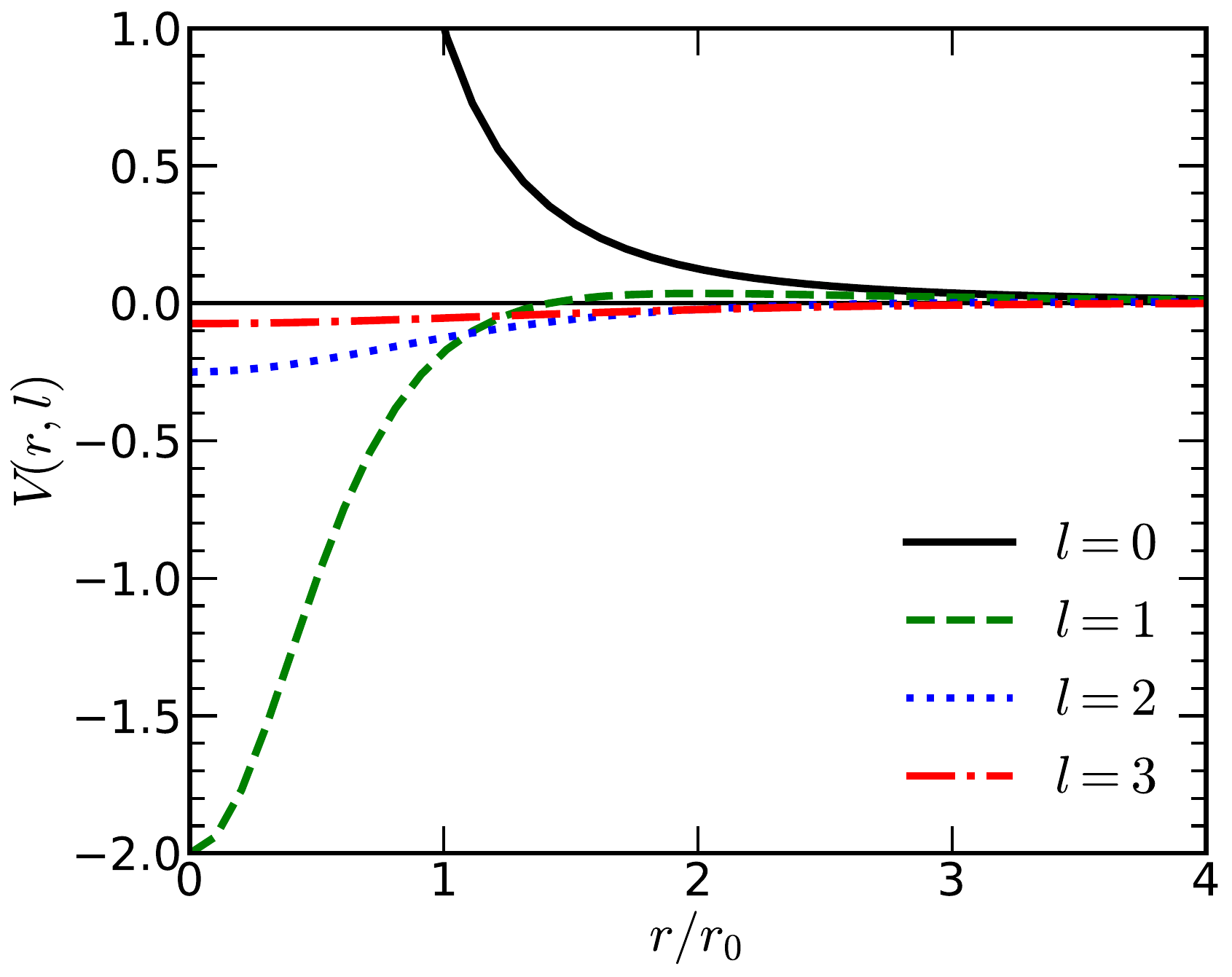}
    \caption{Dipolar potential $V(r,l)$ as a function of $r/r_0$
        for different values of the interlayer separation $lh$,
        with $h/r_0=1.0$.}
    \label{Fig:Ch7_Potential}
\end{figure}

We study an $N$-particle system of dipolar bosons of mass $m$ and dipole
moment $d$ confined to an $M$-layer geometry. The three-dimensional confining structure
is formed of $M$ two-dimensional parallel layers separated by a distance $h$.
The dipolar moment of each boson is considered to be perpendicular to the
layers and there is no interlayer tunneling.
In Fig.~\ref{Fig:Ch7_Multilayer} we show a schematic representation of the multilayer
geometry.
The Hamiltonian of this
system is given by
\begin{equation}
H=\sum_{i=1}^{N}-\frac{\hbar^2}{2m}\nabla^2_i
+\sum_{i<j}^N V_{ij}(|\mathbf{r}_{i}-\mathbf{r}_{j}|,l),
\label{Eq:7.1}
\end{equation}
where $\mathbf{r}_i$ is the position vector of particle $i$.
The first term in the Hamiltonian~(\ref{Eq:7.1}) corresponds
to the dipole kinetic energy. The second term is the 
dipolar interaction between particle $i$ and $j$
\begin{equation}
    V_{ij}({r}_{ij},l)=
    d^2\frac{r_{ij}^2-2l^2h^2}{(r_{ij}^2+l^2h^2)^{5/2}},\quad
    l=0,1,2,\ldots,M-1.
    \label{Eq:DipolarPotential}
\end{equation}
Here, $lh$ denotes the interlayer separation and
$r_{ij}=|{\bf{r}}_{i}-{\bf{r}}_{j}|$ stands for
the distance between the projections onto any of the layers of the positions of 
the $i$-th and $j$-th particles.
For $l\neq0$ the dipolar interaction called interlayer interaction,
is attractive for small values of $r$ but repulsive
for large values of $r$, where $r$ is the in-plane distance between
dipoles. For dipoles in the same layer the dipolar interaction is always repulsive  
\begin{equation}
V_{ij}({r}_{i,j},0)=
\frac{d^2}{r^3_{ij}},
\end{equation}
where we set $l=0$.
The dipolar length $r_0=md^2/\hbar^2$ is used as a unit of length.
In Fig.~\ref{Fig:Ch7_Potential} we show the dipolar potential
as a function of $r/r_0$ for different values of the interlayer distance $lh$,
with $h/r_0=1.0$. Notice that the attractive part of the interlayer
potential for $l=2$ and 3 are much weaker than for $l=1$.

For the Hamiltonian Eq.~(\ref{Eq:7.1}) we anticipate to find three phases: atomic gas,
solid, and gas of chains. This comes from considering previous
studies in a dipolar layer~\cite{Buchler2007,Astrakharchik2007}
and bilayer~\cite{Macia2014,PhysRevA.94.063630,Cinti2017}.
%%%%%%%%%%%%%%%%%%%%%%%%%%%%%%%%%%%%%%%%%%%%%%%%%%%%%%%%%%%%%%%
\section{Details of the Methods}\label{Ch7:Details of the Methods}
To investigate the ground-state properties of 
Hamiltonian Eq.~(\ref{Eq:7.1}) we employ the diffusion Monte Carlo 
(DMC) method, which was described in
Chapter~\ref{Chapter:Quantum Monte Carlo methods}. 
Here, we use different trial wave functions 
appropriate for the description of various 
phases present in the system.
In total we describe three phases with different symmetry: 
gas, solid, and gas of chains. 
%%%%%%%%%%%%%%%%%%%%%%%%%%%%%%%%%%%%%%%%%%%%%%%%%%%%%%%%%%%%%%%
\subsection{Gas Trial Wave Function}
To describe the gas phase we chose a trial wave function of the 
Bijl-Jastrow form Eq.~(\ref{BijlJastrow})
\begin{equation}
    \Psi_{\rm Gas}(\mathbf{r}_1,\dots,\mathbf{r}_N)=\prod_{j<k}^N
f_{2}(|\mathbf{r}_{j}-\mathbf{r}_{k}|).
 \label{TWFGas}
\end{equation}
The wave function $\Psi_{\rm Gas}$ has translational symmetry.
This symmetry is present in the gas phase, while the crystal phase has
broken translational symmetry due to a regular spacing of atoms and,
the gas of chains phase has additional interlayer correlations.
The two-body terms 
$f_2(|\mathbf{r_j}-\mathbf{r_k}|)$ in Eq.~(\ref{TWFGas}) depend
on the distance between a pair of particles.
Here, the two-body terms $f_{2}(r)$ are 
taken as the solution of the two-body problem at short distances.
This solution depends on whether the dipoles are in the
same layer or not. 
For dipoles in the same layer we chose the short distance part of the
two-body correlations term as
\begin{equation}
    f_{2}(r)=C_0 K_0\left(2\sqrt{R_0r_0/r}\right),
\end{equation}
up to $R_{\rm match}$. Here, $K_0(r)$ the modified Bessel function, $R_{\rm match}$
and $R_0$ are 
variational parameters~\cite{PhysRevLett.98.060405}.

For distances larger than $R_{\rm match}$ we chose 
\begin{equation}
    f_{2}(r)=C_1 {\rm exp}\left[ -\frac{C_2}{r} - \frac{C_2}{L-r}\right],
\end{equation}
which take into account the contributions from other particles 
and describe long-range phonons in the form established by Reatto and
Chester~\cite{PhysRev.155.88}. The constants $C_0$, $C_1$ and $C_2$ are fixed
by imposing continuity of the function and its first derivative at the
matching distance $R_{\rm match}$, and also that $f_{2}(L/2)=1$.

For dipoles in different layers, the
interlayer correlations 
are taken as the solution of the two-body problem
$f_{2}(r)$ up to $R_1$. We also impose the boundary condition
$f^{'}_{2}(R_1)=0$. For distances larger than the variational
parameter $0<R_1<L/2$ we set $f_{2}(r)=1$.
%%%%%%%%%%%%%%%%%%%%%%%%%%%%%%%%%%%%%%%%%%%%%%%%%%%%%%%%%%%%%%%
\subsection{Solid Trial Wave Function}
The trial wave function we use to describe the quantum solid phase is of the
Nosanow-Jastrow form~\cite{RevModPhys.89.035003}
\begin{equation}
    \Psi_{\rm Solid}(\mathbf{r}_1,\dots,\mathbf{r}_N;\{\mathbf{R}_I^c\})=
\prod_{j<k}^N
f_{2}(|\mathbf{r}_{j}-\mathbf{r}_{k}|)
\prod_{i=1}^N
f_{1}(|\mathbf{r}_{i}-\mathbf{R}^c_{i}|),
\label{NosanowTWF}
\end{equation}
where $\mathbf{r}_i$ are the positions vectors of the particles,
$\{\mathbf{R}_I^c\}$ are the position vectors defining the equilibrium 
crystal lattice, and 
$f_1(r)$ and $f_2(r)$ are the one-body and two-body correlation factors,
respectively. For a two-dimensional system the equilibrium crystal lattice
is a triangular lattice.
The wave function $\Psi_{\rm Solid}$~(\ref{NosanowTWF}) has a broken
translational symmetry due to particle localization close to the lattice
sites. In two dimensiones the crystal has triangular symmetry.

The two-body correlation functions $f_2(r)$ used in $\Psi_{\rm Solid}$~(\ref{NosanowTWF}) 
are of the same form as
those used to describe the gas phase $\Psi_{\rm Gas}$~(\ref{TWFGas}),
although the specific values of the variational parameters might differ.

The one-body terms $f_1(r)$ used in $\Psi_{\rm Solid}$~(\ref{NosanowTWF})
are modeled by a Gaussian function
\begin{equation}
    f_1(r_i)=e^{-\alpha|\mathbf{r}_{i}-\mathbf{R}^c_{i}|^2},
\end{equation}
where $\alpha$ is the localization strength and $R_i^c$ is the
position of the lattice site. Notice that, $R_i^c$ are fixed by the
triangular lattice. Meanwhile, $\alpha$ is a variational parameter, which
is chosen by minimizing the variational energy.

The trial wave function $\Psi_{\rm Solid}$~(\ref{NosanowTWF}) is
not symmetric under the exchange of particles. Therefore, it does not
give an appropriate description of off-diagonal properties that directly depend
on the Bose-Einstein statistics (for example, one body density matrix and
momentum distribution). However, $\Psi_{\rm Solid}$ leads to an accurate
description of the energy and diagonal properties of quantum solids (for example,
density profile, pair-distribution function, static structure factor).
Examples of symmetric trial wave functions
for modeling quantum solids can be found in Ref.~\cite{RevModPhys.89.035003}.
In this work, we are interested in calculating the ground-state energy
and the pair distributions of the system. Thus we will use 
$\Psi_{\rm Solid}$~(\ref{NosanowTWF}).
We leave for future work the use of symmetric trial wave functions
to describe the properties of our system.
%%%%%%%%%%%%%%%%%%%%%%%%%%%%%%%%%%%%%%%%%%%%%%%%%%%%%%%%%%%%%%%
\subsection{Gas of Chains Trial Wave Function}
To describe the gas of chains phase we propose the following trial wave function
\begin{equation}
    \Psi_{\rm Chains}(\mathbf{r}_1,\dots,\mathbf{r}_N)=\prod_{j<k}^N
f_{2}(|\mathbf{r}_{j}-\mathbf{r}_{k}|)
\prod_{i=1}^N
f_{\rm CM}(|\mathbf{r}_{i}-\mathbf{R}_{i}|).
\label{TWFChains}
\end{equation}
Differently from the Nosanow-Jastrow wave function, the last product
in Eq.~(\ref{TWFChains}) is not a one-body but rather is a many-body
term. That is, the movement of a single particle in that product changes
$M$ terms, with $M$ the number of layers,
while in Nosanow form Eq.~(\ref{NosanowTWF}) such a
movement affects only a single term (i.e. is a one-body correlation).
Here, the two-body term $f_2(|\mathbf{r}_j-\mathbf{r}_k|)$
depends on the distance between a pair of particles.
The many-body term $f_{\rm CM}(|\mathbf{r}_{i}-\mathbf{R}_{i}|)$ depends on the
positions of a single particle $\mathbf{r}_i$ and on the center
of mass of the chains $\mathbf{R}_i$. 
Details of the construction of the trial wave function $\Psi_{\rm Chains}$
and its derivatives can be found in Section~\ref{ConstructionTWF}.

The chain center of mass $\mathbf{R}_i$ is given by
\begin{equation}
    \mathbf{R}_i=\frac{1}{M}\sum_{k\in{\rm C}_i}\mathbf{r}_k.
\end{equation}
Here, 
$k\in{\rm C}_i$ means that the sum is over all
$k$-th particles that belong to the same chain as the $i$-th particle. 
The main difference with the solid case is that
in $\Psi_{\rm Chains}$~(\ref{TWFChains}) the
$\mathbf{R}_i$ values change during the simulation
while the $\mathbf{R}_i^c$ values in $\Psi_{\rm Solid}$~(\ref{NosanowTWF})
are fixed by the triangular lattice. 

The two-body correlation functions $f_2(r)$ used in 
$\Psi_{\rm Chains}$~(\ref{TWFChains}) 
are the same as those for the gas and solid phases.

The terms $f_{\rm CM}(r)$ are described by a Gaussian function
\begin{equation}
    f_{\rm CM}(r_i)=e^{-\alpha|\mathbf{r}_{i}-\mathbf{R}_{i}|^2},
    \label{fcmGaussian}
\end{equation}
where $\alpha$ is the localization strength. 

All the variational parameters that appear in the trial wave functions
$\Psi_{\rm Gas}$, $\Psi_{\rm Solid}$ and $\Psi_{\rm Chains}$,
are chosen by minimizing the variational energy and can have a different value
for each trial function.
%%%%%%%%%%%%%%%%%%%%%%%%%%%%%%%%%%%%%%%%%%%%%%%%%%%%%%%%%%%%%%%
\section{Details of the Gas of Chains Wave Function}
\label{ConstructionTWF}
In this section, we are going to discuss a trial wave function 
which we employ for the description of a gas of chains phase.  
Also, we will give explicit expressions for the gradient and
Laplacian of the trial function.

While the trial Bijl-Jastrow (gas) and Nosanow-Jastrow (crystal) forms
are standard and were extensively studied in the literature, within our
best knowledge this is the first time the gas of chains is calculated
within the DMC method. For methodological reasons we provide a very
detailed description of how such functions are constructed and give
explicit expressions of how the derivatives are calculated.

It is essential that the trial wave function has the same symmetry
as the one present in the phase. In the gas of chains phase, dipoles
belonging to different layers form composite bosons (chains) with
no crystal ordering between them. Each composite bosons is composed
from $M$ dipoles each belonging to a different layer.
Also, this gas fulfills the following properties:
\begin{itemize}
    \item The chains are considered to be a composite objects and
        no exchange of dipoles between them is allowed.
    \item The chains are flexible, each dipole in the chain can move freely in
        the corresponding layer.
    \item The chains will not become tangled because of the repulsive intralayer
        interaction between dipoles.
\end{itemize}
%%%%%%%%%%%%%%%%%%%%%%%%%%%%%%%%%%%%%%%%%%%%%%%%%%%%%%%%%%%%%%%
\subsection{Construction of Trial Wave Function}
A trial wave function to describe the gas of chains phase can be
constructed as a product of many-body and two-body terms as given in
Eq.~(\ref{TWFChains}).
The two-body term $f_2(|\mathbf{r}_j-\mathbf{r}_k|)$
depends on the distance between a pair of particles.
The many-body term $f_{\rm CM}(|\mathbf{r}_{i}-\mathbf{R}_{i}|)$ depends on the
positions of a single particle $\mathbf{r}_i$ and on 
the the chain center of mass $\mathbf{R}_i$. 
The chain center of mass $\mathbf{R}_i$ itself depends
on the positions of $M$ particles. 

The chain center of mass $\mathbf{R}_i$ is given by
\begin{equation}
    \mathbf{R}_i\equiv\frac{1}{M}\sum_{k\in{\rm C}_i}\mathbf{r}_k.
    \label{CCM}
\end{equation}
Here, $M$ is the number of layers,
and the index $k$ runs over all 
$k$-th particles that belong to the same chain as the $i$-th particle. 

In order to implement the QMC algorithm we need to be able to 
calculate the gradient $\nabla_{\mathbf{r}_i}\Psi_{\rm Chains}$ and Laplacian
$\Delta_{\mathbf{r}_i}\Psi_{\rm Chains}$ of the trial wave function.
The expressions for the gradient and Laplacian of the two-body terms
$\prod_{j<k}^Nf_{2}(|\mathbf{r}_{j}-\mathbf{r}_{k}|)$
can be found in the Appendix~\ref{Appendix:Jastrow Trial wave function}.
Here, we are going to obtain the expressions for the many-body terms
$\prod_{i=1}^Nf_{\rm CM}(|\mathbf{r}_{i}-\mathbf{R}_{i}|)$.
The gradient of the product of many-body terms $f_{\rm CM}(|\mathbf{r}_{i}-\mathbf{R}_{i}|)$
with respect to the coordinate $\mathbf{r}_i$ is given by
\begin{equation}
    \label{driftforcef1}
    \vec{F}_{\rm CM,\mathbf{r}_i}=
\frac{\vec{\nabla}_{\mathbf{r}_i}
    \prod_{j=1}^{N}f_{\rm CM}(|\mathbf{r}_j-\mathbf{R}_j|)}
{\prod_{j=1}^{N}f_{\rm CM}(|\mathbf{r}_j-\mathbf{R}_j|)}
=\sum_{k\in{\rm C}_i}\frac{\vec{\nabla}_{\mathbf{r}_i}
    f_{\rm CM}(|\mathbf{r}_k-\mathbf{R}_k|)}
{f_{\rm CM}(|\mathbf{r}_k-\mathbf{R}_k|)}.
\end{equation}
Here, $\vec{F}_{\rm CM,\mathbf{r}_i}$ is called the center of mass drift force
and corresponds to the logarithmic derivative with respect to $\mathbf{r}_i$.

The Laplacian
$\Delta_{\mathbf{r}_i}\prod_{j=1}^Nf_{\rm CM}(|\mathbf{r}_j-\mathbf{R}_j|)$
applied with respect to the coordinate $\mathbf{r}_i$
reads as
\begin{equation}
    \label{Laplacianf1}
\begin{aligned}
\frac{\Delta_{\mathbf{r}_i}\prod_{j=1}^Nf_{\rm CM}(|\mathbf{r}_j-\mathbf{R}_j|)}
{\prod_{j=1}^Nf_{\rm CM}(|\mathbf{r}_j-\mathbf{R}_j|)}=&
\sum_{k\in{\rm C}_i}\left[ 
    \frac{\Delta_{\mathbf{r}_i}f_{\rm CM}(|\mathbf{r}_k-\mathbf{R}_{k}|)}
    {f_{\rm CM}(|\mathbf{r}_k-\mathbf{R}_{k}|)}-
\left(\frac{\vec{\nabla}_{\mathbf{r}_i}f_{\rm CM}(|\mathbf{r}_k-\mathbf{R}_{k}|)}
    {f_{\rm CM}(|\mathbf{r}_k-\mathbf{R}_{k}|)}\right)^2 
\right]\\
&+\left( \sum_{k\in{\rm C}_i}
    \frac{\vec{\nabla}_{\mathbf{r}_i}f_{\rm CM}(|\mathbf{r}_k-\mathbf{R}_{k}|)}
    {f_{\rm CM}(|\mathbf{r}_k-\mathbf{R}_{k}|)}\right)^2.
\end{aligned}
\end{equation}
Now we can write an expression for the center of mass contribution to the kinetic energy
\begin{equation}
    T^{\rm loc}_{\rm CM}=\frac{\hbar^2}{2m}
    \left[
        \sum_{i=1}^N\mathcal{E}_{\rm CM,\mathbf{r}i}^{\rm loc}
        -\sum_{i=1}^N|\vec{F}_{\rm CM,\mathbf{r}_i}|^2
    \right],
    \label{kineticenergyf1}
\end{equation}
where we have used Eq.~(\ref{driftforcef1}) and we have defined
the center of mass local energy as
\begin{equation}
\mathcal{E}_{\rm CM,\mathbf{r}i}^{\rm loc}=-
\sum_{k\in{\rm C}_i}\left[ 
    \frac{\Delta_{\mathbf{r}_i}f_{\rm CM}(|\mathbf{r}_k-\mathbf{R}_{k}|)}
    {f_{\rm CM}(|\mathbf{r}_k-\mathbf{R}_{k}|)}-
    \left(\frac{\vec{\nabla}_{\mathbf{r}_i}f_{\rm CM}(|\mathbf{r}_k-\mathbf{R}_{k}|)}
        {f_{\rm CM}(|\mathbf{r}_k-\mathbf{R}_{k}|)}\right)^2 
\right].
 \label{localenergyf1}
\end{equation}
%%%%%%%%%%%%%%%%%%%%%%%%%%%%%%%%%%%%%%%%%%%%%%%%%%%%%%%%%%%%%%%
\subsection{Many-Body Term: Gaussian Function}
The purpose of the many-body term $f_{\rm CM}(|\mathbf{r}_{i}-\mathbf{R}_{i}|)$
in the trial wave function $\Psi_{\rm Chains}$~(\ref{TWFChains})
is to describe the localization of the particles belonging to one chain around
its center of mass. The simplest form of the localization term is a Gaussian
function Eq.~(\ref{fcmGaussian})
\begin{equation}
    f_{\rm CM}(|\mathbf{r}_j-\mathbf{R}_{j}|)=
 e^{-\alpha|\mathbf{r}_j-\mathbf{R}_{j}|^2},
 \label{Gaussianf1}
\end{equation}
where $\alpha$ is the localization strength, $\mathbf{r}_j$ is the position
vector of the particle $j$ and $\mathbf{R}_j$ is the chain center of mass
to which particle $\mathbf{r}_j$ belongs.
The chain center of mass $\mathbf{R}_j$ is given by Eq.~(\ref{CCM}).

To calculate the center of mass contributions to the drift force
$\vec{F}_{\rm CM,\mathbf{r}_i}$~(\ref{driftforcef1}) and local energy 
$\mathcal{E}_{\rm CM,\mathbf{r}i}^{\rm loc}$~(\ref{localenergyf1}) we need expressions
for the gradient and Laplacian of the function~(\ref{Gaussianf1}).
Let us start by calculating the gradient 
$\vec{\nabla}_{\mathbf{r}_i} e^{-\alpha|\mathbf{r}_k-\mathbf{R}_{k}|^2}$
with respect to the particle $\mathbf{r}_i$ 
\begin{equation}
 \frac{ \vec{\nabla}_{\mathbf{r}_i}   e^{-\alpha|\mathbf{r}_k-\mathbf{R}_{k}|^2}}
 { e^{-\alpha|\mathbf{r}_k-\mathbf{R}_{k}|^2}}=
 \begin{cases}
     &\frac{2\alpha}{M}(\mathbf{r}_k-\mathbf{R}_k) \quad 
     {\rm if} \quad \mathbf{r}_k\neq\mathbf{r}_i, \\
     &2\alpha\left(\frac{1}{M}-1\right)(\mathbf{r}_i-\mathbf{R}_i) \quad
     {\rm if} \quad \mathbf{r}_k=\mathbf{r}_i, 
 \end{cases}    
 \label{GradientGaussian}
\end{equation}
where we assumed that $\mathbf{r}_k\in {\rm C}_i$.

The Laplacian
$\Delta_{\mathbf{r}_i} e^{-\alpha|\mathbf{r}_k-\mathbf{R}_{k}|^2}$
with respect to the particle $\mathbf{r}_i$ reads as
\begin{equation}
    \frac{ \Delta_{\mathbf{r}_i}   e^{-\alpha|\mathbf{r}_k-\mathbf{R}_{k}|^2}}
    { e^{-\alpha|\mathbf{r}_k-\mathbf{R}_{k}|^2}}=
 \begin{cases}
     &\frac{4\alpha^2}{M^2}|\mathbf{r}_k-\mathbf{R}_k|^2-\frac{2\alpha}{M^2}
     \quad {\rm if} \quad \mathbf{r}_k\neq\mathbf{r}_i, \\
     &4\alpha^2\left(1-\frac{1}{M}\right)^2|\mathbf{r}_i-\mathbf{R}_i|^2
    -2\alpha\left(1-\frac{1}{M}\right)^2
    \quad {\rm if} \quad \mathbf{r}_k=\mathbf{r}_i, 
 \end{cases}    
\label{LaplacianGaussian}
\end{equation}
where we assumed that $\mathbf{r}_k\in {\rm C}_i$.

Using Eq.~(\ref{driftforcef1}) and Eq.~(\ref{GradientGaussian}) 
we obtain an expression for the center of mass contribution to the
drift force $\vec{F}_{\rm CM,\mathbf{r}_i}$
\begin{equation}
\vec{F}_{\rm CM,\mathbf{r}_i}=
-2\alpha(\mathbf{r}_i-\mathbf{R}_i).
\end{equation}
From here, the square of the drift force $|\vec{F}_{\rm CM,\mathbf{r}_i}|^2$
reads as 
\begin{equation}
|\vec{F}_{\rm CM,\mathbf{r}_i}|^2=
4\alpha|\mathbf{r}_i-\mathbf{R}_i|^2.
\end{equation}
Using Eq.~(\ref{localenergyf1}) and Eq.~(\ref{LaplacianGaussian}) 
we obtain that 
the center of mass contribution to the local energy $\mathcal{E}_{\rm CM,\mathbf{r}i}^{\rm loc}$ reads as
\begin{equation}
    \mathcal{E}_{\rm CM,\mathbf{r}i}^{\rm loc}=
2\alpha\left(1-\frac{1}{M}\right).
\end{equation}
%%%%%%%%%%%%%%%%%%%%%%%%%%%%%%%%%%%%%%%%%%%%%%%%%%%%%%%%%%%%%%%
\section{Results}
In order to approximate the properties of the systems in the thermodynamic
limit we perform calculations in finite-size systems with periodic boundary
conditions. 
The total density of the system is defined as $n=N/(L_xL_y)$,
where $N$ is the total number of dipoles and
$L_x\times L_y$ is the size of the simulation box. 
In the gas phase, the simulations are performed
in a square box $L_x=L_y$, while in the solid and gas of chains 
phases the simulations are carried out in a rectangular box $L_x\neq L_y$,
commensurate with the crystal lattice.
We ensure that each of the box sides is a multiple
of elementary cell size of a triangular lattice.
%%%%%%%%%%%%%%%%%%%%%%%%%%%%%%%%%%%%%%%%%%%%%%%%%%%%%%%%%%%%%%%
\subsection{Crystallization and Threshold Densities}
%%%%%%%%%%%%%%%%%%%%%%%%%%%%%%%%%%%%%%%%%%%%%%%%%%%%%%%%%%%%%%%
\subsection*{Crystallization Density}
In Ref.~\cite{Astrakharchik2007}, the authors studied the ground-state phase diagram
of a two-dimensional Bose system with dipole-dipole interactions using the
QMC methods. The dipoles were constrained to move in a single plane and
were polarized in the perpendicular direction to the plane.
The authors found a quantum phase transition from a gas to a solid phase
as the density increases.
This transition was estimated to occur at the critical density 
\begin{equation}
    \tilde{n}\tilde{r}_0^2\approx290,
    \label{Eq:criticaldensity}
\end{equation}
with $\tilde{r}_0=r_0=md^2/\hbar^2$ and $\tilde{n}=n$ for a single layer of dipoles. 

\begin{figure}[!t]
	\centering
    \includegraphics[width=0.7\textwidth]{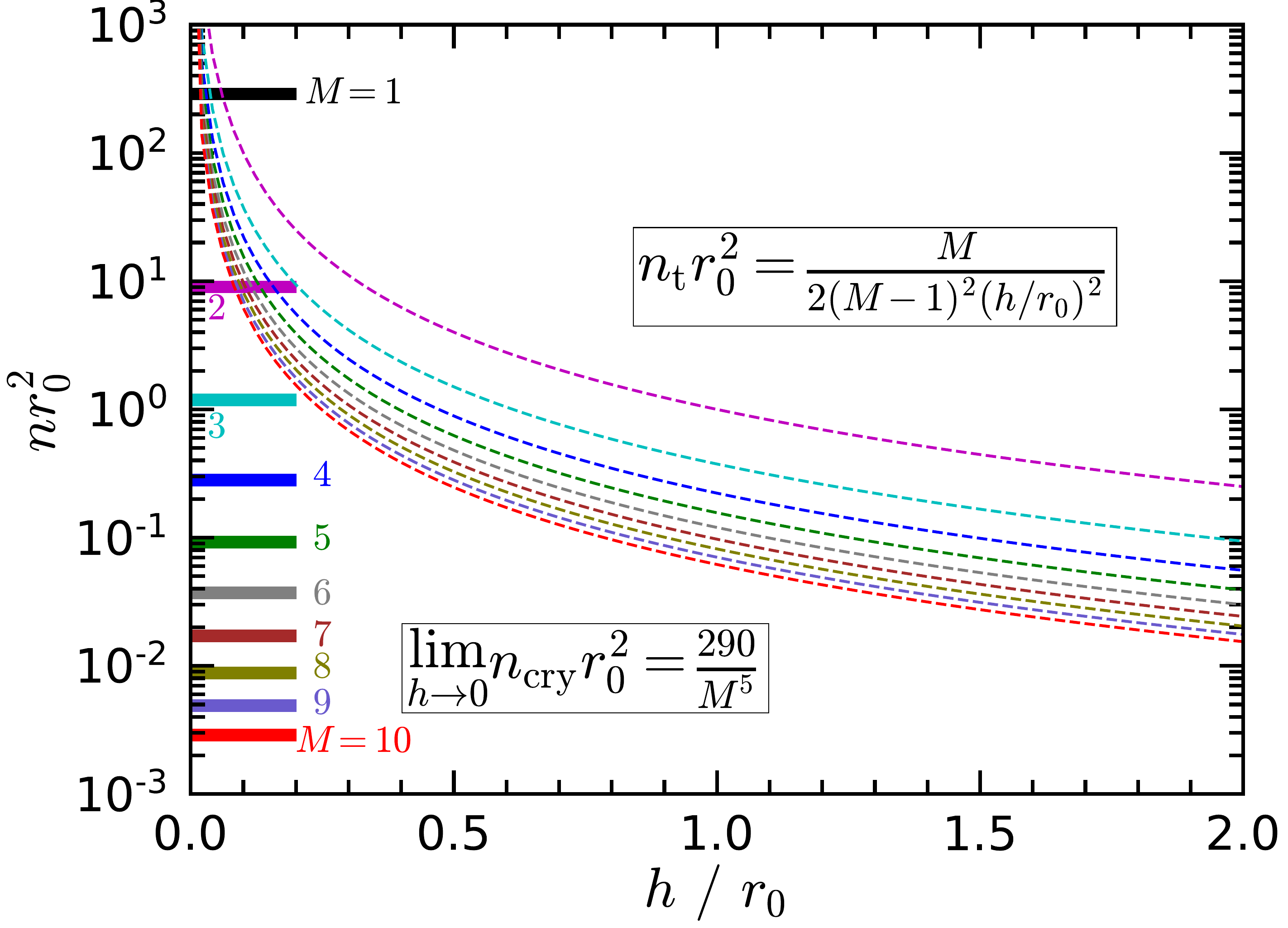}
    \caption{Schematic representation of the phase diagram of dipolar
        bosons confined to an $M$-layer geometry. The phase diagram is shown as
        a function of the total density $nr_0^2$, the separation between layers $h/r_0$,
        and for different number of layers $M$.
        The crystallization densities $n_{\rm cry}r_0^2$, Eq.~(\ref{soliddensity}),
        are shown by the thick lines. The threshold densities $n_{\rm t}r_0^2$,
        Eq.~(\ref{ncritic}), at which classical interaction between the
        top and bottom layers changes the sign are shown with dashed lines.}
\label{Fig:Ch7_PhaseDiagramEquations}
\end{figure}

Our system consists of dipolar bosons confined in an $M$-layer geometry
separated by a distance $h$.
Now, we are going to show how Eq.~(\ref{Eq:criticaldensity}) is
rewritten for the system with $M$ layers in the limit of rigid chains.
To do this,
consider a chain with $M$ dipoles, one in each layer. 
When $h\to0$ the chain becomes a super-dipole with
mass $Mm$, dipolar moment $Md$, and its dipolar length is given by
\begin{equation}
    \tilde{r}_0=\frac{(Mm)(Md)^2}{\hbar^2}=
    M^3\frac{md^2}{\hbar^2}=
    M^3r_0.
    \label{Eq:Sdipoler0}
\end{equation}
Now consider a $M$-layer system with $N$ dipoles and $N/M$ chains evenly distributed.
When $h\to0$ the $M$-layer system effectively becomes to a single-layer one, each chain becomes a super-dipole,
and the number of particles changes from $N$ dipoles to $N/M$ super-dipoles.
As a consequence of the latter the density now is given by
\begin{equation}
    \tilde{n}=\frac{n}{M}.
    \label{Eq:Sdipolen}
\end{equation}
Using Eq.~(\ref{Eq:Sdipoler0}) and Eq.~(\ref{Eq:Sdipolen}) we obtain
\begin{equation}
    \tilde{n}\tilde{r}_0^2=
    \left(\frac{n}{M}\right)\left(M^3r_0\right)^2
    =M^5nr_0^2.
\end{equation}
From the last relation and with Eq.~(\ref{Eq:criticaldensity}) we obtain
\begin{equation}
    nr_0^2=\frac{290}{M^5}.
\end{equation}
Notably, the solidification density has a strong dependence
(fifth power) on the number of layers.
In the limit when $h\to0$ an $M$-layer system of dipolar bosons will
crystallize at the critical density $290/M^5$ 
\begin{equation}
    \lim_{h\to0}  n_{\rm cry}r_0^2=\frac{290}{M^5}.
    \label{soliddensity}
\end{equation}
Such a strong dependence on $M$ makes the multilayer geometry a very
promising setup for future observation of solidification.
In Fig.~\ref{Fig:Ch7_PhaseDiagramEquations} we show a schematic
representation of the phase diagram of dipolar bosons confined to an
$M$-layer geometry. The phase diagram
is shown as a function of the total density $nr_0^2$, the separation
between layers $h/r_0$, and for different number of layers $M$.
The crystallization densities $n_{\rm cry}r_0^2$ Eq.~(\ref{soliddensity})
for different values of the number of layers are shown by the
thick lines. 
Notice that the critical density drops several orders
of magnitude when going from $M=2$ to 10 layers.
This is an important result because it tells us that a very dilute quantum
solid can be obtained just by increasing the number of layers in the system.
%%%%%%%%%%%%%%%%%%%%%%%%%%%%%%%%%%%%%%%%%%%%%%%%%%%%%%%%%%%%%%%
\subsection*{Threshold Density}
\begin{figure}[!b]
	\centering
    \includegraphics[width=0.7\textwidth]{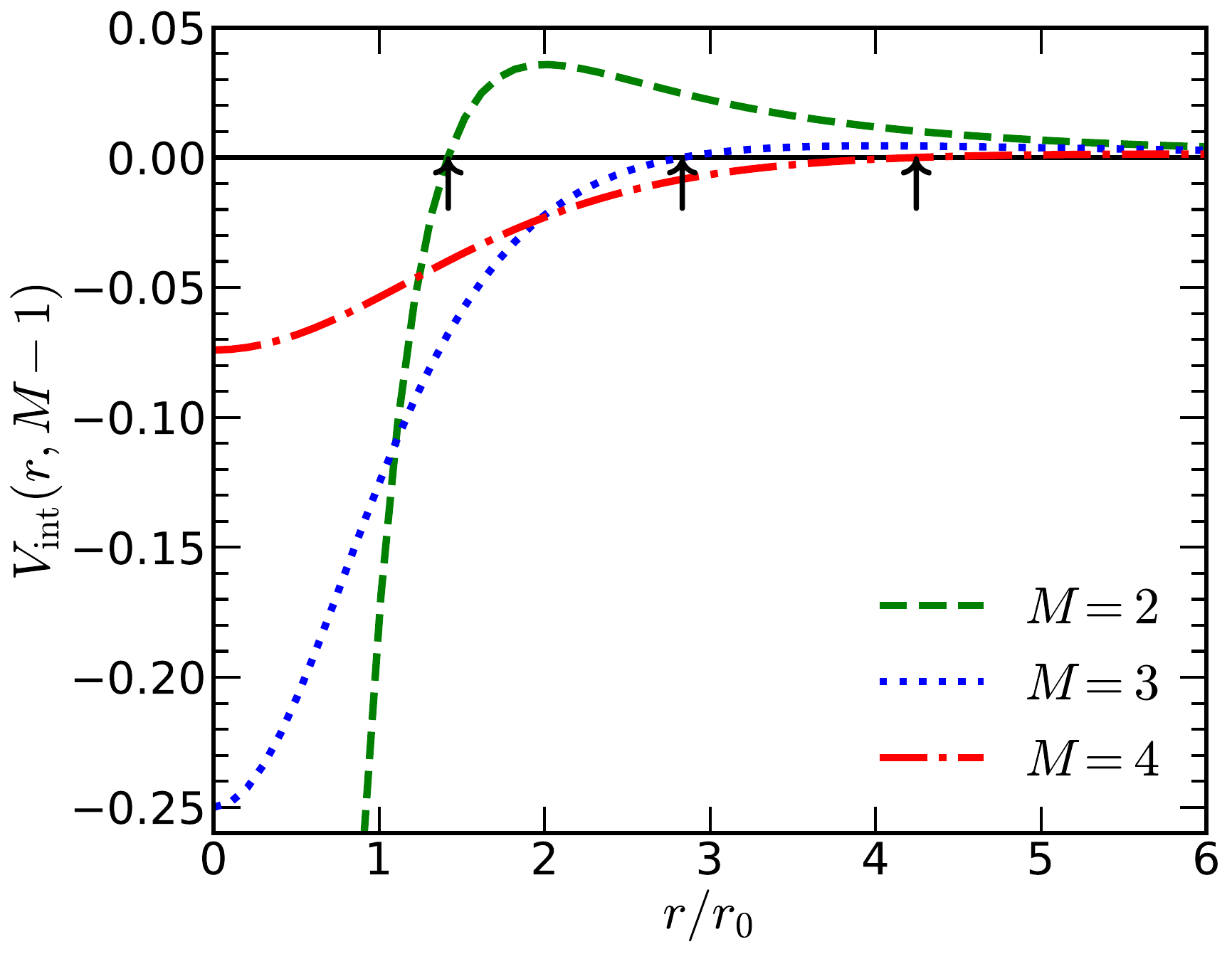}
    \caption{Interlayer potential $V_{\rm int}(r,M-1)$ as a function of $r/r_0$
        for different values of the number of layers $M$,
        with $h/r_0=1.0$. The arrows show the positions of the threshold
        distances $r_{\rm t}$ for $M=$2, 3 and 4.}
    \label{Fig:Ch7_ThresholdDistance}
\end{figure}

The dipolar interlayer potential Eq.~({\ref{Eq:DipolarPotential}})
\begin{equation}
    V_{\rm int}({r},l)=
    d^2\frac{\left(r^2-2l^2h^2\right)}{\left(r^2+l^2h^2\right)^{5/2}},\quad
    l=1,2,\ldots,M-1,
\label{Eq:interlayerpotential}
\end{equation}
is anisotropic. It is attractive for small values of $r$ but it is repulsive
for large values of $r$, where $r$ is the in-plane distance between
dipoles.
Consider an $M$-layer system of dipolar bosons in a solid phase.
Here, the system forms a triangular lattice.
Now, we ask ourselves what is the threshold distance $r_{\rm t}$
between a dipole on the first layer and a dipole on the last
layer and in a next neighbour lattice site such that the dipolar interlayer potential is zero.
This threshold distance $r_{\rm t}$ can be obtained from Eq.~({\ref{Eq:interlayerpotential}})
\begin{equation}
d^2\frac{r_{t}^2-2(M-1)^2h^2}{(r_{t}^2+(M-1)^2h^2)^{5/2}}
=0,
\label{potentialzero}
\end{equation}
with $l=M-1$. Now we solve for $r_{\rm t}$
\begin{equation}
    r_{\rm t}^2=2(M-1)^2h^2.
    \label{rcritic}
\end{equation}
In Fig.~\ref{Fig:Ch7_ThresholdDistance} we show the interlayer
potential $V_{\rm int}(r,M-1)$ as a function of $r/r_0$ for different values
of the number of layers $M$, with $h/r_0=1.0$.
The threshold distances $r_{\rm t}$ are the points where the potential
crosses the horizontal axis.
The values of $r_{\rm t}$ are indicated by arrows
in Fig.~\ref{Fig:Ch7_ThresholdDistance}.

Now, the mean-inter-particle distance in one layer is proportional to  
\begin{equation}
    \langle r/r_0\rangle \sim\frac{1}{\sqrt{nr_0^2/M}}\quad\to\quad
    \frac{nr_0^2}{M}\sim\frac{1}{\langle r/r_0\rangle^2}.
    \label{mpd}
\end{equation}

From the Eq.~(\ref{rcritic}) and Eq.~(\ref{mpd}) we obtain
the threshold density $n_{\rm t}$, which satifies Eq.~(\ref{potentialzero})
\begin{equation}
    n_{\rm t}r_0^2=
    \frac{M}{2(M-1)^2h^2}.
    \label{ncritic}
\end{equation}
The threshold density Eq.~(\ref{ncritic}) is an approximation.
Densities larger than the threshold density $n_{\rm t}r_0^2$
(attractive interlayer potential)
will favor the formation of a gas phase
and lower densities than the threshold density
(repulsive interlayer potential)
will favor the formation of the solid phase.

In Fig.~\ref{Fig:Ch7_PhaseDiagramEquations}
we plot the
threshold density, at which the interaction potential
between the most top and most bottom layers changes
its sign at the mean interparticle distance,
as a function
of the interlayer separation and
for different values of the number of layers. 
Each of the shown lines divides the phase diagram into two regions.
The first region is above the curve, where the formation of a
gas phase is favored. The second region is below the curve
where the formation of the solid phase is facilitated.
According to the prediction Eq.~(\ref{ncritic}),
it is possible to have a solid phase
at very low densities and for a wide range of $h/r_0$.

Finally, we want to emphasize that the equations Eq.~(\ref{soliddensity})
and Eq.~(\ref{ncritic}), which correspond to the crystallization $n_{\rm cry}r_0^2$ 
and threshold $n_{\rm t}r_0^2$ densities, are approximations under some assumptions.
These predictions give us a
picture of how 
the boundaries of the phase diagram are delimited.
To calculate the exact phase diagram it is necessary
to do DMC calculations, we will do this in the next section.

%%%%%%%%%%%%%%%%%%%%%%%%%%%%%%%%%%%%%%%%%%%%%%%%%%%%%%%%%%%%%%%
\subsection{Dipoles within a Four-Layer Geometry}
In this section, we present and discuss our DMC results
for the pair distribution functions and the ground-state
phase diagram of dipolar bosons confined to a four-layer geometry.
%%%%%%%%%%%%%%%%%%%%%%%%%%%%%%%%%%%%%%%%%%%%%%%%%%%%%%%%%%%%%%%
\subsubsection*{Four-Layer Phase Diagram}
The phase diagram of the dipolar multilayer is constructed is the following way. 
We explore the parameter space (total density $nr_0^2$
and interlayer distance $h/r_0$) and we calculate the
ground-state energy with each of the three trial wave functions:
$\Psi_{\rm Gas}$, $\Psi_{\rm Solid}$ and $\Psi_{\rm Chains}$
(see Section~\ref{Ch7:Details of the Methods}).
The phase at each point corresponds to
the phase that yields the lowest energy.

\begin{figure}[t!]
	\centering
    \includegraphics[width=0.7\textwidth]{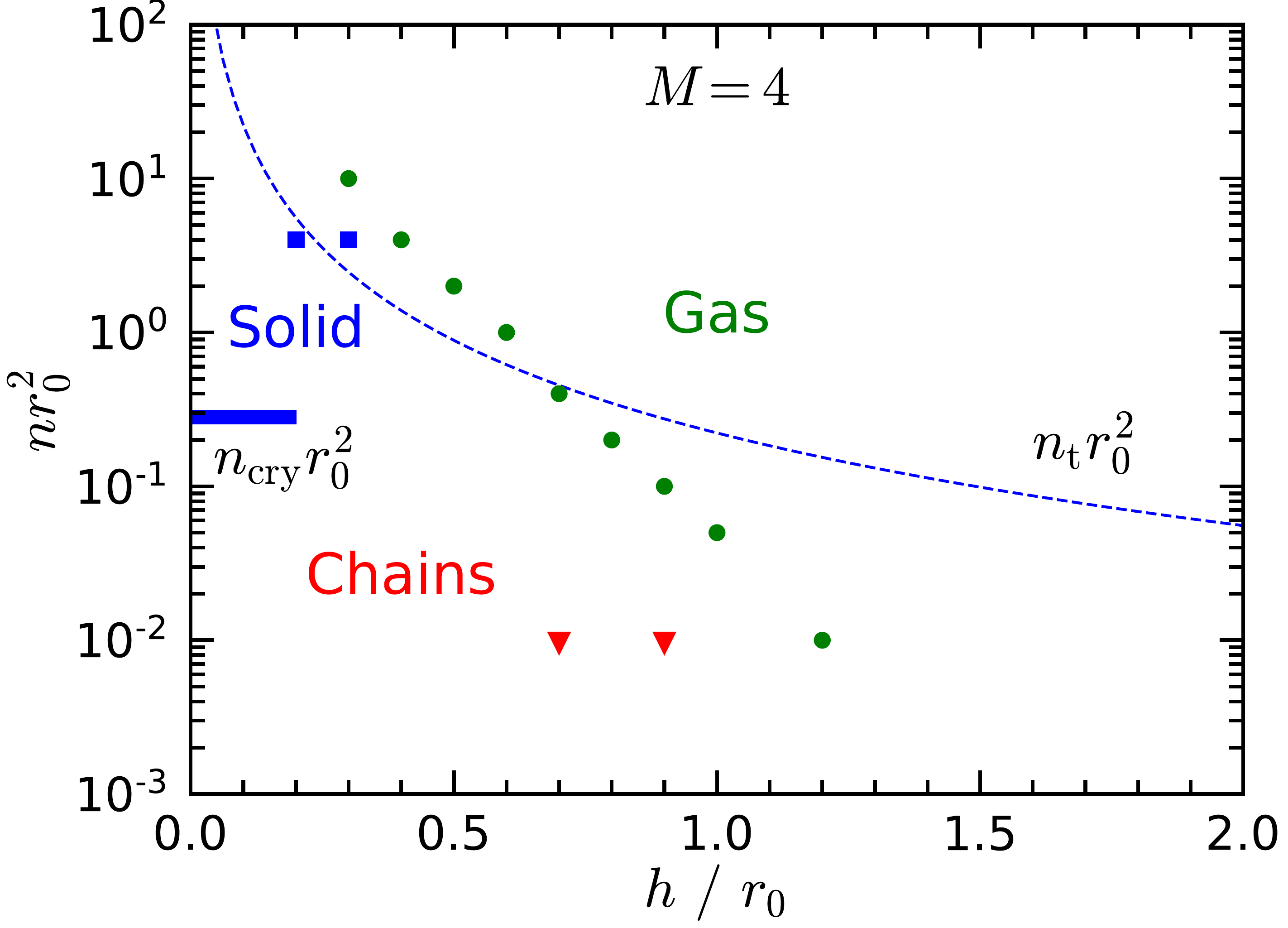}
    \caption{Ground-state phase diagram of dipolar bosons confined to a
        four-layer system at zero temperature.
        The phase diagram is shown as a function of the total density
        $nr_0^2$ and the separation between layers $h/r_0$.
        The green circles correspond to the gas phase, the blue squares
        correspond to the solid phase, and the red triangles correspond
        to the gas of chain phase.
        The crystallization $n_{\rm cry}r_0^2$ 
        and threshold $n_{\rm t}r_0^2$
        densities are shown by the thick and dashed lines,
        respectively.}
    \label{Fig:PDM4}
\end{figure}

The ground-state phase diagram of dipolar bosons within a four-layer
geometry is plotted in Fig.~\ref{Fig:PDM4}. The phase diagram is shown as a function of
the total density $nr_0^2$ and the separation between layers $h/r_0$.
The crystallization Eq.~(\ref{soliddensity}) and threshold Eq.~(\ref{ncritic})
densities for $M=4$ are shown by the thick and dashed lines,
respectively. We found three phases in this system: solid, gas, and gas of chains.
The gas phase and its boundaries were precisely determined,
as well as, the transitions solid-gas and chain-gas.
However, a precise estimation of the solid-chain transition location
is numerally complicated, 
because the energies of these phases are very similar over a wide range of the
parameter space.
We expect the gas of chains phase to appear below the crystallization density
$n_{\rm cry}r_0^2\approx0.283$.

The key result of the phase diagram of Fig.~\ref{Fig:PDM4} is the existence of
three phases: gas, solid, and gas of chains, in a wide range of densities
and interlayer distances. This characteristic
makes the experimental observation of these three phases more feasible.
%%%%%%%%%%%%%%%%%%%%%%%%%%%%%%%%%%%%%%%%%%%%%%%%%%%%%%%%%%%%%%%
\subsubsection*{Pair Distributions}
To quantify how the dipoles are spatially distributed
in the gas, solid, and gas of chains phases, we calculate
the pair distribution function $g(r,l)$ for different values
of the system parameters. The pair distribution $g(r,l)$
is proportional to
the probability of finding two particles at the relative
distance $r$ (see Subsection~\ref{Sec:PairDistribution}). 
Here, $lh$ denotes the interlayer distance between particles.
For example, $g(r,l=0)$ corresponds to the pair distribution of
particles in the same layer, $g(r,l=1)$ corresponds to the pair
distribution of particles at a one-layer distance.
\begin{figure}[!b]
  \centering
  \subfigure{\includegraphics[width=0.49\textwidth]{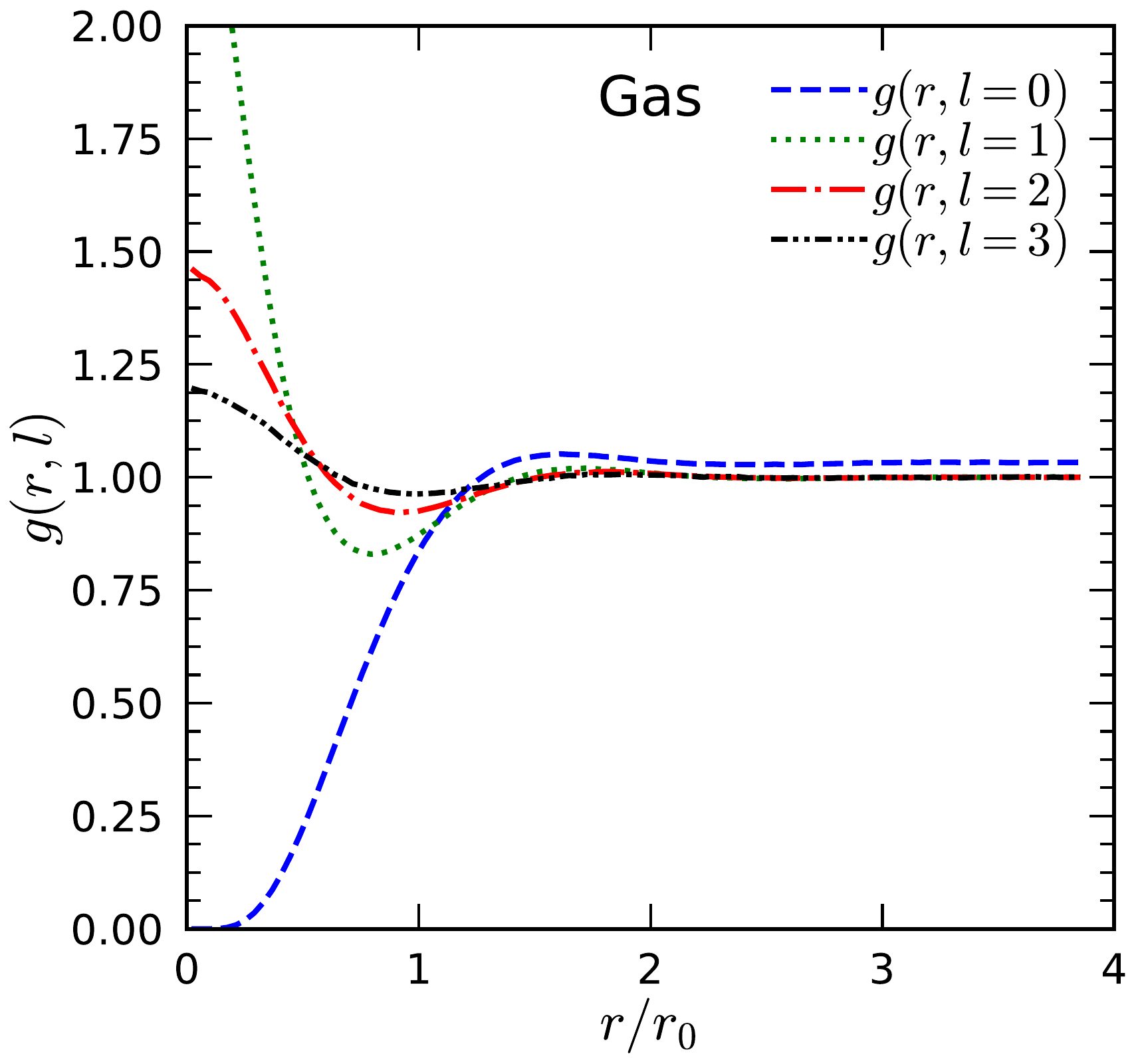}}\quad
  \subfigure{\includegraphics[width=0.46\textwidth]{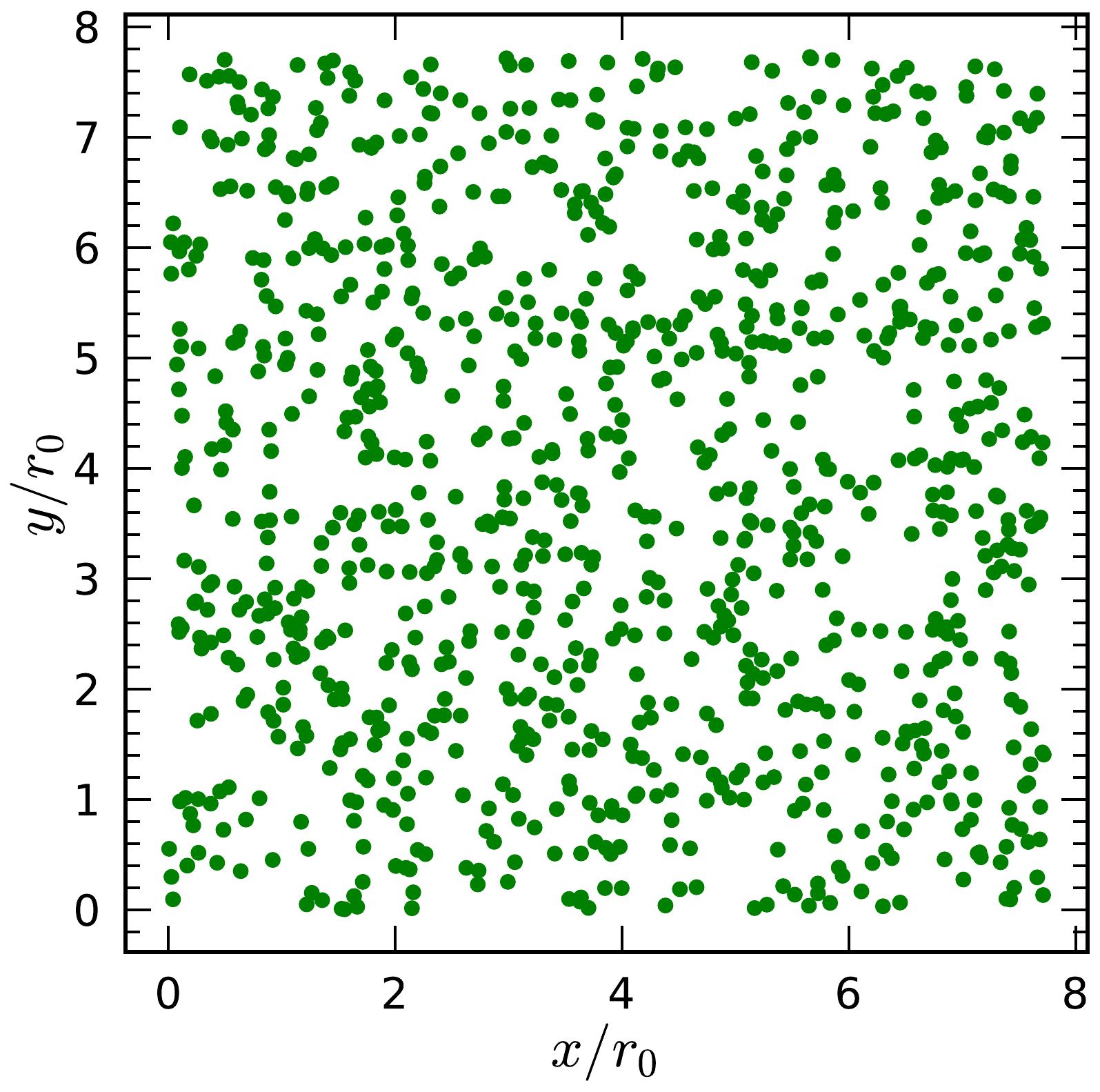}}
  \caption{Left panel: A typical example of the pair distribution functions $g(r,l)$
      for the gas phase within a four-layer geometry.
      The gas parameters are: total density $nr_0^2=2.0$ and interlayer distance $h/r_0=0.5$.
      Here, $lh$ denotes the interlayer distance between particles.
      Right panel: snapshot of the DMC simulations
      for the gas phase.}  
     \label{Fig:PairDistributionsGas}
 \end{figure}

In the left panel of Fig.~\ref{Fig:PairDistributionsGas} we show the pair distribution functions
$g(r,l)$ for the gas phase and for $l=0,1,2,3$. We have set the total density at
$nr_0^2=2.0$ and interlayer distance $h/r_0=0.5$.
The same-layer distribution $g(r,l=0)$
vanishes when $r/r_0\to0$ as a consequence of the strongly repulsive dipolar
intralayer potential. As $r$ increases, $g(r,l=0)$ exhibits
a shallow maximum, next it tends to 1, the asymptotic value of uncorrelated
particles.
The strong-correlation peak of $g(r,l\neq0)$ 
at $r/r_0=0$ is due to the interlayer attraction.
As $r$ increases, $g(r,l\neq0)$ exhibits
a minimum, next it tends to 1.
In the left panel of Fig.~\ref{Fig:PairDistributionsGas} we show
a snapshot of the particle coordinates
during the DMC simulation. Here, the dipoles are uniformly distributed. 
\begin{figure}[!t]
  \centering
  \subfigure{\includegraphics[width=0.49\textwidth]{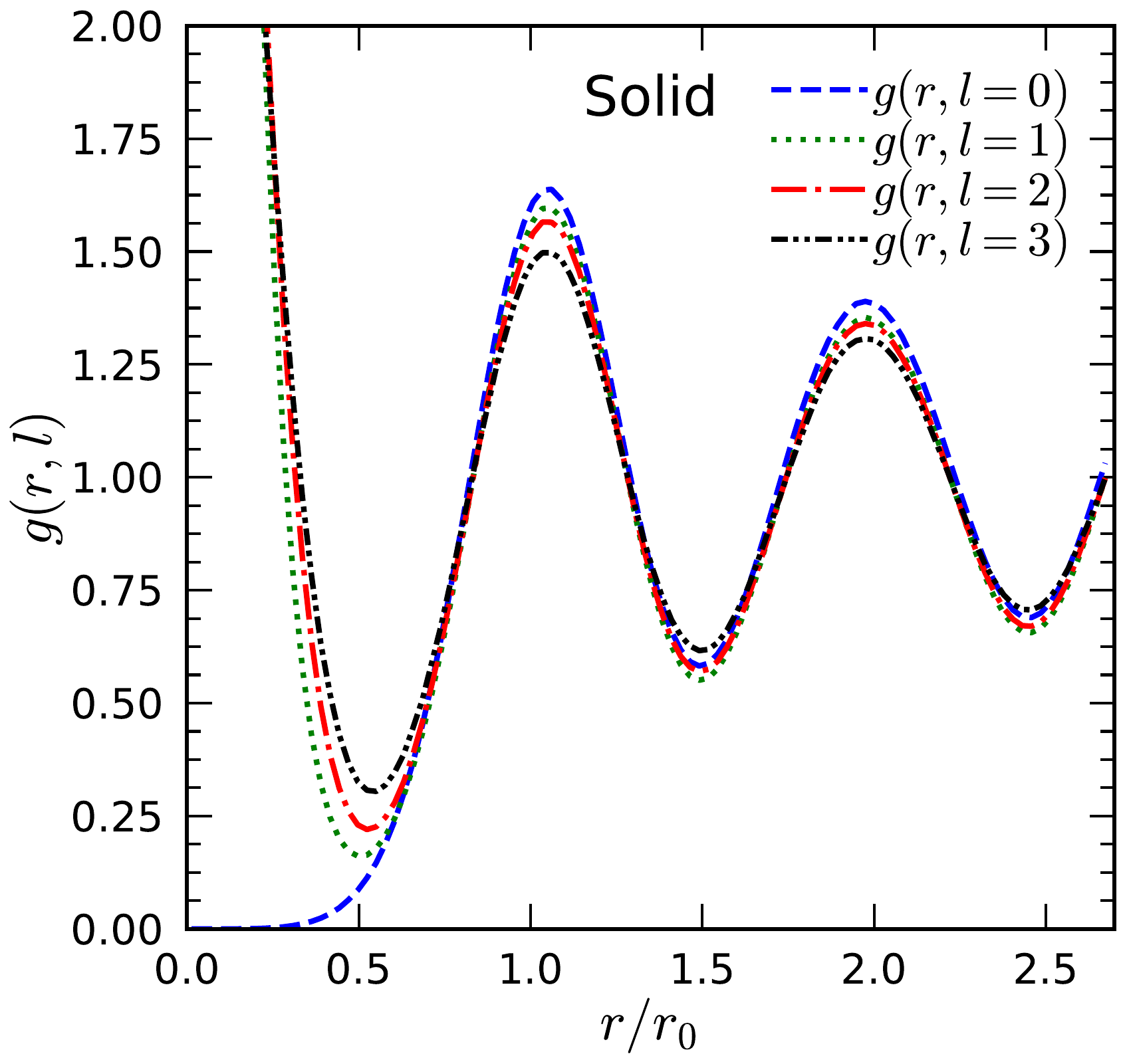}}\quad
  \subfigure{\includegraphics[width=0.46\textwidth]{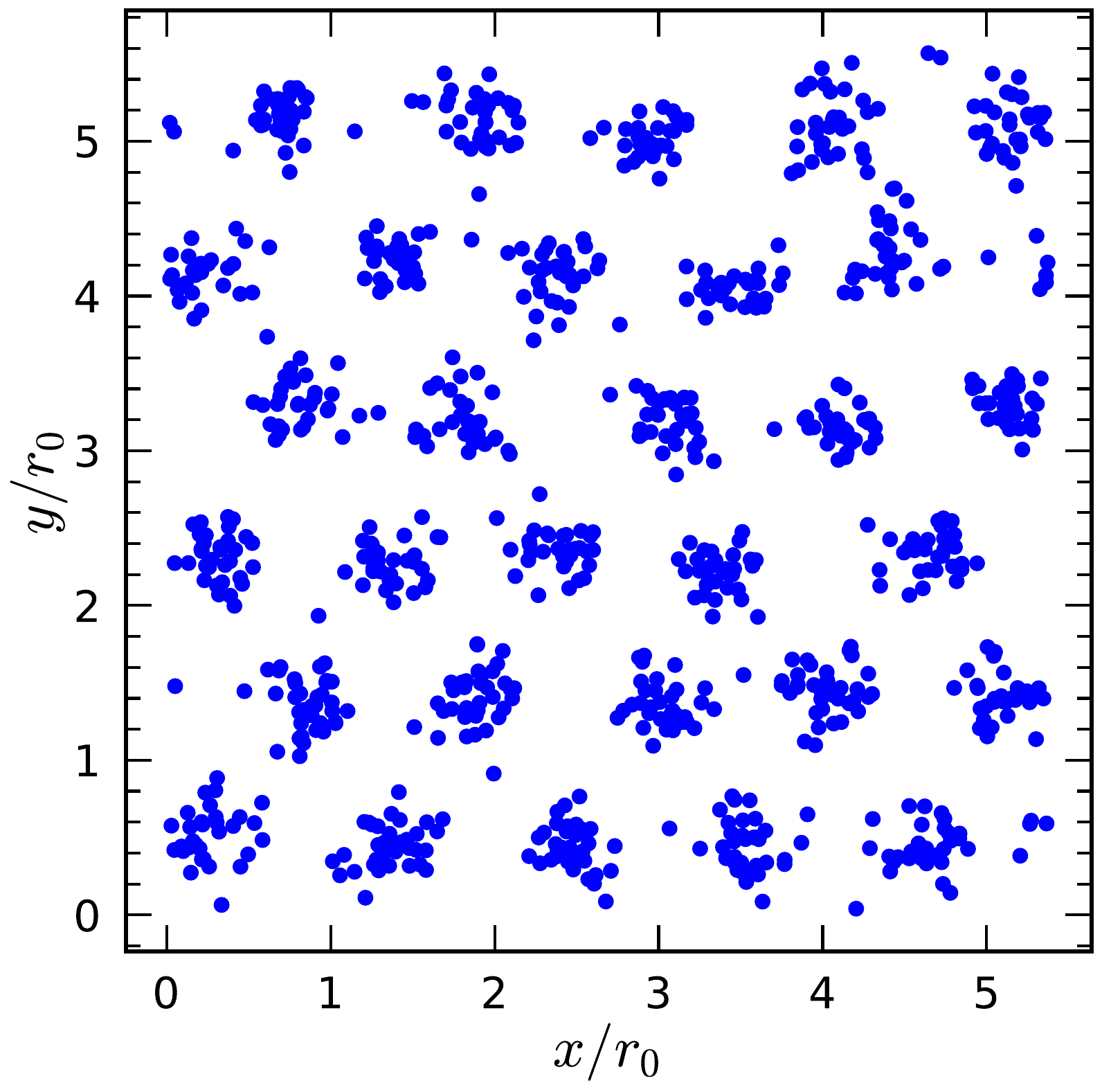}}
  \caption{Left panel: A typical example of the pair distribution functions $g(r,l)$
      for a solid phase within a four-layer geometry.
      The solid parameters are: 
      total density $nr_0^2=4.0$ and interlayer distance $h/r_0=0.3$.
      Here, $lh$ denotes the interlayer distance between particles.
      Right panel: snapshot of the DMC simulations
      for the solid phase.}  
  \label{Fig:PairDistributionsSolid}
\end{figure}

In the left panel of Fig.~\ref{Fig:PairDistributionsSolid} we plot the pair distribution functions
$g(r,l)$ for the solid phase and for $l=0,1,2,3$, with $nr_0^2=4.0$ and $h/r_0=0.3$.
When $r/r_0\to 0$ we observe a behavior that is similar to that previously
reported for the gas phase. The same-layer distribution start from 0 and the different-layer
distributions have a strong correlation peak.
As $r/r_0$ increases the solid pair distributions show a more complex structure than in the case of
the gas phase. Here, all $g(r,l)$ show some local maxima and minima rather than tend to 1.
These extreme values are related to the parameters of the system, $nr_0^2$ and $h/r_0$.
The maxima that appear around $r/r_0=1$ correspond to the mean
interparticle distance and are related to the density of the system.
That is, these maxima apper when $r/r_0\approx 1/\sqrt{nr_0^2/M}=1/\sqrt{4/4}=1$.
In the left panel of Fig.~\ref{Fig:PairDistributionsSolid} we show
a snapshot of the particle coordinates
during the DMC simulation. Here, the dipoles are distributed in a triangular lattice.

\begin{figure}[!t]
  \centering
  \subfigure{\includegraphics[width=0.485\textwidth]{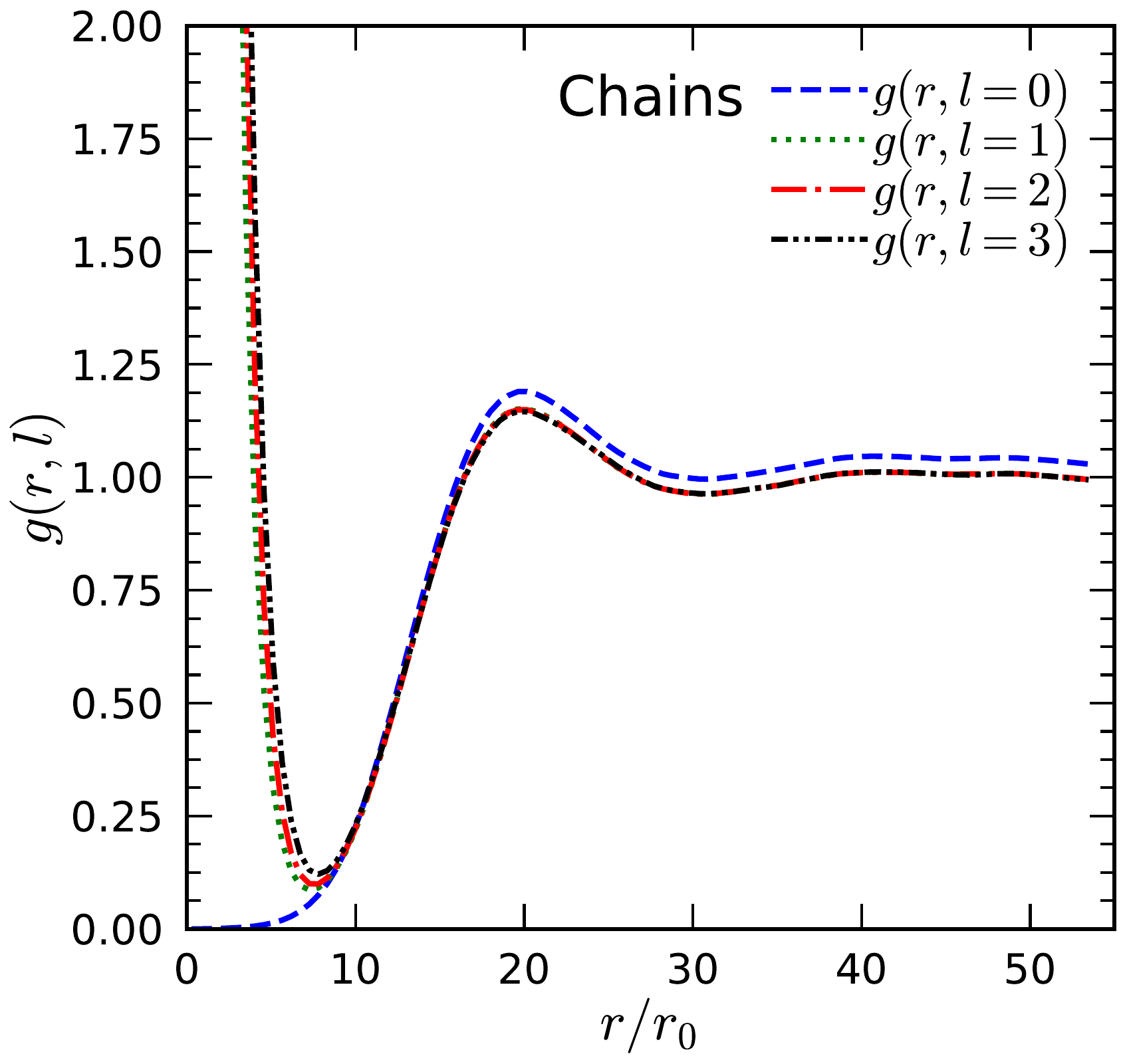}}\quad
  \subfigure{\includegraphics[width=0.48\textwidth]{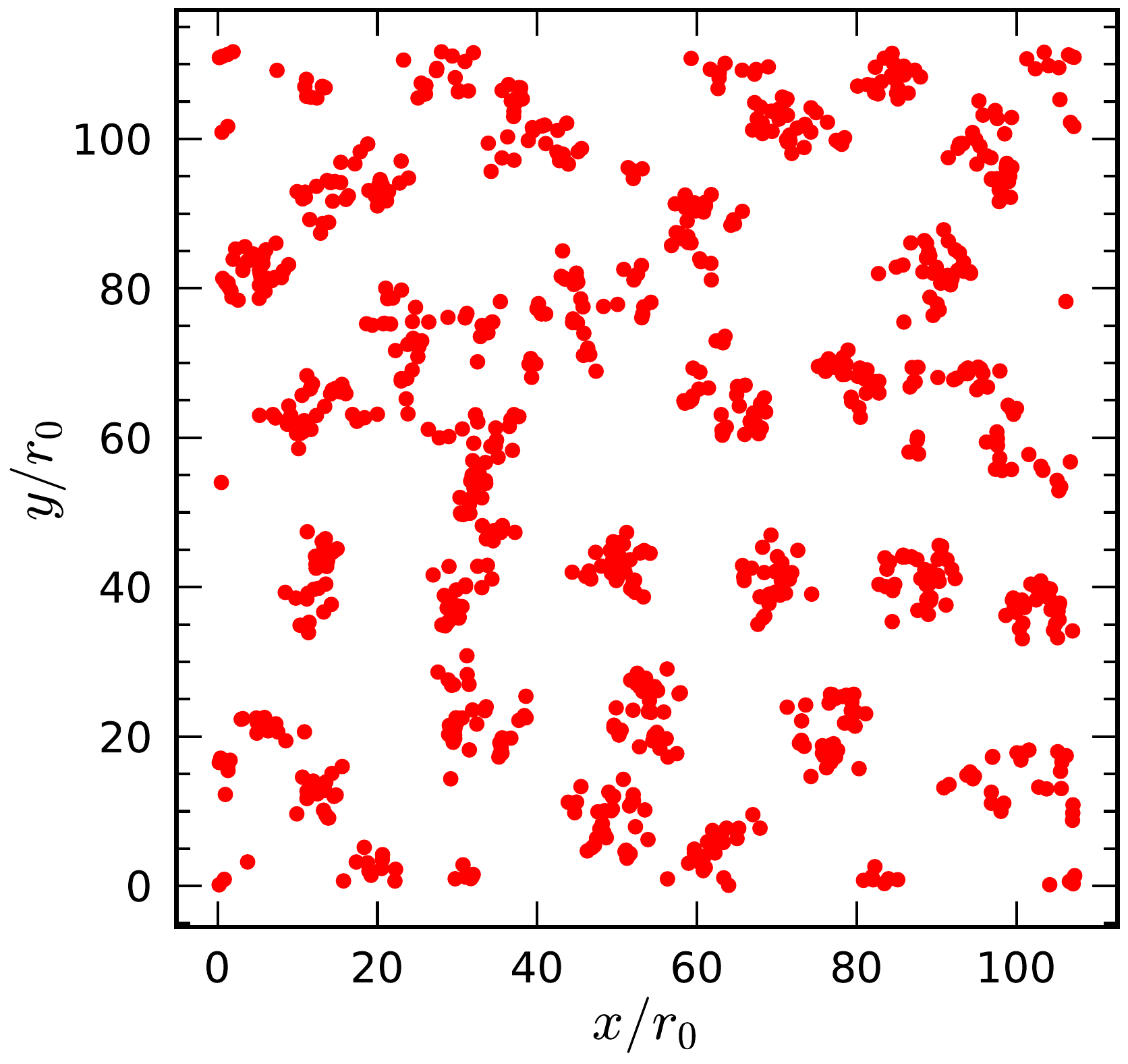}}
  \caption{Left panel: A typical example of the pair distribution functions $g(r,l)$
      for a gas of chains phase within a four-layer geometry.
      The chains parameters are: 
      total density $nr_0^2=0.01$ and interlayer distance $h/r_0=0.9$.
      Here, $lh$ denotes the interlayer distance between particles.
      Right panel: snapshot of the DMC simulations
      for the chains phase.}  
  \label{Fig:PairDistributionsChains}
\end{figure}

In the left panel of Fig.~\ref{Fig:PairDistributionsSolid} we plot the pair distribution functions
$g(r,l)$ for the gas of chains phase and for $l=0,1,2,3$, with $nr_0^2=0.01$ and $h/r_0=0.9$.
When $r/r_0\to 0$ we observe a similar behavior as for the gas and solid phases.
As $r/r_0$ increases $g(r,l)$
shows a maximum, next it tends to 1.
The maxima now appear at$ r/r_0\approx 1/\sqrt{nr_0^2/M}=1/\sqrt{0.01/4}$=20.
In the left panel of Fig.~\ref{Fig:PairDistributionsChains} we show
a snapshot of the particle coordinates
during the DMC simulation. Here, the dipolar chains are
homogeneously distributed.
%%%%%%%%%%%%%%%%%%%%%%%%%%%%%%%%%%%%%%%%%%%%%%%%%%%%%%%%%%%%%%%
\subsection{Dipoles within an M-Layer Geometry}
In this section, we present and discuss our DMC results for the phase
diagram of dipolar bosons confined to an $M$ parallel layers at
zero-temperature.

\begin{figure}[!t]
	\centering
    \includegraphics[width=0.7\textwidth]{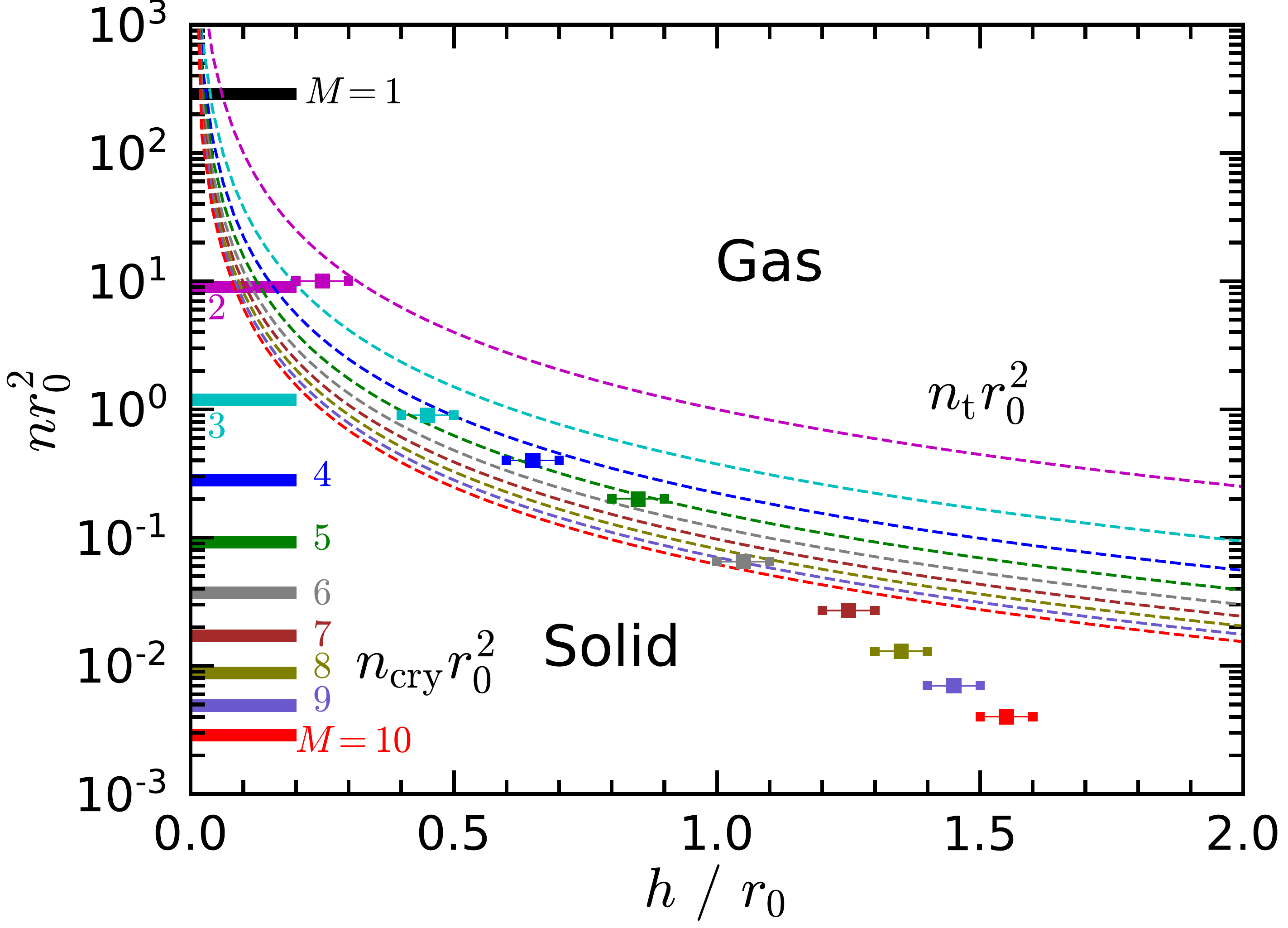}
    \caption{Ground-state phase diagram of dipolar bosons confined to a
        $M$-layer system at zero temperature. The phase diagram is shown as
        a function of the total density $nr_0^2$, the separation between layers $h/r_0$,
        and for different number of layers $M$. The squares indicated the
        interlayer critical value for the transition solid-gas
        for fixed values of density and number of layers $M$. 
        The crystallization $n_{\rm cry}r_0^2$ Eq.~(\ref{soliddensity})
        and threshold $n_{\rm t}r_0^2$ Eq.~(\ref{ncritic})
        densities are shown by the thick and dashed lines,
        respectively.}
    \label{Fig:Ch7_PhaseDiagram}
\end{figure}

The ground-state phase diagram of dipolar bosons within an $M$-layer
geometry is plotted in Fig.~\ref{Fig:Ch7_PhaseDiagram}. The phase diagram
is shown as a function of the total density $nr_0^2$, the separation
between layers $h/r_0$, and for different number of layers $M$.
The crystallization $n_{\rm cry}r_0^2$ and threshold $n_{\rm t}r_0^2$
densities are shown by the thick and dashed lines,
respectively.
The estimated critical interlayer distance for the transition solid-gas
is reported for $M=2$ up to 10. For each value of $M$, the critical
interlayer distance is reported for one density value.
Notice that, as we increase the number of layers the size of the
phase diagram increases for the solid phase. Also that the simple
approximation (dashed lines) discussed previously fails
dramatically when $M$ grows.
%%%%%%%%%%%%%%%%%%%%%%%%%%%%%%%%%%%%%%%%%%%%%%%%%%%%%%%%%%%%%%%
\section{Summary}
We used the diffusion Monte Carlo method to study the ground-state
phase diagram of dipolar bosons in a geometry formed by
equally spaced two-dimensional layers. This system is predicted
to have a rich collection of phases due to the anisotropic
and quasi long-range dipole-dipole interaction between the bosons.
We predicted several quantum phase transitions occurring at zero
temperature as the density and separation between
layers are changed.
First, we have considered the case where there are four layers and
the same number of dipoles in each layer. 
We observe a number of
distinct phases, including atomic gas, solid, and gas of chains.
These phases are present in a wide range of densities and interlayer
distances. 
The solid phase is predicted to be formed for large densities and small
interlayer distances, while the chain phase is presented
for lower densities and for a wide range of interlayer distances.
The atomic gas is observed in a wide range of densities and interlayers
distances.
The transitions of solid-gas and chain-gas were 
determined. Whereas, the exact location of the solid-chain transition
could not be determined.
Furthermore, we have considered the case where the dipoles are confined
up to ten layers.  
We find that the range of densities where the solid is observed decreases
several orders of magnitude with increasing the number of layers in the
system. Our results show that the dipolar multilayer system offers
a highly controllable setup for observing ultra-dilute quantum solids.

A subject of future work is the implementation of the
symmetric trial wave function to describe the solid phase.
Also, the calculation of the static structure factor
and the superfluid fraction for the solid, gas, and gas of
chains. The calculation of these properties will allow us to
accurately characterize the solid-chain transition.

\chapter{Conclusions}\label{Conclusiones}
In this Thesis, we reported a detailed study of the ground-state properties
of a set of quantum few- and many-body systems. In particular, 
we have studied one-dimensional Bose-Bose and Fermi-Bose mixtures
with attractive interspecies and repulsive intraspecies contact interactions.
Here, we characterized a three-dimer repulsion. Also, we have studied
a bosonic dipolar quantum system confined to a bilayer and multilayer
geometries. These setups consist of equally spaced two-dimensional layers,
which can be experimentally realized by imposing tight confinement in one
direction. We have studied the bilayer configuration with few and many dipoles
and the multilayer system with many dipoles only. In all cases, we used Quantum
Monte Carlo simulations to obtain the zero-temperature properties. 

Noticeable, we have found in the bilayer that some properties are inherited
from the few-dipole system to the many-dipole system. In the case of the
few-dipole system, we established the existence of dipolar bound states with
interesting properties. For example, we found that the bound states are
quantum halo states. In the case of the many-dipole system, we demonstrated
the existence of a dilute liquid phase. For the multilayer system we found
an extremely dilute solid phase. Our results show that a dipolar system in
a bilayer and a multilayer geometries offer clean and highly controllable setups 
for observing interesting phases of quantum matter, such as, halo states, and
dilutes liquids and solids. 

Below we present the main conclusions of this Thesis.
%%%%%%%%%%%%%%%%%%%%%%%%%%%%%%%%%%%%%%%%%%%%%%%%%%%%%%%%%%%%%%%%%%%%%%%%
\subsubsection*{One-Dimensional Three-Boson Problem with Two- and Three-Body
Interactions} 
In Chapter~\ref{Chapter:One-dimensional three-boson problem with two- and three-body
interactions}, by using the diffusion Monte Carlo technique we calculated the
binding energy of two and three dimers formed in a one-dimensional Bose-Bose or
Fermi-Bose mixture with attractive interspecies and repulsive intraspecies
interactions. Combining these results with a three-body theory~\cite{Guijarro2018},
we extracted the three-dimer scattering length close to the dimer-dimer zero crossing.
We argued that since in one dimension the three-body energy correction scales
logarithmically with the three-body scattering length $a_3$, three-body effects
are observable even for exponentially small $a_3$, which significantly simplifies
the task of engineering three-body-interacting systems in one dimension. We
demonstrated that Bose-Bose or Fermi-Bose dimers, previously shown to be tunable
to the dimer-dimer zero crossing, exhibit a noticeable three-dimer repulsion. We can
now be certain that the ground state of many such dimers slightly below the
dimer-dimer zero crossing is a liquid in which the two-body attraction is compensated
by the three-body repulsion \cite{Bulgac,PricoupenkoPetrov}.
%%%%%%%%%%%%%%%%%%%%%%%%%%%%%%%%%%%%%%%%%%%%%%%%%%%%%%%%%%%%%%%%%%%%%%%%
\subsubsection*{Few-Body Bound States of Two-Dimensional Bosons}
In Chapter~\ref{Chapter:Few-body bound states of two-dimensional bosons},
we have studied few-body clusters of the type A$_N$B$_M$ with
$N\leq M\leq 3$ in a two-dimensional Bose-Bose mixture of A and B bosons,
with attractive AB and equally repulsive AA and BB interactions.
We considered two very different models: dipolar bosons in a bilayer
geometry and particles interacting via separable Gaussian potentials.
In both cases, the intraspecies scattering length $a_{\1\1}=a_{\2\2}$ is
of the order of the potential ranges, whereas we tune $a_{\1\2}$ by adjusting
the AB attractive potential (or the interlayer distance in the bilayer setup).
We find that for $a_{\1\2}\gg a_{\2\2}$ all considered clusters are (weakly)
bound and their energies are independent of the interaction model. As
the ratio $a_{\1\2}/a_{\2\2}$ decreases, the increasing intraspecies
repulsion pushes the clusters upwards in energy and eventually breaks
them up into $N$ dimers and $M-N$ free B atoms. In the population
balanced case ($N=M$) this happens at $a_{\1\2}/a_{\2\2}\approx 10$
where the dimer-dimer attraction changes to repulsion. The
population-imbalanced ABB trimer, ABBB tetramer, and AABBB pentamer remain
bound beyond the dimer-dimer threshold. In the dipolar model, they break up at
$a_{\1\2}\approx 2 a_{\2\2}$ where the atom-dimer interaction switches to
repulsion. By studying the AAABBB hexamer near the dimer-dimer zero crossing
we find that it very much behaves like a system of three point-like particles
(dimers) characterized by an effective three-dimer repulsion. A
dipolar system in a bilayer geometry can thus exhibit the tunability
and mechanical stability necessary for observing dilute liquids and
supersolid phases.
%%%%%%%%%%%%%%%%%%%%%%%%%%%%%%%%%%%%%%%%%%%%%%%%%%%%%%%%%%%%%%%%%%%%%%%%
\subsubsection*{Quantum Halo States in Two-Dimensional Dipolar Clusters}
In Chapter~\ref{Chapter:Quantum halo states in two-dimensional dipolar clusters},
we have studied the ground-state properties of loosely bound few-body bound states
in a two-dimensional bilayer geometry. We have investigated whether halos, bound
states with a wave function that extends deeply into the classically forbidden region,
can occur in this system. The dipoles are confined to two layers, A and B, with
dipolar moments aligned perpendicularly to the layers. We have studied clusters
composed by up to six particles, for different values of the interlayer distance,
as candidates for quantum halo states. In the case of dimers, we find that for values
of the interlayer separation larger than $h/r_0 = 0.45$ the clusters are halo states
and they follow a universal scaling law. In the cases of trimers up to hexamers, we
find two very different halo structures. For large values of the interlayer separation,
the halo structures are almost radially symmetric and the distances between dipoles
are all of the same scales. In contrast, in the vicinity of the threshold for unbinding,
the clusters are elongated and highly anisotropic. Importantly, our results prove the
existence of stable halo states composed of up to six particles. To the best of our
knowledge this is the first time that halo states with such a large number of
particles are predicted and observed in a numerical experiment. Indeed, the addition
of particles to a two or three body halo states typically makes them shrink towards
a more compact liquid structure. This particular bilayer geometry is the reason of
our distinct results.
%%%%%%%%%%%%%%%%%%%%%%%%%%%%%%%%%%%%%%%%%%%%%%%%%%%%%%%%%%%%%%%%%%%%%%%%
\subsubsection*{Quantum Liquid of Two-Dimensional Dipolar Bosons}
In Chapter~\ref{Chapter:Two-dimensional dipolar liquid},
we have shown that a dipolar bilayer possesses a rich phase
diagram with quantum phase transitions between gas and solid phases
(known before), and a liquid phase (newly predicted). 
Remarkably, the liquid state, which results from the balance of a
dimer-dimer attraction and an effective three-dimer repulsion, exists
in a wide range of densities and interlayer separations which are
experimentally accessible. From the equations of state, we extracted
the spinodal and equilibrium densities, which are controllable through
the interlayer distance. The equilibrium density decreases as the
interlayer distances increases, allowing access to ultra-dilute liquids
in a stable setup.
%%%%%%%%%%%%%%%%%%%%%%%%%%%%%%%%%%%%%%%%%%%%
\subsubsection*{Phases of Dipolar Bosons Confined to a Multilayer Geometry}
In Chapter~\ref{Multilayer System of Dipolar Bosons}, we
have studied the ground-state phase diagram of dipolar bosons in a
multilayer geometry formed by equally spaced two-dimensional layers. 
We predicted several quantum phase transitions occurring at zero
temperature as the density and separation between layers are changed.
We have considered the case where there are four layers and
the same number of dipoles in each layer. When the dipole moment
of the bosons is aligned perpendicular to the layers, we observe 
three distinct phases, namely atomic gas, solid, and gas of chains.
These phases are present in wide range of densities and interlayer
distances. The solid phase is observed for large densities and small
interlayer distances. While the chain phase is presented
for lower densities and for a wide range of interlayer distances.
The atomic gas is observed in a wide range of densities and interlayers
distances. The transitions of solid-gas and chains-gas were precisely
determined. However, the transition solid-chains could not be fully
determined. Furthermore, we have considered the case where the dipoles
are confined to three up to ten layers. We find that the range of densities
where the solid is observed decreases several orders of magnitude with
increasing the number of layers in the system. Our results show that the
dipolar multilayer system offers a highly controllable setup for observing
ultra-dilute quantum solids.

\appendix
\chapter{Trial Wave Function}
\label{Appendix:Jastrow Trial wave function}
One of the most used many-body trial wave functions $\Psi_{\rm T}(\mathbf{R})$
in QMC methods for importance sampling is of the form
\begin{equation}               
\label{Eq:A01}                  
\Psi_{\rm T}(\mathbf{R})=
\prod_{i=1}^{N}f_1(\mathbf{r}_i)
\prod_{k<j}^{N}f_2(|\mathbf{r}_k-\mathbf{r}_j|)
\mathcal{S}(\mathbf{R}).
\end{equation}      
The one-body term $f_1(\mathbf{r}_i)$ depends only on the position of a single
particle $\mathbf{r}_i$. The product of two-body $f_2(|\mathbf{r}_k-\mathbf{r}_j|)$
terms is known as the Bijl-Jastrow factor. The factor $\mathcal{S}(\mathbf{R})$ 
defines the symmetry or antisymmetry of the trial wave function
$\Psi_{\rm T}(\mathbf{R})$ under the exchange of two particles.

In order to implement the QMC algorithm we need to calculate the gradient
$\nabla_{\mathbf{r}_i}\Psi_{\rm T}(\mathbf{R})$ and Laplacian
$\Delta_{\mathbf{r}_i}\Psi_{\rm T}(\mathbf{R})$ of the trial wave function.
Let us start by obtaining expressions for the gradients involved in
Eq.~(\ref{Eq:A01}). The gradient of the product of one-body terms $f_1(\mathbf{r}_j)$
with respect to the coordinate $\mathbf{r}_i$ is given by
\begin{equation}
\label{Eq:A03}
\frac{\vec{\nabla}_{\mathbf{r}_i}\prod_{j=1}^{N}
f_1(\mathbf{r}_j)}{\prod_{j=1}^{N}f_1(\mathbf{r}_j)}=
\frac{\vec{\nabla}_{\mathbf{r}_i}f_1(\mathbf{r}_i)}{f_1(\mathbf{r}_i)},
\end{equation}
and the gradient of the product of two-body terms 
$f_2(|\mathbf{r}_k-\mathbf{r}_j|)$ is given by
\begin{equation}
\label{Eq:A04}
\frac{\vec{\nabla}_{\mathbf{r}_i}
    \prod_{k<j}^{N}f_2(|\mathbf{r}_k-\mathbf{r}_j|)}
{\prod_{k<j}^{N}f_2(|\mathbf{r}_k-\mathbf{r}_j|)}
=\sum_{j}\frac{\vec{\nabla}_{\mathbf{r}_i}
f_2(|\mathbf{r}_i-\mathbf{r}_j|)}
{f_2(|\mathbf{r}_i-\mathbf{r}_j|)}.
\end{equation}
Using the expressions shown in the Eq.~(\ref{Eq:A03}) and
Eq.~(\ref{Eq:A04}) we obtain an expression
for the
gradient of the trial wave function 
$\vec{\nabla}_{\mathbf{r}_i}\Psi_{\rm T}(\mathbf{R})$
with respect to the coordinate $\mathbf{r}_i$
\begin{equation}
\label{Eq:A02} 
\vec{F}_{\mathbf{r}_i}=
\frac{\vec{\nabla}_{\mathbf{r}_i}\Psi_{\rm T}(\mathbf{R})}{\Psi_{\rm T}(\mathbf{R})}=
\frac{\vec{\nabla}_{\mathbf{r}_i}\mathcal{S}(\mathbf{R})}{\mathcal{S}(\mathbf{R})}
+\frac{\vec{\nabla}_{\mathbf{r}_i}f_1(\mathbf{r_i})}{f_1(\mathbf{r_i})}
+\sum_{j}\frac{\vec{\nabla}_{\mathbf{r}_i}
f_2(|\mathbf{r}_i-\mathbf{r}_j|)}
{f_2(|\mathbf{r}_i-\mathbf{r}_j|)}.
\end{equation}
Here, $\vec{F}_{\mathbf{r}_i}$ is called the drift force.
Now we are going to calculate the Laplacian expressions.
The Laplacian of the many-body trial wave function 
$\Delta_{\mathbf{r}_i}\Psi(\mathbf{R})$ with respect to the
coordinate $\mathbf{r}_i$ is given by
\begin{equation}
\label{Eq:A05}
\begin{aligned}
\frac{\Delta_{\mathbf{r}_i}\Psi(\mathbf{R})}{{\Psi(\mathbf{R})}}=&
\frac{{\Delta}_{\mathbf{r}_i}\mathcal{S}}{\mathcal{S}} 
+\frac{{\Delta}_{\mathbf{r}_i}\prod_{j=1}^{N}
    f_1(\mathbf{r_j})}{\prod_{j=1}^{N}
    f_1(\mathbf{r_j})} 
+\frac{{\Delta}_{\mathbf{r}_i}\prod_{k<j}^{N}
    f_2(|\mathbf{r}_k-\mathbf{r}_j|)}{\prod_{k<j}^{N}
    f_2(|\mathbf{r}_k-\mathbf{r}_j|)} \\
&+ 2\frac{\vec{\nabla}_{\mathbf{r}_i}\mathcal{S}}
{\mathcal{S}}\cdot
\left(
\frac{\nabla_{\mathbf{r}_i}
f_1(\mathbf{r_i})}{f_1(\mathbf{r_i})} 
+\sum_{j}\frac{\nabla_{\mathbf{r}_i}
f_2(|\mathbf{r}_i-\mathbf{r}_j|)}
{f_2(|\mathbf{r}_i-\mathbf{r}_j|)} 
\right)\\
&+ 2
\left(\frac{\nabla_{\mathbf{r}_i}
f_1(\mathbf{r_i})}{f_1(\mathbf{r_i})} \right)
\cdot\left(\sum_{j}\frac{\nabla_{\mathbf{r}_i}
f_2(|\mathbf{r}_i-\mathbf{r}_j|)}
{f_2(|\mathbf{r}_i-\mathbf{r}_j|)} \right),
\end{aligned}
\end{equation}
where we have used the expresiones Eq.~(\ref{Eq:A03}) and Eq.~(\ref{Eq:A04}).
The inner products that appear in Eq.~(\ref{Eq:A05}) are difficult to implement
in the QMC code. In order to remove these terms we considered the
square of the drift force,
from Eq.~(\ref{Eq:A02}) we obtain
\begin{equation}
\begin{aligned}
\label{Eq:A06}
|\vec{F_{\mathbf{r}_i}}|^2=&
\left(\frac{\vec{\nabla}_{\mathbf{r}_i}\mathcal{S}}{\mathcal{S}}\right)^2
+\left(\frac{\vec{\nabla}_{\mathbf{r}_i}f_1(\mathbf{r}_i)}{f_1(\mathbf{r}_i)}\right)^2
+\left(\sum_{j=1}^N\frac{\vec{\nabla}_{\mathbf{r}_i}f_2(|\mathbf{r}_i-\mathbf{r}_j|)}
{f_2(|\mathbf{r}_i-\mathbf{r}_j|)}\right)^2\\
&+2\frac{\vec{\nabla}_{\mathbf{r}_i}\mathcal{S}}{\mathcal{S}}\cdot
\left(\frac{\vec{\nabla}_{\mathbf{r}_i}f_1(\mathbf{r}_i)}{f_1(\mathbf{r}_i)}
+\sum_{j}\frac{\vec{\nabla}_{\mathbf{r}_i}f_2(|\mathbf{r}_i-\mathbf{r}_j|)}
{f_2(|\mathbf{r}_i-\mathbf{r}_j|)}\right)\\
&+2\left(\frac{\vec{\nabla}_{\mathbf{r}_i}f_1(\mathbf{r}_i)}{f_1(\mathbf{r}_i)}\right)\cdot
\left(\sum_{j}\frac{\vec{\nabla}_{\mathbf{r}_i}f_2(|\mathbf{r}_i-\mathbf{r}_j|)}
    {f_2(|\mathbf{r}_i-\mathbf{r}_j|)}\right).
\end{aligned}
\end{equation}
Now, we solve for the inner product terms
\begin{equation}
\begin{aligned}
\label{Eq:A07}
&|\vec{F_{\mathbf{r}_i}}|^2-
\left(\frac{\vec{\nabla}_{\mathbf{r}_i}\mathcal{S}}{\mathcal{S}}\right)^2
-\left(\frac{\vec{\nabla}_{\mathbf{r}_i}f_1(\mathbf{r}_i)}{f_1(\mathbf{r}_i)}\right)^2
-\left(\sum_{j}\frac{\vec{\nabla}_{\mathbf{r}_i}f_2(|\mathbf{r}_i-\mathbf{r}_j|)}
{f_2(|\mathbf{r}_i-\mathbf{r}_j|)}\right)^2=\\
&2\frac{\vec{\nabla}_{\mathbf{r}_i}\mathcal{S}}{\mathcal{S}}\cdot
\left(\frac{\vec{\nabla}_{\mathbf{r}_i}f_1(\mathbf{r}_i)}{f_1(\mathbf{r}_i)}
+\sum_{j}\frac{\vec{\nabla}_{\mathbf{r}_i}f_2(|\mathbf{r}_i-\mathbf{r}_j|)}
{f_2(|\mathbf{r}_i-\mathbf{r}_j|)}\right)\\
&+2\left(\frac{\vec{\nabla}_{\mathbf{r}_i}f_1(\mathbf{r}_i)}{f_1(\mathbf{r}_i)}\right)\cdot
\left(\sum_{j}\frac{\vec{\nabla}_{\mathbf{r}_i}f_2(|\mathbf{r}_i-\mathbf{r}_j|)}
{f_2(|\mathbf{r}_i-\mathbf{r}_j|)}\right).
\end{aligned}
\end{equation}
After substituting Eq.~(\ref{Eq:A07}) into Eq.~(\ref{Eq:A05}) we get
an expression for the Laplacian of the trial wave function
$\Delta_{\mathbf{r}_i}\Psi(\mathbf{R})$ with respect to the
coordinate $\mathbf{r}_i$ and without inner product terms 
\begin{equation}
\label{Eq:A08}
\begin{aligned}
\frac{\Delta_{\mathbf{r}_i}\Psi(\mathbf{R})}{{\Psi(\mathbf{R})}}=&
\frac{{\Delta}_{\mathbf{r}_i}\mathcal{S}}{\mathcal{S}} 
+\frac{{\Delta}_{\mathbf{r}_i}\prod_{j=1}^{N}
    f_1(\mathbf{r}_j)}{\prod_{j=1}^{N}
f_1(\mathbf{r}_j)} 
+\frac{{\Delta}_{\mathbf{r}_i}\prod_{k<j}^{N}
    f_2(|\mathbf{r}_k-\mathbf{r}_j|)}{\prod_{k<j}^{N}
    f_2(|\mathbf{r}_k-\mathbf{r}_j|)}
+|\vec{F_{\mathbf{r}_i}}|^2\\
&-\left(\frac{\vec{\nabla}_{\mathbf{r}_i}\mathcal{S}}{\mathcal{S}}\right)^2
-\left(\frac{\vec{\nabla}_{\mathbf{r}_i}f_1(\mathbf{r}_i)}{f_1(\mathbf{r}_i)}\right)^2
-\left(\sum_{j}\frac{\vec{\nabla}_{\mathbf{r}_i}f_2(|\mathbf{r}_i-\mathbf{r}_j|)}
{f_2(|\mathbf{r}_i-\mathbf{r}_j|)}\right)^2.
\end{aligned}
\end{equation}
Now we need to calculate the expressions of the Laplacians
$\Delta_{\mathbf{r}_i}\prod_{j=1}^{N}f_1(\mathbf{r}_j)$ and
$\Delta_{\mathbf{r}_i}\prod_{k<j}^Nf_2(|\mathbf{r}_k-\mathbf{r}_j|)$
and then substitute them in the last equation.
The Laplacian $\Delta_{\mathbf{r}_i}\prod_{j=1}^{N}f_1(\mathbf{r}_j)$
is given by
\begin{equation}
    \label{Eq:A09}
\frac{\Delta_{\mathbf{r}_i}\prod_{j=1}^{N}
    f_1(\mathbf{r}_j)}{\prod_{j=1}^{N}f_1(\mathbf{r}_j)}=
\frac{{\Delta}_{\mathbf{r}_i}f_1(\mathbf{r}_i)}{f_1(\mathbf{r}_i)},
\end{equation}
and the Laplacian
$\Delta_{\mathbf{r}_i}\prod_{k<j}^Nf_2(|\mathbf{r}_k-\mathbf{r}_j|)$
reads as
\begin{equation}
\begin{aligned}
    \label{Eq:A10}
    \frac{\Delta_{\mathbf{r}_i}\prod_{k<j}^Nf_2(|\mathbf{r}_k-\mathbf{r}_j|)}
    {\prod_{k<j}^Nf_2(|\mathbf{r}_k-\mathbf{r}_j|)}=&
\sum_{j=1}^N\left[ 
\frac{\Delta_{\mathbf{r}_i}f_2(|\mathbf{r}_i-\mathbf{r}_{j}|)}
{f_2(|\mathbf{r}_i-\mathbf{r}_{j}|)}-
\left(\frac{\vec{\nabla}_{\mathbf{r}_i}f_2(|\mathbf{r}_i-\mathbf{r}_{j}|)}
{f_2(|\mathbf{r}_i-\mathbf{r}_{j}|)}\right)^2 
\right]\\
&+\left( \sum_{j}
\frac{\vec{\nabla}_{\mathbf{r}_i}f_2(|\mathbf{r}_i-\mathbf{r}_{j}|)}
{f_2(|\mathbf{r}_i-\mathbf{r}_{j}|)}\right)^2.
\end{aligned}
\end{equation}
Substituting the Eq.~(\ref{Eq:A09}) and Eq.~(\ref{Eq:A10})
into Eq.~(\ref{Eq:A08}) we obtain
an expression for the Laplacian of the trial wave function
$\Delta_{\mathbf{r}_i}\Psi(\mathbf{R})$ with respect to the
coordinate $\mathbf{r}_i$ 
\begin{equation}
\label{Eq:A11}
\begin{aligned}
\frac{\Delta_{\mathbf{r}_i}\Psi(\mathbf{R})}{{\Psi(\mathbf{R})}}=&
|\vec{F_{\mathbf{r}_i}}|^2
+\frac{{\Delta}_{\mathbf{r}_i}\mathcal{S}}{\mathcal{S}} 
-\left(\frac{\vec{\nabla}_{\mathbf{r}_i}\mathcal{S}}{\mathcal{S}}\right)^2
+\frac{\Delta_{\mathbf{r}_i}f_1(\mathbf{r}_i)}{f_1(\mathbf{r}_i)} 
-\left(\frac{\vec{\nabla}_{\mathbf{r}_i}f_1(\mathbf{r}_i)}{f_1(\mathbf{r}_i)}\right)^2\\
&+\sum_{j}\left[ 
\frac{\Delta_{\mathbf{r}_i}f_2(|\mathbf{r}_i-\mathbf{r}_{j}|)}
{f_2(|\mathbf{r}_i-\mathbf{r}_{j}|)}-
\left(\frac{\vec{\nabla}_{\mathbf{r}_i}f_2(|\mathbf{r}_i-\mathbf{r}_{j}|)}
{f_2(|\mathbf{r}_i-\mathbf{r}_{j}|)}\right)^2 
\right].
\end{aligned}
\end{equation}
The Eq.~(\ref{Eq:A02}) and Eq.~(\ref{Eq:A11}) 
are the equations that are implemented in the QMC code.
Now we can write an expression for the kinetic energy
\begin{equation}
    T^{\rm loc}=\frac{\hbar^2}{2m}
    \left[
        \sum_{i=1}^N\mathcal{E}_{\mathcal{S},i}^{\rm loc}+
        \sum_{i=1}^N\mathcal{E}_{1,i}^{\rm loc}+
        2\sum_{i<j}^N\mathcal{E}_{2,i}^{\rm loc}
   -\sum_{i=1}^N|\vec{F}_{1,\mathbf{r}_i}|^2
    \right],
    \label{localenergyJastrow}
\end{equation}
where we have used Eq.~(\ref{Eq:A06}) and defined
\begin{equation}
\label{Eq:A12}
\mathcal{E}_{\mathcal{S},i}^{\rm loc}=
-\frac{{\Delta}_{\mathbf{r}_i}\mathcal{S}}{\mathcal{S}} 
+\left(\frac{\vec{\nabla}_{\mathbf{r}_i}\mathcal{S}}{\mathcal{S}}\right)^2,
\end{equation}
\begin{equation}
\label{Eq:A13}
\mathcal{E}_{1,i}^{\rm loc}=
-\frac{\Delta_{\mathbf{r}_i}f_1(\mathbf{r}_i)}{f_1(\mathbf{r}_i)} 
+\left(\frac{\vec{\nabla}_{\mathbf{r}_i}f_1(\mathbf{r}_i)}{f_1(\mathbf{r}_i)}\right)^2,
\end{equation}
\begin{equation}
\label{Eq:A14}
\mathcal{E}_{2,i}^{\rm loc}=-
\sum_{j}\left[ 
\frac{\Delta_{\mathbf{r}_i}f_2(|\mathbf{r}_i-\mathbf{r}_{j}|)}
{f_2(|\mathbf{r}_i-\mathbf{r}_{j}|)}-
\left(\frac{\vec{\nabla}_{\mathbf{r}_i}f_2(|\mathbf{r}_i-\mathbf{r}_{j}|)}
{f_2(|\mathbf{r}_i-\mathbf{r}_{j}|)}\right)^2 
\right].
\end{equation}

\chapter{Symmetric Trial Wave Function}
\label{Appendix:Sym Trial wave function}
An important part of the VMC and DMC methods is the choice of the
trial wave function, which is used for importance sampling.
Here, we consider a symmetric many-body 
trial wave function $\Psi_{\rm S}(\mathbf{R})$, which is given by
\begin{equation}               
\begin{aligned}
\label{Eq:B01}                  
\Psi_{\rm S}(\mathbf{R})=&\prod_{i<j}^{N_{\1}}f_{\1\1}(|\mathbf{r}_i-\mathbf{r}_j|)
\prod_{\alpha<\beta}^{N_{\2}}f_{\2\2}(|\mathbf{r}_{\alpha}-\mathbf{r}_{\beta}|)\\
&\times\left[\prod_{i=1}^{N_{\1}}\sum_{\alpha=1}^{N_{\2}}f_{\1\2}(|\mathbf{r}_i-\mathbf{r}_{\alpha}|)+
\prod_{\alpha=1}^{N_{\2}}\sum_{i=1}^{N_{\1}}f_{\1\2}(|\mathbf{r}_i-\mathbf{r}_{\alpha}|)\right].                
\end{aligned}
\end{equation}      
We use $\Psi_{\rm S}(\mathbf{R})$ to study a mixture of A and B bosons with attractive interspecies AB interactions
and equally repulsive intraspecies AA and BB interactions. 
In Eq.~(\ref{Eq:B01}), $N_\1$ and $N_\2$ are the number of bosons of the species A and B, respectively.
We denote with Latin letters the bosons of the species A and with Greek letters
the bosons of the species B. 

In order to implement the QMC algorithm we need to 
calculate the gradient $\nabla_{\mathbf{r}_i}\Psi_{\rm S}(\mathbf{R})$ and Laplacian
$\Delta_{\mathbf{r}_i}\Psi_{\rm S}(\mathbf{R})$ of the trial wave function. In the
following we are going to obtain the expressions of
$\nabla_{\mathbf{r}_i}\Psi_{\rm S}(\mathbf{R})$ and $\Delta_{\mathbf{r}_i}\Psi_{\rm S}(\mathbf{R})$.
To simplifying the expressions we defined the following quantities  
\begin{equation}
\label{Eq:B02}
\begin{aligned}
&\mathbb{A}\equiv\prod_{i<j}^{N_{\1}}f_{\1\1}(|\mathbf{r}_i-\mathbf{r}_j|),
&\mathbb{P}_1\equiv\prod_{i=1}^{N_{\1}}\sum_{\alpha=1}^{N_{\2}}f_{\1\2}(|\mathbf{r}_i-\mathbf{r}_{\alpha}|),\\
&\mathbb{B}\equiv\prod_{\alpha<\beta}^{N_{\2}}f_{\2\2}(|\mathbf{r}_{\alpha}-\mathbf{r}_{\beta}|),
&\mathbb{P}_2\equiv\prod_{\alpha=1}^{N_{\2}}\sum_{i=1}^{N_{\1}}f_{\1\2}(|\mathbf{r}_i-\mathbf{r}_{\alpha}|).
\end{aligned}
\end{equation}
With the above definitions the trial wave function $\Psi_{\rm S}(\mathbf{R})$
Eq.~(\ref{Eq:B01}) reduces to 
\begin{equation}
\label{Eq:B03} 
\Psi_{\rm S}(\mathbf{R})=\mathbb{A}\mathbb{B}[\mathbb{P}_1+\mathbb{P}_2].
\end{equation}
Let us now obtain expressions for the gradients involved in Eq.~(\ref{Eq:B01}).
The gradient of the trial wave function
$\vec{\nabla}_{\mathbf{r}_i}\Psi_{\rm S}(\mathbf{R})$ with respect to the
coordinate $\mathbf{r}_i$ is given by
\begin{equation}
\label{Eq:B04} 
\vec{F}_{\mathbf{r}_i}=\frac{\vec{\nabla}_{\mathbf{r}_i}
    \Psi_{\rm S}(\mathbf{R})}{\Psi_{\rm S}(\mathbf{R})}=
\frac{\vec{\nabla}_{\mathbf{r}_i}\mathbb{A}}
{\mathbb{A}}+\frac{\vec{\nabla}_{\mathbf{r}_i}\mathbb{P}_1+
\vec{\nabla}_{\mathbf{r}_i}\mathbb{P}_2}{\mathbb{P}_1+\mathbb{P}_2}.
\end{equation}
Here, $\vec{F}_{\mathbf{r}_i}$ is called the drift force.
The term $\vec{\nabla}_{\mathbf{r}_i}\mathbb{B}$ does not appears in
Eq.~(\ref{Eq:B04}), because $\mathbb{B}$ is independent of the 
coordinate $\mathbf{r}_i$.
The expressions of the gradients
$\vec{\nabla}_{\mathbf{r}_i}\mathbb{A}$,
$\vec{\nabla}_{\mathbf{r}_i}\mathbb{P}_1$ and
$\vec{\nabla}_{\mathbf{r}_i}\mathbb{P}_2$ 
with respect to the coordinate $\mathbf{r}_i$ are given by
\begin{equation}
\begin{aligned}
\label{Eq:B05} 
\frac{\vec{\nabla}_{\mathbf{r}_i}\mathbb{A}}{\mathbb{A}}=&\sum_{j}^{N_{\1}}\frac{\vec{\nabla}_{\mathbf{r}_i}f_{\1\1}(|\mathbf{r}_i-\mathbf{r}_j|)}{f_{\1\1}(|\mathbf{r}_i-\mathbf{r}_j|)},\\
\frac{\vec{\nabla}_{\mathbf{r}_i}\mathbb{P}_1}{\mathbb{P}_1}=&\sum_{\alpha=1}^{N_{\2}} \frac{\vec{\nabla}_{\mathbf{r}_i}f_{\1\2}(|\mathbf{r}_i-\mathbf{r}_{\alpha}|)}{\sum_{\alpha=1}^{N_{\2}}f_{\1\2}(|\mathbf{r}_i-\mathbf{r}_{\alpha}|)},\\
\frac{\vec{\nabla}_{\mathbf{r}_i}\mathbb{P}_2}{\mathbb{P}_2}=&\sum_{\alpha=1}^{N_{\2}} \frac{\vec{\nabla}_{\mathbf{r}_i}f_{\1\2}(|\mathbf{r}_i-\mathbf{r}_{\alpha}|)}{\sum_{i=1}^{N_{\1}}f_{\1\2}(|\mathbf{r}_i-\mathbf{r}_{\alpha}|)}. 
\end{aligned}
\end{equation}
Now we are going to calculate the expressions for the Laplacians involved in
Eq.~(\ref{Eq:B01}). 
The Laplacian of the trial wave function $\Delta_{\mathbf{r}_i}\Psi_{\rm S}(\mathbf{R})$
with respect to the coordinate $\mathbf{r}_i$ is given by
\begin{equation}
    \label{Eq:B06}
\frac{\Delta_{\mathbf{r}_i}\Psi_{\rm S}(\mathbf{R})}{{\Psi_{\rm S}(\mathbf{R})}}=
\left[\frac{{\Delta}_{\mathbf{r}_i}\mathbb{A}}{\mathbb{A}} +
2\frac{\vec{\nabla}_{\mathbf{r}_i}\mathbb{A}}
{\mathbb{A}}\cdot\frac{\vec{\nabla}_{\mathbf{r}_i}
\mathbb{P}_1+\vec{\nabla}_{\mathbf{r}_i}\mathbb{P}_2}{\mathbb{P}_1+
\mathbb{P}_2} + \frac{\Delta_{\mathbf{r}_i}\mathbb{P}_1+
\Delta_{\mathbf{r}_i}\mathbb{P}_2}{\mathbb{P}_1+\mathbb{P}_2}\right].
\end{equation}
The second term in the right hand side of Eq.~(\ref{Eq:B06}) is difficult to calculate
in the DMC code, since it involves the inner product of different quantities.
To remove this term let us to calculate the square of the drift force, 
from Eq.~(\ref{Eq:B04}) we obtain \begin{equation}
\label{Eq:B07}
|\vec{F_{\mathbf{r}_i}}|^2=\left(\frac{\vec{\nabla}_{\mathbf{r}_i}\mathbb{A}}{\mathbb{A}}\right)^2+
2\frac{\vec{\nabla}_{\mathbf{r}_i}\mathbb{A}}{\mathbb{A}}\cdot\frac{\vec{\nabla}_{\mathbf{r}_i}\mathbb{P}_1+\vec{\nabla}_{\mathbf{r}_i}\mathbb{P}_2}{\mathbb{P}_1+\mathbb{P}_2}+
\left(\frac{\vec{\nabla}_{\mathbf{r}_i}\mathbb{P}_1+\vec{\nabla}_{\mathbf{r}_i}\mathbb{P}_2}{\mathbb{P}_1+\mathbb{P}_2}\right)^2.
\end{equation}
Now we solve for the inner product term
\begin{equation}
\label{Eq:B08}
2\frac{\vec{\nabla}_{\mathbf{r}_i}\mathbb{A}}{\mathbb{A}}\cdot\frac{\vec{\nabla}_{\mathbf{r}_i}\mathbb{P}_1+\vec{\nabla}_{\mathbf{r}_i}\mathbb{P}_2}{\mathbb{P}_1+\mathbb{P}_2}
=|\vec{F_{\mathbf{r}_i}}|^2-\left(\frac{\vec{\nabla}_{\mathbf{r}_i}\mathbb{A}}{\mathbb{A}}\right)^2-
\left(\frac{\vec{\nabla}_{\mathbf{r}_i}\mathbb{P}_1+\vec{\nabla}_{\mathbf{r}_i}\mathbb{P}_2}{\mathbb{P}_1+\mathbb{P}_2}\right)^2.
\end{equation}
After substituting Eq.~(\ref{Eq:B08}) into Eq.~(\ref{Eq:B06}) we obtain
a expression for the Laplacian of the trial wave function 
$\Delta_{\mathbf{r}_i}\Psi_{\rm S}(\mathbf{R})$
\begin{equation}
\label{Eq:B09}
\begin{aligned}
    \frac{\Delta_{\mathbf{r}_i}\Psi_{\rm S}(\mathbf{R})}{{\Psi_{\rm S}(\mathbf{R})}}=&
|\vec{F_{\mathbf{r}_i}}|^2
+\frac{{\Delta}_{\mathbf{r}_i}\mathbb{A}}{\mathbb{A}}
-\left(\frac{\vec{\nabla}_{\mathbf{r}_i}\mathbb{A}}{\mathbb{A}}\right)^2
 + \frac{\Delta_{\mathbf{r}_i}\mathbb{P}_1+\Delta_{\mathbf{r}_i}\mathbb{P}_2}{\mathbb{P}_1+\mathbb{P}_2}\\
&-\left(\frac{\vec{\nabla}_{\mathbf{r}_i}\mathbb{P}_1+\vec{\nabla}_{\mathbf{r}_i}\mathbb{P}_2}{\mathbb{P}_1+\mathbb{P}_2}\right)^2.
\end{aligned}
\end{equation}
Here, we see that $\Delta_{\mathbf{r}_i}\Psi_{\rm S}(\mathbf{R})$ is given in terms
of the gradients and Laplacians of $\mathbb{A}$, $\mathbb{P}_1$ and 
$\mathbb{P}_2$. We already derived expressions for the gradients, which
are given in Eq.~(\ref{Eq:B05}). Now, we focus on obtaining expressions
for $\Delta_{\mathbf{r}_i}\mathbb{A}$, $\Delta_{\mathbf{r}_i}\mathbb{P}_1$
and $\Delta_{\mathbf{r}_i}\mathbb{P}_2$.
The Laplacian $\Delta_{\mathbf{r}_i}\mathbb{A}$ reads as
\begin{equation}
\label{Eq:B10}
\begin{aligned}
\frac{\Delta_{\mathbf{r}_i}\mathbb{A}}{\mathbb{A}}=&
\sum_{j}^{N_{\1}}
\left[ \frac{\Delta_{\mathbf{r}_i}f_{\1\1}(|\mathbf{r}_i-\mathbf{r}_{j}|)}{f_{\1\1}(|\mathbf{r}_i-\mathbf{r}_{j}|)}-
    \left(  \frac{\vec{\nabla}_{\mathbf{r}_i}f_{\1\1}(|\mathbf{r}_i-\mathbf{r}_{j}|)}{f_{\1\1}(|\mathbf{r}_i-\mathbf{r}_{j}|)}\right)^2 \right]\\
&+\left( \sum_{j}^{N_{\1}}   \frac{\vec{\nabla}_{\mathbf{r}_i}f_{\1\1}(|\mathbf{r}_i-\mathbf{r}_{j}|)}{f_{\1\1}(|\mathbf{r}_i-\mathbf{r}_{j}|)}\right)^2.
\end{aligned}
\end{equation}
The Laplacian $\Delta_{\mathbf{r}_i}\mathbb{P}_1$ is given by the following expression
\begin{equation}
\label{Eq:B11}
\frac{{\Delta}_{\mathbf{r}_i}\mathbb{P}_1}{\mathbb{P}_1}=
\sum_{\alpha=1}^{N_{\2}} \frac{{\Delta}_{\mathbf{r}_i}f_{\1\2}
(|\mathbf{r}_i-\mathbf{r}_{\alpha}|)}{\sum_{\alpha=1}^{N_{\2}}f_{\1\2}(|\mathbf{r}_i-\mathbf{r}_{\alpha}|)}.\\ 
\end{equation}
And finally, the Laplacian $\Delta_{\mathbf{r}_i}\mathbb{P}_2$ is given by
\begin{equation}
\label{Eq:B12}
\begin{aligned}
\frac{\Delta_{\mathbf{r}_i}\mathbb{P}_2}{\mathbb{P}_2}=&\sum_{\alpha=1}^{N_{\2}}
\left[ \frac{\Delta_{\mathbf{r}_i}f_{\1\2}(|\mathbf{r}_i-\mathbf{r}_{\alpha}|)}{\sum_{i=1}^{N_{\1}}f_{\1\2}(|\mathbf{r}_i-\mathbf{r}_{\alpha}|)}-
    \left(  \frac{\vec{\nabla}_{\mathbf{r}_i}f_{\1\2}(|\mathbf{r}_i-\mathbf{r}_{\alpha}|)}{\sum_{i=1}^{N_{\1}}f_{\1\2}(|\mathbf{r}_i-\mathbf{r}_{\alpha}|)}\right)^2 \right]\\
&+\left(\sum_{\alpha=1}^{N_{\2}}   \frac{\vec{\nabla}_{\mathbf{r}_i}f_{\1\2}(|\mathbf{r}_i-\mathbf{r}_{\alpha}|)}{\sum_{i=1}^{N_{\1}}f_{\1\2}(|\mathbf{r}_i-\mathbf{r}_{\alpha}|)}\right)^2.
\end{aligned}
\end{equation}
The full Laplacian of the many-body trial wave function
$\Delta_{\mathbf{r}_i}\Psi_{\rm S}(\mathbf{R})$
with respect to the coordinate $\mathbf{r}_i$ is given by
\begin{equation}
    \label{Eq:B13}
\begin{aligned}
    \frac{\Delta_{\mathbf{r}_i}\Psi_{\rm S}(\mathbf{R})}{\Psi_{\rm S}(\mathbf{R})}=
    &|\vec{F_{\mathbf{r}_i}}|^2 + \sum_{j}^{N_{\1}}\left[\frac{\Delta_{\mathbf{r}_i}f_{\1\1}(|\mathbf{r}_i-\mathbf{r}_j|)}{f_{\1\1}(|\mathbf{r}_i-\mathbf{r}_j|)} -  \left(   \frac{\vec{\nabla}_{\mathbf{r}_i}f_{\1\1}(|\mathbf{r}_i-\mathbf{r}_j|)}{f_{\1\1}(|\mathbf{r}_i-\mathbf{r}_j|)}  \right)^2  \right]\\
+& \frac{\Delta_{\mathbf{r}_i}\mathbb{P}_1+\Delta_{\mathbf{r}_i}\mathbb{P}_2}{\mathbb{P}_1+\mathbb{P}_2}-\left(    \frac{\vec{\nabla}_{\mathbf{r}_i}\mathbb{P}_1+\vec{\nabla}_{\mathbf{r}_i}\mathbb{P}_2}{\mathbb{P}_1+\mathbb{P}_2}\right)^2.
\end{aligned}
\end{equation}

Analogous to the Eq.~(\ref{Eq:B13}), the full Laplacian of the many-body trial
wave function $\Delta_{\mathbf{r}_{\alpha}}\Psi_{\rm S}(\mathbf{R})$
with respect to the coordinate $\mathbf{r}_{\alpha}$ is given by
\begin{equation}
\label{Eq:B14}
\begin{aligned}
    \frac{\Delta_{\mathbf{r}_\alpha}\Psi_{\rm S}(\mathbf{R})}{\Psi_{\rm S}(\mathbf{R})}=
&|\vec{F}_{{\mathbf{r}_\alpha}}|^2 + \sum_{\beta}^{N_{\2}}\left[\frac{\Delta_{\mathbf{r}_\alpha}f_{\2\2}(|\mathbf{r}_\alpha-\mathbf{r}_\beta|)}{f_{\2\2}(|\mathbf{r}_\alpha-\mathbf{r}_\beta|)} -  \left(   \frac{\vec{\nabla}_{\mathbf{r}_\alpha}f_{\2\2}(|\mathbf{r}_\alpha-\mathbf{r}_\beta|)}{f_{\2\2}(|\mathbf{r}_\alpha-\mathbf{r}_\beta|)}  \right)^2  \right]\\
+& \frac{\Delta_{\mathbf{r}_\alpha}\mathbb{P}_1+\Delta_{\mathbf{r}_\alpha}\mathbb{P}_2}{\mathbb{P}_1+\mathbb{P}_2}-\left(    \frac{\vec{\nabla}_{\mathbf{r}_\alpha}\mathbb{P}_1+\vec{\nabla}_{\mathbf{r}_\alpha}\mathbb{P}_2}{\mathbb{P}_1+\mathbb{P}_2}\right)^2.
\end{aligned}
\end{equation}
Here, the drift force with respect to the coordinate
$\mathbf{r}_\alpha$ is given by
\begin{equation}
\label{Eq:B15} 
\vec{F}_{{\mathbf{r}_\alpha}}=\frac{\vec{\nabla}_{\mathbf{r}_\alpha}
\Psi(\mathbf{R})}{\Psi(\mathbf{R})}=
\frac{\vec{\nabla}_{\mathbf{r}_\alpha}\mathbb{B}}{\mathbb{B}}
+\frac{\vec{\nabla}_{\mathbf{r}_\alpha}\mathbb{P}_1
+\vec{\nabla}_{\mathbf{r}_\alpha}\mathbb{P}_2}{\mathbb{P}_1+\mathbb{P}_2}.
\end{equation}
In the following we are going to obtain the expressions for the gradients
and Laplacians involved in Eq.~(\ref{Eq:B14}).
The gradients $\vec{\nabla}_{\mathbf{r}_\alpha}\mathbb{B}$,
$\vec{\nabla}_{\mathbf{r}_\alpha}\mathbb{P}_1$ and
$\vec{\nabla}_{\mathbf{r}_\alpha}\mathbb{P}_2$ are given by
\begin{equation}
\begin{aligned}
\label{Eq:B16} 
\frac{\vec{\nabla}_{\mathbf{r}_\alpha}\mathbb{B}}{\mathbb{B}}=&
\sum_{\beta}^{N_{\2}}\frac{\vec{\nabla}_{\mathbf{r}_\alpha}f_{\2\2}(|\mathbf{r}_\alpha-\mathbf{r}_\beta|)}{f_{\2\2}(|\mathbf{r}_\alpha-\mathbf{r}_\beta|)},\\
\frac{\vec{\nabla}_{\mathbf{r}_\alpha}\mathbb{P}_1}{\mathbb{P}_1}=&\sum_{i=1}^{N_{\1}} \frac{\vec{\nabla}_{\mathbf{r}_\alpha}f_{\1\2}(|\mathbf{r}_i-\mathbf{r}_{\alpha}|)}{\sum_{\alpha=1}^{N_{\2}}f_{\1\2}(|\mathbf{r}_i-\mathbf{r}_{\alpha}|)},\\\frac{\vec{\nabla}_{\mathbf{r}_\alpha}\mathbb{P}_2}{\mathbb{P}_2}=&\sum_{i=1}^{N_{\1}} \frac{\vec{\nabla}_{\mathbf{r}_\alpha}f_{\1\2}(|\mathbf{r}_i-\mathbf{r}_{\alpha}|)}{\sum_{i=1}^{N_{\1}}f_{\1\2}(|\mathbf{r}_i-\mathbf{r}_{\alpha}|)}. 
\end{aligned}
\end{equation}
The Laplacian $\Delta_{\mathbf{r}_\alpha}\mathbb{B}$ is given by
\begin{equation}
\begin{aligned}
\label{Eq:B17} 
\frac{\Delta_{\mathbf{r}_\alpha}\mathbb{B}}{\mathbb{B}}=
&\sum_{\beta}^{N_{\2}}\left[ \frac{\Delta_{\mathbf{r}_\alpha}
f_{\2\2}(|\mathbf{r}_\alpha-\mathbf{r}_\beta|)}
{f_{\2\2}(|\mathbf{r}_\alpha-\mathbf{r}_\beta|)}
-\left(  \frac{\vec{\nabla}_{\mathbf{r}_\alpha}
f_{\2\2}(|\mathbf{r}_\alpha-\mathbf{r}_\beta|)}
{f_{\2\2}(|\mathbf{r}_\alpha-\mathbf{r}_\beta|)}\right)^2 \right]\\
&+\left( \sum_{\beta}^{N_{\2}}
\frac{\vec{\nabla}_{\mathbf{r}_\alpha}
f_{\2\2}(|\mathbf{r}_\alpha-\mathbf{r}_\beta|)}
{f_{\2\2}(|\mathbf{r}_\alpha-\mathbf{r}_\beta|)}\right)^2.
\end{aligned}
\end{equation}
The expression for the Laplacian $\Delta_{\mathbf{r}_\alpha}\mathbb{P}_1$
reads as
\begin{equation}
\begin{aligned}
\label{Eq:B18} 
\frac{\Delta_{\mathbf{r}_\alpha}\mathbb{P}_1}{\mathbb{P}_1}=
&\sum_{i=1}^{N_{\1}}\left[ \frac{\Delta_{\mathbf{r}_\alpha}
f_{\1\2}(|\mathbf{r}_i-\mathbf{r}_{\alpha}|)}{\sum_{\alpha=1}^{N_{\2}}
f_{\1\2}(|\mathbf{r}_i-\mathbf{r}_{\alpha}|)}-
\left(  \frac{\vec{\nabla}_{\mathbf{r}_\alpha}
f_{\1\2}(|\mathbf{r}_i-\mathbf{r}_{\alpha}|)}{\sum_{\alpha=1}^{N_{\2}}
f_{\1\2}(|\mathbf{r}_i-\mathbf{r}_{\alpha}|)}\right)^2 \right]\\
&+\left(\sum_{i=1}^{N_{\1}}\frac{\vec{\nabla}_{\mathbf{r}_\alpha}
f_{\1\2}(|\mathbf{r}_i-\mathbf{r}_{\alpha}|)}
{\sum_{\alpha=1}^{N_{\2}}f_{\1\2}(|\mathbf{r}_i-\mathbf{r}_{\alpha}|)}\right)^2.
\end{aligned}
\end{equation}
Finally, the Laplacian $\Delta_{\mathbf{r}_\alpha}\mathbb{P}_2$ is given by
\begin{equation}
\begin{aligned}
\label{Eq:B19} 
\frac{{\Delta}_{\mathbf{r}_\alpha}\mathbb{P}_2}{\mathbb{P}_2}=
&\sum_{i=1}^{N_{\1}}\frac{{\Delta}_{\mathbf{r}_\alpha}
f_{\1\2}(|\mathbf{r}_i-\mathbf{r}_{\alpha}|)}{\sum_{i=1}^{N_{\1}}
f_{\1\2}(|\mathbf{r}_i-\mathbf{r}_{\alpha}|)}.
\end{aligned}
\end{equation}

\bibliography{bibliography} 

\begin{thebibliography}{100}

\bibitem{Einstein1924}
A.~Einstein.
\newblock Quantentheorie des einatomigen idealen gases.
\newblock {\em Sitzungsberichte der Preussischen Akademie der Wissenschaften},
  page 261–267, 1924.

\bibitem{Anderson198}
M.~H. Anderson, J.~R. Ensher, M.~R. Matthews, C.~E. Wieman, and E.~A. Cornell.
\newblock Observation of {Bose-Einstein} condensation in a dilute atomic vapor.
\newblock {\em Science}, 269(5221):198--201, 1995.

\bibitem{PhysRevLett.75.3969}
K.~B. Davis, M.~O. Mewes, M.~R. Andrews, N.~J. van Druten, D.~S. Durfee, D.~M.
  Kurn, and W.~Ketterle.
\newblock {Bose-Einstein} condensation in a gas of sodium atoms.
\newblock {\em Phys. Rev. Lett.}, 75:3969--3973, Nov 1995.

\bibitem{nobelprize}
www.nobelprize.org.

\bibitem{PhysRevLett.115.155302}
D.~S. Petrov.
\newblock Quantum mechanical stabilization of a collapsing {Bose-Bose} mixture.
\newblock {\em Phys. Rev. Lett.}, 115:155302, Oct 2015.

\bibitem{Cabrera2017}
C.~R. Cabrera, L.~Tanzi, J.~Sanz, B.~Naylor, P.~Thomas, P.~Cheiney, and
  L.~Tarruell.
\newblock Quantum liquid droplets in a mixture of {Bose-Einstein} condensates.
\newblock {\em Science}, 359(6373):301, 2018.

\bibitem{Semeghini2018}
G.~Semeghini, G.~Ferioli, L.~Masi, C.~Mazzinghi, L.~Wolswijk, F.~Minardi,
  M.~Modugno, G.~Modugno, M.~Inguscio, and M.~Fattori.
\newblock Self-bound quantum droplets of atomic mixtures in free space.
\newblock {\em Phys. Rev. Lett.}, 120:235301, Jun 2018.

\bibitem{Ferioli2019}
G.~Ferioli, G.~Semeghini, L.~Masi, G.~Giusti, G.~Modugno, M.~Inguscio,
  A.~Gallem\'{\i}, A.~Recati, and M.~Fattori.
\newblock Collisions of self-bound quantum droplets.
\newblock {\em Phys. Rev. Lett.}, 122:090401, Mar 2019.

\bibitem{Kadau2016}
H.~Kadau, M~Schmitt, M.~Wenzel, C~Wink, T.~Maier, I.~Ferrier-Barbut, and
  T.~Pfau.
\newblock Observing the {Rosensweig} instability of a quantum ferrofluid.
\newblock {\em Nature}, 530(7589):194, February 2016.

\bibitem{Schmitt2016}
M.~Schmitt, M.~Wenzel, F.~B\"ottcher, I.~Ferrier-Barbut, and T.~Pfau.
\newblock Self-bound droplets of a dilute magnetic quantum liquid.
\newblock {\em Nature (London)}, 539:259, 2016.

\bibitem{Ferrier2016}
I.~Ferrier-Barbut, H.~Kadau, M.~Schmitt, M.~Wenzel, and T.~Pfau.
\newblock Observation of quantum droplets in a strongly dipolar {Bose} gas.
\newblock {\em Phys. Rev. Lett.}, 116:215301, 2016.

\bibitem{Chomaz2016}
L.~Chomaz, S.~Baier, D.~Petter, M.~J. Mark, F.~W\"achtler, L.~Santos, and
  F.~Ferlaino.
\newblock Quantum-fluctuation-driven crossover from a dilute {Bose-Einstein}
  condensate to a macrodroplet in a dipolar quantum fluid.
\newblock {\em Phys. Rev. X}, 6:041039, 2016.

\bibitem{PhysRevResearch.2.022008}
Ivan Morera, Grigori~E. Astrakharchik, Artur Polls, and Bruno
  Juli\'a-D\'{\i}az.
\newblock Quantum droplets of bosonic mixtures in a one-dimensional optical
  lattice.
\newblock {\em Phys. Rev. Research}, 2:022008, Apr 2020.

\bibitem{morera2020universal}
Ivan Morera, Grigori~E. Astrakharchik, Artur Polls, and Bruno Juliá-Díaz.
\newblock Universal dimerized quantum droplets in a one-dimensional lattice,
  arXiv 2007.01786 2020.

\bibitem{kohstall2012metastability}
Christoph Kohstall, Mattheo Zaccanti, Matthias Jag, Andreas Trenkwalder, Pietro
  Massignan, Georg~M Bruun, Florian Schreck, and Rudolf Grimm.
\newblock Metastability and coherence of repulsive polarons in a strongly
  interacting {Fermi} mixture.
\newblock {\em Nature}, 485(7400):615--618, 2012.

\bibitem{Massignan_2014}
Pietro Massignan, Matteo Zaccanti, and Georg~M Bruun.
\newblock Polarons, dressed molecules and itinerant ferromagnetism in ultracold
  {Fermi} gases.
\newblock {\em Reports on Progress in Physics}, 77(3):034401, feb 2014.

\bibitem{RaulBombi2019}
Raúl Bombín, Tommaso Comparin, Gianluca Bertaina, Ferran Mazzanti, Stefano
  Giorgini, and Jordi Boronat.
\newblock Two-dimensional repulsive {Fermi} polarons with short and long-range
  interactions.
\newblock {\em Phys. Rev. A}, 100:023608, 2019.

\bibitem{PhysRevX.8.011024}
Shuhei~M. Yoshida, Shimpei Endo, Jesper Levinsen, and Meera~M. Parish.
\newblock Universality of an impurity in a {Bose-Einstein} condensate.
\newblock {\em Phys. Rev. X}, 8:011024, Feb 2018.

\bibitem{PhysRevLett.120.050405}
Nils-Eric Guenther, Pietro Massignan, Maciej Lewenstein, and Georg~M. Bruun.
\newblock Bose polarons at finite temperature and strong coupling.
\newblock {\em Phys. Rev. Lett.}, 120:050405, Feb 2018.

\bibitem{PhysRevResearch.2.023405}
L.~A. Pe\~na Ardila, G.~E. Astrakharchik, and S.~Giorgini.
\newblock Strong coupling {Bose} polarons in a two-dimensional gas.
\newblock {\em Phys. Rev. Research}, 2:023405, Jun 2020.

\bibitem{Lahaye_2009}
T~Lahaye, C~Menotti, L~Santos, M~Lewenstein, and T~Pfau.
\newblock The physics of dipolar bosonic quantum gases.
\newblock {\em Reports on Progress in Physics}, 72(12):126401, nov 2009.

\bibitem{PhysRevLett.94.160401}
Axel Griesmaier, J\"org Werner, Sven Hensler, J\"urgen Stuhler, and Tilman
  Pfau.
\newblock {Bose-Einstein} condensation of chromium.
\newblock {\em Phys. Rev. Lett.}, 94:160401, Apr 2005.

\bibitem{PhysRevLett.95.150406}
J.~Stuhler, A.~Griesmaier, T.~Koch, M.~Fattori, T.~Pfau, S.~Giovanazzi,
  P.~Pedri, and L.~Santos.
\newblock Observation of dipole-dipole interaction in a degenerate quantum gas.
\newblock {\em Phys. Rev. Lett.}, 95:150406, Oct 2005.

\bibitem{PhysRevLett.107.190401}
Mingwu Lu, Nathaniel~Q. Burdick, Seo~Ho Youn, and Benjamin~L. Lev.
\newblock Strongly dipolar {Bose-Einstein} condensate of dysprosium.
\newblock {\em Phys. Rev. Lett.}, 107:190401, Oct 2011.

\bibitem{PhysRevLett.108.215301}
Mingwu Lu, Nathaniel~Q. Burdick, and Benjamin~L. Lev.
\newblock Quantum degenerate dipolar {Fermi Gas}.
\newblock {\em Phys. Rev. Lett.}, 108:215301, May 2012.

\bibitem{Chomaz2018}
Lauriane Chomaz, Rick Bijnen, Daniel Petter, Giulia Faraoni, Simon Baier, Jan
  Becher, Falk Waechtler, Luis Santos, and Francesca Ferlaino.
\newblock Observation of the roton mode in a dipolar quantum gas.
\newblock {\em Nature Physics}, 14, 05 2018.

\bibitem{PhysRevLett.119.250402}
R.~Bombin, J.~Boronat, and F.~Mazzanti.
\newblock Dipolar {Bose} supersolid stripes.
\newblock {\em Phys. Rev. Lett.}, 119:250402, Dec 2017.

\bibitem{PhysRevLett.121.213601}
A.~Trautmann, P.~Ilzh\"ofer, G.~Durastante, C.~Politi, M.~Sohmen, M.~J. Mark,
  and F.~Ferlaino.
\newblock Dipolar quantum mixtures of erbium and dysprosium atoms.
\newblock {\em Phys. Rev. Lett.}, 121:213601, Nov 2018.

\bibitem{PhysRevA.99.043609}
Tommaso Comparin, Ra\'ul Bomb\'{\i}n, Markus Holzmann, Ferran Mazzanti, Jordi
  Boronat, and Stefano Giorgini.
\newblock Two-dimensional mixture of dipolar fermions: Equation of state and
  magnetic phases.
\newblock {\em Phys. Rev. A}, 99:043609, Apr 2019.

\bibitem{Buchler2007}
H.-P. B\"uchler, E.~Demler, M.~Lukin, A.~Micheli, N.~Prokof'ev, G.~Pupillo, and
  P.~Zoller.
\newblock Strongly correlated {2D} quantum phases with cold polar molecules:
  Controlling the shape of the interaction potential.
\newblock {\em Phys. Rev. Lett.}, 98:060404, Feb 2007.

\bibitem{Astrakharchik2007}
G.~E. Astrakharchik, J.~Boronat, I.~L. Kurbakov, and Yu.~E. Lozovik.
\newblock Quantum phase transition in a two-dimensional system of dipoles.
\newblock {\em Phys. Rev. Lett.}, 98:060405, Feb 2007.

\bibitem{Macia2014}
A.~Macia, G.~E. Astrakharchik, F.~Mazzanti, S.~Giorgini, and J.~Boronat.
\newblock Single-particle versus pair superfluidity in a bilayer system of
  dipolar bosons.
\newblock {\em Phys. Rev. A}, 90:043623, Oct 2014.

\bibitem{PhysRevA.94.063630}
G.~E. Astrakharchik, R.~E. Zillich, F.~Mazzanti, and J.~Boronat.
\newblock Gapped spectrum in pair-superfluid bosons.
\newblock {\em Phys. Rev. A}, 94:063630, Dec 2016.

\bibitem{Nespolo_2017}
J.~Nespolo, G.~E. Astrakharchik, and A.~Recati.
\newblock {Andreev{\textendash}Bashkin} effect in superfluid cold gases
  mixtures.
\newblock {\em New J. Phys.}, 19(12):125005, dec 2017.

\bibitem{hammond1994monte}
B.L. Hammond, W.A. Lester, and P.J. Reynolds.
\newblock {\em Monte Carlo Methods in Ab Initio Quantum Chemistry}.
\newblock Lecture and Course Notes In Chemistry Series. World Scientific, 1994.

\bibitem{toulouse2015introduction}
Julien Toulouse, Roland Assaraf, and C.~J. Umrigar.
\newblock Introduction to the variational and diffusion {Monte Carlo} methods,
  2015.

\bibitem{Ceperley}
David Ceperley.
\newblock {An Overview of Quantum Monte Carlo Methods}.
\newblock {\em Reviews in Mineralogy \& Geochemistry - REV MINERAL GEOCHEM},
  71:129--135, 04 2010.

\bibitem{PhysRevA.99.023618}
V.~Cikojevi\ifmmode~\acute{c}\else \'{c}\fi{}, L.~Vranje\ifmmode
  \check{s}\else~\v{s}\fi{} Marki\ifmmode~\acute{c}\else \'{c}\fi{}, G.~E.
  Astrakharchik, and J.~Boronat.
\newblock Universality in ultradilute liquid bose-bose mixtures.
\newblock {\em Phys. Rev. A}, 99:023618, Feb 2019.

\bibitem{LParisi-Thel-2019}
L.~Parisi, G.~E. Astrakharchik, and S.~Giorgini.
\newblock The liquid state of one-dimensional {Bose} mixtures: a quantum
  {Monte-Carlo} study.
\newblock {\em Phys. Rev. Lett.}, 122:105302, 2019.

\bibitem{PhysRevA.99.063607}
L.~A. Pe\~na Ardila, N.~B. J\o{}rgensen, T.~Pohl, S.~Giorgini, G.~M. Bruun, and
  J.~J. Arlt.
\newblock Analyzing a {Bose} polaron across resonant interactions.
\newblock {\em Phys. Rev. A}, 99:063607, Jun 2019.

\bibitem{ThesisOmar}
O.~A. Rodr\'iguez~L\'opez.
\newblock {\em Interacting {Bose} gas in multi-rod lattices : ground state
  properties and quantum phase transitions}.
\newblock PhD thesis, UNAM, Instituto de F\'isica, 2020.

\bibitem{PhysRevA.98.053632}
J.~S\'anchez-Baena, J.~Boronat, and F.~Mazzanti.
\newblock Diffusion {Monte Carlo} methods for spin-orbit-coupled ultracold
  {Bose} gases.
\newblock {\em Phys. Rev. A}, 98:053632, Nov 2018.

\bibitem{Foulkes}
Matthew Foulkes, L.~Mitas, R.~Needs, and Guna Rajagopal.
\newblock Quantum {Monte Carlo} simulation of solids.
\newblock {\em Reviews of Modern Physics}, 73, 02 2001.

\bibitem{RevModPhys.89.035003}
Claudio Cazorla and Jordi Boronat.
\newblock Simulation and understanding of atomic and molecular quantum
  crystals.
\newblock {\em Rev. Mod. Phys.}, 89:035003, Aug 2017.

\bibitem{WDmowski-Obser-2017}
W.~Dmowski, S.~O. Diallo, K.~Lokshin, G.~Ehlers, G.~Ferré, J.~Boronat, and
  T.~Egami.
\newblock Observation of dynamic atom-atom correlation in liquid {Helium} in
  real space.
\newblock {\em Nature Communications}, 8:15294, 2017.

\bibitem{PhysRevLett.124.205301}
M.~C. Gordillo and J.~Boronat.
\newblock Superfluid and supersolid phases of $^{4}\mathrm{He}$ on the second
  layer of graphite.
\newblock {\em Phys. Rev. Lett.}, 124:205301, May 2020.

\bibitem{PhysRev.138.A442}
W.~L. McMillan.
\newblock Ground state of liquid {He}$^4$.
\newblock {\em Phys. Rev.}, 138:A442--A451, Apr 1965.

\bibitem{RafaelGuardiola}
Rafael Guardiola.
\newblock Monte {Carlo} methods in quantum many-body theories.
\newblock In Jes{\'u}s Navarro and Artur Polls, editors, {\em Microscopic
  Quantum Many-Body Theories and Their Applications}, pages 269--336, Berlin,
  Heidelberg, 1998. Springer Berlin Heidelberg.

\bibitem{Reynolds1982}
Peter~J. Reynolds, David~M. Ceperley, Berni~J. Alder, and William~A. Lester.
\newblock Fixed‐node quantum {Monte Carlo} for molecules.
\newblock {\em The Journal of Chemical Physics}, 77(11):5593--5603, 1982.

\bibitem{ThesisGuillem}
G.~Ferr\'e~Porta.
\newblock {\em A Monte Carlo approach to statics and dynamics of quantum
  fluids}.
\newblock PhD thesis, UPC, Departament de F\'isica Aplicada, 2017.

\bibitem{ThesisRaul}
R.~Bomb\'in~Escudero.
\newblock {\em Ultracold {Bose} and {Fermi} dipolar gases: a Quantum {Monte}
  {Carlo} study}.
\newblock PhD thesis, UPC, Departament de F\'isica Aplicada, 2019.

\bibitem{doi:10.1119/1.18168}
Ioan Kosztin, Byron Faber, and Klaus Schulten.
\newblock Introduction to the diffusion {Monte Carlo} method.
\newblock {\em American Journal of Physics}, 64(5):633--644, 1996.

\bibitem{PhysRevA.42.6991}
Siu~A. Chin.
\newblock Quadratic diffusion {Monte Carlo} algorithms for solving atomic
  many-body problems.
\newblock {\em Phys. Rev. A}, 42:6991--7005, Dec 1990.

\bibitem{PhysRevA.10.303}
K.~S. Liu, M.~H. Kalos, and G.~V. Chester.
\newblock Quantum hard spheres in a channel.
\newblock {\em Phys. Rev. A}, 10:303--308, Jul 1974.

\bibitem{CasullerasBoronat1995}
J.~Casulleras and J.~Boronat.
\newblock Unbiased estimators in quantum {Monte Carlo} methods: Application to
  liquid $^{4}\mathrm{He}$.
\newblock {\em Phys. Rev. B}, 52:3654--3661, Aug 1995.

\bibitem{doi:10.1063/1.1446847}
A.~Sarsa, J.~Boronat, and J.~Casulleras.
\newblock Quadratic diffusion {Monte Carlo} and pure estimators for atoms.
\newblock {\em The Journal of Chemical Physics}, 116(14):5956--5962, 2002.

\bibitem{Guijarro2018}
G.~Guijarro, A.~Pricoupenko, G.~E. Astrakharchik, J.~Boronat, and D.~S. Petrov.
\newblock One-dimensional three-boson problem with two- and three-body
  interactions.
\newblock {\em Phys. Rev. A}, 97:061605, Jun 2018.

\bibitem{LiebLiniger}
Elliott~H. Lieb and Werner Liniger.
\newblock Exact analysis of an interacting {Bose} gas. {I}. {The} general
  solution and the ground state.
\newblock {\em Phys. Rev.}, 130:1605--1616, May 1963.

\bibitem{McGuireBosons}
J.~B. McGuire.
\newblock Study of exactly soluble one‐dimensional {N}‐body problems.
\newblock {\em Journal of Mathematical Physics}, 5(5):622--636, 1964.

\bibitem{Nishida}
Yuta Sekino and Yusuke Nishida.
\newblock Quantum droplet of one-dimensional bosons with a three-body
  attraction.
\newblock {\em Phys. Rev. A}, 97:011602, Jan 2018.

\bibitem{PricoupenkoPetrov}
A.~Pricoupenko and D.~S. Petrov.
\newblock Dimer-dimer zero crossing and dilute dimerized liquid in a
  one-dimensional mixture.
\newblock {\em Phys. Rev. A}, 97:063616, Jun 2018.

\bibitem{Bulgac}
Aurel Bulgac.
\newblock Dilute quantum droplets.
\newblock {\em Phys. Rev. Lett.}, 89:050402, Jul 2002.

\bibitem{Petrov3body}
D.~S. Petrov.
\newblock Three-body interacting bosons in free space.
\newblock {\em Phys. Rev. Lett.}, 112:103201, Mar 2014.

\bibitem{Girardeau60}
M.~Girardeau.
\newblock Relationship between systems of impenetrable bosons and fermions in
  one dimension.
\newblock {\em Journal of Mathematical Physics}, 1(6):516--523, 1960.

\bibitem{Falke}
Stephan Falke, Horst Kn\"ockel, Jan Friebe, Matthias Riedmann, Eberhard
  Tiemann, and Christian Lisdat.
\newblock Potassium ground-state scattering parameters and {Born-Oppenheimer}
  potentials from molecular spectroscopy.
\newblock {\em Phys. Rev. A}, 78:012503, Jul 2008.

\bibitem{Zwierlein}
Cheng-Hsun Wu, Ibon Santiago, Jee~Woo Park, Peyman Ahmadi, and Martin~W.
  Zwierlein.
\newblock Strongly interacting isotopic {Bose-Fermi} mixture immersed in a
  {Fermi} sea.
\newblock {\em Phys. Rev. A}, 84:011601, Jul 2011.

\bibitem{Olshanii}
M.~Olshanii.
\newblock Atomic scattering in the presence of an external confinement and a
  gas of impenetrable bosons.
\newblock {\em Phys. Rev. Lett.}, 81:938--941, Aug 1998.

\bibitem{Gora}
A.~Muryshev, G.~V. Shlyapnikov, W.~Ertmer, K.~Sengstock, and M.~Lewenstein.
\newblock Dynamics of dark solitons in elongated {Bose-Einstein} condensates.
\newblock {\em Phys. Rev. Lett.}, 89:110401, Aug 2002.

\bibitem{Kovrizhin}
Subhasis Sinha, Alexander~Yu. Cherny, Dmitry Kovrizhin, and Joachim Brand.
\newblock Friction and diffusion of matter-wave bright solitons.
\newblock {\em Phys. Rev. Lett.}, 96:030406, Jan 2006.

\bibitem{Mazets}
I.~E. Mazets, T.~Schumm, and J.~Schmiedmayer.
\newblock Breakdown of integrability in a quasi-1d ultracold bosonic gas.
\newblock {\em Phys. Rev. Lett.}, 100:210403, May 2008.

\bibitem{PhysRevA.101.041602}
G.~Guijarro, G.~E. Astrakharchik, J.~Boronat, B.~Bazak, and D.~S. Petrov.
\newblock Few-body bound states of two-dimensional bosons.
\newblock {\em Phys. Rev. A}, 101:041602, Apr 2020.

\bibitem{Pricoupenko2018}
A.~Pricoupenko and D.~S. Petrov.
\newblock Dimer-dimer zero crossing and dilute dimerized liquid in a
  one-dimensional mixture.
\newblock {\em Phys. Rev. A}, 97:063616, 2018.

\bibitem{WangLukinDemler2006}
D.-W. Wang, M.~D. Lukin, and E.~Demler.
\newblock Quantum fluids of self-assembled chains of polar molecules.
\newblock {\em Phys. Rev. Lett.}, 97:180413, Nov 2006.

\bibitem{Wang2007}
D.-W. Wang.
\newblock Quantum phase transitions of polar molecules in bilayer systems.
\newblock {\em Phys. Rev. Lett.}, 98:060403, Feb 2007.

\bibitem{Trefzger_2011}
Trefzger C., Menotti C., Capogrosso-Sansone B., and Lewenstein M.
\newblock Ultracold dipolar gases in optical lattices.
\newblock {\em Journal of Physics B: Atomic, Molecular and Optical Physics},
  44(19):193001, sep 2011.

\bibitem{ODell2003}
D.~H.~J. O'Dell, S.~Giovanazzi, and G.~Kurizki.
\newblock Rotons in gaseous {Bose-Einstein} condensates irradiated by a laser.
\newblock {\em Phys. Rev. Lett.}, 90:110402, Mar 2003.

\bibitem{Santos2003}
L.~Santos, G.~V. Shlyapnikov, and M.~Lewenstein.
\newblock {Roton-Maxon} spectrum and stability of trapped dipolar
  {Bose-Einstein} condensates.
\newblock {\em Phys. Rev. Lett.}, 90:250403, Jun 2003.

\bibitem{Lu2015}
Z.-K. Lu, Y.~Li, D.~S. Petrov, and G.~V. Shlyapnikov.
\newblock Stable dilute supersolid of two-dimensional dipolar bosons.
\newblock {\em Phys. Rev. Lett.}, 115:075303, 2015.

\bibitem{Tanzi2019}
L.~Tanzi, E.~Lucioni, F.~Fam\`a, J.~Catani, A.~Fioretti, C.~Gabbanini, R.~N.
  Bisset, L.~Santos, and G.~Modugno.
\newblock Observation of a dipolar quantum gas with metastable supersolid
  properties.
\newblock {\em Phys. Rev. Lett.}, 122:130405, Apr 2019.

\bibitem{Chomaz2019}
L.~Chomaz, D.~Petter, P.~Ilzh\"ofer, G.~Natale, A.~Trautmann, C.~Politi,
  G.~Durastante, R.~M.~W. van Bijnen, A.~Patscheider, M.~Sohmen, M.~J. Mark,
  and F.~Ferlaino.
\newblock Long-lived and transient supersolid behaviors in dipolar quantum
  gases.
\newblock {\em Phys. Rev. X}, 9:021012, 2019.

\bibitem{Bottcher2019}
F.~B\"ottcher, J.-N. Schmidt, M.~Wenzel, J.~Hertkorn, M.~Guo, T.~Langen, and
  T.~Pfau.
\newblock Transient supersolid properties in an array of dipolar quantum
  droplets.
\newblock {\em Phys. Rev. X}, 9:011051, 2019.

\bibitem{Tanzi2019Modes}
L.~Tanzi, S.~M. Roccuzzo, E.~Lucioni, F.~Fam\`a, A.~Fioretti, C.~Gabbanini,
  G.~Modugno, A.~Recati, and S.~Stringari.
\newblock Supersolid symmetry breaking from compressional oscillations in a
  dipolar quantum gas.
\newblock {\em Nature (London)}, 574:382, 2019.

\bibitem{Guo2019}
M.~Guo, F.~B\"ottcher, J.~Hertkorn, J.-N. Schmidt, M.~Wenzel, H.-P. B\"uchler,
  T.~Langen, and T.~Pfau.
\newblock The low-energy goldstone mode in a trapped dipolar supersolid.
\newblock {\em Nature (London)}, 574:386, 2019.

\bibitem{Ferlaino2019}
G.~Natale, R.~M.~W. van Bijnen, A.~Patscheider, D.~Petter, M.~J. Mark,
  L.~Chomaz, and F.~Ferlaino.
\newblock Excitation spectrum of a trapped dipolar supersolid and its
  experimental evidence.
\newblock {\em Phys. Rev. Lett.}, 123:050402, Aug 2019.

\bibitem{Safavi2017}
A.~Safavi-Naini, \c{S}.~G. S\"oyler, G.~Pupillo, H.~R. Sadeghpour, and
  B.~Capogrosso-Sansone.
\newblock Quantum phases of dipolar bosons in bilayer geometry.
\newblock {\em New J. Phys.}, 15:013036, 2013.

\bibitem{bitube}
Note that the analogous integral in the bitube geometry is finite [see, for
  example, B. Wunsch, N.~T. Zinner, I.~B. Mekhov, S.-J. Huang, D.-W. Wang, and
  E. Demler, Few-Body Bound States in Dipolar Gases and Their Detection, Phys.
  Rev. Lett. {\bf 107}, 073201 (2011)].

\bibitem{Yudson1997}
V.~I. Yudson, M.~G. Rozman, and P.~Reineker.
\newblock Bound states of two particles confined to parallel two-dimensional
  layers and interacting via dipole-dipole or dipole-charge laws.
\newblock {\em Phys. Rev. B}, 55:5214, Feb 1997.

\bibitem{Shih2009}
S.-M. Shih and D.-W. Wang.
\newblock Pseudopotential of an interaction with a power-law decay in
  two-dimensional systems.
\newblock {\em Phys. Rev. A}, 79:065603, 2009.

\bibitem{Armstrong2010}
J.~R. Armstrong, N.~T. Zinner, D.~V. Fedorov, and A.~S. Jensen.
\newblock Bound states and universality in layers of cold polar molecules.
\newblock {\em Europhys. Lett.}, 91:16001, 2010.

\bibitem{Klawunn2010}
M.~Klawunn, A.~Pikovski, and L.~Santos.
\newblock Two-dimensional scattering and bound states of polar molecules in
  bilayers.
\newblock {\em Phys. Rev. A}, 82:044701, Oct 2010.

\bibitem{Baranov2011}
M.~A. Baranov, A.~Micheli, S.~Ronen, and P.~Zoller.
\newblock Bilayer superfluidity of fermionic polar molecules: Many-body
  effects.
\newblock {\em Phys. Rev. A}, 83:043602, Apr 2011.

\bibitem{Volosniev2011}
A.~G. Volosniev, D.~V. Fedorov, A.~S. Jensen, and N.~T. Zinner.
\newblock Model independence in two dimensions and polarized cold dipolar
  molecules.
\newblock {\em Phys. Rev. Lett.}, 106:250401, 2011.

\bibitem{Simon1976}
B.~Simon.
\newblock The bound state of weakly coupled {Schr\"{o}dinger} operators in one
  and two dimensions.
\newblock {\em Annals of Physics}, 97(2):279, April 1976.

\bibitem{Volosniev2012}
A.~G. Volosniev, D.~V. Fedorov, A.~S. Jensen, and N.~T. Zinner.
\newblock Few-body bound-state stability of dipolar molecules in two
  dimensions.
\newblock {\em Phys. Rev. A}, 85:023609, Feb 2012.

\bibitem{Ticknor2009}
C.~Ticknor.
\newblock Two-dimensional dipolar scattering.
\newblock {\em Phys. Rev. A}, 80:052702, 2009.

\bibitem{SuzVar98}
Y.~Suzuki and K.~Varga.
\newblock {\em Stochastic variational approach to quantum-mechanical few-body
  problems}.
\newblock Springer, Berlin Heidelberg, 1998.

\bibitem{BazEliKol16}
B.~Bazak, M.~Eliyahu, and U.~van Kolck.
\newblock Effective field theory for few-boson systems.
\newblock {\em Phys. Rev. A}, 94:052502, 2016.

\bibitem{Brodsky2006}
I.~V. Brodsky, M.~Yu. Kagan, A.~V. Klaptsov, R.~Combescot, and X.~Leyronas.
\newblock Exact diagrammatic approach for dimer-dimer scattering and bound
  states of three and four resonantly interacting particles.
\newblock {\em Phys. Rev. A}, 73:032724, 2006.

\bibitem{PricoupenkoPedri2010}
L.~Pricoupenko and P.~Pedri.
\newblock Universal (1+2)-body bound states in planar atomic waveguides.
\newblock {\em Phys. Rev. A}, 82:033625, 2010.

\bibitem{Bellotti2011}
F.~F. Bellotti, T.~Frederico, M.~T. Yamashita, D.~V. Fedorov, A.~S. Jensen, and
  N.~T. Zinner.
\newblock Scaling and universality in two dimensions: three-body bound states
  with short-ranged interactions.
\newblock {\em J. Phys. B: At Mol. Opt. Phys.}, 44:205302, 2011.

\bibitem{BazakPetrov2018}
B.~Bazak and D.~S. Petrov.
\newblock Energy of {N} two-dimensional bosons with zero-range interactions.
\newblock {\em New J. Phys.}, 20(2):023045, 2018.

\bibitem{PRB95_045401}
T.~Ren and I.~Aleiner.
\newblock Three-boson bound states in three dimensions.
\newblock {\em Phys. Rev. B}, 95:045401, 2017.

\bibitem{BazakPetrovPRL2018}
B.~Bazak and D.~S. Petrov.
\newblock Stable $p$-wave resonant two-dimensional {Fermi-Bose} dimers.
\newblock {\em Phys. Rev. Lett.}, 121:263001, 2018.

\bibitem{kagan2013modern}
M.Y. Kagan.
\newblock {\em Modern trends in Superconductivity and Superfluidity}.
\newblock Lecture Notes in Physics. Springer Netherlands, 2013.

\bibitem{inguscio2008ultra}
M.~Inguscio, W.~Ketterle, and C.~Salomon.
\newblock {\em Ultra-cold {Fermi} Gases}.
\newblock International School of Physics “Enrico Fermi”. IOS Press, 2008.

\bibitem{BruchTjon1979}
L.~W. Bruch and J.~A. Tjon.
\newblock Binding of three identical bosons in two dimensions.
\newblock {\em Phys. Rev. A}, 19:425, 1979.

\bibitem{Adhikari1988}
Sadhan~K. Adhikari, A.~Delfino, T.~Frederico, I.~D. Goldman, and Lauro Tomio.
\newblock Efimov and {Thomas} effects and the model dependence of
  three-particle observables in two and three dimensions.
\newblock {\em Phys. Rev. A}, 37:3666--3673, May 1988.

\bibitem{Nielsen1997}
E.~Nielsen, D.~V. Fedorov, and A.~S. Jensen.
\newblock Three-body halos in two dimensions.
\newblock {\em Phys. Rev. A}, 56:3287, 1997.

\bibitem{Nielsen1999}
E.~Nielsen, D.~V. Fedorov, and A.~S. Jensen.
\newblock Structure and occurrence of three-body halos in two dimensions.
\newblock {\em Few-Body Syst.}, 27:15, 1999.

\bibitem{HammerSon2004}
H.-W. Hammer and D.~T. Son.
\newblock Universal properties of two-dimensional boson droplets.
\newblock {\em Phys. Rev. Lett.}, 93:250408, Dec 2004.

\bibitem{KartavtsevMalykh2006}
O.~I. Kartavtsev and A.~V. Malykh.
\newblock Universal low-energy properties of three two-dimensional bosons.
\newblock {\em Phys. Rev. A}, 74:042506, Oct 2006.

\bibitem{PhysRevLett.113.205301}
T.~Takekoshi, L.~Reichs\"ollner, A.~Schindewolf, J.~M. Hutson, C.~R. Le~Sueur,
  O.~Dulieu, F.~Ferlaino, R.~Grimm, and H.-C. N\"agerl.
\newblock Ultracold dense samples of dipolar {RbCs} molecules in the
  rovibrational and hyperfine ground state.
\newblock {\em Phys. Rev. Lett.}, 113:205301, Nov 2014.

\bibitem{PhysRevLett.113.255301}
P.~K. Molony, P.~D. Gregory, Z.~Ji, B.~Lu, M.~P. K\"oppinger, C.~R. Le~Sueur,
  C.~L. Blackley, J.~M. Hutson, and S.~L. Cornish.
\newblock Creation of ultracold $^{87}\mathrm{Rb}^{133}\mathrm{Cs}$ molecules
  in the rovibrational ground state.
\newblock {\em Phys. Rev. Lett.}, 113:255301, Dec 2014.

\bibitem{PhysRevLett.116.205303}
M.~Guo, B.~Zhu, B.~Lu, X.~Ye, F.~Wang, R.~Vexiau, N.~Bouloufa-Maafa,
  G.~Qu\'em\'ener, O.~Dulieu, and D.~Wang.
\newblock Creation of an ultracold gas of ground-state dipolar
  $^{23}\mathrm{Na}^{87}\mathrm{Rb}$ molecules.
\newblock {\em Phys. Rev. Lett.}, 116:205303, May 2016.

\bibitem{PhysRevA.97.020501}
M.~Guo, X.~Ye, J.~He, G.~Qu\'em\'ener, and D.~Wang.
\newblock High-resolution internal state control of ultracold
  $^{23}\mathrm{Na}^{87}\mathrm{Rb}$ molecules.
\newblock {\em Phys. Rev. A}, 97:020501(R), Feb 2018.

\bibitem{Deiglmayr2010}
J.~Deiglmayr, A.~Grochola, M.~Repp, O.~Dulieu, R.~Wester, and M.~Weidem\"uller.
\newblock Permanent dipole moment of {LiCs} in the ground state.
\newblock {\em Phys. Rev. A}, 82:032503, Sep 2010.

\bibitem{Moses2015}
S.~A. Moses, J.~P. Covey, M.~T. Miecnikowski, B.~Yan, B.~Gadway, J.~Ye, and
  D.~S. Jin.
\newblock Creation of a low-entropy quantum gas of polar molecules in an
  optical lattice.
\newblock {\em Science}, 350(6261):659, 2015.

\bibitem{DeMarco2019}
L.~De~Marco, G.~Valtolina, K.~Matsuda, W.~G. Tobias, J.~P. Covey, and J.~Ye.
\newblock A degenerate {Fermi} gas of polar molecules.
\newblock {\em Science}, 363(6429):853, 2019.

\bibitem{Park2015}
J.~W. Park, S.~A. Will, and M.~W. Zwierlein.
\newblock Ultracold dipolar gas of fermionic $^{23}\mathrm{Na}^{40}\mathrm{K}$
  molecules in their absolute ground state.
\newblock {\em Phys. Rev. Lett.}, 114:205302, 2015.

\bibitem{Park2017}
J.~W. Park, Z.~Z. Yan, H.~Loh, S.~A. Will, and M.~W. Zwierlein.
\newblock Second-scale nuclear spin coherence time of ultracold
  $^{23}\mathrm{Na}^{40}\mathrm{K}$ molecules.
\newblock {\em Science}, 357:372, 2017.

\bibitem{Yang2019}
H.~Yang, D.-C. Zhang, L.~Liu, Y.-X. Liu, J.~Nan, B.~Zhao, and J.-W. Pan.
\newblock Observation of magnetically tunable feshbach resonances in ultracold
  $^{23}\mathrm{Na}^{40}\mathrm{K}$ + $^{40}\mathrm{K}$ collisions.
\newblock {\em Science}, 363:261, 2019.

\bibitem{Jensen2004}
A.~S. Jensen, K.~Riisager, D.~V. Fedorov, and E.~Garrido.
\newblock Structure and reactions of quantum halos.
\newblock {\em Rev. Mod. Phys.}, 76:215--261, Feb 2004.

\bibitem{Riisager_2013}
K~Riisager.
\newblock Halos and related structures.
\newblock {\em Physica Scripta}, T152:014001, jan 2013.

\bibitem{EFIMOV1970563}
V.~Efimov.
\newblock Energy levels arising from resonant two-body forces in a three-body
  system.
\newblock {\em Physics Letters B}, 33(8):563 -- 564, 1970.

\bibitem{kraemer2006evidence}
T~Kraemer, Manfred Mark, Philipp Waldburger, Johann~G Danzl, Cheng Chin,
  Bastian Engeser, Almar~D Lange, Karl Pilch, Antti Jaakkola, H-C N{\"a}gerl,
  et~al.
\newblock Evidence for {Efimov} quantum states in an ultracold gas of caesium
  atoms.
\newblock {\em Nature}, 440(7082):315--318, 2006.

\bibitem{Naidon_2017}
Pascal Naidon and Shimpei Endo.
\newblock Efimov physics: a review.
\newblock {\em Reports on Progress in Physics}, 80(5):056001, mar 2017.

\bibitem{sachdev_2009}
Subir Sachdev.
\newblock {\em Quantum phase transitions}.
\newblock Cambridge Univ. Press, 2009.

\bibitem{Bloch2005}
Immanuel Bloch and Immanuel.
\newblock Ultracold quantum gases in optical lattices.
\newblock {\em Nat. Phys.}, 1:23--30, 10 2005.

\bibitem{ReviewFermiGases}
Stefano Giorgini, Lev~P. Pitaevskii, and Sandro Stringari.
\newblock Theory of ultracold atomic {Fermi} gases.
\newblock {\em Rev. Mod. Phys.}, 80:1215--1274, Oct 2008.

\bibitem{Zwerger2012book}
Wilhelm Zwerger.
\newblock {\em The BCS-BEC crossover and the unitary {Fermi} gas}.
\newblock Springer, Heidelberg New York, 2012.

\bibitem{RevModPhys.66.1105}
K.~Riisager.
\newblock Nuclear halo states.
\newblock {\em Rev. Mod. Phys.}, 66:1105--1116, Jul 1994.

\bibitem{TANIHATA1985}
I.~Tanihata, H.~Hamagaki, O.~Hashimoto, S.~Nagamiya, Y.~Shida, N.~Yoshikawa,
  O.~Yamakawa, K.~Sugimoto, T.~Kobayashi, D.E. Greiner, N.~Takahashi, and
  Y.~Nojiri.
\newblock Measurements of interaction cross sections and radii of {He}
  isotopes.
\newblock {\em Physics Letters B}, 160(6):380 -- 384, 1985.

\bibitem{PhysRevLett.55.2676}
I.~Tanihata, H.~Hamagaki, O.~Hashimoto, Y.~Shida, N.~Yoshikawa, K.~Sugimoto,
  O.~Yamakawa, T.~Kobayashi, and N.~Takahashi.
\newblock Measurements of interaction cross sections and nuclear radii in the
  light $p$-shell region.
\newblock {\em Phys. Rev. Lett.}, 55:2676--2679, Dec 1985.

\bibitem{RevModPhys.82.1225}
Cheng Chin, Rudolf Grimm, Paul Julienne, and Eite Tiesinga.
\newblock Feshbach resonances in ultracold gases.
\newblock {\em Rev. Mod. Phys.}, 82:1225--1286, Apr 2010.

\bibitem{RIISAGER1992393}
K.~Riisager, A.S. Jensen, and P.~MÃžller.
\newblock Two-body halos.
\newblock {\em Nuclear Physics A}, 548(3):393 -- 413, 1992.

\bibitem{PhysRevC.49.201}
D.~V. Fedorov, A.~S. Jensen, and K.~Riisager.
\newblock Three-body halos: Gross properties.
\newblock {\em Phys. Rev. C}, 49:201--212, Jan 1994.

\bibitem{PhysRevLett.113.253401}
P.~Stipanovi\ifmmode~\acute{c}\else \'{c}\fi{}, L.~Vranje\ifmmode
  \check{s}\else~\v{s}\fi{} Marki\ifmmode~\acute{c}\else \'{c}\fi{},
  I.~Be\ifmmode \check{s}\else \v{s}\fi{}li\ifmmode~\acute{c}\else \'{c}\fi{},
  and J.~Boronat.
\newblock Universality in molecular halo clusters.
\newblock {\em Phys. Rev. Lett.}, 113:253401, Dec 2014.

\bibitem{Stipanovi2017QuantumHS}
Petar Stipanovi\'c, Leandra~Vranje\v{s} Marki\'c, and Jordi Boronat.
\newblock Quantum halo states in {Helium} tetramers.
\newblock {\em The journal of physical chemistry. A}, 121 1:308--314, 2017.

\bibitem{Stipanovic2019}
Petar Stipanović, Leandra Vranje\v{s}~Marki\'c, Andrii Gudyma, and Jordi
  Boronat.
\newblock Universality of size-energy ratio in four-body systems.
\newblock {\em Scientific Reports}, 9, 12 2019.

\bibitem{LandauLifshitz_iii}
L.~D. Landau and E.~M. Lifshitz.
\newblock {\em Quantum Mechanics : Non-Relativistic Theory}.
\newblock Elsevier Science, Burlington, 1977.

\bibitem{leggett_2015}
Anthony~J. Leggett.
\newblock {\em Quantum liquids: {Bose} condensation and Cooper pairing in
  condensed-matter systems}.
\newblock Oxford Univ. Press, 2015.

\bibitem{PhysRevLett.98.060405}
G.~E. Astrakharchik, J.~Boronat, I.~L. Kurbakov, and Yu.~E. Lozovik.
\newblock Quantum phase transition in a two-dimensional system of dipoles.
\newblock {\em Phys. Rev. Lett.}, 98:060405, Feb 2007.

\bibitem{PhysRev.155.88}
L.~Reatto and G.~V. Chester.
\newblock Phonons and the properties of a {Bose} system.
\newblock {\em Phys. Rev.}, 155:88--100, Mar 1967.

\bibitem{Petrov2016}
D.~S. Petrov and G.~E. Astrakharchik.
\newblock {Ultradilute Low-Dimensional Liquids}.
\newblock {\em Physical Review Letters}, 117(10):1--5, 2016.

\bibitem{Cinti2017}
Fabio Cinti, Daw~Wei Wang, and Massimo Boninsegni.
\newblock {Phases of dipolar bosons in a bilayer geometry}.
\newblock {\em Physical Review A}, 95(2):1--6, 2017.

\bibitem{Landau:1958:SP}
Lev~D. Landau and Evgeny~M. Lifshitz.
\newblock {\em Statistical Physics}, volume~9 of {\em Course of theoretical
  physics}.
\newblock Pergamon Press, Oxford, UK, 1980.
\newblock Translation from Russian by Eugenia and Rudolf F. Peierls.

\bibitem{GUYER1970413}
R.A. Guyer.
\newblock The physics of quantum crystals.
\newblock volume~23 of {\em Solid State Physics}, pages 413 -- 499. Academic
  Press, 1970.

\bibitem{ronen2007radial}
Shai Ronen, Daniele~CE Bortolotti, and John~L Bohn.
\newblock Radial and angular rotons in trapped dipolar gases.
\newblock {\em Physical review letters}, 98(3):030406, 2007.

\bibitem{bohn2009does}
John~L Bohn, Ryan~M Wilson, and Shai Ronen.
\newblock How does a dipolar {Bose-Einstein} condensate collapse?
\newblock {\em Laser Physics}, 19(4):547--549, 2009.

\bibitem{parker2009structure}
NG~Parker, C~Ticknor, AM~Martin, and DHJ O’Dell.
\newblock Structure formation during the collapse of a dipolar atomic
  {Bose-Einstein} condensate.
\newblock {\em Physical Review A}, 79(1):013617, 2009.

\bibitem{blakie2012roton}
PB~Blakie, D~Baillie, and RN~Bisset.
\newblock Roton spectroscopy in a harmonically trapped dipolar {Bose-Einstein}
  condensate.
\newblock {\em Physical Review A}, 86(2):021604, 2012.

\bibitem{jona2013roton}
M~Jona-Lasinio, K~{\L}akomy, and L~Santos.
\newblock Roton confinement in trapped dipolar {Bose-Einstein} condensates.
\newblock {\em Physical Review A}, 88(1):013619, 2013.

\bibitem{wilson2010critical}
Ryan~M Wilson, Shai Ronen, and John~L Bohn.
\newblock Critical superfluid velocity in a trapped dipolar gas.
\newblock {\em Physical review letters}, 104(9):094501, 2010.

\bibitem{natu2014dynamics}
Stefan~S Natu, L~Campanello, and S~Das Sarma.
\newblock Dynamics of correlations in a quasi-two-dimensional dipolar {Bose}
  gas following a quantum quench.
\newblock {\em Physical Review A}, 90(4):043617, 2014.

\bibitem{Chomaz2018observation}
Lauriane Chomaz, Rick~MW van Bijnen, Daniel Petter, Giulia Faraoni, Simon
  Baier, Jan~Hendrik Becher, Manfred~J Mark, Falk Waechtler, Luis Santos, and
  Francesca Ferlaino.
\newblock Observation of roton mode population in a dipolar quantum gas.
\newblock {\em Nature physics}, 14(5):442--446, 2018.

\end{thebibliography}
\prefacesection{Agradecimientos}
Primero quiero agradecer a mis tutores, al Dr. Jordi Boronat y
al Dr. Grigory Astrakharchik, por darme la oportunidad de realizar
este trabajo de investigación bajo su supervisión y por permitirme
formar parte del grupo de investigación Barcelona Quantum Monte Carlo.
Además, agradezco a los miembros de este grupo, los cuales 
ayudaron a mi formación académica. 

En especial agradezco a mis compañeros
y amigos: Huixia Lu, Juan Sanchéz, Raúl Bombín, Viktor 
Cikojević, y Giulia De Rosi, por hacer más amenas las comidas en la
universidad. También agradezco a mis amigos en México, que a pesar de
los más de 9000 Km de distancia nuestra amistad continúa. 

Agradezco al Consejo Nacional de Ciencia y Tecnología (CONACyT) y al pueblo de México
por haberme otorgado una beca para realizar mis estudios de doctorado en
la Universitat Politècnica de Catalu\~na.

Dedico este trabajo a mi familia, por su cari\~no y apoyo incondicional.
También lo dedico a Esteban, por su apoyo en todo momento, y por los buenos
momentos que hemos pasado juntos en estos a\~nos.

\end{document}